\documentclass[a4paper,twoside,12pt]{elsarticle}
\usepackage{epsfig}
 
\usepackage{eurosym}

\usepackage{fancyhdr}
\begin{document}
\thispagestyle{empty}
\pagenumbering{roman}

\noindent
%\epsfig{width=\textwidth,file=CTA_concept_tit.eps}
%\vfill
 
%\begin{center}
%{\bf
%\huge
%Design Concepts for the   \\[2mm]
%Cherenkov Telescope Array \\[2mm]
%CTA \\[1cm]
%\large
%An Advanced Facility for Ground-Based\\
%High-Energy Gamma-Ray Astronomy\\
% 
%\vfill
% 
%\huge  The CTA Consortium\\
%\vfill
% 
%\large May 2010
%}
%\end{center}
%\vfill
%\noindent
%\epsfig{width=\textwidth,file=CTA_concept_tit2.eps}
% 
%\clearpage
%\thispagestyle{empty}
%\vspace*{19cm}
%\normalsize
%\noindent
%contact:\\[2mm]
%W. Hofmann (Werner.Hofmann@mpi-hd.mpg.de)\\
%M. Martinez (martinez@ifae.es) \\[3mm]
%http://www.cta-observatory.org/
% 
%\cleardoublepage
%\thispagestyle{empty}
% 
%\setcounter{page}{1}
%\pagenumbering{roman}
%
%%%%%%%%%%%%%%%%%%%%%%%%%%%%%%%%%%%%%%%%%%%%%%%%%%%%%%%%%%%%%%%%%%%
%%%%%%%%%%%%%%%%%%%%%%%%%%%%%%%%%%%%%%%%%%%%%%%%%%%%%%%%%%%%%%%%%%%
%%%%%%%%%%%%%%%%%%%%%%%%%%%%%%%%%%%%%%%%%%%%%%%%%%%%%%%%%%%%%%%%%%%
%%%%%%%%%%%%%%%%%%%%%%%%%%%%%%%%%%%%%%%%%%%%%%%%%%%%%%%%%%%%%%%%%%%

\begin{frontmatter}
\title{Design Concepts for the Cherenkov Telescope Array CTA,\\
An Advanced Facility for Ground-Based High-Energy Gamma-Ray Astronomy}
 
\begin{abstract}
Ground-based gamma-ray astronomy has had a major breakthrough with the impressive 
results obtained using systems of imaging atmospheric Cherenkov telescopes. 
Ground-based gamma-ray astronomy has a huge potential in astrophysics, particle physics 
and cosmology. CTA is an international initiative to build the next generation instrument, 
with a factor of 5-10 improvement in sensitivity in the 100 GeV to 10 TeV range and the 
extension to energies well below 100 GeV and above 100 TeV. 
CTA will consist of two arrays (one in the north, one in the south) for 
full sky coverage and will be operated as open observatory. 
The design of CTA is based on currently available technology. 
This document reports on the status and presents the major design concepts of CTA.
\end{abstract}

\author[cta]{{\bf CTA Consortium}\footnote{contact: 
W. Hofmann (Werner.Hofmann@mpi-hd.mpg.de), M. Martinez (martinez@ifae.es),
http://www.cta-observatory.org}}
\author[arg1c]{M.~Actis}
\author[inaf1]{G.~Agnetta} 
\author[ire]{F.~Aharonian} 
\author[arm]{A.~Akhperjanian}
\author[sp6]{J.~Aleksi{\'c}} 
\author[barn]{E.~Aliu}
\author[dur1]{D.~Allan} 
\author[arg1a]{I.~Allekotte}
\author[arg1c]{F.~Antico}
% ...

\address[cta]{CTA Project Office, Landessternwarte, Universit\"at Heidelberg, K\"onigstuhl, D 69117 Heidelberg, Germany}
\address[arg1a]{Centro At\'omico Bariloche (CNEA-CONICET-IB/UNCuyo), Av.~E.~Bustillo 9500, (8400) San Carlos de Bariloche, Rio Negro, Argentina}
% ...
\address[arg1c]{UID GEMA - Departamento de Aeron\'autica (Facultad de Ingeniería, UNLP), Calle 116 entre 47 y 48, La Plata (1900), Buenos Aires, Argentina}
% ...
\address[arm]{Alikhanyan National Science Laboratory, 2 Alikhanyan Brothers St., 0036, Yerevan, Armenia}

% ...
\address[ire]{Dublin Institute for Advanced Studies, 31 Fitzwilliam Place, Dublin 2, Ireland}

\address[inaf1]{INAF - Istituto di Astrofisica Spaziale e Fisica Cosmica di Palermo, Via U. La Malfa 153, I-90146 Palermo, Italy}
% ...
\address[sp6]{IFAE, Edifici Cn, Campus UAB, E-08193 Bellaterra, Spain}
% ...
\address[dur1]{Dept. of Physics, Durham University, Science Laboratories, South Road, Durham DH1 3LE, UK}
% ...
\address[barn]{Barnard College, Columbia University, Dept. of Physics \& Astronomy, 3009 Broadway, New York, NY 10027}
% ...

\end{frontmatter}

 \clearpage
 
%\input{ExecSummary.tex}
%%%%% HEADER: ExecSummary.tex
%%%%%%%%%%%%%%%%%%%%%%%%%%%%%%%%%%%%%%%%%%%%%%%%%%%%%%%%%%%%%%%%%%%
 
\section*{Executive Summary}
 \label{sec:exec}
 
The present generation of imaging atmospheric Cherenkov telescopes
(H.E.S.S., MAGIC and VERITAS)
has in recent years opened the realm of ground-based
gamma ray astronomy for energies above a few tens of GeV.
The Cherenkov Telescope Array (CTA) will explore in depth our
Universe in very high energy gamma-rays and investigate
cosmic processes leading to relativistic particles,
in close cooperation with observatories
of other wavelength ranges of the electromagnetic spectrum,
and those using cosmic rays and neutrinos.
 
Besides guaranteed high-energy astrophysics results, CTA will have a large
discovery potential in key areas of astronomy, astrophysics and fundamental physics research.
These include the study of the origin of cosmic rays and their impact
on the constituents of the Universe through the investigation of galactic particle accelerators,
the exploration of the nature and variety of black hole particle accelerators
through the study of the production and propagation of extragalactic gamma rays,
and the examination of the
ultimate nature of matter and of physics beyond the Standard Model through searches
for dark matter and the effects of quantum gravity.
 
With the joining of the US groups of the Advanced Gamma-ray Imaging System (AGIS) project,
and of the Brazilian and Indian groups in Spring 2010,
and with the strong Japanese participation,
CTA represents a genuinely world-wide effort,
extending well beyond its European roots.
 
CTA will consist of two arrays of Cherenkov telescopes, which aim to:
(a) increase sensitivity by another order of magnitude for deep observations around 1~TeV,
(b) boost significantly the detection area and hence detection rates,
particularly important for transient phenomena and at the highest energies,
(c) increase the angular resolution and hence the ability to resolve
the morphology of extended sources,
(d) provide uniform energy coverage for photons from some tens of GeV to beyond 100 TeV, and
(e) enhance the sky survey capability, monitoring capability and flexibility of operation.
CTA will be operated as a proposal-driven open observatory, with a Science
Data Centre providing transparent access to data, analysis tools and user training.
 
To view the whole sky, two CTA sites are foreseen. The main site will be
in the southern hemisphere, given the wealth of sources in the central
region of our Galaxy and the richness of their morphological features.
A second complementary northern site will be primarily devoted
to the study of Active Galactic Nuclei (AGN) and cosmological galaxy and star formation and evolution.
The performance and scientific potential of arrays of Cherenkov telescopes
have been studied in significant detail, showing that the performance goals can be reached.
What remains to be decided is the exact layout of the telescope array.
Ample experience exists in constructing
and operating telescopes of the 12 m class (H.E.S.S., VERITAS). Telescopes of
the 17 m class are operating (MAGIC) and
one 28 m class telescope is under construction (H.E.S.S. II).
These telescopes will serve as prototypes for CTA.
The structural and optical properties of such
telescopes are well understood, as many have been built for applications from radio astronomy to
solar power installations.
The fast electronics needed in gamma ray astronomy to capture the
nanosecond-scale Cherenkov pulses have long been mastered,
well before such electronics became commonplace with the Gigahertz
transmission and processing used today in telephony, internet, television,
and computing.
 
The extensive experience of members of the consortium in the area of conventional
photomultiplier tubes (PMTs) provides a solid foundation for the design of
cameras with an optimal cost/performance ratio.
Consequently, the base-line design relies on conventional PMTs.
Advanced photon detectors with improved quantum efficiency are under development
and test and may well be available when the array is constructed.
In short, all the technical solutions needed to carry out this
project exist today. The main challenge lies in the industrialisation of all
aspects of the production and the exploitation of economies of scale.
 
Given the large amounts of data recorded by the instrument
and produced by computer simulations of the experiment,
substantial efforts in e-science and grid computing are envisaged
to enable efficient data processing.
Some of the laboratories involved in CTA are Tier 1 and 2 centres
on the LHC computing grid and the Cosmogrid.
Simulation and analysis packages for CTA are developed
for the grid. The consortium has set up a CTA-Virtual Organisation
within the EGEE project (Enabling Grids for E-sciencE; funded by the European Union)
for use of grid infrastructure and the sharing of computing
resources, which will facilitate worldwide collaboration for
simulations and the processing and analysis of scientific data.
 
Unlike current ground-based gamma-ray instruments, CTA will be an open observatory, with
a Science Data Centre (SDC) which provides pre-processed data to the user,
as well as the tools necessary for the most common analyses. The software tools
will provide an easy-to-use and well-defined
access to data from this unique observatory. CTA data will be accessible through
the Virtual Observatory, with varying interfaces matched to different levels of
expertise. The required toolkit is being developed by
partners with experience in SDC management from, for example, the
INTEGRAL space mission.
 
Experiments in astroparticle physics have proven to be an excellent training ground
for young scientists, providing a highly interdisciplinary work environment
with ample opportunities to acquire not only physics skills but also to learn
data processing and data mining techniques, programming of complex control and monitoring
systems and design of electronics. Further, the environment of the large multi-national
CTA Collaboration, working across international borders, ensures that
presentation skills, communication ability and management and leadership
proficiency are enhanced. Young scientists frequently participate in outreach activities
and, thus,  hone also their skills in this increasingly important area.
With its training and mobility opportunities for young scientists, CTA will
have a major impact on society.
 
Outreach activities will be an important part of the CTA operation.
Lectures and demonstrations
augmented by web-based non-expert tools for viewing CTA data  will be
offered to pupils and lay audiences.
Particularly interesting objects
will be featured on the CTA web pages,
along the lines of the ``Source of the Month'' pages of the H.E.S.S. collaboration.
CTA is expected to make highly visible
contributions towards popularising science and generating enthusiasm for research
at the cosmic frontier and to create interest in the technologies applied in this field.

\cleardoublepage
 
\tableofcontents
\cleardoublepage

\setcounter{page}{1}
\pagenumbering{arabic}
 
%% 1
%\input{CTAScienceInfra.tex}
%%%%% HEADER: CTAScienceInfra.tex
%%%%%%%%%%%%%%%%%%%%%%%%%%%%%%%%%%%%%%%%%%%%%%%%%%%%%%%%%%%%%%%%%%%
 
\section{CTA, a New Science Infrastructure}
 \label{sec:sciinfra}
 
\begin{figure}[hbtp]
\centering
\epsfig{width=0.87\textwidth,file=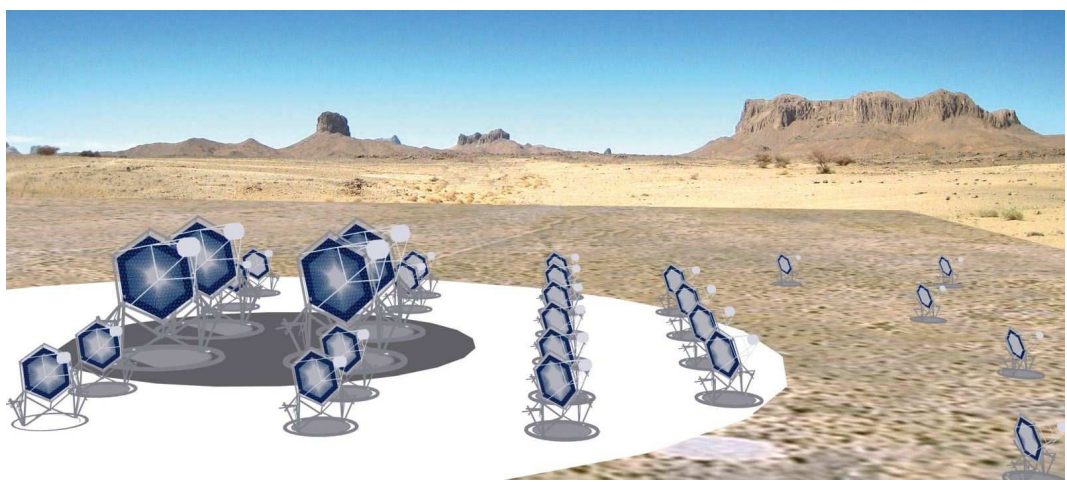}
\caption{\small  Conceptual layout of a possible Cherenkov Telescope Array  (not to scale).}
\label{fig_cta}
\end{figure}

In the field of very high energy gamma-ray astronomy 
(VHE, energies $>$100~GeV\footnote{1 GeV = 10$^9$ eV;
1 TeV = 10$^{12}$ eV; 1 PeV = 10$^{15}$ eV}),
the instruments 
H.E.S.S. \cite{hess}, MAGIC \cite{magic} and VERITAS \cite{veritas}
have been driving the development in recent years.
The spectacular astrophysics results from the current Cherenkov instruments
have generated considerable interest in both the astrophysics and particle
physics communities and have
created the desire for a next-generation, more sensitive and more flexible facility, able to serve a
larger community of users. The proposed CTA\footnote{CTA was first publicly presented to
an ESFRI panel in autumn 2005.} \cite{cta} is
a large array of Cherenkov telescopes of different sizes, based on proven
technology and deployed on an unprecedented scale (fig. \ref{fig_cta}).
It will allow significant extension of our current knowledge in high-energy astrophysics.
CTA is a new facility, with capabilities well beyond those of conceivable
upgrades of existing instruments such as H.E.S.S., MAGIC or VERITAS. The CTA project
unites the main research groups in this field in a common strategy, resulting in an
unprecedented convergence of efforts, human resources, and know-how. Interest in and support for the project
is coming from scientists in Europe, America, Asia and Africa, all of whom wish to use such a facility
for their research and are willing to contribute to its design and construction. CTA will offer worldwide
unique opportunities to users with varied scientific interests. The number of in particular young scientists
working in the still evolving field of gamma-ray astronomy is growing at a steady rate, drawing from
other fields such as nuclear and particle physics. In addition, there is increased interest by other
parts of the astrophysical community, ranging from radio to X-ray and satellite-based gamma-ray
astronomers. CTA will, for the first time in this field,
provide open access via targeted observation proposals and
generate large amounts of public data, accessible using Virtual Observatory tools.
CTA aims to become a cornerstone in a networked multi-wavelength,
multi-messenger exploration of the
high-energy non-thermal universe.
 
\clearpage
 
%% 2
%\input{Science.tex}
%%%%% HEADER: Science.tex
%%%%%%%%%%%%%%%%%%%%%%%%%%%%%%%%%%%%%%%%%%%%%%%%%%%%%%%%%%%%%%%%%%%

%%%%%%%%%%%%%%%%%%%%%%%%%%%%%
%%%%%%%%%%%%%%%%%%%%%%%%%%%%%
%%%%%%%%%%%%%%%%%%%%%%%%%%%%%
%%%%%%%%%%%%%%%%%%%%%%%%%%%%%
%%%%%%%%%%%%%%%%%%%%%%%%%%%%%

\section{The Science Case for CTA}
 \label{sec:phys}
 
\subsection{Science Motivation in a Nutshell}
 
\subsubsection{Why Observing in Gamma-Rays?}
 
Radiation at gamma-ray energies differs fundamentally from that detected
at lower energies and hence longer wavelengths: GeV to TeV gamma-rays cannot
conceivably be generated by thermal emission from hot celestial objects. The energy of thermal
radiation reflects the temperature of the emitting body, and apart from the Big Bang there is and has been nothing
hot enough to emit such gamma-rays in the known Universe. Instead, we find that high-energy gamma-rays
probe a non-thermal Universe, where other mechanisms allow the
concentration of large amounts of energy onto a single quantum of radiation. In a bottom-up fashion,
gamma-rays can be generated when highly relativistic particles --
accelerated for example in the gigantic shock waves of stellar explosions --
collide with ambient gas, or interact with photons and magnetic fields. The flux and energy
spectrum of the gamma-rays reflects the flux and spectrum of the high-energy particles.
They can therefore be used to trace these cosmic rays and electrons
in distant regions of our own Galaxy or even in other galaxies.
High-energy gamma-rays can also be produced in a top-down fashion by decays of heavy particles such
as hypothetical dark matter particles or cosmic strings, both of which might be relics of the Big Bang.
Gamma-rays therefore provide a window on the discovery of the
nature and constituents of dark matter.
 
High-energy gamma-rays, as argued above, can be used to trace the populations of
high-energy particles in distant regions of our own
or in other galaxies. Meandering in interstellar magnetic fields, cosmic rays will usually
not reach Earth and thus cannot be observed directly. Those which do arrive have lost all directional
information and cannot be used to pinpoint their sources, 
except for cosmic-rays of extreme energy $> 10^{18}$ eV. However, such high-energy particle populations
are an important aspect of the dynamics of galaxies. Typically, the energy content in cosmic rays
equals the energies in magnetic fields or in thermal radiation. The pressure generated by
high-energy particles drives galactic outflows and helps balance the gravitational collapse of
galactic disks.
Astronomy with high-energy gamma-rays is so far the only way to directly probe and
image the cosmic particle accelerators responsible for these particle populations,
in conjunction with studies of the synchrotron radiation resulting form
relativistic
electrons moving in magnetic fields and giving rise to non-thermal 
radio and X-ray emission.

\subsubsection{A First Glimpse of the Astrophysical Sources of Gamma-Rays}
 
\begin{figure}[tbtp]
	\centering
\epsfig{file=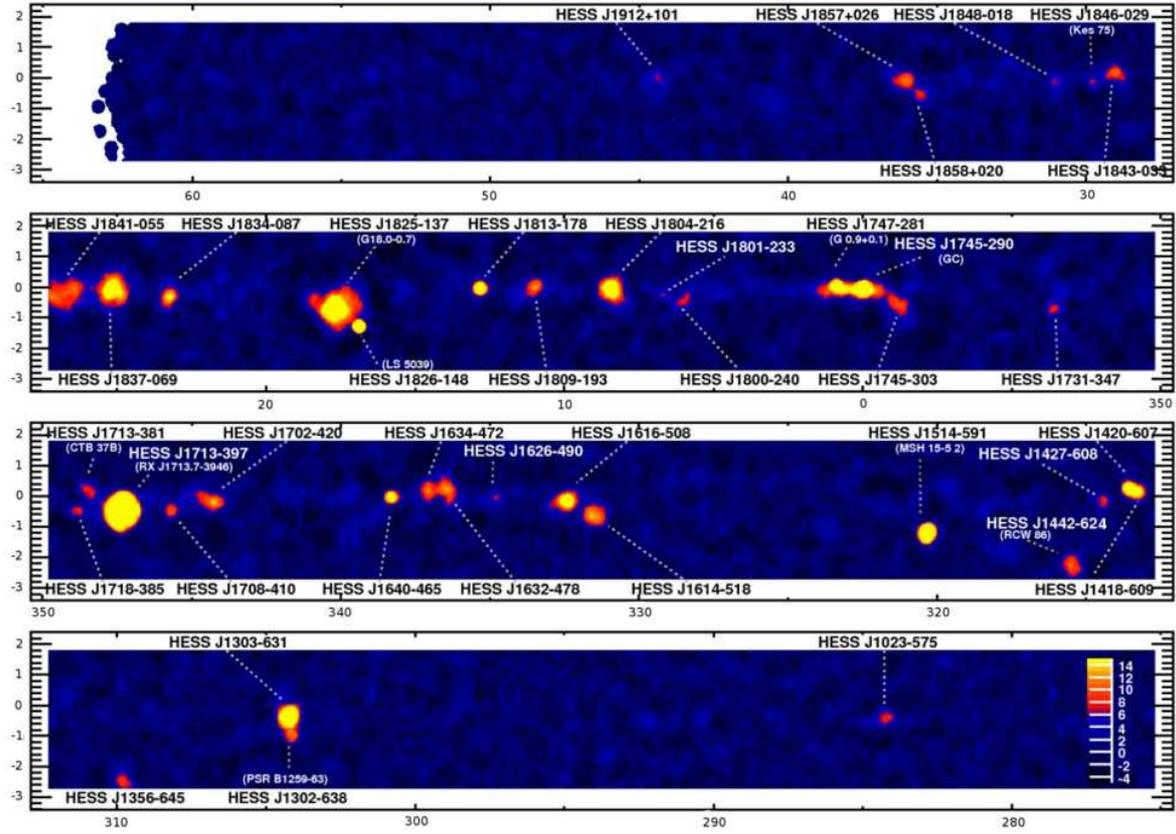,width=\textwidth}
  \caption{\small The Milky Way viewed in VHE gamma-rays, in four bands of Galactic longitude 
  \cite{hd1-survey}.}
  \label{fig_hess_scan}
\end{figure}

The first images of the Milky Way in VHE
gamma-rays have been obtained in the last few years.
These reveal a chain of gamma-ray emitters situated along the Galactic equator
(see fig.~\ref{fig_hess_scan}), demonstrating that sources of high-energy radiation are ubiquitous
in our Galaxy. Sources of this radiation include supernova shock waves, where presumably atomic nuclei
are accelerated and generate the observed gamma-rays. Another important class of objects
are ``nebulae'' surrounding pulsars, where giant rotating magnetic fields give rise to a
steady flow of high-energy particles. Additionally, some of the objects discovered to emit at such
energies are binary systems, where a black hole or a pulsar orbits a massive star. Along the elliptical
orbit, the conditions for particle acceleration vary and hence the intensity of the radiation is modulated
with the orbital period. These systems are particularly interesting in that they enable the study of how
particle acceleration processes respond to varying ambient conditions. One of several surprises
was the discovery of ``dark sources'', objects which emit VHE gamma rays, but have no obvious
%(that is, significantly shiny)
counterpart in other wavelength regimes. In other words, there are
objects in the Galaxy which might in fact be only detectable in high-energy gamma-rays.
Beyond our Galaxy, many extragalactic sources of high-energy radiation have been discovered, located
in active galaxies, where a super-massive black hole at the centre of the galaxy is fed by a steady stream
of gas and is releasing enormous amounts of energy. Gamma-rays are believed to be emitted from the
vicinity of these black holes, allowing the study of the processes occurring in this violent and as
yet poorly understood environment.
 
\subsubsection{Cherenkov Telescopes}
 
The recent breakthroughs in VHE gamma-ray astronomy were achieved with ground-based
Cherenkov telescopes.
When a VHE gamma-ray enters the atmosphere, it interacts with atmospheric nuclei
and generates a shower of secondary electrons, positrons and photons. Moving through the
atmosphere at speeds higher than the speed of light in air, these electrons and positrons
emit a beam of bluish light, the Cherenkov light.
For near vertical showers this Cherenkov light illuminates a circle with a diameter of about 250~m
on the ground. For large zenith angles the area can increase considerably.
This light can be captured with optical elements and be used to image the
shower, which vaguely resembles a shooting star. Reconstructing the shower
axis in space and tracing it back onto the sky allows the celestial origin of the gamma-ray to
be determined. Measuring many gamma-rays  enables an image of the gamma-ray sky, such as that shown
in fig.~\ref{fig_hess_scan}, to be created. Large optical
reflectors with areas in the 100~m$^2$ range and beyond are required to collect enough light,
and the instruments can only be operated in dark nights at clear sites. 
With Cherenkov telescopes, the effective area of the detector is about the size of the Cherenkov
pool at ground. As this is a circle with 250 m diameter this 
is about 10$^5\times$ larger than the size that
can be achieved
with satellite-based detectors. Therefore much lower fluxes at
higher energies can be investigated with Cherenkov Telescopes, enabling the study of
short time scale variability.
 
The Imaging Atmospheric Cherenkov
Technique was pioneered by the Whipple Collaboration in the United States.
After more than 20 years of
development, the Crab Nebula, the first source of VHE gamma-rays,  was discovered in 1989.
The Crab Nebula is among the strongest sources of very high energy gamma-rays, and is often
used as a ``standard candle''. Modern instruments, using multiple telescopes to track the cascades
from different perspectives and employing fine-grained photon detectors for improved imaging, can
detect sources down to  1\% of the flux of the Crab Nebula. Finely-pixellated imaging was first
employed in the French CAT telescope \cite{cat}, and the use of 
``stereoscopic'' telescope systems to provide
images of the cascade from different viewing points was pioneered by the European HEGRA 
IACT system \cite{hegra}.
For summaries of the achievements in recent years and the science case
for a next-generation very high energy gamma ray observatory
see Refs. \cite{hd1, hd2, hiho, aha, agis_white_paper}.
 
In March 2007, the High Energy Stereoscopic System (H.E.S.S.) project was awarded the Descartes
Research Prize of the European Commission for offering  ``A new glimpse at the highest-energy
Universe''.  Together with the instruments MAGIC and VERITAS (in the northern hemisphere) and
CANGAROO (in the southern hemisphere), a new wavelength domain was opened for astronomy, the
domain of very high energy gamma-rays with energies between about 100~GeV
and about 100 TeV, energies which are a million million times higher than the energy
of visible light.
 
At lower energies, in the GeV domain,
the launch of
a new generation of gamma-ray telescopes (like AGILE, but in particular
Fermi, which was launched in 2008) has opened a new era in gamma-ray discoveries \cite{fermisymp}.
The Large Area Telescope (LAT), the main instrument
onboard Fermi,
is sensitive to gamma-rays with energies in the range from 20
MeV to about 100 GeV.
The energy range covered by CTA will smoothly connect to that
of Fermi-LAT and overlap with that of the current generation of ground based instruments
and extends to the higher energies, while providing an improvement in both sensitivity and angular resolution.
\clearpage
 
\subsection{The CTA Science Drivers}
 \label{sec:drivers}
 
The aims of the CTA can be roughly grouped into three main themes,
serving as key science drivers:
 
\begin{enumerate}
\item Understanding the origin of cosmic rays and their
%impact onto the constituents of
role in the Universe
\item Understanding the nature and variety of particle acceleration
around black holes
\item Searching for the ultimate nature of matter and physics beyond the Standard Model
\end{enumerate}
 
Theme 1 comprises the study of the physics of galactic particle
accelerators, such as pulsars and pulsar wind nebulae, supernova remnants,
and gamma-ray binaries. It deals with the impact of
the accelerated particles on their environment (via the emission
from particle interactions with the interstellar medium and radiation
fields), and the cumulative effects seen at various scales,
from massive star forming regions to starburst galaxies.
 
Theme 2 concerns particle acceleration near super-massive and stellar-sized black holes.
Objects of interest include microquasars at the Galactic scale, and 
blazars, radio galaxies and other classes of AGN
that can potentially be studied in high-energy gamma rays. The fact
that CTA will be able to detect a large number of these objects
enables population studies which will be a major step forward in this area.
Extragalactic background light (EBL), Galaxy clusters and Gamma Ray Burst (GRB)
studies are also connected to this field.
 
Finally, Theme 3 covers what can be called ``new physics'', with
searches for dark matter through possible annihilation signatures,
tests of Lorentz invariance, and any other observational signatures that may challenge
our current understanding of fundamental physics.
 
CTA will be able to generate significant advances in all these areas.

\subsection{Details of the CTA Science Case}
 
We conclude this chapter with a few examples of physics issues
that could be significantly advanced with an instrument like CTA.
The list is certainly not exhaustive. The physics of the CTA is being
explored in detail by many scientists and their findings indicate
the huge potential for numerous interesting discoveries with CTA.

\subsubsection{Cosmic Ray Origin and Acceleration}
 
A tenet of high-energy astrophysics is that cosmic rays (CRs) are accelerated
in the shocks of supernova explosions. However, while particle acceleration up to energies well beyond
$10^{14}$ eV has now clearly been demonstrated with the current
generation of instruments, it is by no means proven that supernovae accelerate the bulk of cosmic
rays. The large sample of supernovae which will be observable with CTA -- in some scenarios several
hundreds of objects -- and in particular the increased energy coverage at lower and higher
energies, will allow sensitive tests of acceleration models and determination of their parameters.
Improved angular resolution (arcmin) will help to resolve fine structures in supernova
remnants which are essential for the study of particle acceleration and particle interactions.
Pulsar wind nebulae surrounding the pulsars (created in supernova explosions) are another abundant
source of high-energy particles, including possibly high-energy nuclei.
Energy conversion within pulsar winds and the interaction of the wind with the
ambient medium and the surrounding supernova shell challenge current ideas in plasma physics.
 
The CR spectrum observed near the Earth can be described by a pure power law up to an energy of a
few PeV, where it slightly steepens. The feature is called the ``knee''. The absence
of other features in the spectrum suggests that, if supernova remnants (SNRs) 
are the sources of galactic CRs, they
must be able to accelerate particles at least up to the knee.
For this to happen, the acceleration in diffusive shocks
has to be fast enough for particles to reach PeV energies before the
SNR enters the Sedov phase, when the shock slows down and consequently becomes
unable to confine the highest energy CRs \cite{Reynolds}
Since the initial free expansion velocity of
SNRs does not vary much from object to object, only the amplification of magnetic
fields can increase the acceleration rate to the required level.
Amplification factors of 100-1000 compared to the interstellar medium value
and small diffusion coefficients are needed \cite{MalD}.
The non-linear theory of diffusive shock acceleration suggests that such an
amplification of the magnetic field might be induced by the CRs themselves, and
high resolution X-ray observations of SNR shocks seem to support this scenario,
though their interpretation is debated. Thus, an accurate determination
of the intensity of the magnetic field at the shock is of crucial importance for
disentangling the origin of the observed gamma-ray emission and
understanding the way diffusive shock acceleration works.
 
Even if a SNR can be detected by Cherenkov telescopes
during a significant fraction of its lifetime (up to several 10$^4$ years),
it can make 10$^{15}$ eV CRs only for a much shorter time (several hundred years),
due to the rapid escape of PeV particles from the SNR.
This implies that the number of SNRs which
have currently a gamma-ray spectrum extending up to hundreds of TeV is very roughly
of the order of $\sim$10. The actual number of detectable objects will depend on the distance
and on the density of the surrounding interstellar medium. The detection of such objects
(even a few of them) would be extremely important, as it would be clear evidence
for the acceleration of CRs up to PeV energies in SNRs. A sensitive scan of the galactic plane with CTA
would be an ideal way of searching for these sources. In general, the spectra of radiating particles
(both electrons and protons) and therefore also the spectra of gamma-ray radiation,
should show characteristic curvature, reflecting acceleration
at CR modified shocks. However, to see such curvature,
one needs a coverage of a few decades in energy,
far from the cutoff region. CTA will provide this coverage.
If the general picture of SNR evolution described above is correct, the position of
the cutoff in the gamma-ray spectrum depends on the age of the SNR and on the
magnetic field at the shock. A study of the number of objects detected as a function of
the cutoff energy will allow tests of this hypothesis and constraints to be placed on the physical
parameters of SNRs, in particular of the magnetic field strength.
\clearpage
%Among the breakthrough topics in this rich area for CTA research,
%we can {\it additionally} emphasize the following
%tip-of-the-iceberg examples:
 
CTA offers the possibility of real breakthroughs in the understanding of cosmic rays; as
there is the potential to directly observe their diffusion (see, e.g., \cite{AhaAto})
The presence of a massive molecular cloud located in the proximity of a SNR (or
any kind of CR accelerator) provides a thick target for CR hadronic interactions and thus
enhances the gamma-ray emission. Hence, studies of molecular clouds in gamma-rays can
be used to identify the sites where CRs are accelerated.
While travelling from the accelerator to the target, the spectrum of cosmic rays is a strong
function of time, distance to the source, and the (energy-dependent) diffusion coefficient.
Depending on the values of these parameters varying proton, and therefore gamma-ray, spectra may be expected.
CTA will allow the study of emission depending on these three quantities, which is
impossible with current experiments. A determination, with high sensitivity, of spatially
resolved gamma-ray sources related to the same accelerator would lead to the experimental
determination of the local diffusion coefficient and/or the local injection spectrum of cosmic rays.
Also, the observation of the penetration of cosmic rays into molecular clouds will be possible.
If the diffusion coefficient inside a cloud is significantly smaller than the average
in the neighbourhood,
low energy cosmic rays cannot penetrate deep into the cloud, and part of the gamma-ray
emission from the cloud is suppressed, with the consequence that its
gamma-ray spectrum appears harder than the cosmic-ray spectrum.
 
Both of these effects are more pronounced in the
denser central region of the cloud. Thus, with an angular resolution of the order of $\le$1 arcmin
one could resolve the inner part of the clouds and measure the degree of penetration of cosmic rays \cite{Gab}.
 
More information on general aspects of cosmic rays and their relationship to VHE gamma
observations is available in the review talks and papers presented at the
International Cosmic Ray Conference 2009 held in \L{}\'od\'z
and the online proceedings are a good source of information \cite{icrc_lodz}.

\subsubsection{Pulsar Wind Nebulae}
 \label{sec:pwn}
Pulsar wind nebulae (PWNe) currently constitute the most populous class of
identified Galactic VHE gamma-ray sources.
As is well known, the Crab Nebula is a very effective accelerator (shown by
emission across more than 15 decades in energy) but not an effective inverse
Compton gamma-ray emitter. Indeed, we see gamma rays from the Crab because of its large spin-down
power ($\sim 10^{38}$ erg s$^{-1}$), although the gamma-ray luminosity is much less than the
spin-down power of its pulsar. This can be understood as resulting from a large (mG) magnetic field,
which also depends on the spin-down power. A less powerful pulsar would imply a
weaker magnetic field, which would allow a higher gamma-ray efficiency (i.e. a more efficient
sharing between synchrotron and inverse Compton losses).
For instance, HESS J1825-137 has
a similar TeV luminosity to the Crab, but a spin-down power that is 2 orders of magnitude smaller, and its magnetic
field has been constrained to be in the range of a few, instead of hundreds, of $\mu$G.
The differential gamma-ray spectrum of the whole emission region from the latter object
has been measured over more than two orders of magnitude, from 270 GeV to 35 TeV, and shows indications
of a deviation from a pure power law that CTA could confirm and investigate in detail.
Spectra have also been determined for spatially separated regions of HESS J1825-137 \cite{1825}. Another example is
HESS J1303-61 \cite{Dal}
The photon spectra in the different regions show a softening
with increasing distance from the pulsar and therefore an energy dependent morphology. If the emission
is due to the inverse Compton effect, the pulsar power is not sufficient to generate the gamma-ray luminosity,
suggesting that the pulsar had a higher injection power in the past. Is this common for other PWNe and
what can that tell us about the evolution of pulsar winds?
In the case of Vela X \cite{vel},  the first detection of what appears to be a VHE inverse Compton peak in the
spectral energy distribution (SED) was found.
Although a hadronic interpretation has also been put forward it is as yet unclear how large the contribution
of ions to the pulsar wind could be.
CTA can be used to test leptonic vs. hadronic models of
gamma-ray production in PWNe.

The return current problem for pulsars have not
been solved to date, but if we detect a clear hadronic signal, this will show that ions
are extracted from the pulsar surface, which may lead to a solution of the most
fundamental question in pulsar magnetospheric physics: how do we close the
pulsar current? In systems where we see a clear leptonic signal, it is important to
measure the magnetisation (or ``sigma'') parameter of the PWNe. Are the
magnetic fields and particles in these systems in equipartition (as in the Crab
Nebula) or do have particle dominated winds?
This will contribute significantly
to the understanding of the magnetohydrodynamic flow in PWNe.
Understanding the time evolution of the multi-wavelength
synchrotron and inverse Compton (or hadronic) intensities is also an aim of CTA. Such evolutionary
tracks are determined by the nature of the progenitor stellar wind, the properties
of the subsequent composite SNR explosion and the surrounding interstellar
environment. Finally, the sensitivity and angular resolution achievable with CTA
will allow detailed
multi-wavelength studies of large/close PWNe, and the understanding of particle propagation,
the magnetic field profile in the nebula, and inter-stellar medium (ISM) feedback.

The Evolution and Structure of Pulsar Wind Nebulae is discussed in a
recent review \cite{gaensler}.
Many key implications for VHE gamma ray measurements,
and an assessment of the current observations
can be found in ref. \cite{dejager}.

\subsubsection{The Galactic Centre Region}
 \label{sec:galcen}
 
It is clear that the galactic centre region itself will
be one of the prime science targets
for the next generation of VHE instruments \cite{cen1, cen2}. The galactic centre hosts the nearest super-massive black
hole, as well as a variety of other objects likely to generate high-energy radiation, including
hypothetical dark-matter particles which may annihilate and produce gamma-rays. Indeed, the
galactic centre has been detected as a source of high-energy gamma-rays, and indications for
high-energy particles diffusing away from the central source and interacting with the dense gas
clouds in the central region have been observed.
In observations with improved sensitivity and
resolution, the Galactic Centre can potentially yield a variety of interesting results
on particle acceleration and gamma-ray production in the vicinity of black holes,
on particle propagation in central molecular clouds, and, possibly,
on the detection of dark matter annihilation or decay.
\clearpage
 
The VHE gamma-ray view of the galactic centre region is dominated by two point
sources, one coincident with a PWN inside SNR G0.9+0.1, and one
coincident with the super-massive black hole Sgr A* and another putative PWN (G359.95-0.04).
After subtraction of these sources diffuse emission along the galactic centre ridge is
visible, which shows two important features:  it appears correlated with molecular
clouds (as traced by the CS (1--0) line), and it exceeds by a factor of 3 to 9
the gamma-ray emission that would be produced if the same target material was
exposed to the cosmic-ray environment in our local neighbourhood.
The striking correlation of diffuse gamma-ray emission with the
density of molecular clouds within  $\sim$150 pc of the galactic centre favours a
scenario in which cosmic rays interact with the cloud material and
produce gamma-rays via the decay of neutral pions. The
differential gamma-ray flux is stronger and harder than expected from just
``passive'' exposure of the clouds to the average galactic cosmic ray flux,
suggesting one or more nearby particle accelerators are present. In a first
approach, the observed gamma-ray morphology can be explained by cosmic rays
diffusing away from an accelerator near the galactic centre into the surroundings.
Adopting a diffusion coefficient of
$D=O(10^{30})$ cm$^2$/s, the lack of VHE gamma-ray emission beyond 150 pc
in this model points to an accelerator age of no more than 10$^4$ years.
Clearly, improved sensitivity and angular resolution would permit the
study of the diffusion process in great detail, including any possible energy
dependence. An alternative explanation (which CTA will address) is the
putative existence of a number of electron sources (e.g. PWNe) along the
galactic centre ridge, correlated with the density of molecular clouds.
Given the complexity and density of the source population in the galactic centre region, CTA's improved
sensitivity and angular resolution is needed to map the morphology of the diffuse emission,
and to test its hadronic or leptonic origin.
 
CTA will also measure VHE absorption in the interstellar radiation field (ISRF). This is
impossible for other experiments, like Fermi-LAT, as their energy coverage is too small, and very
hard or perhaps impossible for current air Cherenkov experiments, as they lack the required sensitivity.
At 8 kpc distance, VHE gamma-ray attenuation due to the CMB is
negligible for energies $<$500 TeV. But the attenuation due to the ISRF (which has a comparable
number density at wavelengths 20 $\mu$m to 300 $\mu$m) can produce absorption at about 50 TeV \cite{Zhang}.
Observation of the cutoff energy for different sources will provide independent
tests and constraints of ISRF models.
CTA will observe sources at different distances and
thereby independently measure the absorption model and the ISRF. Due to their smaller distances
there is less uncertainty in identifying intrinsic and extrinsic features in the spectrum
than is the case for EBL studies.
 
\subsubsection{Microquasars, Gamma-ray-, and X-ray Binaries}
 \label{sec:microquas}
 
Currently, a handful of VHE gamma-ray emitters are known to be binary systems,
consisting of a compact object, a neutron star or a black hole, orbiting a massive star. Whilst
many questions on the gamma-ray emission from such systems are still open (in some cases it is not
even clear if the energy source is a pulsar-driven nebula around a neutron star or accretion onto a black hole)
it is evident that they offer a unique chance to ``experiment'' with cosmic accelerators.
Along the eccentric orbits of the compact objects, the environment (including the radiation field)
changes, resulting in a periodic modulation of the gamma-ray emission, allowing the study of how
particle acceleration is affected by environmental conditions. Interestingly, the physics
of microquasars in our own Galaxy resembles the processes occurring around super-massive black
holes in distant active galaxies, with the exception of the much reduced time scales, providing insights
in the emission mechanisms at work. The following are key questions in this area which CTA
will be able to address, because of the extension of the accessible energy domain,
the improvement in sensitivity, and the superior angular resolution it provides:
 
a) {\it Studies of the formation of relativistic outflows from highly magnetised, rotating objects.}
If gamma-ray binaries are pulsars, is the gamma-ray emission coming mostly from processes within
the pulsar wind zone or rather from particles accelerated in the wind
collision shock? Is the answer to this question a function of energy?
What role do the inner winds play, particularly with regard to particle injection?
Gamma-ray astronomy
can provide data that will help to answer these questions,  but which will also throw light on
the particle energy distribution within the pulsar wind zone itself.
Recent Fermi-LAT
results on gamma-ray binaries, such as LS I +61 303 and LS 5039
(which are found to be periodic at GeV
and TeV energies, although anti-correlated \cite{fer}), show the existence
of a cutoff in the SED at a few GeV (a feature that was not predicted by any models).
Thus, the large energy coverage of CTA is an essential prerequisite in disentangling of the pulsed and continuous components of the radiation and the exploration of the
processes leading to the observed GeV-TeV spectral differences.
 
b) {\it Studies of
the link between accretion and ejection around compact objects and
transient states associated with VHE emission.}
It is known that
black holes display different spectral states in X-ray emission,
with transitions between a low/hard state,
where a compact radio jet is seen,
to a high/soft state, where the radio emission is reduced by large factors
or not detectable at all \cite{fen}.
Are these spectral changes related to changes in the gamma-ray emission?
Is there any
gamma-ray emission during non-thermal radio flares (with increased flux by up to a
factor of 1000)?
Indeed, gamma-ray emission via the inverse Compton effect
is expected when flares occur in the radio to X-ray region,
due to synchrotron radiation of relativistic
electrons and radiative, adiabatic and energy-dependent escape losses in
fast-expanding plasmoids (radio clouds).
Can future gamma-ray observations put
constraints on the  magnetic fields in plasmoids?\\
Continued observations of key objects (such as Cyg X-1)
with the sensitivity of current instruments (using sub-arrays of CTA) can provide
good coverage. Flares of less than 1 hour at a flux of 10\% of the Crab could be detected
at the distance of the Galactic Centre.
Hence variable sources could be monitored and triggers provided
for observations with all CTA telescopes or with other instruments.
For short flares, energy coverage in the 10-100 GeV band is not possible with
current instruments (AGILE and Fermi-LAT lack sensitivity).
Continuous coverage at higher energies is also impossible, due to lack of sensitivity
with the current generation of Imaging Atmospheric Cherenkov Telescopes (IACTs).
CTA will provide improved access to both regions.
 
c) {\it Collision of the jet with the ISM, as a non-variable source of gamma-ray emission.}
Improved angular resolution at high energies will provide
opportunities for the study of microquasars, particularly if their jets contain a sizeable fraction of
relativistic hadrons. While inner engines will still remain unresolved with future Cherenkov
telescope arrays, microquasar jets and their interaction with the ISM might become resolvable,
leading to the distinction of emission
from the central object (which may be variable) and
from the jet-ISM interaction (which may be stable).
Microquasars, Gamma-ray-, and X-ray binaries,
and high-energy aspects of astrophysical jets and binaries are discussed
in ref. \cite{microquas}.
 
\subsubsection{Stellar Clusters, Star Formation, and Starburst Galaxies}
 \label{sec:clusters}
 
While the classical paradigm has supernova explosions as the dominant source of cosmic rays,
it has been speculated that cosmic rays are also accelerated in stellar winds around massive young
stars before they explode as supernovae, or around star clusters \cite{par}.
Indeed, there is growing evidence
from gamma-ray data for a population of sources related to young stellar clusters and
environments with strong stellar winds. However, lack of sensitivity currently
prevents the detailed study and clear identification of these sources of gamma radiation.
CTA aims at a better understanding of the relationship between star formation processes
and gamma-ray emission. CTA can experimentally establish whether there is a direct correlation
between star formation rate and gamma-ray luminosity when convection and absorption processes at
the different environments are taken into account.  Both the VERITAS and H.E.S.S. arrays have done deep
observations of the nearest starburst galaxies, and have found them to be emitting TeV gamma-rays
at the limit of their sensitivity. Future observations, with improved sensitivity at higher and lower energies,
will reveal details of this radiation which in turn will
help with an understanding of the spectra,
provide constraints on the physical emission scenarios
and extend the study of the relationship between star
formation processes and gamma-ray emission to
extragalactic environments.
A good compendium of the current status of this topic can be found in
the proceedings of a recent conference \cite{jaen}.

\subsubsection{Pulsar Physics}
 \label{sec:pulsphys}
 
Pulsar magnetospheres are known to act as efficient cosmic accelerators, yet there is no complete
and accepted
model for this acceleration mechanism, a process which involves electrodynamics with very high
magnetic fields as well as the effects of general relativity. Pulsed gamma-ray emission allows
the separation of processes occurring in the magnetosphere from the emission in the surrounding
nebula. That pulsed  emission at tens of GeV can be detected with
Cherenkov telescopes was recently demonstrated by MAGIC with the Crab pulsar \cite{crab-mag}
(and the sensitivity for pulsars with known pulse frequency is nearly an order of magnitude higher 
than for standard sources).
Current Fermi-LAT results provide some support for models in which gamma-ray emission occurs far out in the
magnetosphere, with reduced magnetic field absorption (i.e. in outer gaps).
In these models, exponential cut-offs in the spectral energy distribution are
expected at a few GeV, which have already been found in several Fermi pulsars.
To make further progress in understanding the emission mechanisms in pulsars
it is necessary to
study their radiation at extreme energies.  In particular,  the
characteristics of pulsar emission in the GeV domain
(currently best examined by the Fermi-LAT) and at VHE will tell us more about the electrodynamics
within their magnetospheres.
Studies of interactions of magnetospheric particle winds with external
ambient fields (magnetic, starlight, CMB) are equally vital.
Between $\sim$10 GeV and $\sim$50 GeV (where the LAT performance is limited)
CTA, with a special low-energy trigger for pulsed sources,
will allow a closer look at
unidentified Fermi sources and deeper analysis of Fermi pulsar candidates. Above 50 GeV
CTA will explore the most extreme energetic processes in millisecond pulsars.
The VHE domain will be particularly important for the study of millisecond pulsars, very
much as the HE domain (with Fermi) is for classical pulsars.
On the other hand, the high-energy emission mechanism from magnetars is essentially unknown.
For magnetars, we do not expect polar cap emission. Due to the large magnetic field, all
high-energy photons would be absorbed if emitted close to the neutron star, i.e.,
CTA would be testing outer-gap models, especially if large X-ray flares are accompanied by
gamma-emission.
 
CTA can study the GeV-TeV emission related to short-timescale pulsar phenomena, which is
beyond the reach of currently working instruments. CTA can observe
possible high-energy phenomena related to timing noise (in which the pulse phase and/or
frequency of radio pulses drift stochastically) or to sudden increases in the pulse
frequency (glitches) produced by apparent changes in the momentum of inertia of neutron stars.
 
Periodicity measurements
with satellite instruments, which require very long integration times,
may be compromised by such glitches, while
CTA, with its much larger detection area and correspondingly shorter
measurement times, is not.
 
A good compendium of the current status of this topic can be found in
the proceedings and the talks presented at the
``International Workshop on the High-Energy Emission from Pulsars and
their Systems'' \cite{pulsarphysics}.

\subsubsection{Active Galaxies, Cosmic Radiation Fields and Cosmology}
 \label{sec:agn}
 
Active Galactic Nuclei (AGN) are among the largest storehouses of energy known in our cosmos.
At the intersection of powerful low-density plasma inflows and outflows, they offer excellent
conditions for efficient particle acceleration in shocks and turbulences. AGN represent one
third of the known VHE gamma-ray sources, with most of the detected objects belonging to the BL Lac
class. The fast variability of the gamma-ray flux (down to minute time scales) indicates that gamma-ray
production must occur close to the black hole, assisted by highly relativistic motion resulting in
time (Lorentz) contraction when viewed by an observer on Earth. Details of how these jets are
launched or even the types of particles of which they consist are poorly known. Multi-wavelength
observations with high temporal and spectral resolution can help to distinguish between
different scenarios, but this is at the limit of the capabilities of
current instruments.
The sensitivity of CTA, combined with simultaneous observations in other wavelengths, will
provide a crucial advance in understanding the mechanisms driving these sources.
 
Available surveys of BL Lacs suffer several biases at all wavelengths, further complicated by
Doppler boosting effects and high variability. The big increase of sensitivity of CTA will provide large
numbers of VHE sources of different types
and opens the way to statistical studies of the VHE blazar and AGN populations.
This will enable the exploration of the relation between different types of blazars,
and of the validity of
unifying AGN schemes. The distribution in redshift of known and relatively nearby BL Lac objects peaks around
$z \sim 0.3$. The large majority of the population is found within $z < 1$, a range
easily accessible with CTA. CTA will therefore be able to analyse in detail blazar populations
(out to $z \sim 2$) and
the evolution of AGN with redshift and to start a genuine ``blazar cosmology''.
 
Several scenarios have been proposed to explain the VHE emission of 
blazars\footnote{There are several clear cases of blazar SEDs where the X-ray peak and the 
$\gamma$-ray peak, with their correlated luminosity and spectral changes, are 
interpreted within a synchrotron-self-Compton (SSC) model \cite{11} 
as the synchrotron and IC peak respectively, produced by a time-varying 
population of particles (e.g. \cite{12,13}). A variant 
of the Compton scenario considers that the soft photons produced externally to 
the jet may be more effective than the internal ones (external Compton, EC, 
model; e.g. \cite{14,15}). Models based on 
hadronic acceleration (e.g. \cite{16,17}) can also 
reproduce the blazar SEDs and lightcurves.}. However, none of
them is fully self-consistent, and the current data are not sufficient to firmly rule out or confirm a
particular mechanism. In the absence of a convincing global picture, a first goal for CTA will be to
constrain model-dependent parameters of blazars within a given scenario. This is achievable due
to the wide energy range, high sensitivity and high spectral resolution of CTA combined with
multi-wavelength campaigns. Thus, the physics of basic radiation models will
be constrained by CTA, and some of the models will be ruled out. A second more difficult goal
will be to distinguish between the different remaining options and to firmly identify the dominant
radiation mechanisms. Detection of specific spectral features, breaks, cut-offs, absorption or
additional components, would be greatly helpful for this.
The role of CTA as a timing explorer will be decisive for constraining both the radiative
phenomena associated with, and the
global geometry and dynamics of, the AGN engine. Probing variability down to the shortest time scales
will significantly constrain acceleration and cooling times, instability growth rates, and the time
evolution of shocks and turbulences. For the brightest blazar flares,
current instruments are able to detect variability
on the scales of several minutes. With CTA, such flares
should be detectable within seconds, rather than minutes. A study of the minimum variability times of
AGN with CTA would allow the localisation of VHE emission regions (parsec distance scales in the jet,
the base of the jet, or the central engine) and would provide stringent constraints
on the emission mechanisms
as well as the intrinsic time scale connected to the size of the central super-massive black hole.
 
Recently, radio galaxies have emerged as a new class of VHE emitting 
AGN \cite{ragal}. Given the proximity of the
sources and the larger jet angle to the line of sight compared to BL Lac objects, the outer and
inner kpc jet structures will be spatially resolved by CTA. This will allow
precise location of the main emission site and
searches for
VHE radiation from large-scale jets and hot spots besides the central core and
jets seen in very long baseline interferometry images.
 
The observation of VHE emission from distant objects and their surroundings will also offer the
unique opportunity to study extragalactic magnetic fields at large distances.  If the fields are
large, an $e^+$ $e^-$ pair halo forms around AGNs, which CTA, with its high sensitivity and
extended field of view, should be capable of detecting. For smaller magnetic field values, the
effect of $e^+$ $e^-$ pair formation along the path to the Earth is seen through energy-dependent
time-delays of variable VHE emission, which CTA with its excellent time resolution will be
ideally suited to measure.
 
CTA will also have the potential to
deliver for the first time significant results on extragalactic diffuse emission at VHE, and offers
the possibility of probing the integrated emission from all sources at these energies.
While well measured at GeV energies with the EGRET and Fermi-LAT instruments, the diffuse emission at VHE
is extremely challenging to measure due to its faintness and the difficulty of adequately
subtracting the background. Here, the improved sensitivity coupled with the large field of view puts
detection in reach of CTA.
 
VHE gamma-rays traveling from remote sources interact with the EBL
via $e^+$ $e^-$ pair production and are absorbed. Studying such effects as a function of the energy
and redshift will provide unique information on the EBL density, and thereby on the history of
the formation of stars and galaxies in the Universe. This approach is complementary to
direct EBL measurements, which are hampered by strong foreground emission from our planetary
system (zodiacal light) and the Galaxy.
 
We anticipate that MAGIC II and H.E.S.S. II will at least double the number of
detected sources, but this is unlikely to resolve the ambiguity between intrinsic
spectral features and effects due to the EBL. It would still be very difficult to extract
spectral information beyond $z >0.5$,
if our current knowledge of the EBL is correct. Only CTA will be able to provide a sufficiently
large sample of VHE gamma-ray sources, and high-quality spectra for individual objects. For
many of the sources, the SED will be determined at GeV energies,
which are much less affected by the absorption and, thus, more suitable for the study of the
intrinsic properties
of the objects. We therefore anticipate that with CTA it will be possible to make robust predictions
about the intrinsic spectrum above 40 to 50 GeV, for individual sources and for particular source classes.
 
The end of the dark ages of the Universe, the epoch of reionisation, is 
a topic of great interest \cite{darkages}.
Not (yet) fully accessible via direct observations, most of our knowledge comes from simulations and from
integral observables like the cosmic microwave background. The first (Population III) and second
generations of stars are natural candidates for being the source of reionisation. If the first stars are hot
and massive, as predicted by simulations, their UV photons emitted at $z > 5$ would be redshifted to
the near infrared and could leave a unique signature on the EBL spectrum. If the EBL contribution
from lower redshift galaxies is sufficiently well known (for example, as derived from source counts) upper
limits on the EBL density can be used to probe the properties of early stars and galaxies. Combining
detailed model calculations with redshift-dependent EBL density measurements could allow the probing of the
reionisation/ionisation history of the Universe. A completely new wavelength region of the EBL will
be opened up by observations of sources at very high redshifts ($z > 5$), which will most likely be
gamma-ray bursts. According to high-redshift UV background models, consistent with our current knowledge
of cosmic reionisation, spectral cut-offs are expected in the few GeV to few tens of GeV range at $z > 5$.
Thus, CTA could have the unique potential to probe cosmic reionisation models through gamma-ray
absorption in high-z GRBs. We analyse the GRB prospects in more detail in the following.
 
A good compendium of the current state of this topic can be found in
the talks and the proceedings of the meeting,
High-energy phenomena in relativistic outflows II \cite{agn}.
 
\subsubsection{Gamma-Ray Bursts}
 \label{sec:grb}
 
Gamma-Ray Bursts are the most powerful explosions in the Universe, and
are by far the most electromagnetically luminous sources known to us.
The peak luminosity of GRBs, equivalent to the light from millions of
galaxies, means they can be detected up to high redshifts, hence act
as probes of the star formation history and reionisation of the
Universe. The highest measured GRB redshift is $z=$8.2 but GRBs have
been observed down to z=0.0085 (the mean redshift is $z\sim$2.2). GRBs occur
in random directions on the sky, briefly outshining the rest of the hard
X-ray and soft gamma-ray sky, and then fade from view.  The rapid
variability seen in gamma- and X-rays indicates a small source size,
which together with their huge luminosities and clearly non-thermal
spectrum (with a significant high-energy tail) require the emitting
region to move toward us with a very large bulk Lorentz factor of
typically $>$100, sometimes as high as $>$1000 \cite{22,23,24}

Thus, GRBs are
thought to be powered by ultra-relativistic jets produced by rapid
accretion onto a newly formed stellar-mass black hole or a rapidly
rotating highly-magnetised neutron star (i.e. a millisecond magnetar).
The prompt gamma-ray emission is thought to originate from dissipation
within the original outflow by internal shocks or magnetic
reconnection events. Some long duration GRBs are clearly associated
with core-collapse supernovae of type Ic (from very massive Wolf-Rayet
stars stripped of their H and He envelope by strong stellar winds),
while the progenitors of short GRBs are much less certain: the
leading model involves the merger of two neutron stars or a
neutron star and a black hole. \cite{44,45}
 
Many of the details of GRB explosions remain unclear. Studying them
requires a combination of rapid observations to observe the prompt
emission before it fades, and a wide energy range to properly capture
the spectral energy distribution. Most recently, GRBs have been
observed by the Swift and Fermi missions, which have revealed an even
more complex behaviour than previously thought, featuring significant
spectral and temporal evolution. As yet, no GRB has been detected at
energies $>$100 GeV due to the limited sensitivity of
current instruments and the large typical redshifts of these events.
In just over a
year of operation, the Fermi-LAT has detected emission above 10 GeV (30 GeV) from 4 (2)
GRBs. In many cases, the LAT detects emission $>$0.1 GeV
for several hundred seconds in the GRB rest-frame. In
GRB090902B a  photon of energy $\sim$33.4 GeV was detected, which translates to an
energy of $\sim$94 GeV at its redshift of $z = 1.822$.
Moreover, the observed
spectrum is fairly hard up to the highest observed energies.
 
Extrapolating the Fermi spectra to CTA energies
suggests that a good fraction of the bright LAT GRBs could be detected
by CTA even in $\sim$minute observing times, if it could
be turned to look at the prompt emission fast enough. The
faster CTA could get on target, the better the scientific
return. Increasing the observation duty cycle by observing for a
larger fraction of the lunar cycle and at larger zenith angles could also
increase the return.
 
Detecting GRBs in the CTA energy range would greatly enhance our
knowledge of the intrinsic spectrum and the particle acceleration
mechanism of GRBs, particularly when combined with data from Fermi and other observatories.
As yet it is unclear what the relative importance is of
the various proposed emission processes, which divide mainly into
leptonic (synchrotron and inverse-Compton, and in particular
synchrotron-self-Compton) and hadronic processes
(induced by protons or nuclei at very high energies which either radiate synchrotron
emission or produce pions with subsequent electromagnetic cascades).
CTA may help to determine the identity of the distinct high-energy component
that was
observed so far in 3 out of the 4 brightest LAT GRBs. The origin of
the high-energy component may in turn shed light on the more familiar
lower-energy components that dominate at soft gamma-ray energies. The bulk
Lorentz factor and the composition (protons, $e^+$ $e^-$ pairs,
magnetic fields) of the outflows
are also highly uncertain and may be probed by CTA.
The afterglow emission which
follows the prompt emission is significantly fainter, but should also
be detectable in some cases. Such detections would be expected from
bright GRBs at moderate redshift, not only from the afterglow synchrotron-self-Compton
component, but perhaps also from inverse-Compton emission
triggered by bright, late (hundreds to thousands of seconds)
flares that are observed in about half of all Swift GRBs.
 
The discovery space at high energies is large and readily
accessible to CTA. The combination of GRBs being extreme astrophysical
sources and cosmological probes make them prime
targets for all high-energy experiments. With its large
collecting area, energy range and rapid response,
CTA is by far the most
powerful and suitable VHE facility for GRB research and
will open up a new energy range for their study.
 
\subsubsection{Galaxy Clusters}
\label{sec:galclust}
 
Galaxy clusters are storehouses of cosmic rays, since all cosmic rays produced in the
galaxies of the cluster since the beginning of the Universe
will be confined there. Probing the density of
cosmic rays in clusters via their gamma-ray emission thus provides a calorimetric measure of the total
integrated non-thermal energy output of galaxies. Accretion/merger shocks outside cluster galaxies
provide an additional source of high-energy particles. Emission from galaxy clusters is predicted at
levels just below the sensitivity of current instruments \cite{66}.
 
Clusters of galaxies are the largest, gravitationally-bound objects in the Universe.
The observation of mainly radio (and in some cases X-ray) emission
proves the existence of non-thermal phenomena therein,
but gamma-rays have not yet been detected.
A possible
additional source of non-thermal radiation from clusters is the annihilation of dark matter (DM). The
increased sensitivity of CTA will help to
establish the DM signal, and CTA could possibly be the first instrument to map DM at the scale of
galaxy clusters.
 
\subsubsection{Dark Matter and Fundamental Physics}
 \label{sec:DM}
The dominant form of matter in the Universe is the as yet unknown
dark matter, which is most likely to exist in the form of a new class of particles such
as those predicted in supersymmetric or extra dimensional extensions to the
standard model of particle physics. Depending on the model, these DM particles can
annihilate or decay to produce detectable Standard Model particles, in
particular gamma-rays. Large dark matter densities due to the accumulation in
gravitational potential wells leads to detectable fluxes, especially for
annihilation, where the rate is proportional to the square of the
density. CTA is a discovery instrument with unprecedented sensitivity for this radiation
and also an ideal tool
to study the properties of the dark matter
particles. If particles beyond the standard model are discovered
(at the Large Hadron Collider or in underground experiments),
CTA will be able to verify whether they actually form the dark matter in the Universe.
Slow-moving dark matter particles could give rise to a striking,
almost mono-energetic photon emission.
The discovery of such line emission would be conclusive evidence
for dark matter.
CTA might have the capability
to detect gamma-ray lines even if the cross-section is loop-suppressed,
which is the case for the most popular candidates of dark matter,
i.e. those
inspired by the minimal supersymmetric extensions to the standard model (MSSM) and
models with extra dimensions, such as Kaluza-Klein theories. Line radiation
from these candidates is not detectable by Fermi, H.E.S.S. II
or MAGIC II, unless optimistic assumptions on the dark matter density
distribution are made. Recent updates of
calculations regarding the gamma-ray spectrum from the annihilation of MSSM
dark matter indicate the possibility of final-state contributions
giving rise to distinctive spectral features (see the reviews in \cite{fund})

The more generic
continuum contribution (arising from pion production) is more ambiguous
but, with its curved shape, potentially distinguishable from the usual
power-law spectra exhibited by known astrophysical sources.
 
Our galactic centre is one of the most promising regions to
look for dark matter annihilation radiation due to its predicted very high dark
matter density. It has been observed by many experiments so far (e.g.
H.E.S.S., MAGIC and VERITAS) and high-energy gamma emission
has been found. However, the identification of dark matter in
the galactic centre is complicated by the presence of many conventional
source candidates and the difficulties of modelling the diffuse
gamma-ray background adequately. The angular and energy resolution of CTA, as
well as its enhanced sensitivity will be crucial to disentangling the
different contributions to the radiation from the galactic centre.
 
Other individual targets for dark matter searches are dwarf spheroidals and dwarf galaxies.
They exhibit large mass-to-light ratios, and allow
dark matter searches with low astrophysical backgrounds. With
H.E.S.S., MAGIC and Fermi-LAT, some of these objects were observed and
upper limits on dark matter annihilation calculated, which are currently
about an order of magnitude above the prediction of the most relevant cosmological models.
CTA will have good sensitivity for Weakly Interacting Massive Particle (WIMP) annihilation searches in the low
and medium energy domains. An improvement in flux sensitivity of
1-2 orders of magnitude over current instruments is expected. Thus
CTA will allow tests in significant regions of the
MSSM parameter space.
 
Dark matter would also cause spectral and spatial signatures in extra-galactic and
galactic diffuse emission.
While the emissivity of conventional astrophysical
sources scale with the local matter density, the emissivity of annihilating dark matter
scales with the density squared, causing differences in the small-scale anisotropy
power spectrum of the diffuse emission.
 
Recent measurements of the
positron fraction presented by the PAMELA Collaboration \cite{PAMELA} point towards a
relatively local source of positrons and electrons, especially if
combined with the measurement of the $e^+$ $e^-$ spectrum by
Fermi-LAT \cite{fermi-elec}. The main candidates being put forward are either pulsar(s)
or dark matter annihilation. One way to distinguish between these two hypotheses is the
spectral shape. The dark matter spectrum exhibits a sudden drop at an energy
which corresponds to the dark matter particle mass, while the pulsar spectrum falls
off more smoothly. Another hint is a small anisotropy, either in the
direction of the galactic centre (for dark matter) or in the direction of the
nearest mature pulsars. The large effective area of CTA, about 6
orders of magnitudes larger than for balloon- and satellite-borne
experiments,
and the greatly improved performance compared to existing Cherenkov observatories,
might allow the measurement of the spectral shape and even the tiny
dipole anisotropy.
 
If the PAMELA result originated from dark matter, the DM particle's mass would be
$>$1~TeV/c$^2$, i.e. large in comparison to most dark matter candidates in
MSSM and Kaluza-Klein theories.
With its best sensitivity at 1 TeV, CTA would be well suited to detect
dark matter particles of TeV/c$^2$ masses. The best sensitivity
of Fermi-LAT for dark matter is at masses of the order of 10 to 100 GeV/c$^2$.
 
Electrons and positrons originating from dark matter annihilation or decay also
produce synchrotron radiation in the magnetic fields present in the
dense regions where the annihilation might take place. This opens up the
possibility of multi-wavelength observations.
Regardless of the
wavelength domain in which dark matter will be detectable using present or future
experiments, it is evident that CTA will provide coverage for the
highest-energy part of the multi-wavelength spectrum necessary to
pinpoint, discriminate and study dark matter indirectly.
 
Due to their extremely short wavelength and long propagation
distances, very high-energy gamma-rays are sensitive to the
microscopic structure of space-time. Small-scale perturbations of the
smooth space-time continuum should manifest themselves in an (extremely small)
energy dependence of the speed of light.
%photon propagation speeds.
Such a violation of Lorentz invariance,
on which the theory of special relativity is
based, is present in some quantum gravity (QG) models. Burst-like events in which
gamma-rays are  produced, e.g. in active galaxies, allow this
energy-dependent dispersion of gamma-rays to be probed and can be used to
place limits on certain classes of quantum gravity scenarios, and may
possibly lead to the discovery of effects associated with Planck-scale
physics. 
%Short delays in arrival times (measured within minute-long timescale variability), large source distances, and large differences in photon energy can then lead to significant astrophysical (lower) limits on the symmetry-breaking scale.
 
CTA has the sensitivity to detect characteristic time-scales and QG effects
in AGN light curves (if indeed any exist) on a routine basis
without exceptional source flux states and in small observing
windows. CTA can resolve time scales as small as few
seconds in AGN light curves and QG effects down to 10 s. Very good
sensitivity at energies $>$1~TeV is especially important
to probe the properties of QG effects at higher orders. Fermi
recently presented results based on observations of a GRB which
basically rule out linear-in-energy variations of the speed of light
up to 1.2$\times$ the Planck scale \cite{1-2}
To test
quadratic or higher order dependencies the sensitivity provided by CTA
will be needed.
 
This topic is thoroughly discussed in the book ``Particle dark matter''
edited by G. Bertone \cite{fund},
and aspects of the fundamental physics implications
of VHE gamma-ray observations are covered in a recent review
\cite{martinez}.
 
\subsubsection{Imaging Stars and Stellar Surfaces}
%Spinoff scientific output of CTA:
 \label{sec:imagstars}
The quest for better angular resolution in astronomy is driving much of the
instrumentation developments throughout the world,
from gamma-rays through low-frequency radio waves.  The optical region is
optimal for studying objects with stellar temperatures, and the current frontier in angular resolution
is represented by optical interferometers such as ESO's VLTI in Chile or the CHARA array in California.
Recently, these have produced images of giant stars surrounded by ejected gas shells and revealed the
oblate shapes of stars deformed by rapid rotation.  However, such phase interferometers are limited
by atmospheric turbulence to baselines of no more than some 100 metres, and to wavelengths longer
than the near infrared.  Only very few stars are large enough to be imaged by current facilities. To
see smaller details (e.g. magnetically active regions, planet-forming disks obscuring parts of the stellar
disk) requires interferometric baselines of the order of 1 km. It has been proposed to
incorporate such instruments on
ambitious future space missions (Luciola Hypertelescope for the ESA Cosmic Vision; Stellar Imager as a
NASA vision mission), or to locate them on the Earth in regions with the best-possible seeing,
e.g. in Antarctica (KEOPS array).  However, the complexity and cost of these concepts seems to put
their realisation beyond the immediate planning horizon.
 
An alternative that can be realised much sooner is offered by CTA, which could become the first
kilometre-scale optical imager.  With many telescopes distributed over a square km or more, its
unprecedented optical collecting area forms an excellent facility for ultrahigh angular
resolution (sub-milliarcsecond) optical imaging through long-baseline intensity interferometry.
This method was originally developed by Hanbury Brown and Twiss in the 1950s \cite{intensityinterf}
for measuring the sizes
of stars. It has since been extensively used in particle physics (``HBT interferometry'') but it has
had no recent application in astronomy because it requires large telescopes spread out over large distances,
which were not available until the recent development of atmospheric Cherenkov telescopes.
 
The great observational advantages of intensity interferometry are its lack of sensitivity to
atmospheric disturbances and to imperfections in the optical quality of the telescopes.
This is because of the
electronic (rather than optical) connection of telescopes. The noise relates to electronic
timescales of nanoseconds (and light-travel distances of centimetres or metres) rather than to
those of the light wave itself (femtoseconds and nanometres).
 
The requirements are remarkably similar to those for studying Cherenkov light:
large light-collecting telescopes,
high-speed optical detectors with sensitivity extending into the blue, and
real-time handling of the signals on nanosecond levels.
The main difference to ordinary Cherenkov Telescope operation
lies in the subsequent signal analysis which digitally synthesises
an optical telescope. From the viewpoint of observatory operations, it is worth
noting that bright stars can be measured for interferometry during bright-sky periods of
full Moon, which would hamper Cherenkov studies.
 
Science targets include studying the disks and surfaces of hot and bright stars \cite{77,88}
Rapidly rotating stars
naturally take on an oblate shape, with an equatorial bulge that, for stars rotating close to their
break-up speed, may extend into a circumstellar disk, while the regions with higher effective gravity
near the stellar poles become overheated, driving a stellar wind.  If the star is observed from
near its equatorial plane, an oblate image results. If the star is instead observed from near its
poles, a radial temperature gradient should be seen.  Possibly, stars with rapid and strong differential
rotation could take on shapes, midway between that of a doughnut and a sphere. The method permits studies
in both broad-band optical light and in individual emission lines, and enables the mapping of gas
flows between the components in close binary stars.
 
\subsubsection{Measurements of Charged Cosmic Rays}
%Spinoff scientific output of CTA:
\label{sec:CR}
Cherenkov telescopes can contribute to cosmic ray physics by detecting
these particles directly \cite{iron_kieda}.
CTA can provide measurements of the spectra of cosmic-ray electrons and nuclei
in the energy regime where balloon- and space-borne instruments run out
of data.
The composition of cosmic rays has been measured by balloon- and space-borne instruments
(e.g. TRACER) up to $\approx 100$~TeV. Starting at about 1 PeV instruments can detect
air showers at ground level (e.g. KASCADE). Such air shower experiments have, however, difficulties in
identifying individual nuclei, and consequently their composition results are of lower resolution
than direct measurements.
Cherenkov telescopes are the most
promising candidates to close the experimental gap between the TeV and PeV domains,
and will probably achieve better mass resolution than ground based particle arrays.
Additionally, CTA can perform crucial measurements of the spectrum of cosmic-ray electrons.
TeV electrons have very short lifetimes and thus propagation distances due to their rapid energy loss.
The upper end of the electron spectrum (which is not accessible by current balloon and satellite experiments)
is therefore expected to be dominated by local electron accelerators and the cosmic-ray electron
spectrum can provide valuable information about characteristics of the contributing sources and of
the electron propagation.
While such measurements involve analyses that differ from the conventional gamma-ray studies,
a proof-of-principle has already been performed with the H.E.S.S. telescopes.
Spectra of electrons and iron nuclei have been published \cite{99}. The
increase in sensitivity expected from CTA will provide significant improvements in such measurements.

\subsection{The CTA Legacy}
\label{sec:legacy}
 
The CTA legacy will most probably not be limited to individual observations addressing the
issues mentioned above, but also comprise a survey of the inner Galactic
plane and/or, depending on the final array capabilities, a deep survey of all
or part of the extragalactic sky.
Surveys provide coverage of large parts of the sky,
maximise serendipitous detections, allow for optimal use of telescope time,
and thereby ensure the legacy of the project for the future scientific community.
Surveys of different extents and depths are among
the scientific goals of all major facilities planned or in operation at all wavelengths.
In view of both H.E.S.S. (see Fig. \ref{fig_hess_scan}) and Fermi-LAT survey results,
the usefulness of surveys is
unquestioned, and many of the scientific cases discussed above can
be encompassed within such an observational strategy.
 
Two possible CTA survey schemes have been studied to date:
\begin{itemize}
\item All-sky survey: With an effective field-of-view of 5$^\circ$, 500 pointings of
0.5 hours would cover a survey area of a quarter of the sky at the target sensitivity of
0.01 Crab. Hence, using about a quarter of the observing time in a year,
a quarter of the sky can be surveyed
down to a level of $<$0.01 Crab, which is equivalent to the flux level of the faintest
AGN currently detected at VHE energies.
\item Galactic plane survey: The H.E.S.S. Galactic plane survey covered 1.5\% of the sky,  at a
sensitivity of 0.02 Crab above 200 GeV, using about 250 hours of observing time. The increase in CTA
sensitivity means that a similar investment in time can be expected to result in a sensitivity of
2-3 mCrab over the accessible region of the Galactic plane.
\end{itemize}
The high-energy phenomena which can be studied with CTA span a wide field of galactic and
extragalactic astrophysics, of plasma physics, particle physics, dark matter studies, and
investigations of the fundamental physics of space-time.
They carry information on the birth and death of stars, on the matter
circulation in the Galaxy, and on the history of the Universe.
Optimisation of the layout of CTA with regards to these different science goals is a
difficult task and detailed studies of the response of different array configurations
to these scientific problems being conducted during the Design Study and
the Preparatory Phase.
 
\clearpage
 
%% 3
%\input{CTAStateOfTheArt.tex}
%%%%% HEADER: CTAStateOfTheArt.tex
%%%%%%%%%%%%%%%%%%%%%%%%%%%%%%%%%%%%%%%%%%%%%%%%%%%%%%%%%%%%%%%%%%%
 
\section{Advancing VHE Gamma-Ray Astronomy with CTA}
\label{sec:state}
 
The latest generation of ground-based gamma-ray instruments (H.E.S.S.,
MAGIC, VERITAS, Cangaroo III \cite{canga3} and MILAGRO \cite{milagro}) allow
the imaging, photometry and spectroscopy of sources of high energy radiation and have ensured that
VHE gamma ray studies have grown to become a genuine branch of astronomy.
The number of known sources of VHE gamma rays is
exceeding 100, and source types include supernovae, pulsar wind nebulae, binary systems, stellar
winds, various types of active galaxies and unidentified sources without obvious counterparts.
H.E.S.S. has conducted a highly successful survey of the Milky Way covering about 600 square degrees,
which resulted in the detection of tens of new sources. However, a survey of the full visible sky would
require at least a decade of observations, which is not feasible.
 
Due to the small fluxes, instruments for detection of high-energy gamma rays (above some 10 GeV)
require a large effective detection area, eliminating space-based instruments which directly detect
the incident gamma rays. Ground-based instruments allow much larger detection areas. They measure the
particle cascade induced when a gamma ray is absorbed in the atmosphere, either by using
arrays of particle detectors to record the cascade particles which reach the ground (or mountain altitudes), or
by using Cherenkov telescopes to image the Cherenkov light emitted by secondary electrons and positrons
in the cascade.
 
Compared to Cherenkov telescopes, air shower arrays (such as MILAGRO, AS-gamma or ARGO)
have the advantage of a large duty cycle -- they can observe during the daytime -- and of a large solid angle
coverage. However, their current sensitivity is such that they can only detect sources with a flux around
the level of the flux from the Crab Nebula, the strongest known steady source of VHE gamma
rays. Results from air shower arrays demonstrate that there are relatively few sources emitting at this
level. The recent rapid evolution of VHE gamma-ray astronomy was therefore primarily driven
by Cherenkov instruments, which reach sensitivities of 1\% of the Crab flux for typical observing times
of 25 h, and which provide significantly better angular resolution. While there are proposals for
better air shower arrays with improved sensitivity (e.g. the HAWC project), which will certainly offer
valuable complementary information, such approaches will not be able to compete
in sensitivity with next-generation Cherenkov telescopes.
 
The properties of the major current and historic Cherenkov instruments are listed in tab.~\ref{table_cts}.
The instruments consist of up to 4 Cherenkov telescopes (or 5 for the H.E.S.S. II upgrade). They reach
sensitivities of about 1\% of the flux of the Crab Nebula at energies in the 100\,GeV to 1\,TeV range.
Sensitivity degrades towards lower energies, due to threshold effects, and towards higher energies, due
to the limited detection area. A typical angular resolution is $0.1^\circ$
or slightly better for single gamma rays. Sufficiently intense sources can be located with a
precision of 10-20''.
\begin{table}
\centering
{\footnotesize
  \begin{tabular}{|l|c|c|c|c|c|c|c|c|c|c|} \hline
    Instrument\rule[3mm]{0mm}{1mm} & Lat & Long & Alt & \multicolumn{3}{c|}{Telescopes} & Pixels & FoV & Thresh & Sensi- \\
               &     &      &     & \#  & Area & Total&     & FoV &        & tivity \\
    & ($^{\circ}$) & ($^{\circ}$) & (m) &  & (m$^{2})$  & (m$^{2}$)  &   &($^{\circ}$) & (TeV) & (\% Crab) \\
    \hline
    \hline
    H.E.S.S. \rule[3mm]{0mm}{1mm}& -23 & 16 & 1800 & 4 & 107 & 428 & 960 & 5 & 0.1 & 0.7 \\
    VERITAS & 32 & -111 & 1275 & 4 & 106 & 424 & 499 & 3.5 & 0.1 & 0.7 \\
    MAGIC I$^{\dagger}$+II & 29 & 18 & 2225 & 2 & 234 & 468 & 576/1039 &  3.5 & 0.03 & 1.0 \\
    CANGAROO-III & -31 & 137 & 160 & 3 & 57.3 & 172 & 427 & 4 & 0.4 & 15 \\
    \hline
    Whipple$^{\dagger}$ \rule[3mm]{0mm}{1mm} & 32 & -111 & 2300 & 1 & 75 & 75 & 379 & 2.3 & 0.3 & 15 \\
    HEGRA & 29 & 18 & 2200 & 5 & 8.5 & 43 & 271 & 4.3 & 0.5 & 5 \\
    CAT$^{\dagger}$ & 42 & 2 & 1650 & 1 & 17.8 & 17.8 & 600 & 4.8 & 0.25 & 15 \\
    \hline
  \multicolumn{11}{l}{$^{\dagger}$: These instruments have pixels of two different sizes.}
  \end{tabular}}
  \caption{\small Properties of selected air-Cherenkov instruments,
  including two of historical interest (HEGRA and CAT).
    Adapted from ref. \cite{hinton09}.}
\label{table_cts}
\end{table}
 
All these instruments are operated by the groups who built them, with very limited access for
external observers and no provision for open data access. Such a mode is appropriate for current
instruments, which detect a relatively limited number of sources,
and where the analysis and interpretation
can be handled by the manpower and experience accumulated in these consortia. However, a
different approach is called for in next-generation instruments, with their expected ten-fold
increase in the number of detectable objects.
CTA will advance the state of the art in astronomy at the highest energies
of the electromagnetic spectrum in a number of decisive areas,
all of which are unprecedented in this field:
\begin{description}
\item [European and international integration:] CTA will for the first time bring together and
combine the experience of all virtually all groups world-wide
working with atmospheric Che\-ren\-kov telescopes.
\item[Performance of the instrument:] CTA aims to provide 
full-sky view, from a southern and a northern site, with
unprecedented sensitivity, spectral coverage,
angular and timing resolution, combined with a high degree of flexibility of operation. Details are
addressed below.
\item[Operation as an open observatory:] The characteristics listed above imply that CTA will, for the
first time in this field, be operated as a true observatory, open to the entire astrophysics (and
particle physics) community, and providing support for easy access and analysis of data. Data will be
made publicly available and will be accessible through Virtual Observatory tools. Service to
professional astronomers will be supplemented by outreach activities and interfaces for laypersons to the
data.
\item[Technical implementation, operation, and data access:] While based on existing and proven
techniques, the goals of CTA imply significant advances in terms of efficiency of construction and
installation, in terms of the reliability of the telescopes, and in terms of data preparation and dissemination.
With these characteristics, the CTA observatory is qualitatively different from experiments such as
H.E.S.S., MAGIC or VERITAS and the increase in capability goes well beyond anything that could ever be
achieved through an expansion or upgrade of existing instruments.
\end{description}
 
\begin{figure}[htbp]
	\centering
  {\epsfig{width=\textwidth,file=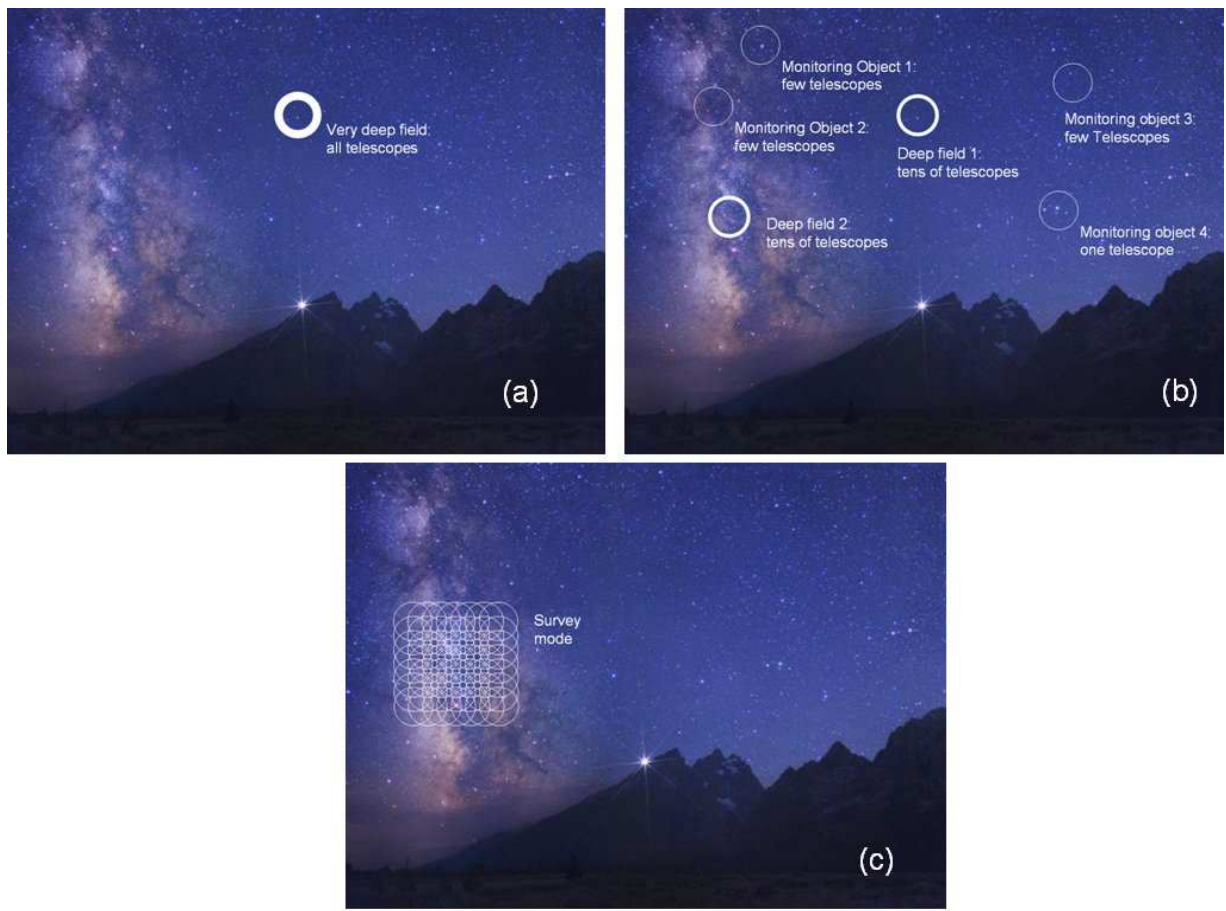}}
  \caption{\small Some of the possible operating modes of CTA:
  (a) very deep observations, (b) combining monitoring
  of flaring sources with deep observations, (c) a survey mode allowing full-sky surveys.}
  \label{fig:opmodes}
\end{figure}

\noindent
Science performance goals for CTA include in particular:
\begin{description}
\item[Sensitivity:] CTA will be about a factor of 10 more sensitive than any existing instrument.
It will therefore for the first time allow detection and in-depth study of large samples of known
source types, will explore a wide range of classes of suspected gamma-ray emitters beyond the
sensitivity of current instruments, and will be sensitive to new phenomena. In its core energy
range, from about 100 GeV to several TeV, CTA will have milli-Crab sensitivity, a factor of 1000 below
the strength of the strongest steady sources of VHE gamma rays, and a factor of 10000 below
the highest fluxes measured in bursts. This dynamic range will not only allow study of weaker sources
and of new source types, it will also reduce the selection bias in the taxonomy of known types of sources.
\item[Energy range:] Wide-band coverage of the electromagnetic spectrum is crucial for understanding
the physical processes in sources of high-energy radiation. CTA is aiming to cover, with a single
facility, three to four orders of magnitude in energy range. Together with the much improved precision
and lower statistical errors, this will enable astrophysicists to distinguish between key hypotheses
such as the leptonic or hadronic origin of gamma rays from supernovae. Combined with the Fermi
gamma-ray observatory in orbit, an unprecedented seamless coverage
of more than seven orders of magnitude in energy can be achieved.
\item[Angular resolution:] Current instruments are able to resolve extended sources, but they cannot
probe the fine structures visible in other wavebands. In supernova remnants, for example, the exact
width of the gamma-ray emitting shell would provide a sensitive probe of the acceleration mechanism.
Selecting a subset of gamma-ray induced cascades detected simultaneously by many of its telescopes, CTA can
reach angular resolutions in the arc-minute range, a factor of 5 better than the typical values for current instruments.
\item[Temporal resolution:] With its large detection area, CTA will resolve flaring and time-variable
emission on sub-minute time scales, which are currently not accessible. In gamma-ray emission from
active galaxies, variability time scales probe the size of the emitting region. Current instruments
have already detected flares varying on time scales of a few minutes, requiring a paradigm shift concerning the
phenomena in the vicinity of the super-massive black holes at the cores of active galaxies, and concerning
the jets emerging from them. CTA will also enable access to episodic and 
periodic phenomena such as emission
from inner stable orbits around black holes or from pulsars and other objects where frequent variations
and glitches in period smear the periodicity when averaging over longer periods.
\item[Flexibility:] Consisting of a large number of individual telescopes, CTA can be operated in a
wide range of configurations, allowing on the one hand the in-depth study of individual objects with
unprecedented sensitivity, and on the other hand the simultaneous monitoring of tens of potentially flaring objects,
and any combination in between (see fig.~\ref{fig:opmodes}).
\item[Survey capability:] A consequence of this flexibility is the dramatically enhanced survey capability of CTA.
Groups of telescopes can point at adjacent fields in the sky, with their fields of view overlapping,
providing an increase of sky area surveyed per unit time by an order of magnitude, and for the first
time enabling a full-sky survey at high sensitivity.
\item[Number of sources:] Extrapolating from the intensity distribution of known sources, CTA is
expected to enlarge the catalogue of objects detected from currently several tens of objects to
about 1000 objects.
\item[Global coverage and integration:] Ultimately, CTA aims to provide full sky coverage
from multiple observatory sites, using transparent access and identical tools to extract and analyse data.
\end{description}
 
\noindent
The feasibility of the performance goals listed above is borne out by detailed simulations of arrays of telescopes,
using currently available technology (details are given below). The implementation of CTA
does requires significant advances in the engineering, construction and operation of the array,
and the data access. These issues are addressed in the design study and the preparatory phase of CTA.
Issues include:
\begin{description}
\item[Construction, installation and commissioning of the telescopes:] To reach the performance
targets, tens of telescopes of 2-3 different types will be required, and the design of the
telescopes must be optimised in terms of their construction cost, making best use of the economics of
large-scale production. In current instruments, consisting at most of a handful of identical telescopes,
design costs were a substantial fraction of total costs, enforcing a different balance between design
and production costs. The design of the telescopes will have to concentrate on modularity and ease of
installation and commissioning.
\item[Reliability:] The reliability of current instruments is far from perfect, and down-times
of individual telescopes due to hardware or software problems are non-negligible. For CTA, telescope
design and software must provide significantly improved reliability. Frequent down-times of individual
telescopes in the array or of pixels within a telescope not only require substantial technical on-site
support and cause higher operating costs, but in particular they make the data analysis much more
complicated, requiring extensive simulations for each configuration of active telescopes, and inevitably
result in systematic errors which are likely to limit the achievable sensitivity.
\item[Operation scheduling and monitoring:] The large flexibility provided by the CTA array also raises
new challenges concerning the scheduling of observations, taking into account the state of the array and
the state of the atmosphere. For example, sky conditions may allow ``discovery observations'' in certain
parts of the sky, but may prevent precise, deep observations of a source. Availability of a given telescope
may be critical for certain types of observations, but may not matter at all in modes where the array
is split up in many sub-arrays tracking different sources at somewhat reduced sensitivity. To make optimum
use of the facility, novel scheduling algorithms will need to be developed, and the monitoring of the
atmosphere over the full sky needs to be brought to a new level of precision.
\item[Data access:] So far, none of the current Cherenkov telescopes has made data publicly available,
or has tools for efficient non-expert data access. Cherenkov telescopes are inherently more complicated
than, say, X-ray satellite instruments in that they do not directly take images of the sky, but
rather require extensive processing to go from the Cherenkov images to the parameters of the primary
gamma ray. Depending on the emphasis in the data analysis - maximum detection rate, lowest energy
threshold, best sensitivity, or highest angular resolution - there is a wide range of selection
parameters, all resulting in different effective detection areas and instrument characteristics.
Effective detection areas also depend on zenith angle, orientation relative to the Earth's magnetic
field, etc. Background subtraction is critical in particular for extended sources which may cover
a significant fraction of the field of view. Providing efficient data access and analysis tools
represents a major challenge and requires significant lead times and extensive software prototyping and tests.
\end{description}

\clearpage
 
%% 4
%\input{Performance.tex}
%%%%% HEADER: Performance.tex
%%%%%%%%%%%%%%%%%%%%%%%%%%%%%%%%%%%%%%%%%%%%%%%%%%%%%%%%%%%%%%%%%%%
 
\section{Performance of Cherenkov Telescope Arrays}
 \label{sec:perf}
In order to achieve improvements of a factor of 10 in several areas, it is
essential to understand and review the factors limiting the performance,
and to establish the extent to which limitations are of technical
nature which can be overcome with sufficient effort (e.g. due to a given size
of the camera pixels or point spread function (PSF)
of the reflector), and to which extent they represent fundamental limitations of the
technique (e.g. due to unavoidable fluctuations in the development of air showers).
 
To detect a cosmic gamma-ray source in a given energy band, three conditions have to
be fulfilled:
\begin{description}
\item[The number of detected gamma rays N$_\gamma$] has to exceed a minimum value,
usually taken to be between 5 and 10 gamma rays.
The number of gamma rays is the product of flux $\phi_\gamma$, effective detection area $A$, observing time
$T$ (usually for sensitivity evaluation taken as between 25 and 50\,h) and a detection
efficiency $\epsilon_\gamma$ which is typically not too far below unity. The number of detected gamma rays
and hence the effective area $A$ are virtually always the limiting factor at the high-energy end of the useful
energy range. For example, to detect a 1\% Crab source above 100\,TeV,
which equivalent to a flux of $2 \cdot 10^{-16}$/cm$^2$s$^{-1}$,
in 50\,h, an area $A$ of $\ge$ 30\,km$^2$ is required.
 
\item[The statistical significance of the gamma ray excess] has to exceed a certain number
of standard deviations, usually taken to be 5. For background dominated observations of faint sources, significance
can be approximated as $N_\gamma/\sqrt{N_{bg}}$ where the background events $N_{bg}$
arise from cosmic ray nuclei, cosmic
ray electrons, local muons, or random images caused by night-sky background (NSB). Background events are usually
distributed more or less uniformly across the useful field of view of the instrument. Their number is given by
the flux per unit solid angle, $\phi_{bg}$, the solid angle $\Omega_{src}$ over which gamma rays from
a candidate source (and hence background)
are accumulated, the effective detection area $A_{bg}$, the observation time and a background
rejection factor $\epsilon_{bg}$. The sensitivity limit $\phi_\gamma$ is hence proportional to
$\sqrt{\epsilon_{bg} A_{bg} T\Omega_{src}}/(\epsilon_\gamma A_\gamma T) \sim
\sqrt{\Omega_{src}}/\sqrt{\epsilon_{bg} A T}$ (assuming $A_{bg} \sim A_\gamma$). In current instruments,
electron and
cosmic nucleon backgrounds limit the sensitivity in the medium to lower part of their energy range.
 
\item[The systematic error on the number of excess gamma rays] due to un\-cer\-tain\-ties
in background estimates
and background subtraction has to be sufficiently small, and has to be accounted for in the
calculation of the significance.
Fluctuations in the background rates due to changes in voltages,
pulse shapes, calibration, in particular when non-uniform over the field of
view, or in the cut efficiencies, e.g. due to non-uniform
NSB noise, will
result in such background systematics.
Effectively, this means that a minimal signal-to-background ratio
is required to safely detect a source. The systematic limitation becomes important in the limit
of small statistical errors, when event numbers are very large
due to large detection areas, observation times, or low energy thresholds resulting in high count rates.
Since both signal and background scale with $A$ and $T$, the systematic sensitivity limit is proportional
to the relative background rate, $\phi_\gamma \sim (\epsilon_{bg} \Omega_{src})/\epsilon_\gamma$.
For current
instruments, background uncertainties at a level of a few \% have been reported \cite{background_paper}.
High reliability and availability of telescopes and pixels as well as improved schemes for calibration
and monitoring will be crucial in controlling systematic errors and exploiting the full sensitivity
of the instrument. An accuracy of the
background modelling and subtraction of 1\% seems reasonable and is assumed
in the following. Systematic errors may still limit sensitivity in the sub-100\,GeV range.
\end{description}
 
\begin{figure}[htbp]
\centering
%  \resizebox{10cm}{!}{\epsfig{file=toymodel.eps}}
%  \epsfig{width=0.49\textwidth, file=sens_toy.eps} \hfill
  \epsfig{width=0.8\textwidth, file=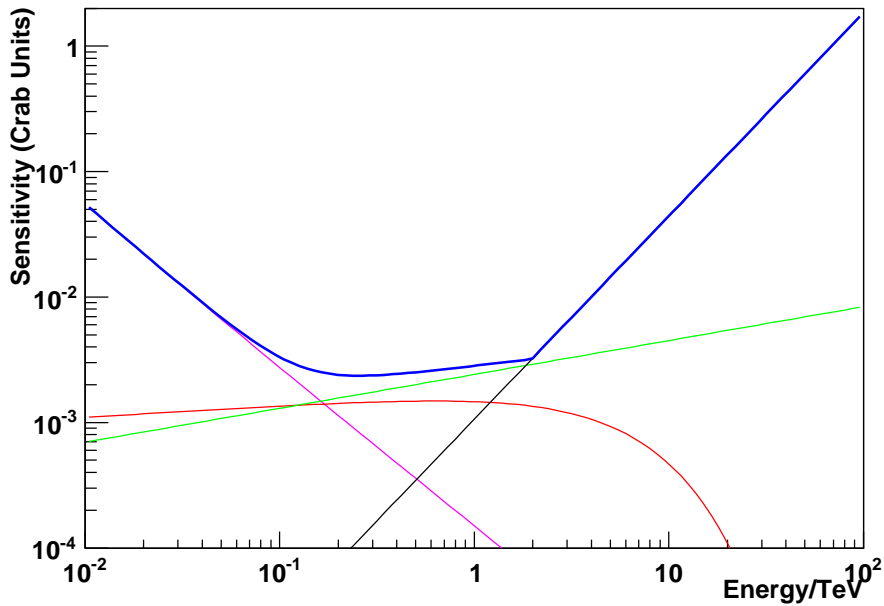}
  \caption{\label{fig_toy_model} \small
  Toy model of a telescope array to illustrate limiting sensitivity, quoted as the minimal
  detectable fraction of the Crab flux per energy band $\Delta \log_{10}(E)=0.2$, (assuming a simple
  power law for the Crab flux and ignoring the change in spectral index at low energy). The model
  assumes an energy-independent effective detection area of 1\,km$^2$, a gamma-ray efficiency
  of $\epsilon_\gamma$ of 0.5, the same efficiency for detection of cosmic-ray electrons, a
  cosmic-ray efficiency after cuts of $\epsilon_{bg} = 0.01$, an angular resolution $\delta$ of
  $0.1^\circ$ defining the integration region $\Omega_{src}$, and a systematic background uncertainty
  of 1\%. The model takes into account that cosmic-ray showers generate less Cherenkov light
  than gamma-ray showers, and are hence reconstructed at lower equivalent gamma-ray energy.
  At high energy, the sensitivity is limited by the gamma-ray count rate (black line),
  at intermediate energies by electron (red) and cosmic-ray backgrounds (green), and
  at low energies, in the area of high statistics, by systematic background uncertainty (purple).
  The plot includes also the effect of the PSF improving like 1/$\sqrt{\rm E}$
  (with PSF = 0.1$^\circ$ for 80\% containment at 200 GeV).}
\end{figure}

\noindent
Fig.~\ref{fig_toy_model} illustrates the various sensitivity limitations in the context of a simple toy model.
Obviously, sensitivity is boosted by large effective area $A$, efficient rejection of background, i.e.
small $\epsilon_{bg}$, and in the case of point-like structures by good angular resolution $\delta$ with
$\Omega_{src} \propto \delta^2$. Sensitivity gains can furthermore be achieved with a large field of view
of the instrument, observing multiple sources at a time and effectively multiplying the attainable
observation time $T$.
 
The annual exposure time amounts to about 1000\,h of useful moonless
observation time per year, varying by maybe 20\% between good and excellent sites.
Observations with
partial moon may increase this by a factor of 1.5, at the expense of reduced performance, depending
on the amount of stray light. Some instruments, such as MAGIC, routinely operate under
moonlight \cite{magic_moon}. While in principle more than 500\,h per year can be dedicated
to a given source (depending on its RA, and the maximum zenith angle under which observations
are carried out), in practice rarely more than 50\,h to at most 100\,h are dedicated to a
given source per year. With the increased number of sources detectable for CTA, there will be pressure
to reduce the time per source compared to current observations.
 
In real systems, the effective area $A$, background rejection $\epsilon_{bg}$ and angular resolution
$\delta$
depend on gamma-ray energy, since a minimal number of detected Cherenkov photons (around
50 to 100) are required to detect and analyse an image, and since the quality of
shower reconstruction depends both on the statistics of detected photons
and shower particles. The performance of the instrument depends on whether gamma-ray
energies are in the sub-threshold regime, near the nominal energy threshold, or well
above threshold.
 
In the sub-threshold regime, the amount of Cherenkov light is below the level needed for the
trigger logic, at a sufficiently low rate of random triggers due to
NSB photons. Only showers with upward fluctuations in the amount of Cherenkov light
will occasionally trigger the system. At GeV energies these fluctuations are large and
there is no sharp trigger threshold. Energy measurement in this domain is strongly biased.
 
In the threshold regime, there is usually enough Cherenkov light for triggering the system but
the signal in each telescope may still be too low for a) location of the image
centroid, b) determination of the direction of the image major axis, or c) accurate
energy assignment. Frequently, a higher
threshold than that given by the trigger is imposed in the data analysis. Most showers with upward fluctuations
will be reconstructed in a narrow energy range at the trigger (or analysis) threshold.
Sources with cut-offs below the analysis threshold may be detectable but
only at very high flux levels. Good imaging and spectroscopic performance of the
instrument is only available at energies $\ge 1.5\times$ the trigger threshold.
 
High sensitivity over a wide energy range, therefore, requires an instrument which
is able to detect a sufficient number of Cherenkov photons for low energy showers,
which covers a very large area for high-energy showers, and
which provides high angular resolution and background rejection.
High angular resolution is also crucial to resolve fine structures in
extended sources such as supernova remnants. On the other hand, for the detection of extended sources, the integration
region $\Omega_{src}$ is determined by the source size rather than the angular resolution and
cosmic-ray rejection becomes a most critical parameter in minimising statistical and systematic
uncertainties.
 
\begin{figure}[htbp]
\centering
\epsfig{file=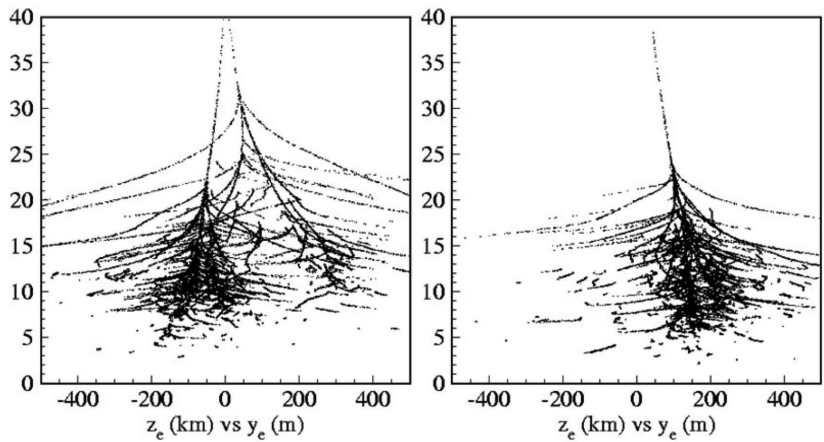, width=0.7\textwidth}
\caption{\small Two low-energy gamma-ray showers developing
in the atmosphere. Both gamma rays were incident vertically.
The difference in shower direction results from the energy sharing between electron and positron in the
first conversion and the subsequent deflection in the Earth's magnetic field.}
\label{fig_magn_field}
\end{figure}

\begin{figure}[htbp]
\centering
\epsfig{file=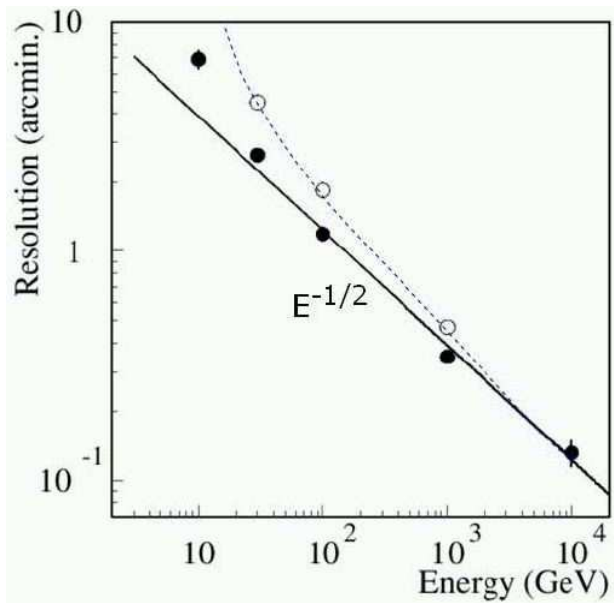, width=0.5\textwidth}
\caption{\small Limiting angular resolution of Cherenkov instruments as a function of gamma-ray energy,
derived from a likelihood fit to the directions of all Cherenkov photons reaching the ground, and
assuming perfect measurement of photon impact point and direction. At low energies, the resolutions
differ in the bending plane of the Earth's magnetic field (open symbols) and in the orthogonal
direction (closed symbols). The simulations assume near-vertical incidence at the H.E.S.S. site in Namibia.}
\label{fig_limiting_resolution}
\end{figure}

A crucial question is therefore to which extent angular resolution and cosmic-ray rejection can be
influenced by the design of the instrument, by parameters such as the number of Cherenkov photons detected
or the size of the photo-sensor pixels. Simulation studies assuming an ideal instrument
\cite{Hofmann_limits}, one which detects
all Cherenkov photons reaching the ground with perfect resolution for impact point
and photon direction, show that achievable
resolution and background rejection are ultimately limited by fluctuations in the shower development.
Angular resolution is in addition influenced by the deflection of shower particles
in the Earth's magnetic field, making
the reconstructed shower direction dependent on the energy sharing between electron and positron in
the first conversion of a gamma ray (fig.~\ref{fig_magn_field}). However, these resolution limits
(fig.~\ref{fig_limiting_resolution}) are
well below the resolution achieved by current instruments. At 1\,TeV, a
resolution below one arc-minute is in principle achievable.
Similar conclusions appear to hold for cosmic-ray background rejection.
There is a virtually irreducible background due to events in which, in the first interaction of a cosmic ray,
almost all the energy is transferred to one or a few neutral pions and, therefore, to electromagnetic cascades
(see, e.g. \cite{Maier}).
However, with their typical cosmic-ray rejection factors of $>$10$^3$ at TeV energies,
current instruments still seem
1-2 orders of magnitude away from this limit, offering space for improvement. Such improvements
could result from improved imaging of the air shower, both in terms of resolution and photon statistics,
and from using a large and sensitive array to veto cosmic-ray induced showers based on the debris
frequently emitted at relatively large angles to the shower axis.
 
At low energies, cosmic-ray electrons become the dominant background, due to their steep spectrum.
Electrons and gamma-rays cannot be distinguished efficiently using shower characteristics, as both induce
electromagnetic cascades. The height of the shower
maximum differs by about one radiation length \cite{PDG}, but this height also fluctuates from shower to
shower by about one radiation length, rendering an efficient rejection impossible.
A technique which is beyond the capability of current instruments but might become possible with
future arrays is to detect Cherenkov radiation from the primary charged particle and use it as a veto
\cite{Hofmann_limits}. Detection of the ``direct Cherenkov light'' has been proposed
\cite{iron_kieda} and successfully applied \cite{hess_iron} for highly charged primary nuclei such as iron,
where Cherenkov radiation is enhanced by a factor of $Z^2$. While in a 100\,m$^2$ telescope, an iron nucleus
generates O(1000) detected photons, a charge-1 primary will provide at most a few photons, not far from night
sky noise levels. Larger telescopes, possibly with improved photo-sensors, fine pixels and high temporal
resolution, could enable detection of primary Cherenkov light from electrons, at the expense of gamma-ray
efficiency, since gamma-rays converting at high altitude will be rejected, too,
and since unrelated nearby cosmic rays may generate fake vetos.
Nevertheless, this approach (not yet studied in detail)
may help at the lowest energies where event numbers are high but there are large
uncertainties in the background systematics. Sakahian et al. \cite{Sahakian_electrons} note that
at energies $<$20\,GeV, deflection of electrons in the Earth's magnetic field is sufficiently
large to disperse Cherenkov photons over a larger area on the ground, reducing light density and therefore
the electron-induced trigger rate. The effect is further enhanced by a dispersion in photon arrival times.
 
In summary, it is clear that the performance of
Cherenkov telescope arrays can be improved significantly,
before fundamental limitations are reached.

\clearpage
 
%% 5
%\input{CTAArray.tex}
%%%%% HEADER: CTAArray.tex
%%%%%%%%%%%%%%%%%%%%%%%%%%%%%%%%%%%%%%%%%%%%%%%%%%%%%%%%%%%%%%%%%%%%%%%%%%%%%%%%%%%%%%%
 
\section{The Cherenkov Telescope Array}
\label{sec:cta}
 
The CTA consortium plans to operate from one site in the southern and one in the northern hemisphere,
allowing full-sky coverage.
The southern site will cover the central part of the galactic plane and see most of the galactic sources and will
therefore be designed to have sensitivity over the full energy range.
The northern site will be optimised for extragalactic astronomy, and will not require
coverage of the highest energies.
 
Determining the arrangement and characteristics of the CTA telescopes in the two arrays is a complex optimisation
problem, balancing cost against performance in different bands of the spectrum. This section will address
the general criteria and considerations for this optimisation, while the technical implementation
is covered in the following sections.
 
\subsection{Array Layout}
\label{sec:arraylayout}
Given the wide energy range to be covered, a uniform
array of identical telescopes, with fixed spacing, is not the most efficient solution for the CTA.
For the purpose of discussion, separation into three energy ranges, without sharp boundaries,
is appropriate:
 
\begin{description}
\item[The low-energy range $\le$100\,GeV:] To detect showers down to a few tens
of GeV, the Cherenkov light needs to be sampled and detected efficiently, with the fraction of
area covered by light collectors being of the order of 10\%
(assuming conventional PMT light sensors). Since event rates
are high and systematic background uncertainties are likely to limit the achievable sensitivity,
the area of this part of the array can be relatively small,
being of order of a few $10^4$\,m$^2$. Efficient photon detection can be achieved either with few large telescopes
or many telescopes of modest size.
For very large telescopes,
the cost of the dish structures dominates, for small telescopes the photon detectors and electronics
account for the bulk of the cost. A (shallow) cost optimum in terms of cost per telescope area is
usually reached for medium-sized telescopes in the 10 to 15\,m diameter range.
However, if small to medium-sized telescopes are used in this energy range, the challenge
is to trigger the array, since no individual telescope detects enough Cherenkov photons to provide a
reliable trigger signal. Trigger systems which combine and superimpose images at the pixel level in real time,
with a time resolution of a few ns, can address this issue \cite{tbd_timeres} but represent a significant
challenge, given
that a single 1000-pixel telescope sampled at (only) 200 MHz and 8 bits per pixel generates a data stream
of more than one Tb/s. CTA designs conservatively assume a small number
of very large telescopes, typically with about a 20 to 30\,m dish diameter, to cover the low energy range.
 
\item[The core energy range from about 100\,GeV to about 10\,TeV:] shower detection\\
and reconstruction
in this energy range are well understood from current instruments, and an appropriate solution seems a
grid of telescopes of the 10 to 15\,m class, with a spacing of about 100\,m.
Improved sensitivity is obtained both by the increased area covered, and by the higher quality of shower
reconstruction, since showers are typically imaged by a larger number of telescopes than is the case for current
few-telescope arrays. For the first time, array sizes will be larger than the Cherenkov light pool,
ensuring that images will be uniformly sampled across the light pool, and that a number of images
are recorded close to the optimum distance from the shower axis (about 70 to 150\,m), where the light intensity
is large and intensity fluctuations are small, and where the shower axis is viewed under a
sufficiently large angle for efficient reconstruction of its direction.
At H.E.S.S. for example, events which are seen and triggered by all four telescopes
provide significantly improved resolution
and strongly reduced backgrounds, but represent only a relatively small fraction of events. Unless energies
are well
above trigger threshold, only events with shower core locations within the telescope square can trigger
all telescopes. A further advantage is that an extended telescope grid operated with a two-telescope trigger condition will
have a lower threshold than a small array, since there are always telescopes sufficiently close to the shower core.
 
\item[The high-energy range above 10\,TeV:] Here, the key limitation is the number of detected gamma-ray
showers and the array needs to cover multi-km$^2$ areas. At high energies the light yield is large, so
showers can be detected well beyond
the 150 m radius of a typical Cherenkov light pool.
Two implementation options can be considered:
either a large number
of small telescopes with mirror areas of a few m$^2$ and spacing matched to the size of the light
pool of 100 to 200\,m,
or a smaller number of larger telescopes with some 10 m$^2$ area which can see showers up
to distance of $\ge$500\,m, and can hence be deployed with a spacing of several 100\,m,
or in widely separated
subclusters of a few telescopes.
%\cite{tbd_subcluster}.
While it is not immediately obvious which options
offers best cost/performance ratio at high energies, the subcluster concept with larger telescopes has
the advantage of providing additional high-quality shower detection towards lower energies,
for impact points near the subcluster.
\end{description}
 
\begin{figure}[htbp]
\centering
 \resizebox{10cm}{!}{\epsfig{file=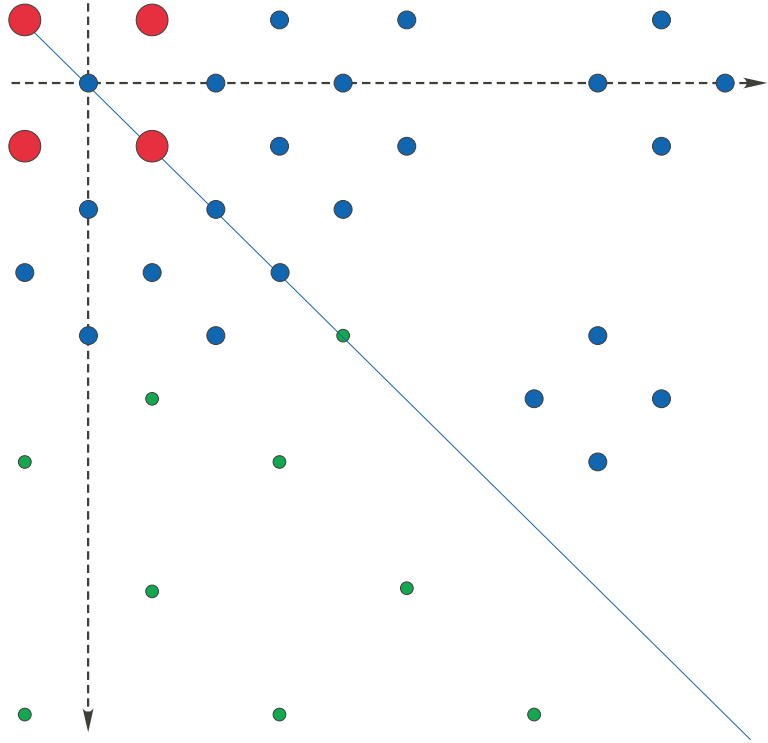}}
 \caption{\small A quadrant of possible array schemes promising excellent
  sensitivity over an extended energy range, as suggested by the Monte Carlo studies.
  The centre of the installation is near the upper left corner. Telescope
  diameters are not drawn to scale.
  In the upper right part, clusters of telescopes of the 12 m class are shown at the perimeter,
  while in the lower left part an option with wide-angle telescopes of the 3-4 m class is shown.}
\label{fig_array_scheme}
 \end{figure}
 
\noindent
Fig.~\ref{fig_array_scheme} shows possible geometries of arrays with separate regions optimized for low,
intermediate and high energies.
 
\subsection{Telescope Layout}
\label{sec:tellayout}
 
Irrespective of the technical implementation details, as far as its performance is concerned,
a Cherenkov telescope is primarily characterised by its light collection capability, i.e.
the product of mirror area, photon collection efficiently and photon detection efficiency,
by its field of view and by its pixel size, which limits the size of image features which
can be resolved. The optical system of the telescope should obviously be able at achieve
a point spread function matched to the pixel size. The electronics for signal capture and
triggering should provide a bandwidth matched to the length of Cherenkov pulses of a
few nanoseconds. The performance of an array is also dependent on the triggering strategy;
Cherenkov emission from air showers
has to be separated in real time from the high flux of night sky background photons,
based on individual images and global array information. The
huge data stream from Cherenkov telescopes does not allow untriggered recording.
 
The required light collection capability in the different parts of the array is
determined by the energy thresholds, as outlined in the previous section. In the following,
field of view, pixel size and the requirements on the readout system and trigger system are reviewed.
 
\subsubsection{Field of View}
\label{sec:telfov}
 
Besides mirror area, an important telescope design parameter is the field of view.
A relatively large field of view is mandatory for the widely spaced telescopes of the high-energy
array, since the distance of the image from the camera centre scales with the distance
of the impact point of the air shower from the telescope.
For the low- and intermediate-energy arrays, the best choice of the field of view is not trivial
to determine. From the science point of view, large fields of view are
highly desirable, since they allow:
\begin{itemize}
\item the detection of high-energy showers at large impact distance without image truncation;
\item the efficient study of extended sources and of diffuse emission regions; and
\item large-scale surveys of the sky and the parallel study of many clustered sources, e.g. in the
band of the Milky Way.
\end{itemize}
In addition, a larger field of view generally helps in improving the uniformity
of the camera and reducing background systematics.
 
\begin{figure}[b]
\centering
  \epsfig{width=0.92\textwidth,file=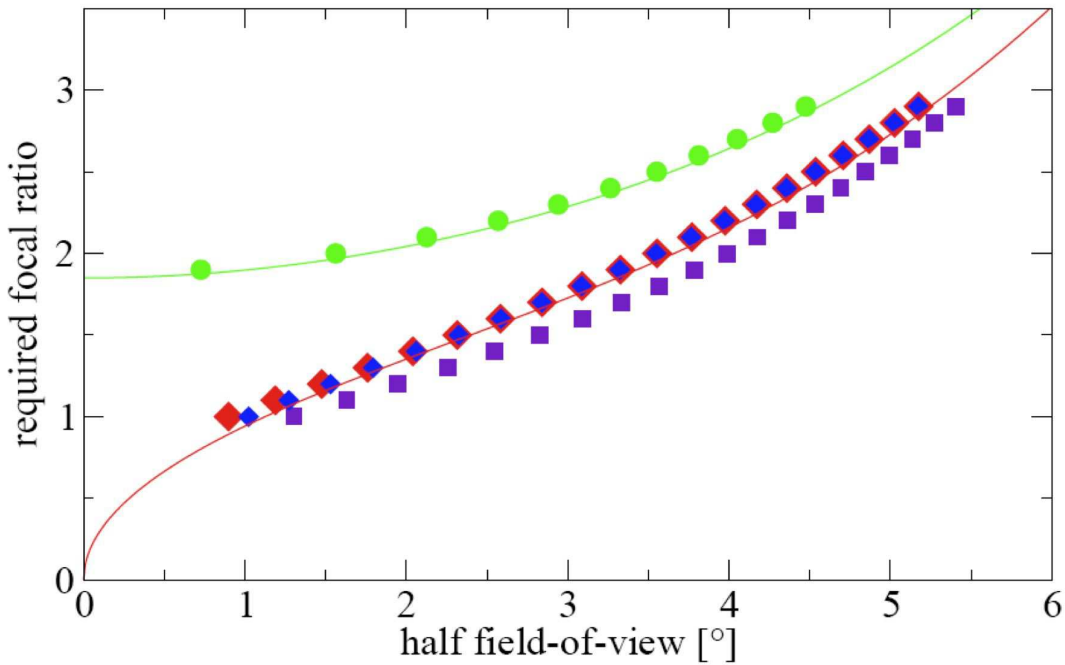}
  \caption{\label{fig_fov} \small Focal ratio required for sufficiently precise shower imaging,
  as a function of the half angle of the field of view \cite{schliesser}.
  Points: simulations for spherical design (green), parabolic design
with constant radii (red), Davies-Cotton design (violet),
parabolic design with adjusted radii (blue). Lines: third-order approximation for
a single-piece paraboloid (red) and a single-piece sphere (green).}
\end{figure}

However, larger fields of view for a given pixel size, result in rapidly growing numbers
of photo-sensor pixels and electronics channels. Large fields of view also require technically
challenging telescope optics. With the current single-mirror optics and $f/d$ ratios in
the range up to 1.2, an acceptable point spread function is obtained out to $4-5^\circ$.
Larger fields of view with single-mirror telescopes require increased $f/d$ ratios,
in excess of 2 for a $10^\circ$
field of view (see fig.~\ref{fig_fov}, \cite{schliesser}),
which are mechanically difficult to realise,
since a large and heavy focus box needs to be supported at a long distance from the dish.
Also, the single-mirror optics solutions which
provide the best imaging use Davies-Cotton or elliptical dish geometries, which in turn result in a time
dispersion of shower photons which seriously impacts on the trigger performance once
dish diameters exceed 15\,m. An alternative solution is the use of secondary mirrors. Using non-spherical
primaries and secondaries, good imaging over fields of up to $10^\circ$ diameter can be achieved
\cite{vassiliev}. Disadvantages are the increased cost and complexity, significant shadowing of the
primary mirror by the secondary, and complex alignment issues if faceted primary and
secondary mirrors are used. With the resulting large range of incidence angles of photons onto the
camera, can imply that baffling of albedo also becomes an issue.
 
The choice of the field of view therefore requires that the
science gains and the cost and increased complexity be carefully balanced.
When searching for unknown source types which are not associated with
non-thermal processes in other, well-surveyed wavelength domains,
a large field of view helps, as several sources may appear in
typical fields of view. This increases the effective
observation time per source by a corresponding factor
compared to an instrument which can look only at one source at a time.
An instrument with CTA-like sensitivity is expected to detect
of the order of 1000 sources. In the essentially one-dimensional galactic plane, there will always be
multiple sources in a field of view. In extragalactic space, the average angular distance between (an
estimated 500) sources would be
about $10^\circ$, implying that even for the maximum conceivable fields of view the gain is modest.
Even in the
galactic plane, a very large field of view will not be the most cost-effective solution, since the gain
in terms of the number of sources viewed simultaneously scales essentially with the diameter of the field of view,
given that sources are likely to cluster within a fraction of a degree from the plane, whereas camera
costs scale with the diameter squared. A very rough estimate based on typical dish costs and per-channel
pixel and readout costs suggests an economic optimum in the cost per source-hour at around a diameter of
6-8$^\circ$ field of view.
 
The final choice of the field of view will have to await detailed studies related to dish and mirror technology
and costs, and the per-channel cost of the detection system.
 
Sensitivity estimates given below do not include an enhancement factor accounting for multiple
sources in the field of view, but effective exposure time should increase by factors of $\ge$4
for Galactic sources, and sensitivity correspondingly by factors of $\ge$2.
 
\subsubsection{Pixel Size}
 \label{sec:pixelsize}
 
\begin{figure}[htbp]
\centering
\epsfig{width=\textwidth, file=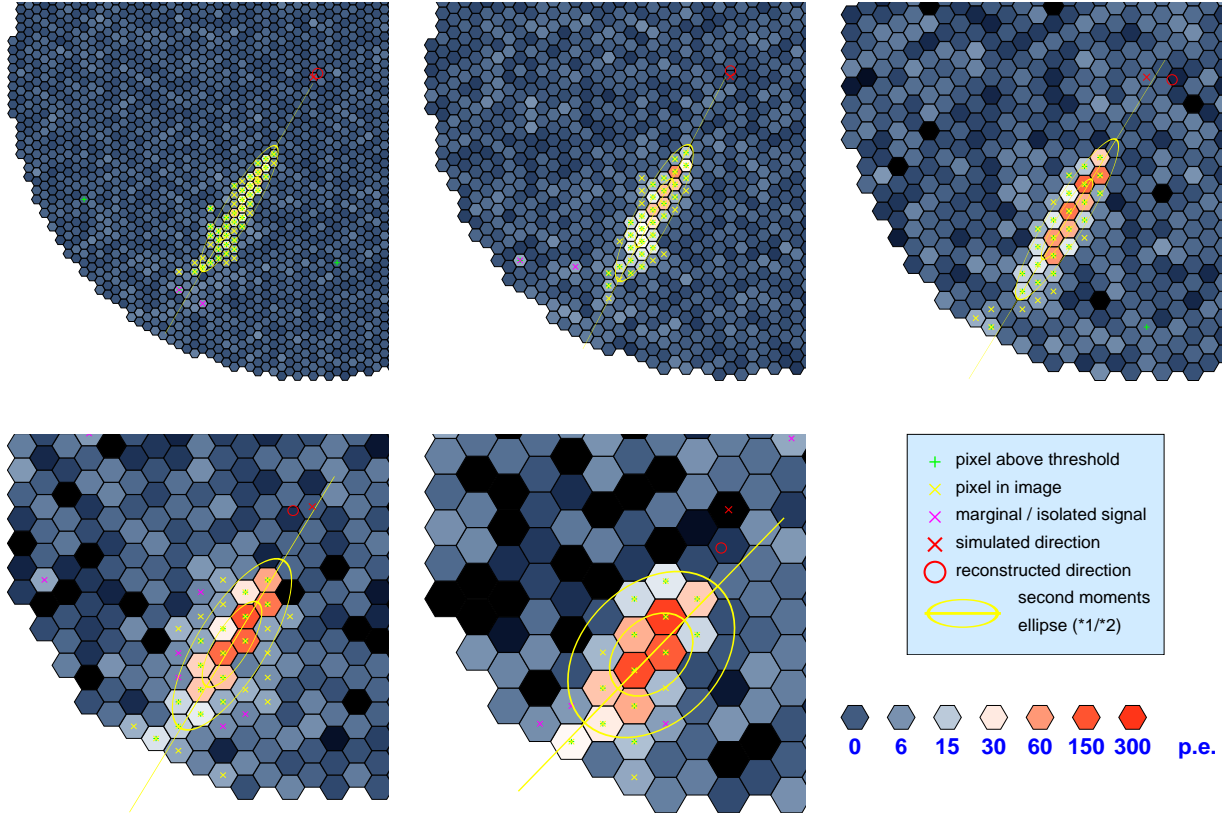}
\caption{\label{fig:pixel-size} \small
Part of the field of view of cameras
with different pixel sizes (0.07, 0.10, 0.14, 0.20, and 0.28$^{\circ}$) but
identical field-of-view (of about 6$^\circ$), viewing the same shower (460 GeV gamma-ray
at 190 m core distance) with a 420 m$^2$ telescope.
Low-energy showers would be difficult to register, both with very small
pixels (signal not contiguous in adjacent pixels) and with very large pixels (not
enough pixels triggered above the increased
thresholds, due to high NSB rates).}
\end{figure}

The size of focal plane pixels is another parameter which requires careful optimisation.
Fig.~\ref{fig:pixel-size} illustrates how a shower image is resolved at pixel sizes
ranging from $0.28^\circ$ (roughly the pixel size of the HEGRA telescopes) down to
pixel sizes of $0.07^\circ$, as used for example in the large H.E.S.S. II telescope. The cost of focal
plane instrumentation is currently driven primarily by the number of pixels and, therefore,
scales like the square of the inverse pixel size. The gain due to the use of small pixels depends strongly
on the analysis technique. In the classical second-moment analysis, performance seems to saturate
for pixels smaller than $0.2-0.15^\circ$ \cite{Aharonian_1995}. Analysis techniques which use the full
image distribution (e.g. \cite{Model-analysis}),
on the other hand, can extract the information contained in the well-collimated
head part of high-intensity images, as compared to the more diffuse tail, and benefit
from pixel sizes as small as
$0.06-0.03^\circ$ \cite{Hofmann_limits,vassiliev}.
Pixel size also influences trigger strategies. For large pixels, gamma-ray
images are contiguous, allowing straight-forward topological triggers, whereas for small pixels,
low-energy gamma-ray images may have gaps between triggered pixels.
 
The final decision concerning pixel size (and telescope field of view)
will to a significant extent be driven by the cost per pixel. Current simulations
favour pixel sizes of 0.07-0.1$^\circ$ for the large telescopes, allowing the resolution of compact
low-energy images and reducing the rate of NSB photons in each pixel,
0.15-0.2$^\circ$ for the medium size telescopes, similar to the pixel sizes used by
H.E.S.S. and VERITAS, and 0.2-0.3$^\circ$ for the pixels of the telescopes in the halo of
the array, where large fields of view are required but shower images also tend to be long due
to the large impact distances and the resulting viewing angles.
Studies to determine the benefits of smaller pixels, as are proposed for 
AGIS-type dual-mirror telescopes \cite{agis},
are underway for the medium-sized telescopes.
 
\subsubsection{Signal Recording}
 \label{sec:sigrec}
\begin{figure}[htbp]
\centering
\epsfig{width=\textwidth,file=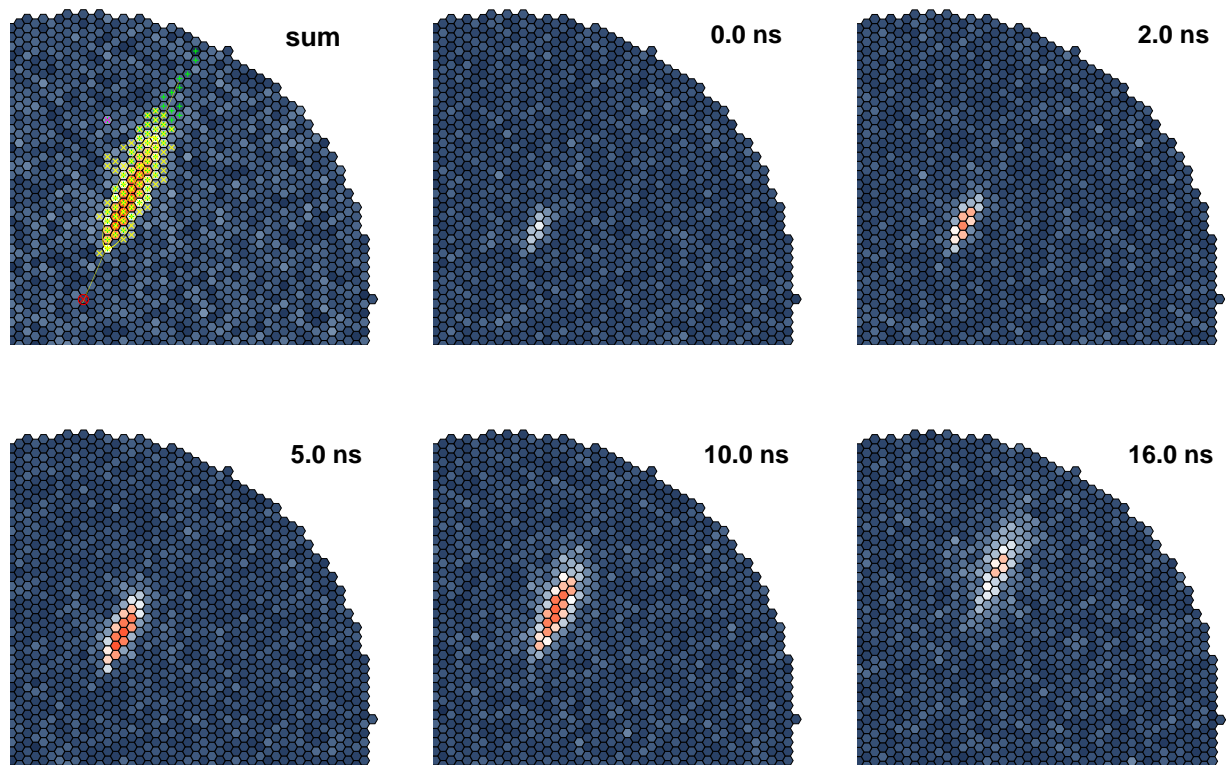}
  \caption{\label{fig-time-development}
  \small Integrated
  signal (upper left) and 1 ns
  samples of the development of a 10 TeV gamma shower at 250 m core distance
  as seen in a telescope with optics and pixels similar to a H.E.S.S.-1
  telescope but with a FoV of 10$^\circ$ diameter. Pixels near the ``head'' of the
  shower have a pulse width dominated by the single photoelectron  pulse width, while
  those in the ``tail'' of the shower see longer pulses. The shower image
  moves across almost half the FoV in about 25~ns.}
\end{figure}

Most modern telescopes use some kind of transient recorders to capture
pixel signals, either with analogue switched-capacitor systems or with
fast digitisers \cite{fadc}, so that, at least in principle, signal shape and timing
can be used in the image analysis. Signal shape and timing can be employed
in two ways: (a) to reject backgrounds such as hadronic showers and local muons; and
(b) to reduce the signal integration windows and hence the amount of NSB
noise in the shower image. For example, muon rejection based on
signal waveform is discussed in \cite{muon_timing}. Quantifying how much
background rejection can be improved using these techniques is non-trivial. The effect of signal-shape
image selection is correlated with other cuts imposed in the analysis.
For single telescopes, signal shape and timing can provide significant improvements.
For telescope systems, the cuts on image shapes in multiple telescopes are already
very powerful and background events passing these cuts will have images and signal shapes
that look very much like those of gamma-rays,
so that less improvement is expected, if any.
The second area where signal waveform recording can improve performance concerns
the signal amplitudes. In particular for larger shower impact parameters,
photon arrival times are not isochronous across the image (fig.~\ref{fig-time-development}),
and photons in the ``tail'' end of the image arrive with significant delays compared
to those from its ``head''. Use of variable and matched integration windows across the image
allows the extraction of shower signals with minimal contamination from NSB
noise.
Use of signal shape and timing information is already used in the current
MAGIC \cite{magic_waveform} and VERITAS systems, and these results will help to
guide final design choices for CTA.
 
The performance numbers quoted for the simulations described below are conservative in that they
are based on fixed (and relatively large) signal integration windows. Improvements
can be expected once the use of image shape information is fully understood.
\clearpage
 
\subsubsection{Trigger}
 \label{sec:trigchar}
 
The trigger scheme and readout electronics are  closely related and fundamentally
influence the design and performance of the telescope array.
For most applications, multi-telescope trigger coincidence is required to reject
backgrounds at the trigger level and to reduce the load on the data acquisition system.
The main issue here is how much information is exchanged between telescopes, and how
image information is stored while the trigger decision is made.
 
One extreme scenario
is to let each telescope trigger independently and only exchange a trigger flag with neighbouring
telescopes, allowing identification of coincident triggers (e.g. \cite{hess_trigger}). The energy
threshold of the system is then determined by the
minimum threshold at which a telescope can trigger. The other extreme is to combine signals
from different telescopes at the pixel level, either in analogue or digital form, and to extract
common image features. % \cite{star_trigger}.
In this case, the system energy threshold could be well below the thresholds
of individual telescopes, which is important when the array is made up of many small or medium-sized
telescopes. However, the technical complexity of such a solution is significant.
There is a wide range of intermediate
solutions, where trigger pre-processors extract image features, such as the image centroid, on a
telescope basis and the system trigger decision includes this information.
 
In cases where individual telescopes generate a local trigger, pixel signals
need to be stored while a global trigger decision is made. The time for which signals can be
stored without introducing deadtime,
is typically ms in the case of digital storage and $\mu$s if analogue storage is used,
which strongly influences the design of higher level triggers.
 
Trigger topology is another important issue. Triggers can either be derived locally within the array by
some trigger logic connecting neighbouring telescopes, or all trigger information
can be routed to a central station where a global decision is made, which
is then propagated back to the telescopes. The first approach
requires shorter signal storage at the telescopes and is more easily scaled up to large arrays, the
second provides maximum flexibility.
Whether local or global, trigger schemes will employ a multi-level hierarchy, with a first trigger
level acting on pixels and pixel groups, and higher levels using information on image topology and/or
the topology of triggered telescopes in the array. As in modern high-energy physics experiments,
trigger decisions will, to the extent possible, be performed using programmable rather than
``hardwired'' processors. If the signal is recorded using fast digitisers, even the first-level
discrimination of pixel signals could be implemented digitally in the gate array controlling the
digitiser, instead of applying analogue thresholds.
 
Whatever implementation is chosen, it is important that the trigger system is
very flexible and software-configurable, since operation modes vary from deep observations,
where all telescopes follow the same source,
to monitoring or survey applications, where groups
of a few telescopes or even single telescopes point in different directions.
 
The simulations discussed below assume a very conservative approach. Each telescope
makes an independent trigger decision with thresholds defined such that the telescope trigger rate
is in the manageable range of a few to some tens of kHz. This is followed by a global decision based on the
number of triggered telescopes.
 
\subsection{CTA Performance Summary}
 
Section~\ref{sec:MC} gives a detailed description of the layout and
performance studies conducted so far for CTA. Many candidate layouts have been
considered. Here we provide a brief description of the nature and
performance of one promising configuration (E), which is
illustrated in fig.~\ref{fig-configurations}. This configuration
utilises three telescope types:
four 24~m telescopes with 5$^{\circ}$
field-of-view and 0.09$^{\circ}$ pixels,
23  telescopes of 12~m diameter
with 8$^{\circ}$ field-of-view and 0.18$^{\circ}$ pixels,
and 32 telescopes
of 7~m diameter with a 10$^{\circ}$ field-of-view and
0.25$^{\circ}$ pixels. The telescopes are distributed over $\sim$3
km$^{2}$ on the ground and the effective collection area of the array
is considerably larger than this at energies beyond 10 TeV. The
sensitivity of array E from detailed calculations and using standard
data analysis techniques is shown in fig.~\ref{fig-sens-ebc}. More sophisticated
analyses result in sensitivities that are $\sim$20\% better across the
whole energy range. As fig.~\ref{fig-sens-ebc} shows, such an array
performs an order of magnitude better than an instrument like
H.E.S.S. over most of required energy range. Fig.~\ref{fig_angres_2}
shows the angular resolution of this array, which approaches one
arcminute at high energies. The energy resolution of layout E is
better than 10\% above a few hundred GeV.
 
Array layout E has a nominal construction cost of 80 M\euro{} and meets
the main design goals of CTA. Given that the configuration itself, and
the analysis methods used, have not yet been optimised, it is likely
that a significantly better sensitivity can be achieved with an array of
this nominal cost which follows the same basic concept. Therefore,
despite the uncertainties in the cost model employed (see sec. \ref{sec:sched}),
we are confident that the design goals of CTA can be realised at close to the
envisaged cost.

\clearpage
 
%% 6
%\input{RealizingCTA.tex}
%%%%% HEADER: RealizingCTA.tex
%%%%%%%%%%%%%%%%%%%%%%%%%%%%%%%%%%%%%%%%%%%%%%%%%%%%%%%%%%%%%%%%%%%
 
\section{Realizing CTA}
\label{sec:real}
 
This section provides a brief overview
of the position of CTA in the European and global context,
the organisation of CTA during the various stages,
of its operation as an open observatory,
of the potential sites envisaged for CTA,
and of the schedule for and cost of CTA design, construction and operation.
 
\subsection{CTA and the European Strategy in Astrophysics and Astro\-particle Physics}
 
CTA, as a major future facility for astroparticle physics, is firmly embedded in the European
processes guiding science in the fields of astronomy and astroparticle physics.
\begin{description}
\item[The European Strategy Forum on Research Infrastructures (ESFRI):]
ESFRI\\
is a strategic organisation whose objective is
to promote the scientific integration of Europe, to streng\-then
the European Research Area and to increase its international impact.
A first Roadmap for pan-European research infrastructures was released in 2006, listing CTA as an
``emerging project''. In the December 2008 update of this Roadmap, CTA was included as one of
eight  Physical Sciences and Engineering projects, together with facilities such as
E-ELT, KM3Net and SKA. As such, CTA is eligible for FP7  Preparatory
Phase funding. The CTA application for this funding was successful, providing up to 5.2~M\euro{}
for the preparation of the construction of the observatory in 3 years time.
The contracts with the EC are in the process of being finalised and signed.
 
\item[The Astroparticle Physics European Coordination (ApPEC) group:] ApPEC\\ was cre\-a\-ted to enhance
coordination in astroparticle physics across Europe. It has stimulated cooperation and convergence
between competing groups in Europe, and has initiated the production of a European roadmap in
astroparticle physics, on which CTA is one of the key projects.
 
\item[ASPERA:] ASPERA is a network of national government agencies responsible for
coordinating and funding national research efforts in Astroparticle Physics. One of the tasks of ASPERA
is to create a scientific roadmap for Astroparticle Physics \cite{aspera_roadmap}
and link it with the more general
European scientific infrastructure roadmap. A Phase I roadmap has been published, presenting the
overarching science
questions and the new instruments planned to address these questions.
Phase II saw the release of the resulting ``European Strategy for Astroparticle Physics''
in September 2008, prioritising the projects under consideration.
In this roadmap, CTA emerges as a
near-term high-priority project. The roadmap states: \\
{\em The priority project for VHE gamma-ray astrophysics
is the Cherenkov Telescope Array, CTA. We recommend design and prototyping of CTA,
the selection of sites, and proceeding rapidly towards start of deployment in 2012.}\\
CTA was one of the two projects targeted by the 2009 ASPERA Common Call for
cross-national funding and received in total 2.7 M\euro{}
from national funding agencies.
 
\item[The ASTRONET Eranet:] ASTRONET was created by a group of European funding agencies
to establish comprehensive long-term planning for the development of European astronomy.
The objective of this effort is to consolidate and reinforce the world-leading position that
European astronomy attained at the beginning of this century. Late in 2008, ASTRONET
released ``The ASTRONET Infrastructure Roadmap: A Strategic Plan for European Astronomy''.
CTA is one of the three medium-scale facilities recommended on this roadmap,
together with the neutrino telescope KM3Net and the solar telescope EST.
\end{description}
 
\subsection{CTA in the World-Wide Context}
 \label{sec:worldwide}
Ground-based gamma-ray astronomy has attracted considerable attention world-wide, and while CTA
is the key project in Europe, other projects have been considered elsewhere.
These include primarily:
\begin{description}
\item[The Advanced Gamma-ray Imaging System (AGIS):] In both science an instrumentation,
AGIS \cite{agis} followed a very similar plan to that of CTA.
The AGIS project was presented in a White Paper prepared for the Division of Astrophysics of the
American Physical Society \cite{agis_white_paper}.
AGIS proposed a square-kilometre array of mid-sized telescopes, similar to the core array
of mid-sized telescopes in CTA but without the additional large telescopes to cover the very lowest
energies, and an extended array of small telescopes to provide large detection area at the
very highest energies.
The baseline configuration of AGIS consisted of 36 two-mirror
Schwarzschild-Couder telescopes with an 11.5 m diameter primary mirror.
These have a large field of view and a very good angular resolution.
Close contacts were established between AGIS and CTA,
during the design study phase;
information was openly exchanged and common developments undertaken.
After a US review panel recommended that AGIS join forces with CTA,
the US members of the AGIS Collaboration have joined CTA in spring 2010.
Within the overall context of CTA, development of Schwarzschild-Couder
telescopes will be continued to investigate their potential for further
improving CTA performance.
Significant intellectual, technological and financial contributions
to CTA from the US groups are anticipated.
%In the recent US Decadal Survey in Astronomy and Astrophysics,
%which prioritised projects for the decade from 2010 to 2020,
%a substantial US participation in CTA was recommended,
%as one of four top priority projects in the category of large ground-based facilities.
Strong US participation in CTA was endorsed by
PASAG\footnote{Particle Astrophysics Scientific Assessment Group
of the High Energy Physics Advisory Panel}
and the Decadal Survey in Astronomy and Astrophysics (Astro-2010).
 
\item[The High-Altitude Water-Cherenkov Experiment (HAWC):] HAWC \cite{hawc}
builds on the technique developed by the MILAGRO group,
which detects shower particles on the ground using water Cherenkov detectors, and
reconstructs the shower direction using
timing information. It is proposed to construct the new detector on a site at 4100\,m a.s.l.
in the Sierra Negra, Mexico. HAWC will
provide a tenfold increase in sensitivity over MILAGRO and detection capability down to
the lower energy of 100\,GeV,
largely due to
its increased altitude. While it will have lower sensitivity, poorer angular resolution and
a higher energy threshold compared to CTA, HAWC has the advantage of a large
field of view ($\approx 2\pi$ sr) and nearly
100\% duty cycle. HAWC therefore complements imaging Cherenkov instruments.
In fact, it would be desirable to construct and
operate a similar instrument in the southern hemisphere, co-located with CTA.
 
\item[The Large High Altitude Air Shower Observatory (LHAASO):] LHAASO is an extensive (km$^2$) cosmic ray
experiment. The proposal is to locate this near the site of the ARGO and AS-Gamma experiments
in Tibet, at 4300\,m a.s.l.
The array includes large-scale water Cherenkov detectors (~90000\,m$^2$), ground scintillation
counter arrays for detecting both muons and electromagnetic particles,
fluorescence/Cherenkov telescope arrays and a shower core detector array.
The science goals encompass a survey of gamma-ray sources in the energy range $\ge$100 GeV,
measurement of gamma-ray energy spectra of sources above 30\,TeV to identify cosmic ray sources, and
the measurement of cosmic ray spectra and composition at energies above 30\,TeV.
If realised, LHASSO will complement the northern CTA array, as it
concentrates primarily on the detection of low-energy gamma-rays in the energy range from
a few times 10\,GeV to some 100\,GeV.
\end{description}
 
\noindent
In summary, the other large-scale instruments for ground-based gamma-ray astronomy that are being discussed
outside Europe (e.g. HAWC, LHAASO), are complementary to CTA in their capabilities.
 
\subsection{Operation of CTA as an Open Observatory}
 \label{sec:operation}
CTA is to address a wide range of astroparticle physics and
astrophysics questions. The majority of studies will be based on observations
of specific astronomical sources. The scientific programme will hence be steered
by proposals to conduct measurements of specific objects.
CTA will be operated as an open observatory.
Beyond a base programme, which will include for example a survey of the Galaxy and deep observations of
``legacy sources'',
observations will be conducted according to observing
proposals selected for scientific excellence by peer-review among suggestions
received from the community. Following the general procedures developed for and by other major
astrophysical facilities, a substantial number of outstanding proposals from
scientists working in institutions outside the CTA-supporting countries will be
executed. All data obtained by the CTA will be made available in an archive
that is accessible to scientists outside the proposing team.
 
Following the experience of currently operating Cherenkov telescope observatories,
the actual observations will normally be conducted over an extended period in
time, with several different projects being scheduled each night. The operation
of the array will be fairly complex. CTA observations will not, therefore, be
conducted by the scientists whose individual proposals were selected, but by a
dedicated team of operators.
 
CTA observatory operation involves
%the oversight of construction and site development,
%testing and acceptance of telescopes, cameras, and auxiliary equipment during
%the construction phase, maintenance and upgrades of the array following
%completion,
proposal handling and evaluation, managing observation and
data-flow, and maintenance.
The actual work may be conducted in a central location or in
decentralised units (e.g. a data centre and an operations centre) with a coordinating
office.
 
\subsubsection{Observatory Logistics}
\label{sec:logistics}
 
\begin{figure}[htbp]
\centering
  \epsfig{file=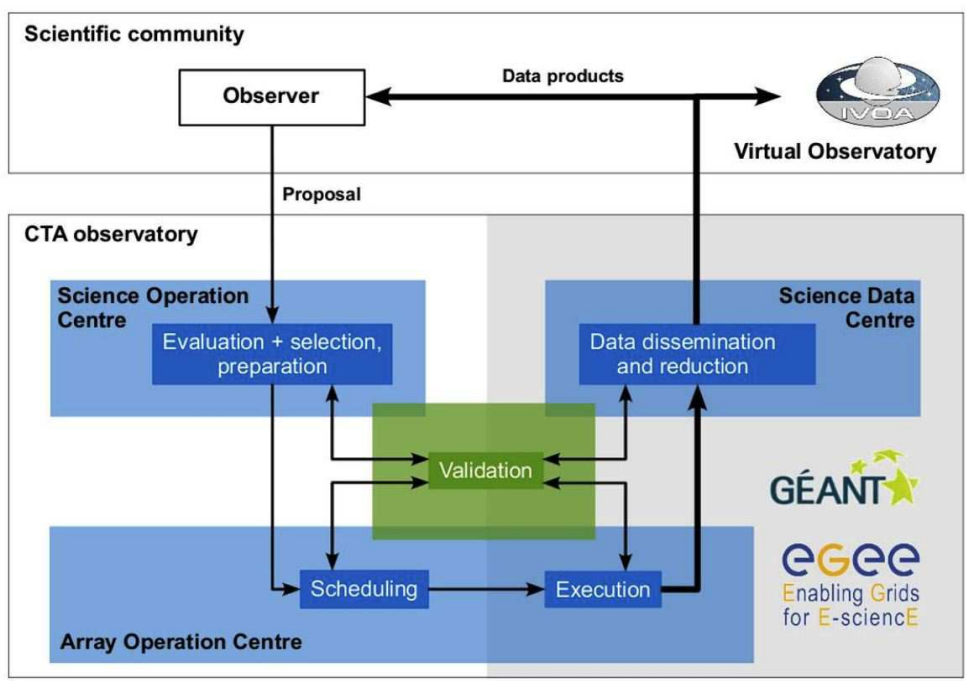,width=0.8\textwidth}
  \caption{\label{fig_dataflow} \small Work flow diagram of the CTA observatory. The three main
 elements which guarantee the functionalities of the observatory are the Science Operation Centre,
 the Array Operation Centre and the Data Centre.
Data handling and dissemination will build on existing infrastructures,
such as EGEE and G\'{E}ANT.}
\end{figure}

The main logistic elements of the CTA observatory are:
the Science Operation Centre (SOC), which is in charge of the organisation of observations;
the Array Operation Centre (AOC), which looks after the operation and monitoring of the telescopes,
and the Science Data Centre (SDC), which provides and disseminates
data and analysis software to the science community at large,
and using the standards of the International Virtual Observatory Alliance
(see fig.~\ref{fig_dataflow}).
 
The use of existing infrastructures, such as EGEE and G\'{E}ANT, and the use of a
Virtual Observatory
is recommended for all data management tasks in the three elements of the CTA observatory.
The high data rate of CTA, together with the large computing power required for
data analysis, demand dedicated resources. Hence, EGEE-Grid infrastructures
and middleware for distributed data storage, analysis and data access
are considered the most efficient solution for CTA.
The CTA observatories will very probably be placed in remote locations
in southern Africa, Latin or Central America, and/or the Canary Islands.
Thus, high-bandwidth networking is critical for remote diagnostics and
instant transfer of the data to well-connected European data centres.
As for other projects in astronomy, a CTA Virtual Organisation, will provide access to the data.
CTA aims to support a wide scientific community, providing access
to all levels of data that is archived in a standardised way.
 
It is envisaged to start CTA operations already during the construction phase as soon
as the first telescopes are ready to conduct competitive science operations.
 
\subsubsection{Proposal Handling}
\label{sec:proposals}
 
The world-wide community of scientists actively exploiting the results from
ground-based VHE gamma-ray experiments currently consists of about
600 physicists
(about 150 in each of the H.E.S.S. and MAGIC Collaborations,
about 100 in VERITAS,
50 in Cangaroo and
50 in Indian gamma ray activities,
plus about 100 scientists either associated, or regularly collaborating, with
these experiments).
Planning and designing CTA involves about
another 100 scientists not currently participating in either of the currently
running experiments. Proposals for observations with CTA are hence expected to
serve a community of at least 700 scientists, larger than that of any
national astronomical facility in Europe, and comparable to the size of the
community using the ESO observatory in the 1980s. CTA must therefore
efficiently deal with a
large number of proposals for a facility which, based on experience with current experiments,
is expected to be
oversubscribed by a large factor.
CTA plans to follow the practice of other major, successful
observatories (e.g. ESO), and announce calls for proposals at regular
intervals. These proposals will be peer-reviewed by a group of international
experts which will change on a regular basis.
Different classes of proposals (targeted, surveys,
time-critical, target of opportunity, and regular programmes) are foreseen,
as is common for current experiments and other ground-based
observatories.
Depending on the science under investigation, subarray operation may be required.
Each site may therefore be conducting several different observation programmes
concurrently.
 
\subsubsection{Observatory Operations}
\label{sec:obsop}
 
The observing programme of the CTA  will be driven by the best
proposals from the scientific community, which will be selected
in a peer-review process. Successful applicants will provide all the information
required for the optimum completion of their measurements. An observing
programme will be compiled by the operations centre, taking the requirements
of individual projects into account. The programme will be
conducted in robotic fashion with a minimum amount of professional staff
on site. Proposers are not expected to participate in measurements.
Quicklook analysis will enable triggers and on-the-fly modification of
projects, if required. Data and calibration files will be provided to
the user.
Frequent modifications to the scheduled observing programme can be
expected for several reasons.
Openness of triggers is essential given the transitory and variable nature
of many of the phenomena to be studied by CTA.
CTA must adapt its schedule to changing atmospheric conditions to ensure
the science programme is optimised.
The flexibility to pursue several potentially very different programmes at the
same time may increase the productivity of the CTA observatory.
Routine calibrations and monitoring of the array and of environmental data
must be scheduled as needed to ensure
the required data quality.
 
Observatory operations covers day-to-day use of the arrays, including
measurements and continuous hardware and software maintenance, proposal
handling and evaluation, automated analysis and user support, as well as
the long-term programme for upgrades and improvements to ensure continued
competitiveness over the lifetime of the observatory.
 
\subsubsection{Data Dissemination}
 \label{sec:datadiss}
 
The measurements made with CTA will be subject to on-line analysis, including
event-selection and calibration for instrumental effects. The analysis of
data obtained with Cherenkov telescopes differs from the procedures typical in
other wavelength ranges in that extended Monte-Carlo simulations are used
to determine the effects of, and calibrate for, the influence of a large range of factors
on the measurements. The necessary simulations
will be carried out by CTA, used in calibrating standard pipline-processed
data and will also be made
available to the community for use in proposal planning etc.  The principal investigators
of accepted proposals will be provided with the results of standard
processing and access to the standard MC simulations and the analysis pipelines
used in data processing. Storage of data and
archiving of scientific and calibration data, programs, and MC simulations used
in the processing will be organised through the distributed computing resources made available
in support of the CTA EGEE Virtual Organisation.
\begin{figure}[htbp]
\centering
  \epsfig{width=0.7\textwidth,file=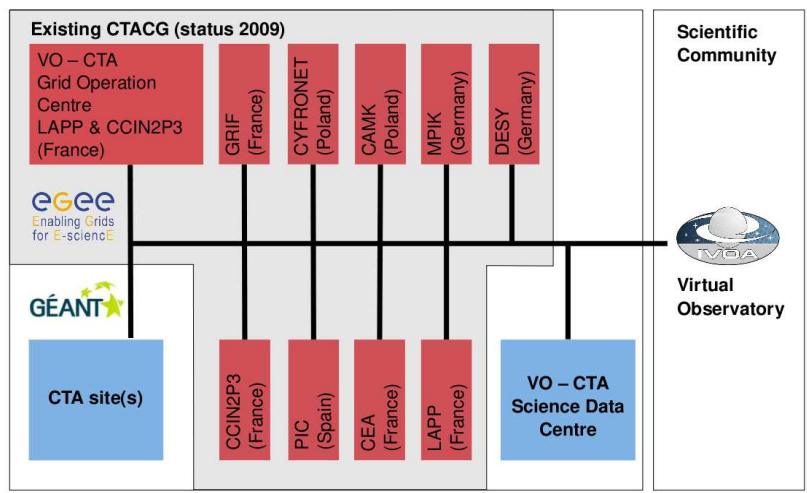}
  \caption{\label{fig_GRIDCTAVO} \small Schematic of the integrated application of
  e-infrastructures like EGEE-GRID, G\'{E}ANT and VO for the CTA
  observatory, together with the 2009 status of the CTACG (CTA Computing Grid)
  project~\cite{lappwiki}.
  The VO-CTA Grid Operation Centre houses the EGEE services.}
\end{figure}
 
The processing of CTA data represents a major computational challenge.
It will be necessary to
reduce a volume of typically 10 TBytes of raw data per observation to a
few tens of MBytes of high-level data within a couple of hours.
This first-level data processing will make heavily use of Grid technology
by running hundreds of processes within a global pipeline.
Data processing requires also the production and analysis of the MC simulations needed for calibration.
The integrated services and infrastructures dedicated to the MC production,
analysis and dissemination have to be taken into account in the CTA data pipeline.
 
All levels of data will be archived in a standardised way, to allow access and
re-processing by the scientific community.
Access to all levels of data and Grid infrastructures will be provided through a single access point,
the  ``VHE gamma-ray Science Gateway''.
 
Fig.~\ref{fig_GRIDCTAVO} shows an overview of the integrated application
e-infrastructures such as EGEE-Grid, G\'{E}ANT and CTA VO.
 
It is foreseen that the high level analysis of CTA data can be conducted by
individual scientists using the analysis software made available by CTA. This software
will follow the standards used by other high-energy observatories and will be provided
free of charge to the scientific community.
 
\subsection{CTA Organisation}
 \label{sec:org}
 
The organisation of the CTA consortium will evolve over the various stages of the project.
These include:
\begin{description}
\item[The design study phase.]
Definition of the layout of the arrays,
specification of the telescope types,
design of the telescopes and small-scale prototyping.
\item[The prototyping and preparatory phase.]
Prototyping and deployment of full-scale telescopes,
preparation of the construction and installation including
solving technical, organisational and legal issues,
site preparation.
\item[Construction phase.]
Construction, deployment and commissioning  of the telescopes.
\item[Operation Phase.]
Operation as an open observatory, with calls for proposals and scheduling,
operation and maintenance of the facility,
processing of the data and provision of analysis tools.
\end{description}
For the design study phase, the organisation of the consortium was defined
in a {\em Memorandum of Understanding} modelled on those proven by
large experiments
in particle and astroparticle physics.
The governing body is the Consortium Board and operational decisions are taken and work is coordinated
by the Spokespersons and the Executive Board.
Work Package Convenors organise and drive the work on essential parts of the project.
The work packages and the area they cover are:
\begin{description}
\item[PHYS] The astrophysics and astroparticle physics that will be studied using CTA.
\item[MC]   Development of simulations for optimisation of the array layout and analysis algorithms,
and for performance studies.
\item[SITE] Evaluation of possible sites for CTA and infrastructure requirements.
\item[MIR]  Design of telescope optics and mirror construction.
\item[TEL]  Design of telescope structure and associated drive and control systems.
\item[FPI]  Development of focal plane instrumentation.
\item[ELEC] Design and development of the readout electronics and trigger.
\item[ATAC] Development of atmospheric monitoring and calibration techniques and associated instrumentation.
\item[OBS]  Development of observatory operation and access strategies.
\item[DATA] Studies of data handling, processing, management and data access.
\item[QA]   Quality assurance and risk assessment strategies.
\end{description}
The CTA design study phase
was organised in terms of
scientific/technical topics, rather than in terms of telescope types, to ensure that,
as far as possible, common technical solutions are employed across the array,
maximising economies of scale and simplifying array operation.
 
For the preparatory phase, the organisation will be adapted to the needs of the project.
The Project Office will be extended, and work packages for each telescope type will be established
to steer prototyping and preparations for construction.
External advisors will assist in guiding and reviewing the project.
 
A significant task for the preparatory phase will be the definition of the legal framework
and governance structure of the
CTA Collaboration and observatory.
Different models exist, each of which has its own advantages and disadvantages.
CTA could for example be realised within an existing international organisation such as CERN or ESO.
CTA could also be operated by a large national
laboratory which has sufficient administrative and technical infrastructure. Suitable national
laboratories exist e.g. in Germany, France, or the UK, for example.
On a smaller scale, H.E.S.S. and MAGIC are operated in this mode.
CTA could be established as an independent legal entity under the
national law of some country, following the example of IRAM.
The definition of the legal structure of CTA will
be determined in close interaction with ASPERA
(a group of European Research Area funding agencies which coordinates astroparticle physics in Europe).
One of their main tasks is the
``Implementation of new European-wide procedures for large infrastructures''.
 
Regardless of the legal implementation, CTA management will be assisted by an
international scientific and technical Advisory Board,
and a Resource Board, composed of representatives of
the national funding organisations supporting CTA.
 
Close contacts between CTA and the funding agencies (via the Resource Board)
during all stages of the project are vital to
secure sufficient and timely funding for the construction of the facility.
 
\subsection{Time Schedule and Costs}
 \label{sec:sched}
 
CTA builds largely on proven technologies and Cherenkov telescopes of sizes similar to those needed for
CTA have already been built or are in the advanced stages of construction. Remaining challenges
are:
(a) optimisation of the cost of telescope components;
(b) improvement of the reliability of telescope components, requiring extensive prototyping;
(c) establishment of the formal framework for building and operating the instrument,
and the selection and provision of sites; and
(d) the funding of the infrastructure.
 
These challenges will be addressed during the Preparatory Phase (2010-2013) which will be
supported by an FP7 grant of up to 5.2 M\euro{} from the European Community
and by grants from various national funding agencies.
 
After a successful Preparatory Phase, and provided the funding has been secured,
construction and deployment will then take from 2013 until 2018.
 
A detailed evaluation of the required construction and running costs
is part of the Preparatory Phase studies.
%A reliable estimate of CTA construction costs will only be
%possible once the telescope design is in a more advanced stage
%and the choice of the sites has been made.
Current design efforts are conducted within an envelope of investment costs
for the CTA construction and site infrastructure  of
100 M\euro{} for the southern site, featuring full energy coverage, and
50 M\euro{} for the more specialised northern site  (all in 2005 \euro{}).
CTA aims to keep running costs below 10\% of the total investment,
in line with typical running costs for other astrophysical facilities.
 
%(with a fraction of 7.2 \%  of total investment costs).
%The running costs will be covered by the member countries and organizations.
%========

Estimates for the costs of all major components of CTA are required for
any optimisation of the array design.
The current model makes the following assumptions:
\begin{itemize}
\item The investment required to construct CTA
(according to European accounting schemes)
is 100 M\euro{} for CTA-South and 50 M\euro{} for CTA-North.
\item For both sites 20\% of the budget is required for infrastructure and a
central processing farm. Therefore, for example,
telescope construction for CTA-South is anticipated to cost
80 M\euro{}.
\item The construction of the telescope foundation, optical support structure, drive/safety system and
camera masts will cost 450 k\euro{} for a 12 m telescope and the cost scales as dish area$^{1.35}$.
\item Mirrors, mounts and actuators will cost $\approx$ 1.7 k\euro{}/m$^2$.
\item Camera mechanics, photo-sensor and electronics costs will be 400 \euro{}/pixel,
including lightcones, support structures and cooling systems.
\item  Miscellaneous additional costs of about 20 k\euro{}/telescope will be incurred.
\end{itemize}
This cost model will evolve as the design work on the different components of CTA progresses.

\clearpage
 
%% 7
%\input{ProofMC.tex}
%%%%% HEADER: ProofMC.tex
%%%%%%%%%%%%%%%%%%%%%%%%%%%%%%%%%%%%%%%%%%%%%%%%%%%%%%%%%%%%%%%%%%%
 
\section{Monte Carlo Simulations and Layout Studies}
\label{sec:MC}
 
The performance of an array of imaging atmospheric Cherenkov
telescopes such as CTA depends on a large number of technical and
design parameters. These include the general layout of the
installation, with telescope sizes and locations,
%but also many other aspects such as
telescope optics, camera field-of-view and pixel size,
signal shapes and trigger logic. In searching for the optimum
configuration of a Cherenkov telescope array, one finds that most of these parameters
are intimately related, either technically or through constraints on the
total cost. For many of these parameters there is experience from
previous gamma-ray installations such as HEGRA, CAT, H.E.S.S., and MAGIC that
provide reasonable starting points for the optimisation of CTA
parameters. Whilst the full optimisation of CTA has not yet been
completed, extensive simulation studies have been performed and
demonstrate that an array of $\ge$60 Cherenkov telescopes can achieve
the key performance targets for CTA, within the cost envelope
described earlier. This section gives a summary of the most
important simulation studies performed so far.
 
\subsection{Simulation Tools}
 \label{sec:tools}
 
Only a modest number of candidate configurations has been simulated in
full detail during the design study, but this still required the simulation
of close to 10$^{11}$ proton, gamma, and electron induced showers,
with full treatment of every interaction, tracking all the particles generated in these
showers through the atmosphere, simulating emission of Cherenkov light,
propagating
the light down to the telescopes, reflecting it on multi-faceted
mirrors, entering photomultiplier tubes, generating pulses in
complex trigger electronics, and having them registered in
analogue-to-digital circuits.  Simulations include not only Cherenkov
photons but also NSB light resulting in the registration of photons at
rates of $\sim$100~MHz in a typical photo-sensor.
%%% Too technical? , also generating after-pulses with a substantial rate.
 
Since the discrimination between $\gamma$-ray and hadron showers in
CTA will surpass that of the best current instruments by a significant
factor, huge numbers of background showers must be simulated before
conclusions on the performance of a particular configuration can be
drawn. Work is underway to reduce the CPU-time requirement by
preferentially selecting proton showers early in their development if
they are more likely to appear {\it $\gamma$-like}. This should lead to a
substantial speed improvement in future studies. Early results from
{\it Toy models}, which parametrize shower detection characteristics and
are many orders of magnitude faster, are encouraging, but
cannot yet be seen as adequate replacements for the detailed simulation process.
 
The air-shower simulation results presented here are based on the
CORSIKA  program \cite{CORSIKA}, which is widely used in the
community and very well tested. Cross-checks with the KASCADE-C$^{++}$
air-shower code \cite{KASCADE} have been performed as part of this
study. Simulations of the instrument response have been carried out
with three codes. Two packages initially developed for H.E.S.S.
(sim\_telarray \cite{sim_telarray} and SMASH \cite{SMASH}),
and one for MAGIC simulations \cite{magicMC},
were cross-checked using an initial {\it benchmark} arrays configurations.
 
The large volume of simulations, dominated by those of proton-induced
showers needed for background estimations, has motivated the use of EGEE
(Enabling Grids for E-sciencE) for the massive production of shower
and detector simulations. A Virtual
Organisation has been founded and a first set of CORSIKA
showers has been generated on the GRID, while a specific interface for
job submission and follow-up for simulations and analysis is currently
under development.
 
The detailed simulations described here, result in data equivalent to
experimental raw data (ADC counts for each time-slice for each
pixel). Analysis tools are needed to reconstruct shower parameters (in
particular energy and direction) and to identify $\gamma$-ray
showers against the background from hadron-initiated showers
(note that the additional background from electron-induced showers is
important at intermediate energies despite the much lower electron flux as
electron showers are extremely difficult to differentiate from those initiated by
photons). The analysis methods currently used are based on experience with past
and current instruments, but are being developed to make full use of
the information available for CTA, in particular to exploit the large
number of shower images that CTA will provide for individual events.
 
The analyses in this study are based on several independent codes, all
of which start with cleaning of images to identify signal pixels, and a
parametrisation of images by second-moment Hillas parameters
\cite{HillasParam}, augmented by parameters such as the height of
shower maximum as reconstructed from stereo images. Background
rejection is achieved both by direct cuts on (suitably normalised)
image parameters, and more general multivariate analysis tools such as
a {\it Random Forest} \cite{RandomForest} classifier and {\it
Boosted Decision Trees} within the open source software package TMVA
\cite{TMVA, OhmTMVA, TMVA_LAPP}.
%TMVA framework (see e.g. \cite{OhmTMVA}).
There are also other analysis methods in use
for the analysis of Cherenkov telescope data, such as the {\it 3-D-model} analysis
\cite{3D-model} the {\it Model++} analysis
\cite{Model-analysis}, and analytical combinations of probability density functions
of discriminating variables
%(e.g. Xeff \cite{Xeff} {\red leave out or explain})
which have advantages over the standard second-moments analysis in at
least some energy ranges. Some of these alternative
methods have been used for a subset of the studies presented
here.
%%Could also mention Xeff
 
\subsection{Verification of Simulation Tools}
\label{sec:verif}
 
The optimisation of CTA relies heavily on detailed simulations to
predict signal and background rates, angular resolution and overall
sensitivity. To demonstrate that the simulation tools in use accurately describe
reality, we show here some key data/simulation comparisons, taking H.E.S.S. as an example.
 
% Verify - 1) Optics
%          2) Gammas - Clementina's figure
%          3) BG - model uncertainty
%          4) BG - absolute rate  - trigger
%          5) Background - electrons plot
 
\begin{figure}[htbp]
\centering
\epsfig{width=0.6\textwidth,file=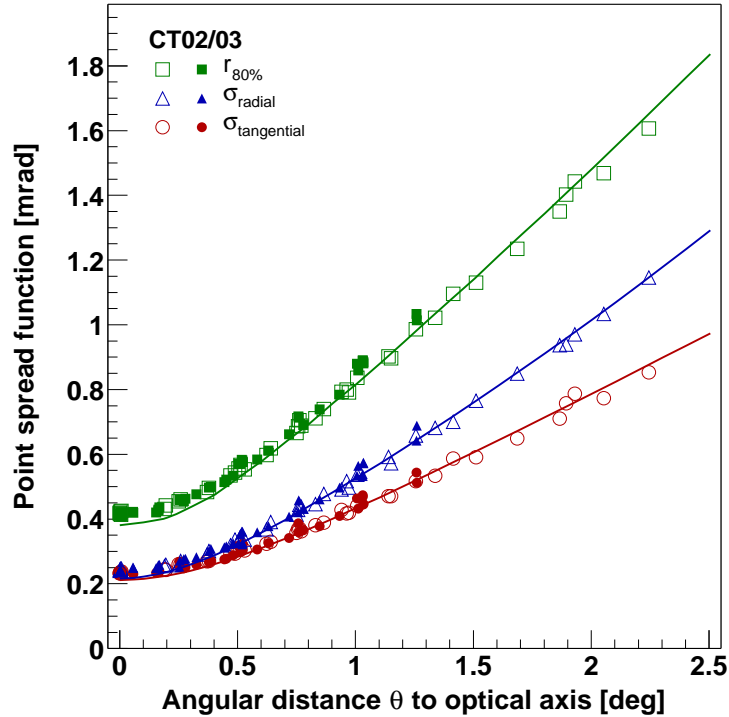}
\caption{\small  Optical point spread function of two
H.E.S.S. telescopes as a function of angle of incidence,
measured using stars, and compared to simulations. Data points are
shown for the radial and tangential width of the PSF, and the 80\%
containment radius. Lines represent the results of simulations of the
telescope optics using sim\_telarray. See \cite{hess_optics2} for details.}
\label{fig_psf}
\end{figure}

A key aspect of the simulation of the detector response to Cherenkov
light from an air-shower is the ray-tracing of light through the
optical system of an individual telescope. An understanding of the
typical misalignments of all components is needed at this stage, as
is the ideal performance. The optical performance of a telescope
is described by its point spread function (PSF), which degrades for off-axis rays. Fig.~\ref{fig_psf}
illustrates that the modelling of the optical system of, in this case,
a H.E.S.S. telescope reproduces the width and shape of the PSF in all
details, and that essentially identical imaging is achieved for
different telescopes in the system.
%for all four telescopes.
 
\begin{figure}[htbp]
\centering
\epsfig{file=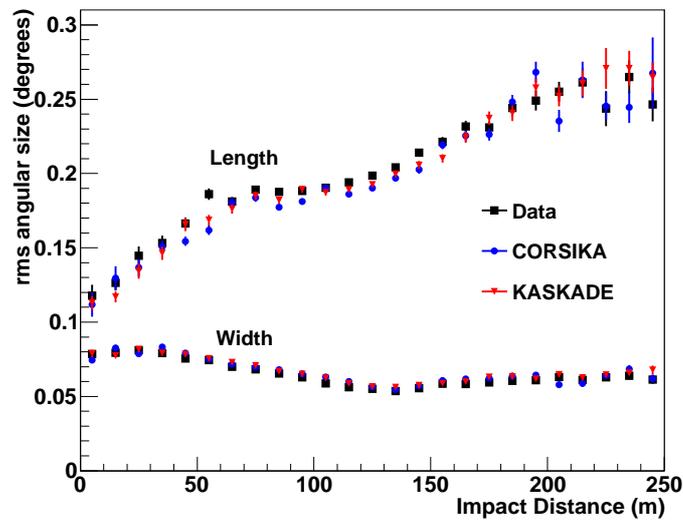, width=0.6\textwidth}
\caption{\label{fig_hillas_comp} \small
Comparison of measured (black squares) and simulated (red triangles and blue circles)
image parameters for the H.E.S.S. telescopes.
Measured data are taken from a flare of the blazar PKS~2155-304 \cite{hess_pks2155} for
which the signal/noise ratio was very high and large gamma-ray statistics are available.}
\end{figure}

An end-to-end test of the correct simulation of gamma-ray induced showers can be made
using the signal from a strong source under very high signal/background conditions.
The giant flare from the blazar
PKS~2155-304 observed with H.E.S.S. in 2006 provides an excellent opportunity for such a test.
Fig.~\ref{fig_hillas_comp} shows the satisfactory agreement (typically at the 5\% level)
between the simulated and detected shape of the shower image  as characterised by their Hillas width and
length parameters. Gamma-ray showers were simulated with the CORSIKA and KASKADE-C$^{++}$
programs and have been
passed through one of the H.E.S.S. detector simulation and analysis chains. The measured
spectrum, optical efficiency, zenith angle and other runtime parameters were used as inputs
to this simulation.
 
\begin{figure}[htbp]
  \centering
\epsfig{file=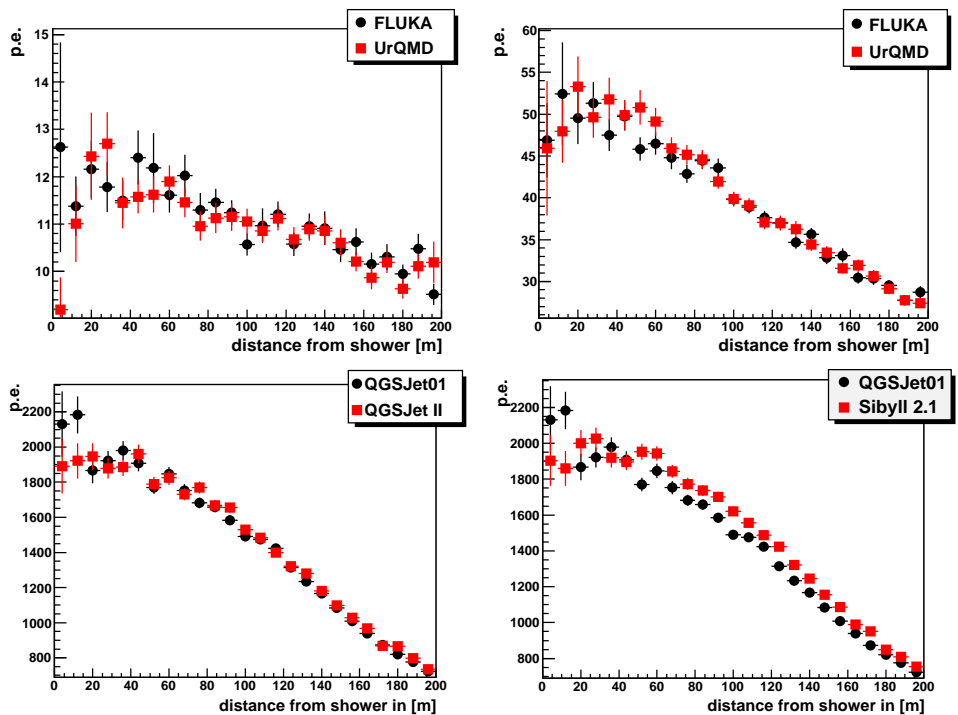,width=0.8\textwidth}
\caption{\label{fig_had_models} \small Comparison of the Cherenkov light profiles for proton-induced
showers generated with different hadronic interaction models.
The profiles for FLUKA and UrQMD at 50 GeV (left) and 100 GeV (right) are shown in the top panel.
Two QGSJet versions and SIBYLL at 1~TeV are compared in the bottom panels.}
\end{figure}
\begin{figure}[htbp]
  \centering
\epsfig{file=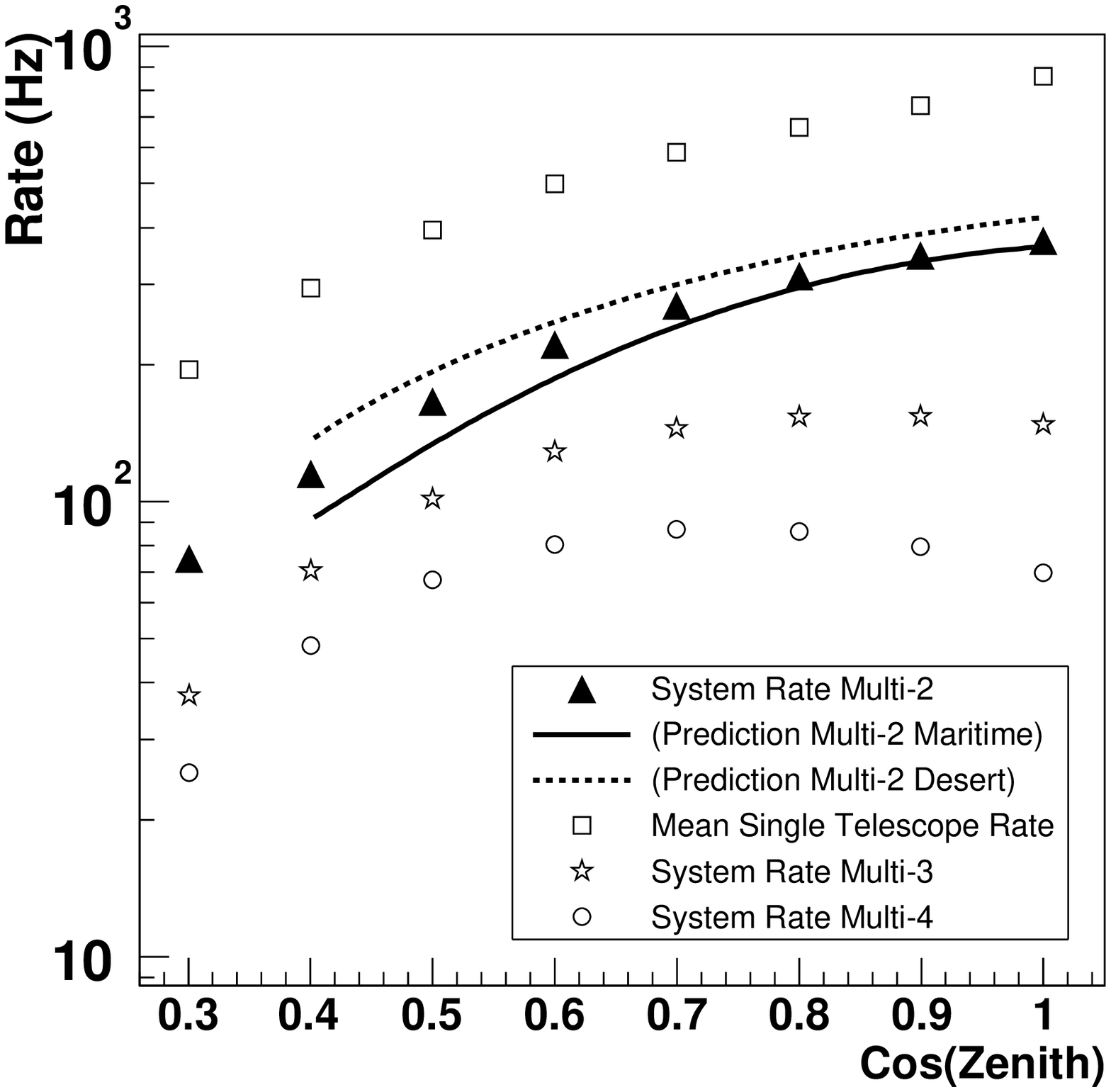,width=0.6\textwidth}
  \caption{\label{fig_rate_vs_zenith} \small Dependence of H.E.S.S. system
  trigger rate on zenith angle, for data and simulations. The
  simulations assume two different model atmospheres, with the
  atmosphere at the H.E.S.S. site representing an intermediate case. See \cite{hess_trigger} for more details.}
\end{figure}
 
In the analysis of experimental data, it is sufficient for simulations to describe the characteristics of
gamma-ray detection, since the cosmic-ray background can (except for very diffuse sources) be modelled
and subtracted using
measurements in regions without gamma-ray emission. However, for the design of new instruments,
simulations must also provide a
reliable modelling of all relevant backgrounds. Experience with existing systems shows that this is
indeed possible, provided that background events are simulated over a very wide area, up to an impact
distance of around
a kilometre from any telescope and over a large solid angle, well beyond the direct field of view
of the instrument, so that far off-axis shower particles are properly included.
%\clearpage
 
An inherent uncertainty in the simulation of the hadronic background is given by
the currently limited knowledge of
hadronic interaction processes at very high energies. The impact of this uncertainty on the Cherenkov
light profile has been
studied using CORSIKA simulations with different interaction models. As can be seen in
fig.~\ref{fig_had_models}, the
low energy ($<$80 GeV) models FLUKA~\cite{fluka} and UrQMD~\cite{urqmd} do not exhibit
significant differences, whereas
the known discrepancy between the high-energy models QGSJet-01~\cite{qgsjet01},
QGSJet-II~\cite{qgsjet2a, qgsjet2b} and SIBYLL~2.1~\cite{sibyll} leads to an uncertainty of about 5\%
in the Cherenkov light profile at 1 TeV.

\begin{figure}[htbp]
\centering
 \epsfig{file=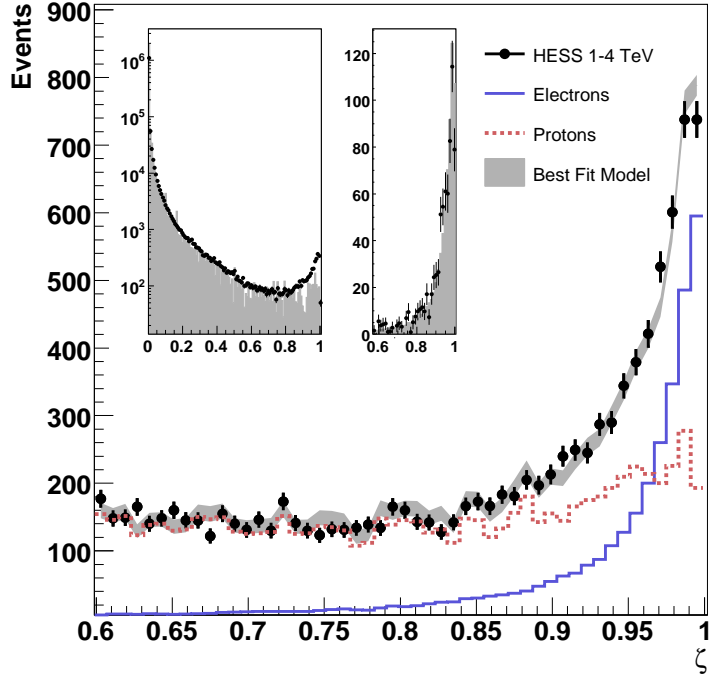,width=0.6\textwidth}
  \caption{\label{fig_electrons_zeta} \small Measured distribution of the proton/electron separation
  parameter $\zeta$ for 239 hours of H.E.S.S. data on sky fields without gamma emission,
  compared to simulations of proton- and electron-induced showers. The shape of the background
  is very well reproduced by simulations across the full range of $\zeta$.
Gamma-ray signals appear close to $\zeta=1$. The electron background is therefore important
despite the relatively low flux of electrons in comparison
to hadrons. See \cite{hess_electrons} for more details.}
\end{figure}

As can be seen in fig.~\ref{fig_rate_vs_zenith}, the raw cosmic-ray detection rate as a function of
zenith angle is described to within about 20\%. Given the uncertainties on cosmic-ray flux, composition above
the atmosphere and in the hadronic interaction models, better agreement cannot be expected.
In the background-limited regime this
uncertainty corresponds to a 10\% uncertainty in sensitivity, assuming that the fraction of
{\it $\gamma$-like} events is understood.
Fig.~\ref{fig_electrons_zeta} demonstrates that the fraction of such events, and the distributions of
separation parameters, are indeed well understood for instruments such as H.E.S.S. using the
simulation and analysis tools applied here to CTA.

\subsection{Energy Range and Sensitivity of Telescope Arrays}
 \label{sec:sens}
 
Three methods of representing the sensitivity of a Cherenkov telescope
are used in the following discussion. All three have merits and
emphasise different features. The traditional way to represent the
sensitivity of Cherenkov Telescope systems is in terms of integral sensitivity,
including all events reconstructed above a given energy (and often
multiplied by the threshold energy to flatten the curves and give more
useful units of erg/(cm$^2$s).  An observation time of 50
hours (typical for the first generation of IACTs) is assumed
for comparison to published sensitivity curves of historical and
current instruments. Integral sensitivities depend on the assumed
source spectrum and can be deceptive in that much of the detection
power quoted for a given threshold may actually be derived from events
well above that threshold. A more useful, but less common, way to
represent the sensitivity of IACTs
%possible CTA configurations
is in terms
of differential sensitivity, where a significant detection (above
5\% of the background level, with $\ge 5 \sigma$ statistical
significance and at least 10 events) is required in each energy
bin. Five bins per decade in energy are used for the following
results for possible CTA configurations.
The differential flux sensitivity is sometimes multiplied by
$E^{2}$ to show the minimum source flux in terms of power per
logarithmic frequency interval and given in units of erg
cm$^{-2}$s$^{-1}$ for ease of comparison with other wavebands.
Alternatively, the Crab nebula, as a strong and non-variable gamma-ray
source with a rather typical spectral shape, can be used as a
reference. Here we use the VHE spectrum as measured with the HEGRA
telescope array as a reference, i.e.  1 Crab Unit (CU) =
$2.79\,\times\,10^{-11} E^{-2.57}$ cm$^{-2}$ s$^{-1}$ TeV$^{-1}$.
(Note that the true spectrum
of the Crab nebula falls below this expression at the highest and lowest energies.)
 
Several different telescope configurations have been investigated in
simulation studies for CTA so far.  The first simulations were used
to cross-check the different simulation packages and to begin the investigation
of the dependence of performance on telescope and array parameters.
Selected results from one of these, an array of 9 telescopes
with 24\,m diameter (the ``benchmark'' array),
are discussed below. Following these studies
%to establish a reasonable altitude and pixel size for the CTA telescopes,
a series of simulations were conducted
with larger telescope arrays (including $41\times$ 12~m telescopes and a
97-telescope array with two different telescope sizes) to
demonstrate that the goals of CTA are attainable with a large
telescope array (see \cite{icrc_sims}). More recently, a 275 telescope ``production
configuration'' has been simulated, subsets of which constitute
CTA candidate configurations. So far 11 candidate configurations have
been defined with an approximately equal construction cost of about
80 M\euro{} (in 2005 \euro{}) with the current CTA cost model.
 
\begin{figure}[htbp]
  \centering
\epsfig{file=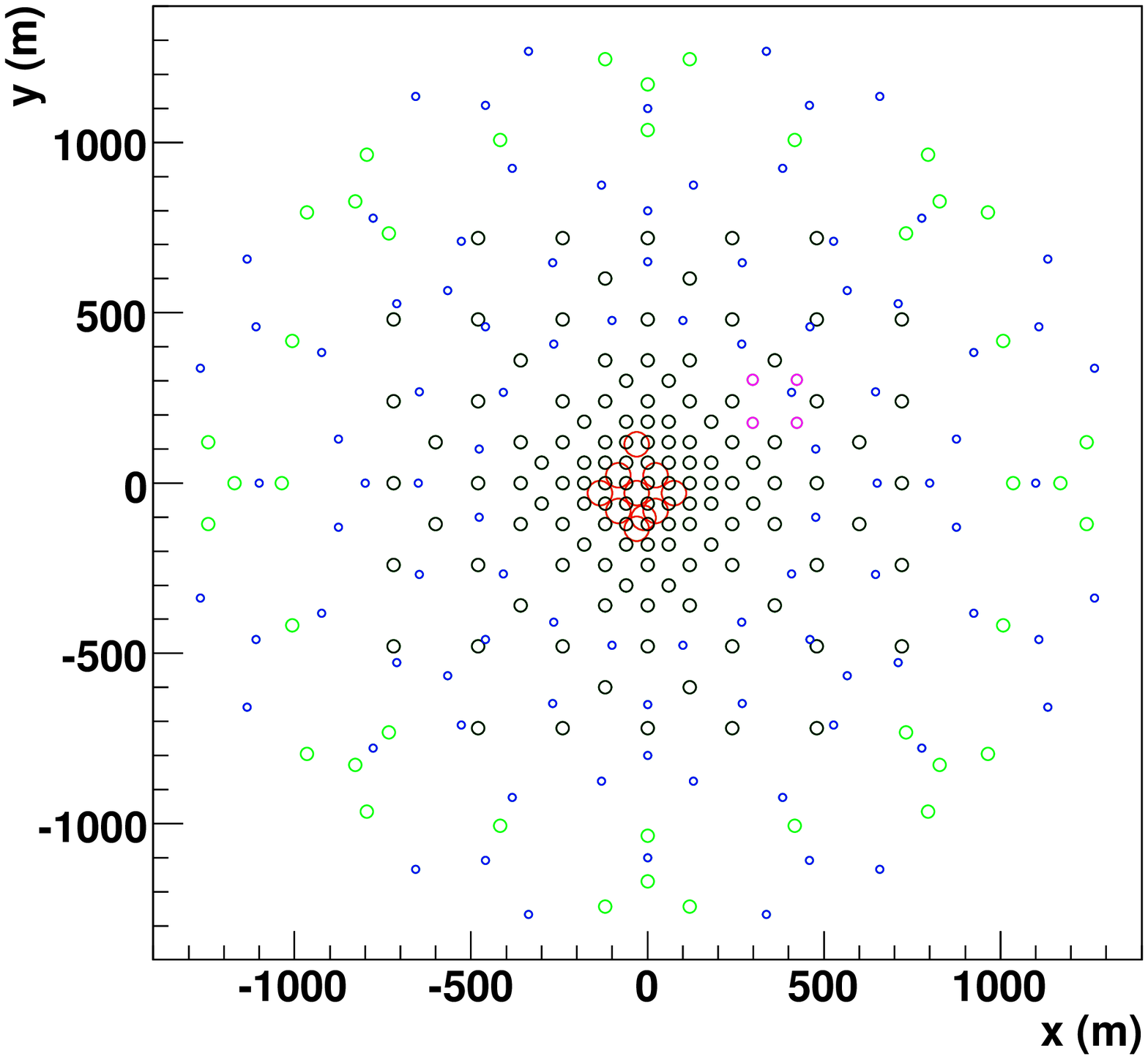,width=0.8\textwidth}
\epsfig{file=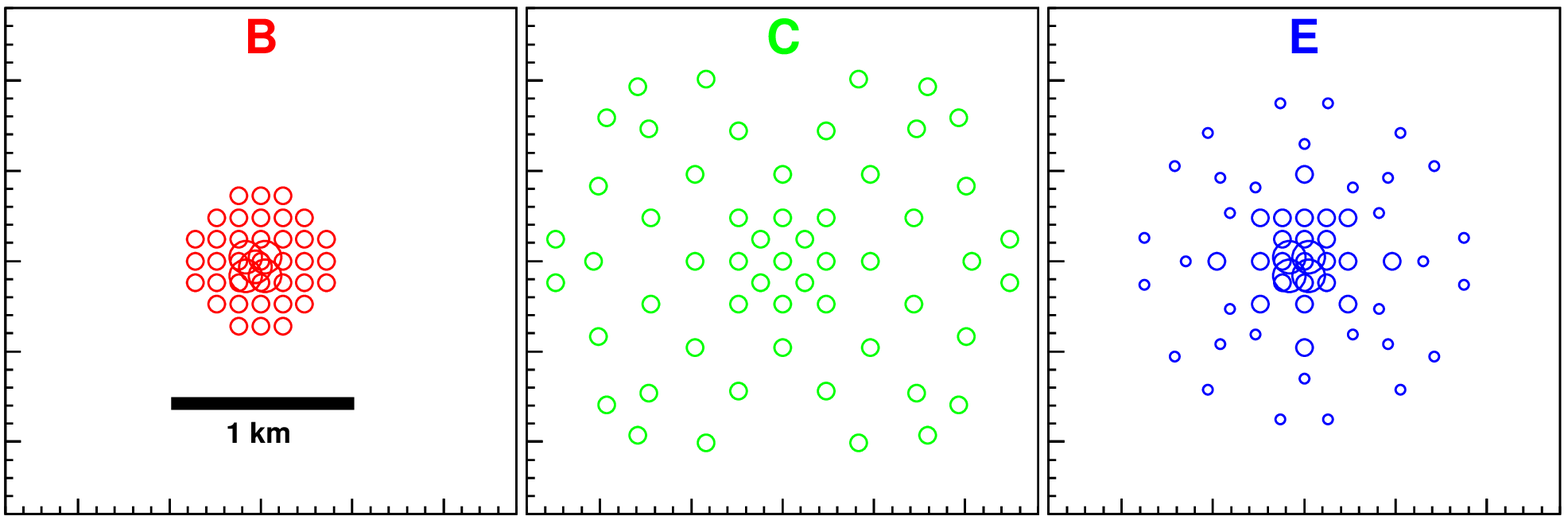,width=\textwidth}
  \caption{\label{fig-configurations} \small {\bf Top:} 275 telescope
    super-configuration for the MC mass production.  5 telescope types
    are simulated (red: 24~m diameter telescopes, black and green: 12~m, pink: 10~m, blue: 7~m),
    with the circle size
    proportional to the mirror area.  {\bf Bottom:} Three example candidate
    configurations (B, C \& E) which are subsets of the 275 telescope
    array and would all have an approximate construction cost of
    80 M\euro{}.}
\end{figure}

The evaluation of the
performance of these candidate arrays is a first step towards the
optimisation of the CTA design. Fig.~\ref{fig-configurations} shows
some of the telescope layouts used. All systems assume conventional technology
for mirrors, PMTs and read-out electronics. Standard analysis techniques are used in general, with
the results from more sophisticated methods shown for comparison in specific cases.
 
\begin{figure}[tbp]
  \centering
  \epsfig{file=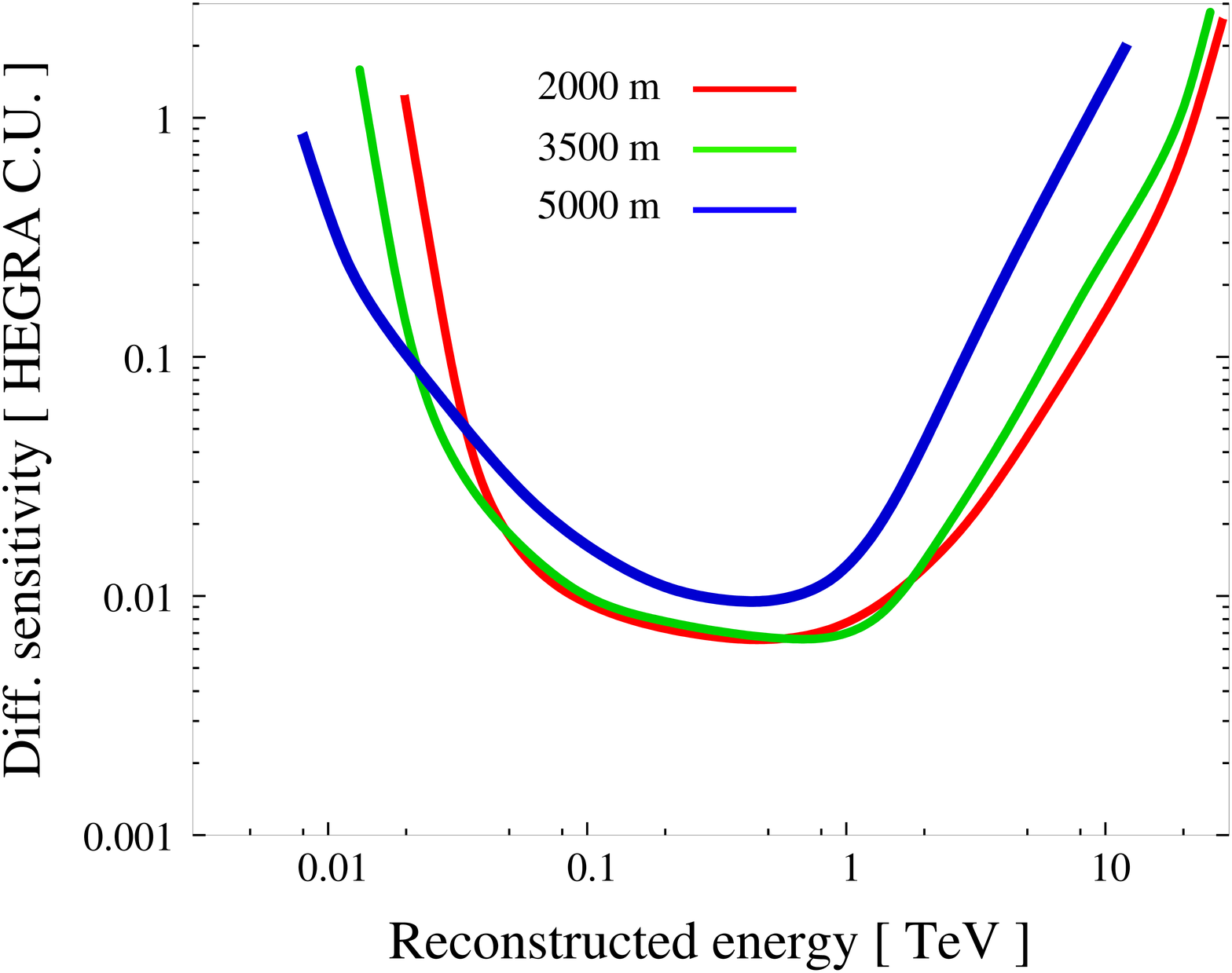,width=0.6\textwidth}
  \caption{\label{fig_9_2000_5000} \small
Differential sensitivity (with 5 independent bins per decade in energy) of the
9-telescope benchmark array placed at 2000\,m, 3500\,m and at 5000\,m elevation,
for point sources observed for 50 hours at a zenith angle of
20$^\circ$. A 5$\sigma$ significance, at least 10 signal events,
and a signal exceeding 5\% of the remaining background is required
for a detection. The image cleaning method applied uses dual threshold
5/10 photoelectron.}
\end{figure}

The 9-telescope benchmark array has been used to test several
aspects of array performance, in particular the desirable altitude
range and best pixel size for the lower part of the CTA energy range.
Fig.~\ref{fig_9_2000_5000} compares arrays
located at different elevations (2000, 3500 and 5000\,m) and also
illustrates the influence of systematic errors in the background
determination at low energies. The spacing of telescopes is adjusted to compensate for the
changing radius of the Cherenkov light-pool with altitude. For 2000\,m elevation, the array has
useful sensitivity above $\approx$20~GeV and at higher energies dips
below the 1\% Crab level.
An equivalent system at high elevation (5000\,m) provides a lower
threshold but worse performance at high energies, at least partly
reflecting the smaller diameter of the light pool at high altitude and
hence the reduced detection area. Another potential problem at very high altitudes is the contamination
of the signal by Cherenkov light from individual shower particles which reach the observation level.
Sensitivities cross at about
30\,GeV, implying that a high-altitude installation is mainly relevant
for specialised very-low-energy instruments, such as the 5@5 array
\cite{5at5}.  Similar conclusions were reached in earlier simulations
by Plyasheshnikov \cite{Plya_alt} and Konopelko \cite{Konopelko_alt}.
A 3500~m altitude array delivers a somewhat lower energy threshold than one at 2000~m and
comparable performance at 0.1-1 TeV for the benchmark array.
However, it is not clear that this result on relative performance at intermediate energies
can be generalised to the much larger telescope array of smaller telescopes with which CTA plans to
cover this energy range.
Simulations of the 275-telescope array at 3700~m altitude are underway to address this question.
\begin{figure}[bp]
  \centering
  \epsfig{file=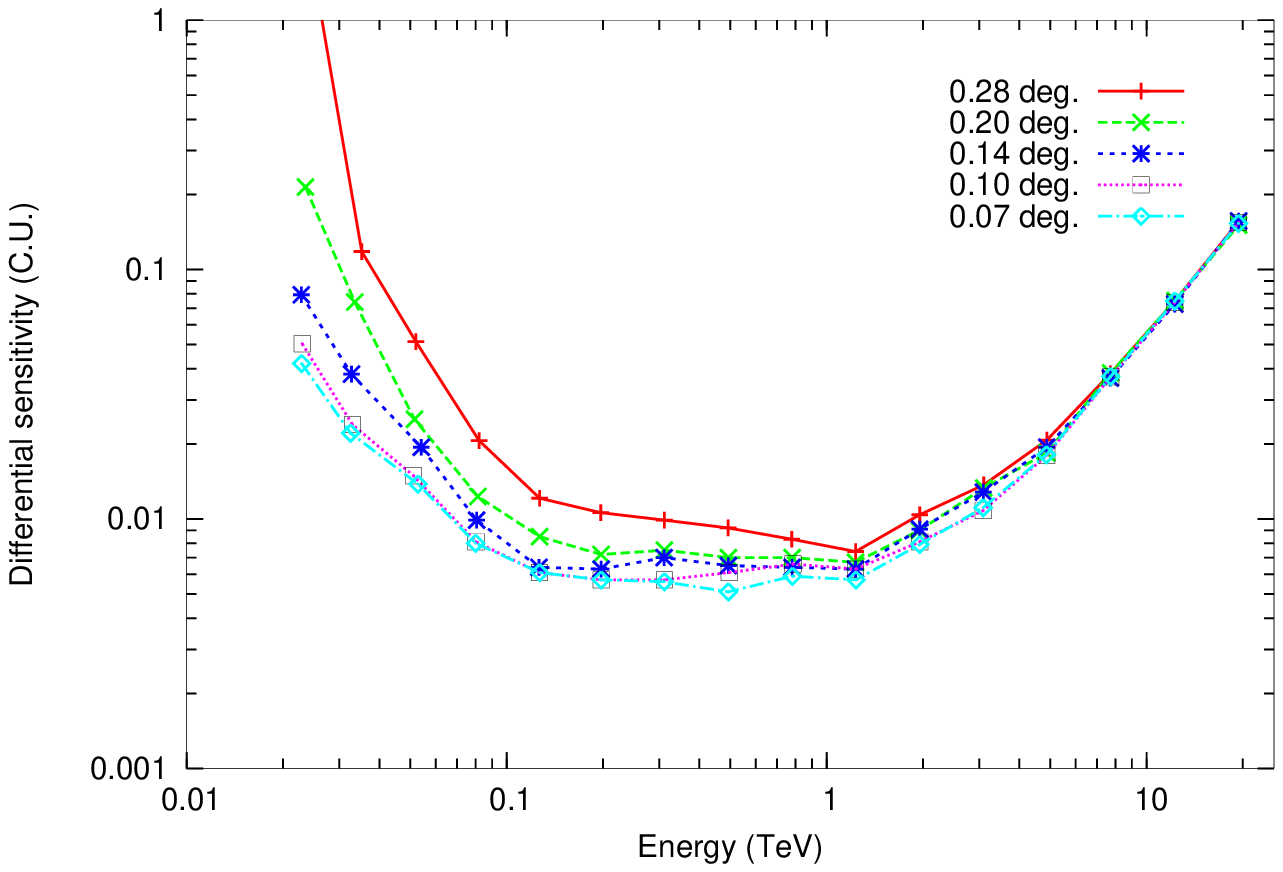,width=0.6\textwidth}
  \caption{\label{fig_pixsize_sens} \small Differential sensitivity curves
  for the  9-telescope benchmark array for several different pixel
  sizes using the same criteria as for the previous figure.
  Image cleaning is adapted to the respective noise levels in each case.
  The impact of reduced pixel size is mainly visible close to the threshold energy.}
\end{figure}
Fig.~\ref{fig_pixsize_sens} shows the impact of changing the (angular) pixel
diameter ($\Theta_{p}$) on the sensitivity of the benchmark array at
2000~m altitude. It can be seen that only modest improvements are possible
with pixels below 0.1$^{\circ}$ diameter.  As the camera cost
increases as $1/\Theta_{p}^{2}$, smaller pixels sizes are
strongly disfavoured.  The improvement of angular resolution at
smaller pixel size is also found to be modest in our studies (see also
\cite{funk_pixsize}). Alternative analyses may lead to
significant benefits from smaller pixel sizes, but this has not yet been demonstrated.
\clearpage
 
\begin{figure}[htbp]
  \centering
  \epsfig{file=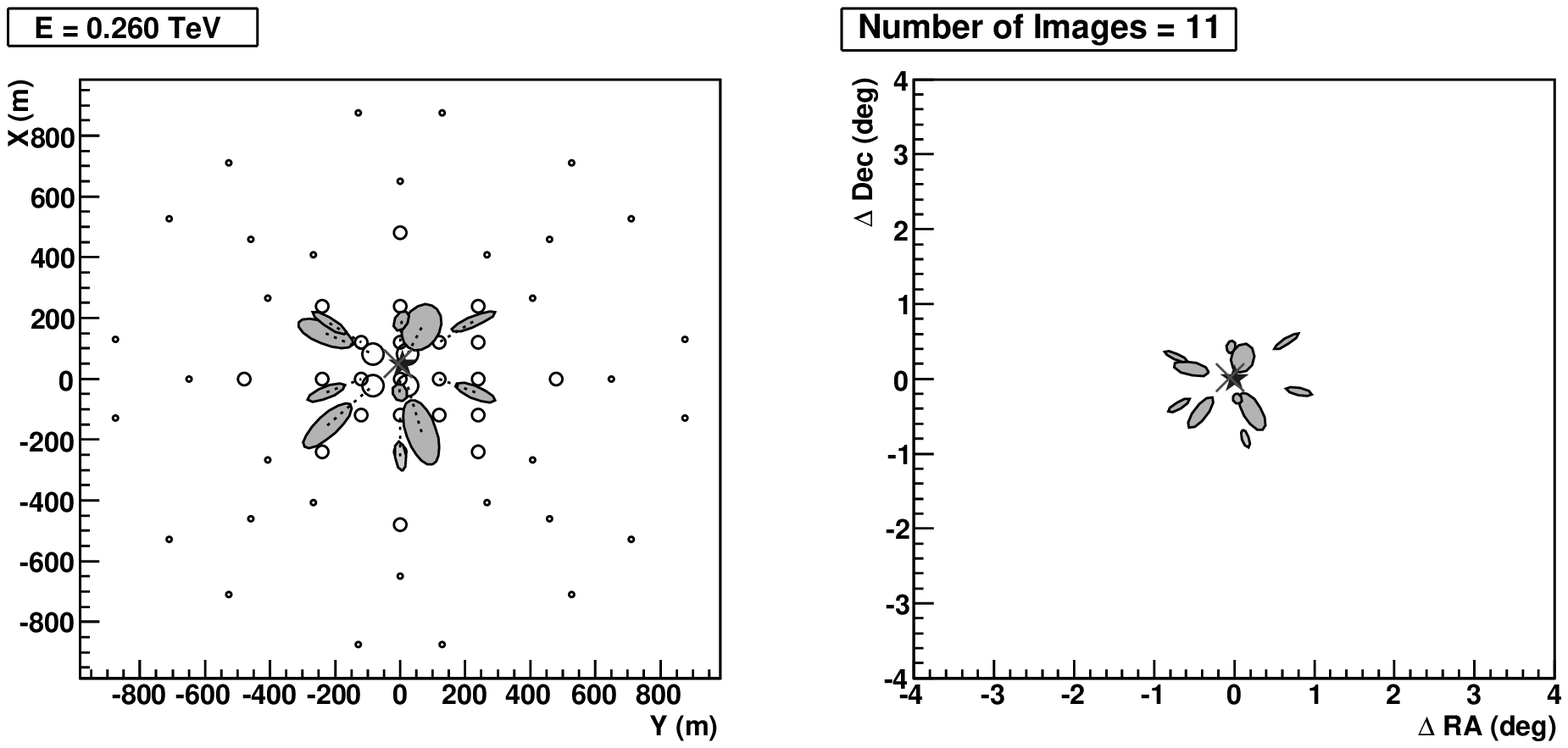,width=0.73\textwidth}
  \epsfig{file=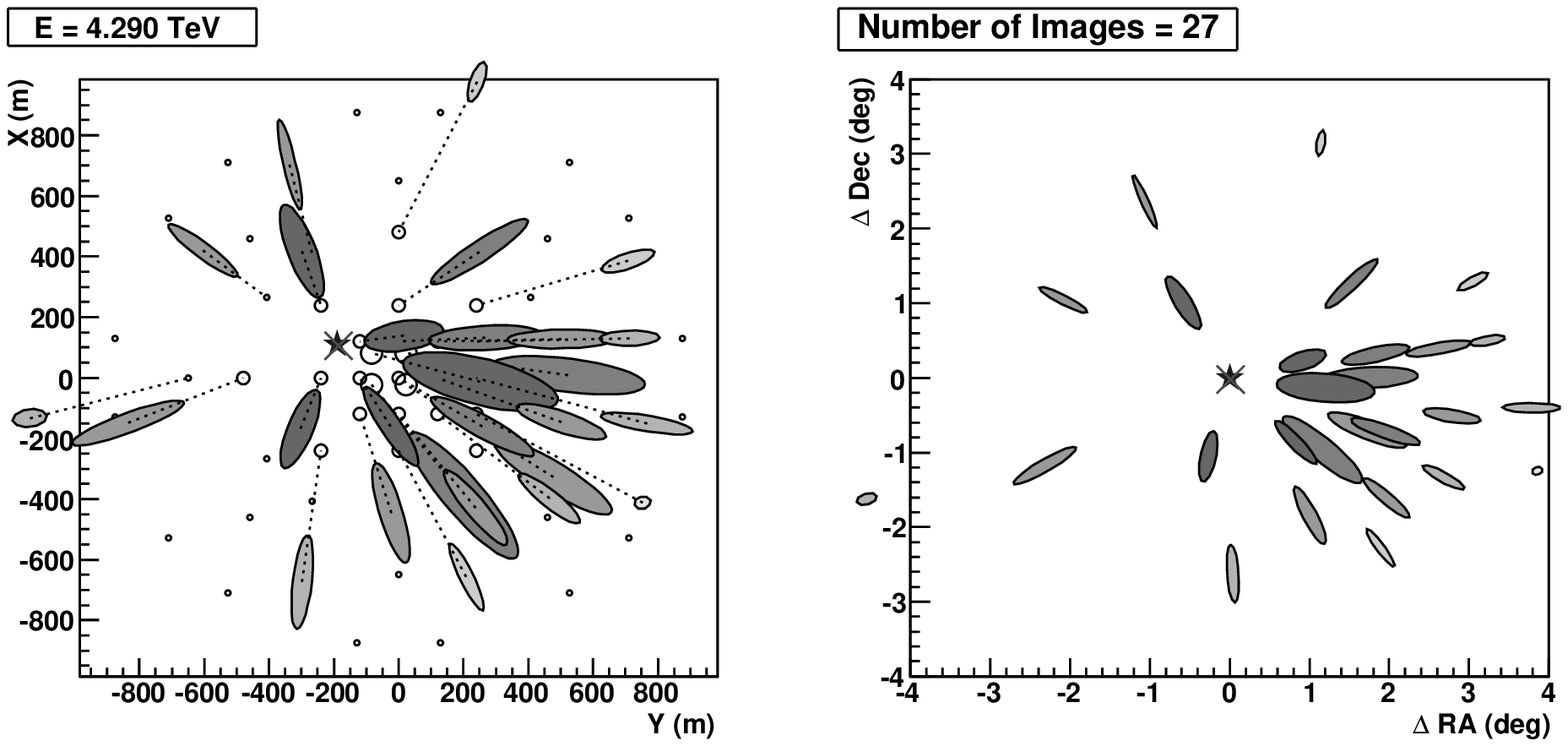,width=0.73\textwidth}
  \epsfig{file=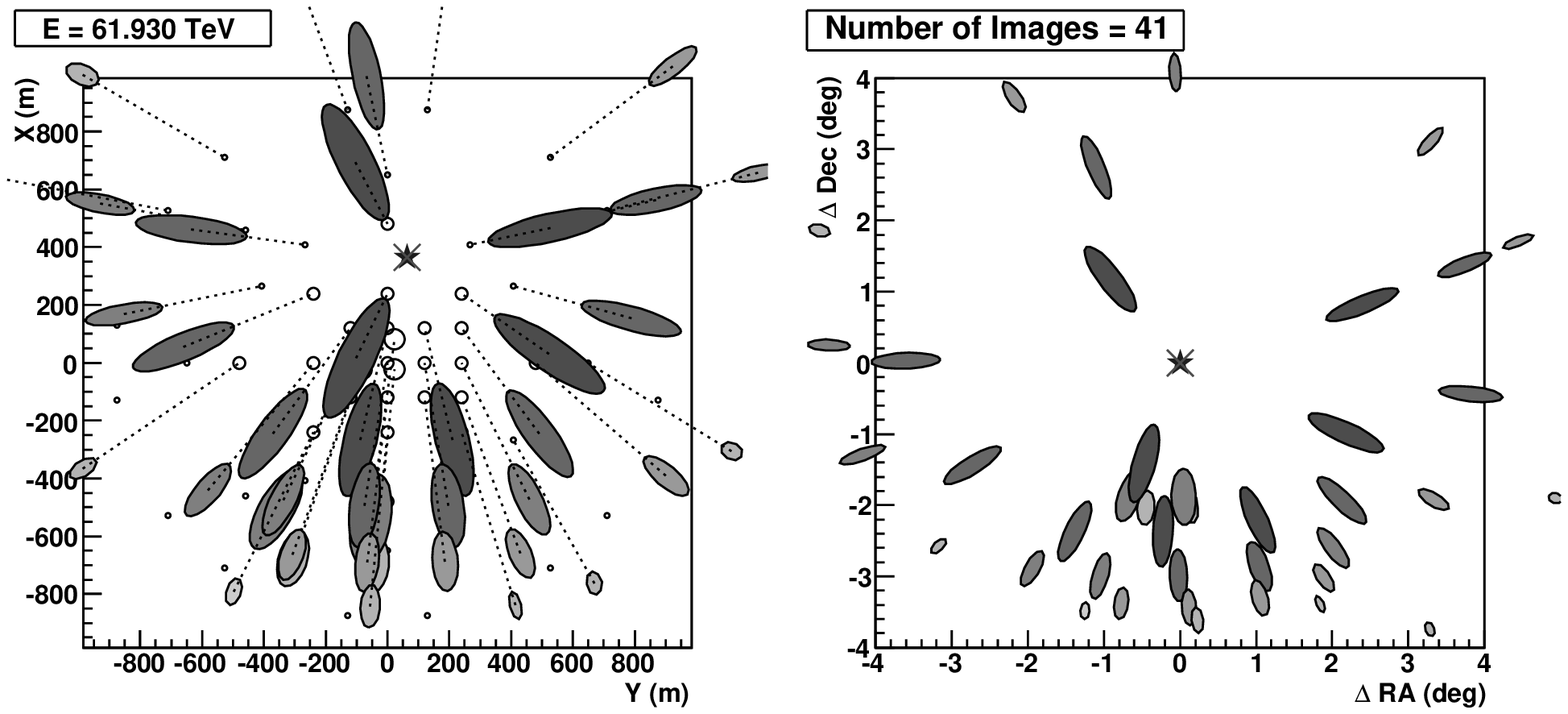,width=0.73\textwidth}
  \caption{
    \label{fig-E-events} \small
Three events as seen by the 59-telescope candidate array E.
The gamma-ray energy and number of images seen are shown in each instance.
The left-hand plots show the telescopes on the ground (the three sizes
of circles for the telescopes of diameters 7~m, 12~m and 24~m, respectively),
with projected Hillas ellipses drawn relative to each telescope position
for each triggered telescope.
Higher amplitude images are filled with darker grey.
The point of intersection of the primary trajectory with the ground is marked with a star.
It is found in a simultaneous fit of both core and direction.
The truncation of images at large impact distances is clearly visible.
The right-hand plots shows the same ellipses in the camera plane, with
the gamma-ray source position marked with a star.
(In the most rudimentary analysis one can reconstruct
the impact point on ground by the intersection of the directions
from image centroids to each of their telescope positions (dotted lines on the left),
and the gamma-ray direction in the sky from the intersection of the image axes (right).)}
%  \epsfig{file=event600gev,width=0.9\textwidth}
%  \epsfig{file=event6tev,width=0.9\textwidth}
%  \epsfig{file=event60tev,width=0.9\textwidth}
%  \caption{
%    \label{fig-E-events} \small Three events as seen by the 59-telescope candidate
%    array E. The number of images recorded are  16, 24 and 36, respectively.
%    The left-hand plots show the telescopes on
%    the ground (triangles, squares and circles indicated telescopes of
%    diameters 7~m, 12~m and 24~m, respectively) with projected Hillas
%    ellipses drawn for selected telescopes. The point of intersection
%    of the primary trajectory with the ground is marked with a
%    star. It is found in a simultaneous fit of both core and direction.
%   Higher amplitude images are filled with darker grey. The
%    right-hand plot shows the same ellipses in the camera plane, with
%    the gamma-ray source position marked with a star.}
\end{figure}

The 275-telescope production configuration described above is the focus of the current
work within CTA and has been used to demonstrate the validity of the
CTA concept. Fig.~\ref{fig-E-events} shows some example events as
seen in a candidate sub-configuration of this production array,
demonstrating the high telescope multiplicity (and event quality)
which is a key element of the CTA design.
 
\begin{figure}[htbp]
  \centering
  \epsfig{file=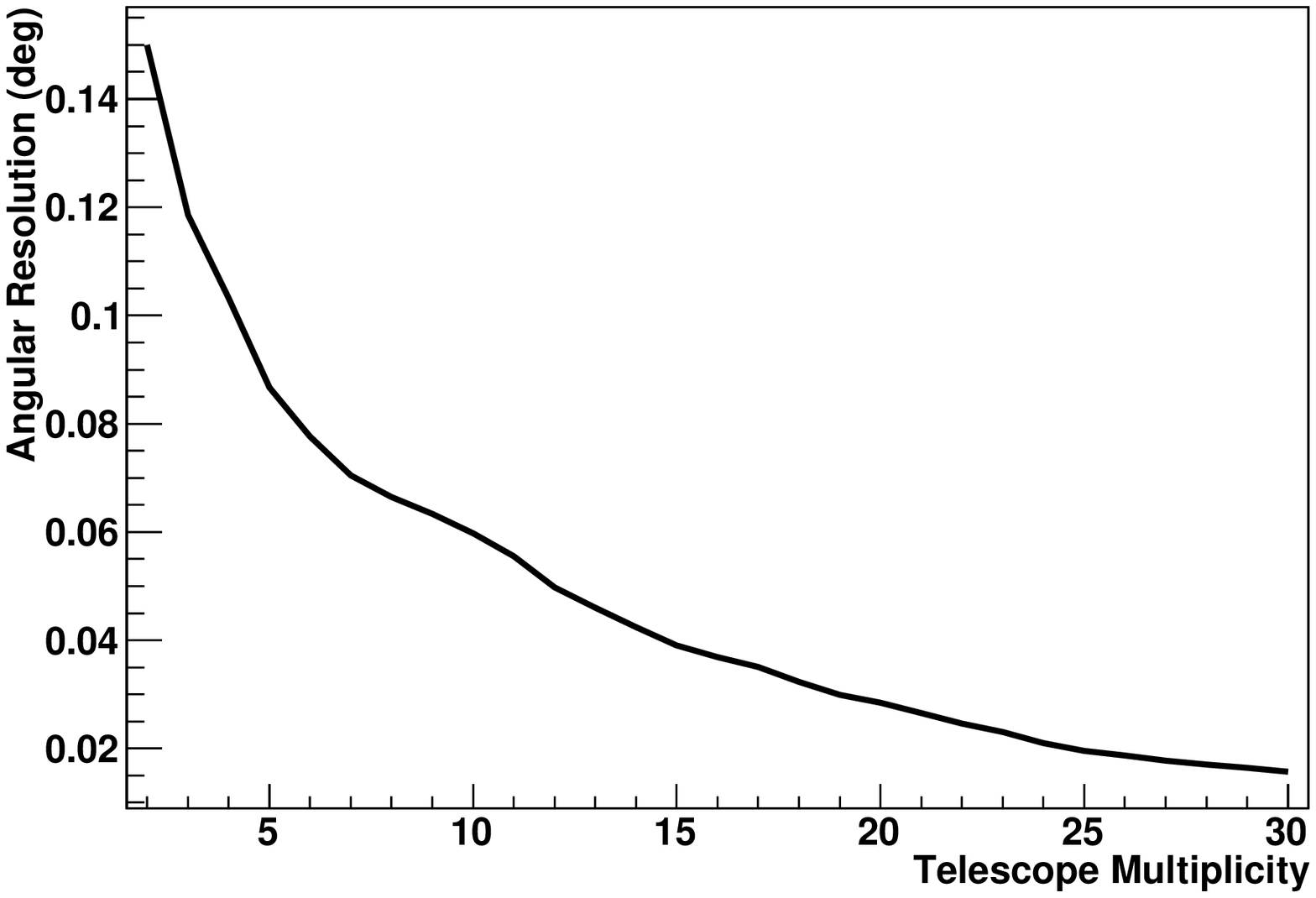,width=0.7\textwidth}
  \caption{
    \label{fig-angresmult} \small
    Angular resolution (68\% containment radius) for array configuration E,
    as a function of the number of telescopes with good shower images.}
 
\end{figure}
Fig. \ref{fig-angresmult} shows how the angular resolution  defined as the 
68\% containment radius, improves with the number of telescopes
that record a shower image. With 4 images (as for instruments like H.E.S.S. or VERITAS)
a resolution of about 0.1$^\circ$ is reached,
while with $\ge$ 12 images the resolution is $\le$0.05$^\circ$.
For the most energetic showers, resolutions of $<$0.02$^\circ$ are reached.
Analogous simulations for AGIS \cite{agis_sims}
give a very similar angular resolution.
The telescopes simulated
include one type of 12~m diameter, 8$^{\circ}$ field-of-view and
0.18$^{\circ}$ pixels (squares in fig.~\ref{fig-E-events}, used in
configurations B, C and E), one type of 7~m diameter, 10$^{\circ}$
field-of-view and 0.25$^{\circ}$ pixels (triangles in
fig.~\ref{fig-E-events}, used in configuration E) and a 24~m
telescope type with 5$^{\circ}$ field-of-view and 0.09$^{\circ}$
pixels (circles in fig.~\ref{fig-E-events},
used in configurations B
and E).  The 24~m telescopes use parabolic optics, all other
telescopes are based on the Davies-Cotton design. Optical designs intermediate between
parabolic and Davies-Cotton are now under consideration to optimise the trade-off between
time-dispersion and off-axis performance. For the cameras,
a quantum efficiency curve of similar spectral shape (blue-sensitive)
to that of current bi-alkali PMTs is assumed. This is a conservative assumption as
$\sim$50\% higher efficiency cathodes have recently been announced by
several major manufacturers (albeit with larger after-pulsing rates, which may
limit the advantage gained in terms of trigger threshold).
 
\begin{figure}[htbp]
  \centering
  \epsfig{file=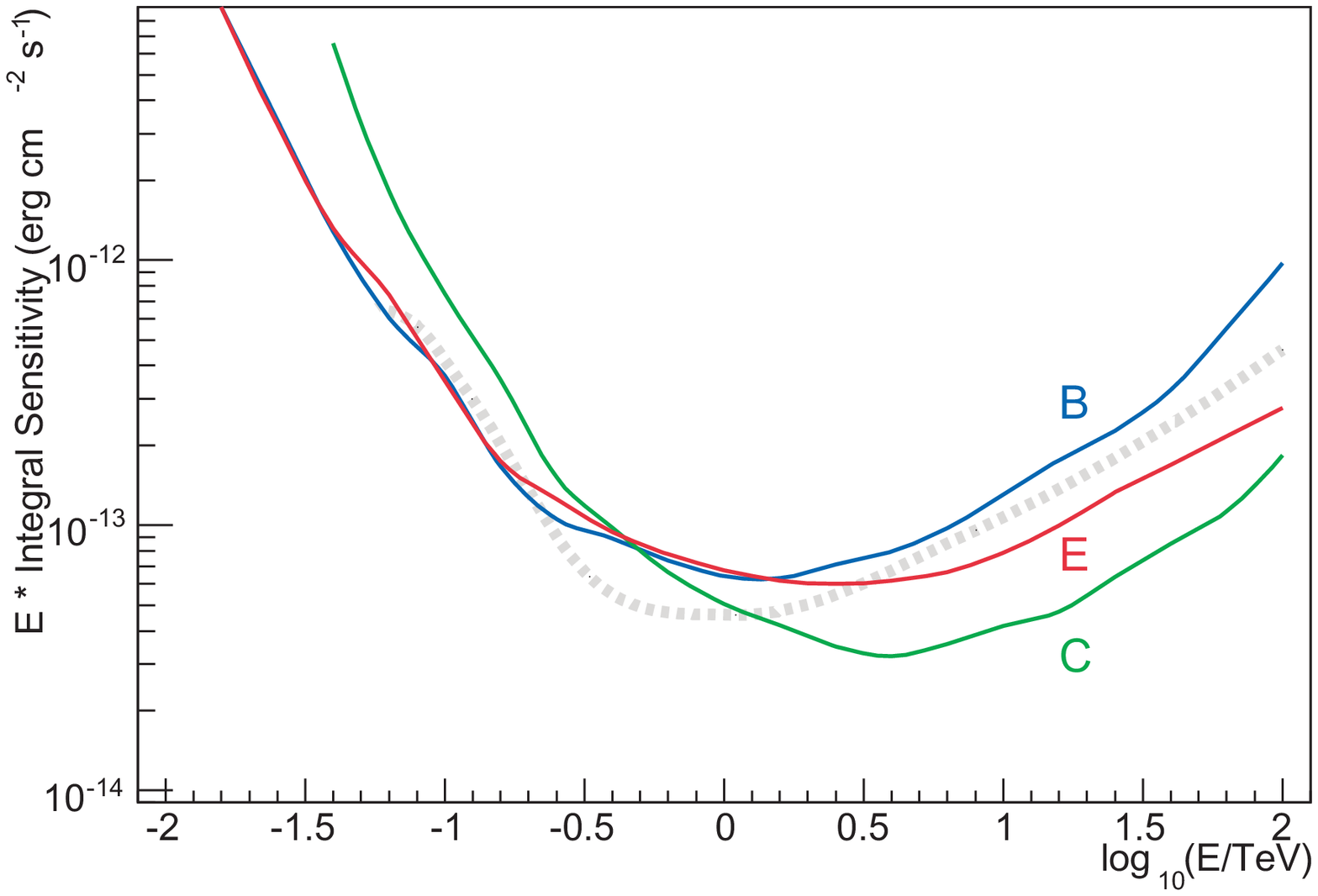,width=0.7\textwidth}
  \caption{\label{fig-sens-ebc} \small Integral sensitivity (multiplied by
  $E$) for the candidate configurations B, C and E, for point sources
  observed for 50 hours at a zenith angle of 20$^\circ$. The goal
  curve for CTA (dashed line) is shown for comparison.
%  {\red really show the low energy part of the dashed curve ???}
}
\end{figure}

Fig.~\ref{fig-sens-ebc} illustrates the integral flux sensitivity
achieved with the three candidate CTA configurations shown above.  The
goal sensitivity curve for CTA
%(as discussed earlier in this document)
is shown for comparison.
It can be seen that these configurations (even with
rather basic analysis methods) are close to achieving the goal
performance in most energy ranges.
%The goal is most distant at
%energies below 30~GeV, where background systematics are the limiting
%factor.
%Significantly better sensitivity is claimed in this energy range for
% simulations by Konopelko \cite{Konopelko_stereo} using only
%two telescope of  larger size (30\,m diameter) than the 24\,m telescopes used here.
%The reason for the large difference in sensitivity estimates is not clear at this time.
At very high energies it seems to be possible to exceed
the original goal performance by a significant factor within the nominal project
budget. As the three configurations B, C and E have roughly equal cost, they can be
used to show the impact of changing the energy emphasis of the
observatory on the performance achieved. Configuration C covers a very
large area ($\sim$5~km$^{2}$) but lacks any telescopes larger than 12~m and
hence has very little sensitivity below 100~GeV. Configuration B has a
low-energy core of 24~m telescopes surrounded by a closely spaced
12~m telescope array. This configuration provides superior hadron
rejection and angular resolution (see later) but provides a more
modest effective collection area at multi-TeV energies. Configuration E
is a compromise array, which attempts to do well in all energy ranges
using multiple telescope types and spacings. As can be seen from
fig.~\ref{fig-sens-ebc}, such an array comes closest to achieving the CTA
performance goals.
 
\begin{figure}[htbp]
  \centering
  \epsfig{file=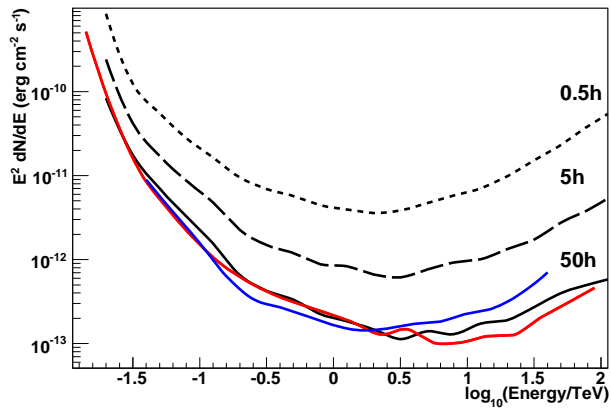,width=0.55\textwidth}
  \caption
      {\label{fig-sens-exposures} \small Time and energy dependence of the
	differential sensitivity (for 5 independent measurements per decade in energy, multiplied by
	$E^{2}$) for configuration E. Exposure times of 0.5, 5 and 50
	hours are shown.  Selection cuts were optimised separately for
	each exposure time. For the 50 hour curve two alternative
	analysis methods are also shown. The red curve is for an
	analysis procedure with an %independent
	image cleaning
	procedure and a Random Forest-based method for hadron
	rejection.  An independent analysis using TMVA for hadron
	rejection is shown as a blue curve.}
\end{figure}

It is  important to study the potential sensitivity of CTA at
much shorter observation times than the 50 hours used for reference.
Fig.~\ref{fig-sens-exposures} shows how the sensitivity changes for
5 hour and 0.5 hour observations. The sensitivity scales linearly with
time $t$ in the regime limited by gamma-ray statistics and approximately with
$\sqrt{t}$ in the background limited regime at lower energies.
For candidate array E, the detection of a source with 2\% of the
Crab Nebula flux (the flux level of the weakest known sources of VHE
gamma-rays until 2007) would be possible in just over 30 minutes.
%Differential sensitivity as a function of observation time
%is plotted in Fig.~ref{}. Already for 5\,h of observation, differential sensitivity
%at intermediate energies is below
%1\% of the Crab flux. The total detection rate for a Crab-level source is about 50\,Hz.
Extreme AGN outbursts, which in the past have reached flux levels $>$10$\times$ the Crab flux,
could be studied with a time resolution of seconds, under virtually background-free
conditions. Fig.~\ref{fig-sens-exposures} also shows 50~hour sensitivity
curves calculated using two independent analyses, illustrating a) that
the conclusions on sensitivity presented here are robust and b) that
the sensitivity can be improved using more advanced methods for
background suppression over much of the CTA energy range.
 
\begin{figure}[htbp]
  \centering
  \epsfig{file=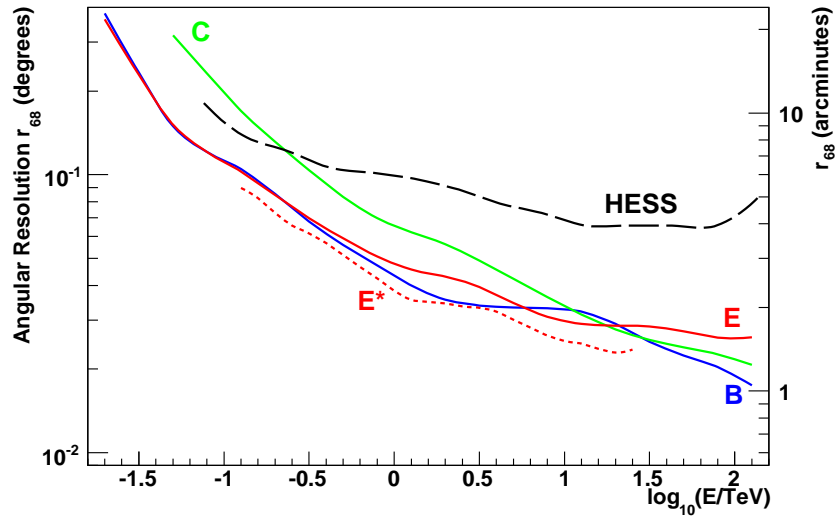,width=0.7\textwidth}
  \caption{\small   \label{fig_angres_2}
  Angular resolution (68\% containment radius of the
  gamma-ray PSF) versus energy for the candidate
  configurations B, C and E. The resolution for a more sophisticated
  shower axis reconstruction method for configuration E is shown for
  comparison (dashed red line - E*). The angular resolution of
  H.E.S.S. (basic Hillas analysis, standard cuts) is shown as a reference \cite{funk_thesis}}.
\end{figure}

The angular resolution for the CTA candidate systems is summarised in
fig.~\ref{fig_angres_2}.  Resolution at 1\,TeV is in the 0.04-0.05$^\circ$
range for configurations B and E, and somewhat worse
for the larger area configuration C, illustrating the
trade-off between collection area and precision at fixed cost. A
simultaneous minimisation to find the best shower core and direction,
using pixel timing information, provides a significant improvement
over the traditional intersection of image axes technique (see dashed
line in fig.~\ref{fig_angres_2}). The resolution approaches 1
arcminute at high energies. Fiducial cuts on core location and/or
harder telescope multiplicity cuts improve this performance, at the
expense of collection area.
\begin{figure}[htbp]
  \centering
  \epsfig{file=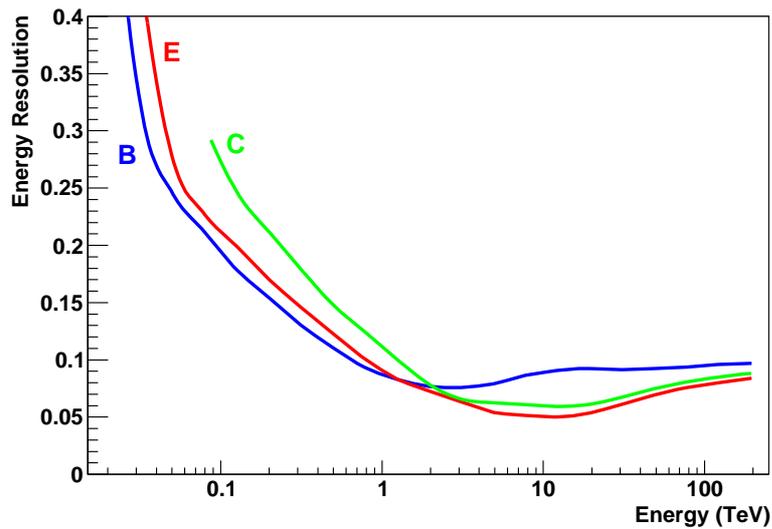,width=0.7\textwidth}
  \caption{\small Energy resolution versus energy for the candidate
  configurations B, C and E.}
  \label{fig_enres}
\end{figure}
The energy resolution (for photon showers) as a function of energy
is shown in fig. \ref{fig_enres}
for the candidate arrays B, C and E.
The energy resolution is below 30\% in almost the whole range of interest and $\le$10\%
above about 1 TeV.
 
In summary, whilst the final optimisation of the CTA design will
require accurate cost models and input from quantitative ``key science
projects'', it is clear from our current studies that an array of
$\sim$60 wide field of view Cherenkov telescopes can achieve the
key performance goals of CTA within the envisaged level of investment.

\clearpage
 
%% 8
%\input{CTAteltech.tex}
%%%%% HEADER: CTAteltech.tex
%%%%%%%%%%%%%%%%%%%%%%%%%%%%%%%%%%%%%%%%%%%%%%%%%%%%%%%%%%%%%%%%%%%
 
\section{CTA Telescope Technology}
\label{sec:teltech}
 
A particular size of Cherenkov telescope is only optimal for covering about 1.5 to 2 decades in energy.
Three sizes of telescope are therefore needed to cover the large energy range
CTA proposes to study (from a few tens of GeV to above 100 TeV).
The current baseline design consists of three single-mirror telescopes:
{\bf SST:} Small size telescopes of 5-8\,m diameter;
{\bf MST:} Medium size telescopes of 10-12\,m diameter; and
{\bf LST:} Large size telescopes of 20-30\,m diameter.
 
While telescope optics involving multiple reflectors or optical correctors have been proposed
\cite{schwarzschild, RC, vassiliev} and do provide improved and more uniform imaging
across large fields of view,
these designs are also more complicated than the classical
single-reflector Cherenkov telescopes. Single-reflector designs are adequate
for the fields of view necessary for CTA and provide a PSF well-matched to the proposed PMT-based camera.
Imaging is improved by choosing relatively large $f/d$ values, in the range of
1.2 to 1.5. A second variable is the dish shape: a Davies-Cotton layout provides good imaging
over wide fields, but introduce a time dispersion.
For small dish diameters this dispersion is smaller than the intrinsic width of the photon distribution,
and therefore insignificant. For large dish diameters, the difference
in photon path length from different parts of the reflector
becomes larger than the intrinsic spread of photon arrival times, broadening the light pulse.
A parabolic shape, which does not introduce this dispersion, is therefore preferred for very large telescopes.
The transition between the two regimes is at about the size of the
MST. Other alternative dish shapes %are discussed in \cite{tbd_dishshapes}, but
face the same general trade-off between time dispersion and imaging quality.

%\input{MountAndDish.tex}
%%%%% HEADER: MountAndDish.tex
%%%%%%%%%%%%%%%%%%%%%%%%%%%%%%%%%%%%%%%%%%%%%%%%%%%%%%%%%%%%%%%%%%%
 
\subsection{Telescope Mount and Dish}
\label{sec:mount}
 
One of the most important mechanical components of a telescope is the mount, with its
associated drive systems.
This must allow the slewing of the dish and the tracking of celestial objects.
The dish structure supports the segmented
reflector and the camera support which holds the camera at the focus on the reflector.
Critical properties for the structural components of a telescope include:
\begin{description}
\item[Positioning of mirror facets.] The dish structure supports mirror facets forming a
pa\-ra\-bo\-lic or Davies-Cotton reflector. Its prime task is to keep the relative orientation
of the mirror facets stable at the arcminute level.
%{\red Approximations concerning the shape of
%the dish, e.g. using planar segments, are tolerable as discussed below. ???}
\item[Mechanical stability of the optical system]
Stability must be achieved
under observing and ``survival'' conditions.
Typical camera pixel sizes are 5' to 10'. To achieve a stable focus, independent of pointing,
modest wind loads and temperature variations, mirror facets have to be kept stable to well
below 1', either by a suitably stiff structure and/or by active mirror attitude control.
Survival conditions refer to high wind and snow loads, which the telescope must
tolerate without suffering damage.
\item[Pointing and tracking precision.] The effective optical pointing of a telescope,
i.e. the location of images on the camera, is determined by the precision of the tracking
system, the overall deformations of the dish and the deformations of the camera support.
Given the extremely short exposure times (ns), the pointing does not need to be stable or precise to
more than a few arcminutes, provided that the effective pointing is monitored with
sufficient precision.
\item[Slewing speed.] A slewing speed that allows repointing to any location in the sky within a minute
%of 1.5-3$^\circ$/sec
is normally sufficient, given that objects are usually tracked over tens of minutes
before repositioning. Only for one special class of targets, the GRB alert follow-ups,
is the fastest possible slewing desirable. Faster slewing of 180$^\circ$ in
20 s is planned for the large-sized telescopes, which are most suited for such follow-ups,
given their low energy threshold.
\item[Efficiency of construction, transport, and installation.] This is a key factor
in reducing costs. For mass production of telescopes, it may be most efficient to
set up a factory for assembly of structural components at the instrument
site, avoiding shipment of large parts and minimising tooling.
\item[Minimal maintenance requirements.] Reducing on-site maintenance to a minimum
aids high efficiency operation
and minimises  the requirements for on-site technical staff.
\item[Safety considerations.]  All procedures for installation and maintenance
have to ensure a high level of safety for workers. The telescopes must also be constructed so that
even in the case of failures of the drive systems or power they can be returned
to their parking positions.
\end{description}

\subsubsection{Mounting System and Drives}
 \label{sec:drives}
 
While some of the very first Cherenkov telescopes were equipped with equatorial mounts,
alt-azimuth mounts offer obvious advantages and have been adopted for all modern instruments.
Two main types of mounts are in use (fig.~\ref{fig:mounting}):
\begin{figure}[hbtp]
\centering
\epsfig{width=0.51\textwidth,file=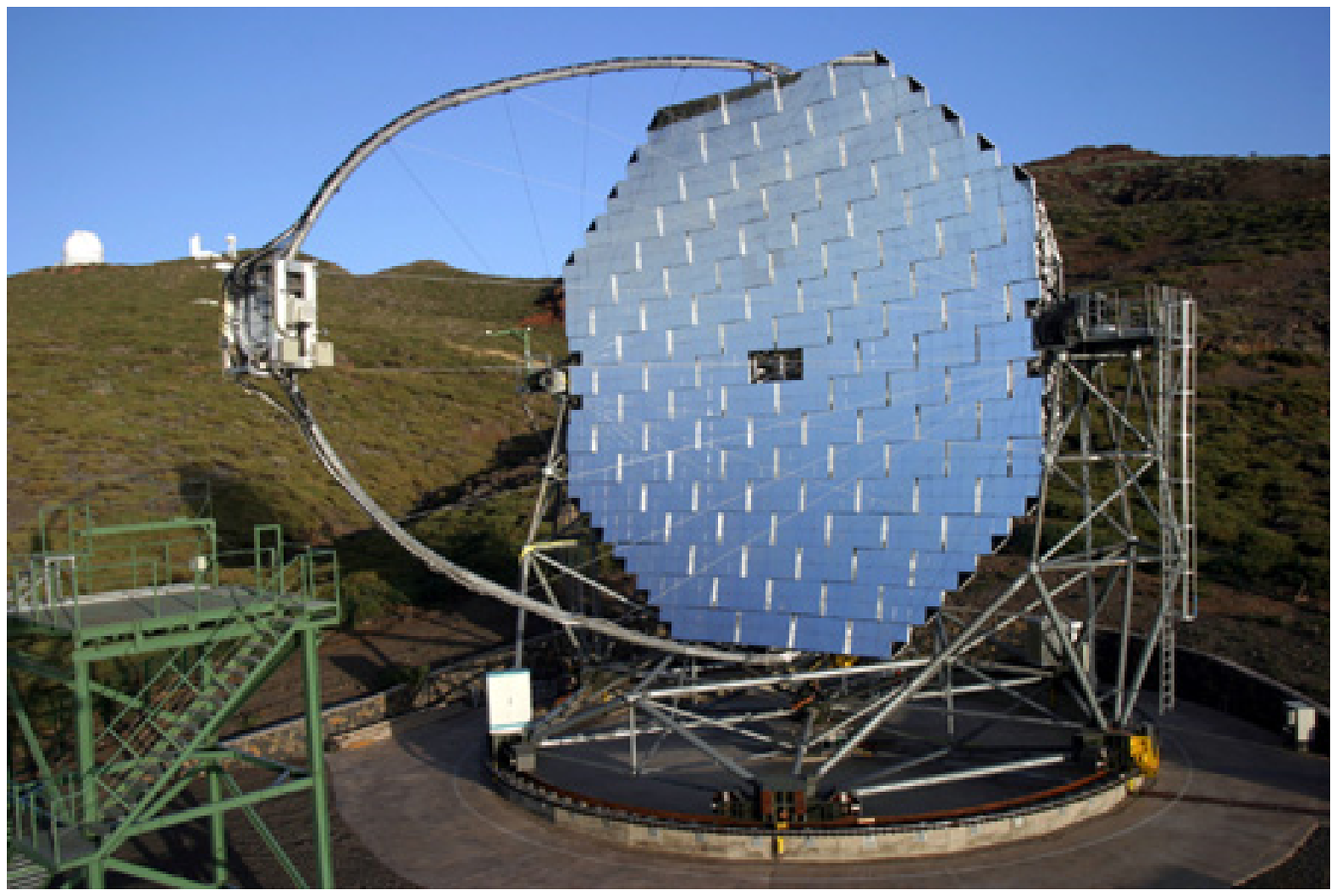} \hfill
\epsfig{width=0.45\linewidth,file=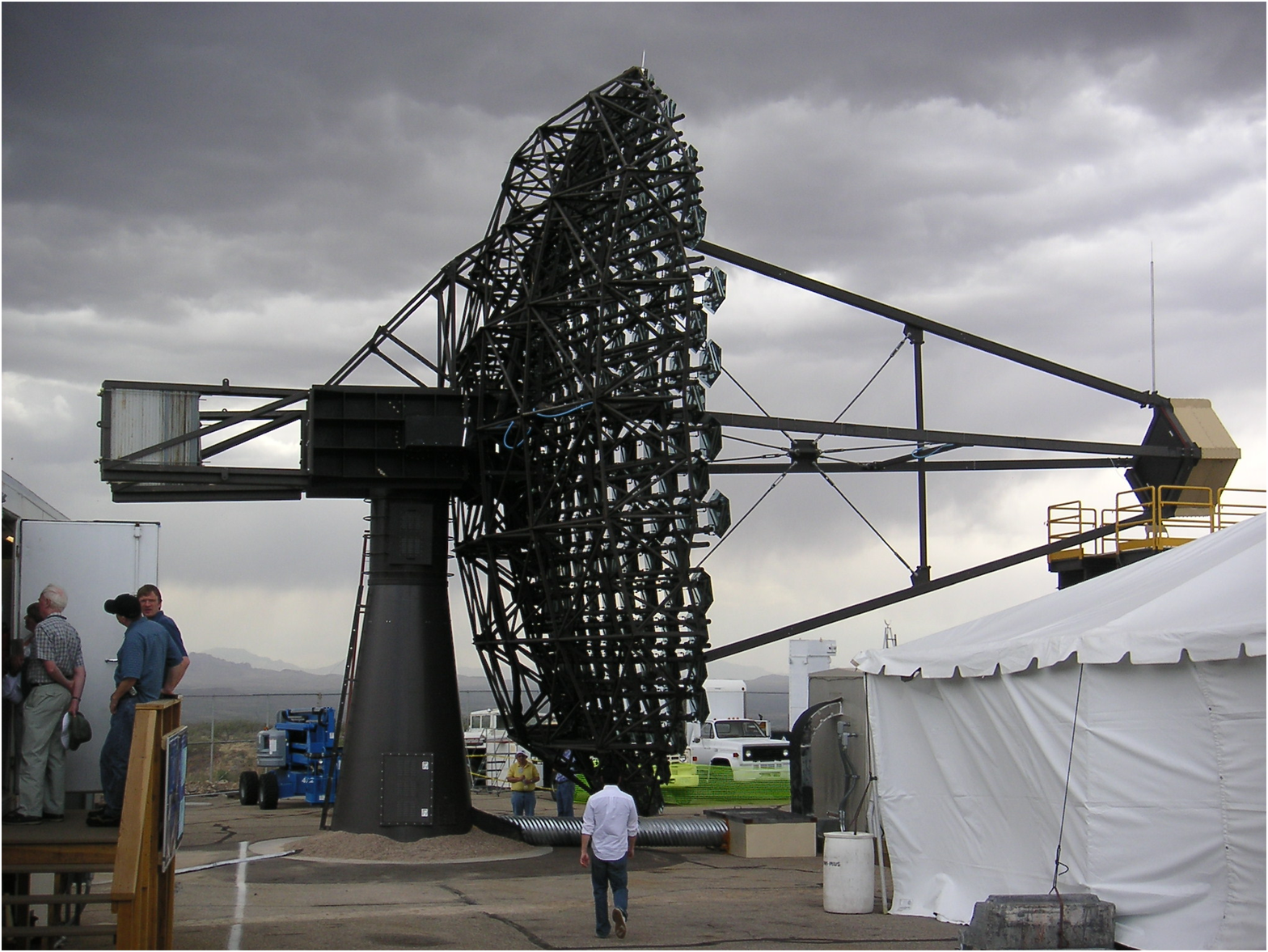}
\caption{\small Examples of alt-azimuth mount, as used for H.E.S.S. and MAGIC (left)
and a central positioner design, used for the Whipple and VERITAS telescopes (right).}
\label{fig:mounting}
\end{figure}
 \begin{description}
\item[Circular rail system] for azimuthal motion, supporting the dish between two elevation
towers, as is used by H.E.S.S. and MAGIC. The elevation axis is positioned such that the dish
is balanced and little or no counterweight is
required. This support scheme will in general permit a large
movement range in elevation,
allowing the positioning of the camera near ground level for easy access, and
the tracking of sources which go through the zenith without repositioning by 180$^\circ$ in azimuth.
A disadvantage of a rail system is the considerable
on-site effort required:  a large ring foundation must be constructed,
the azimuth rail needs to be carefully levelled, and drive systems have to be mounted and
cabled on-site.
 
\item[The central positioner] as used by VERITAS, in which the dish is supported from
near its center in the back.
The central positioner construction
is often used for radio and radar antennae and mirrors for solar power concentrators.
The construction of the foundation is considerably simplified
and the on-site installation work reduced which can be of importance
at sites with poor access or difficult terrain.
In addition, maintenance tends to be simplified since all bearings and drive
components are contained and protected within a compact positioner unit, as opposed
to rails and wheels which are more exposed. While these advantages make the choice obvious
for antennae and solar concentrators, for which focal plane instrumentation is generally of low weight
and $f/d$ is normally very short,
the trend for Cherenkov telescopes is now towards
large $f/d$ ratios, well above 1, to provide  improved image quality.
More and more components
are also being installed in the camera, resulting in increased weight.
Large counterweights are then required to balance the elevation axis in the central positioner design,
as is visible in the VERITAS case. Without these counterweights, the
elevation mechanism has to handle large torques and
the desired positioning speeds require much larger drive power than needed for balanced systems.
Access to the camera at ground level is also
possible in these designs if one locates the elevation axis away from the centre of the tower.
\end{description}
 
Alternative mounting schemes have been considered. For example, a hexpod mount
was investigated for the H.E.S.S. II telescope,
but was abandoned as the initially assumed cost advantages over
conventional mounts turned out to be marginal due to the complexity of the hydraulic drive system and the
extensive safety features required. In addition, a hexpod mount requires a mirror cover during day time,
when the dish is parked facing up. Camera access is also non-trivial.
 
\begin{figure}[hbtp]
\centering
\epsfig{width=0.49\linewidth,file=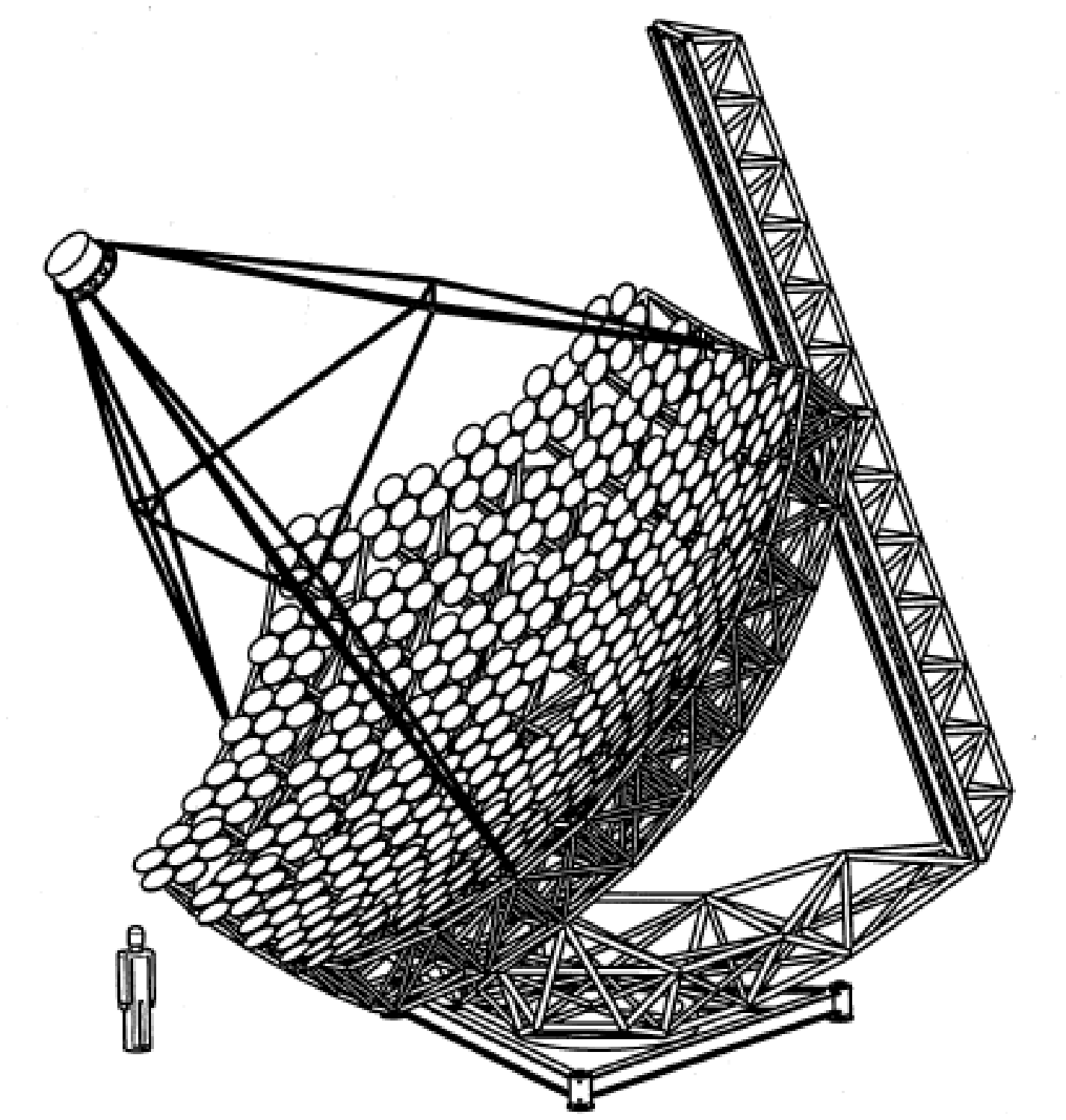} \hfill
\epsfig{width=0.49\linewidth,file=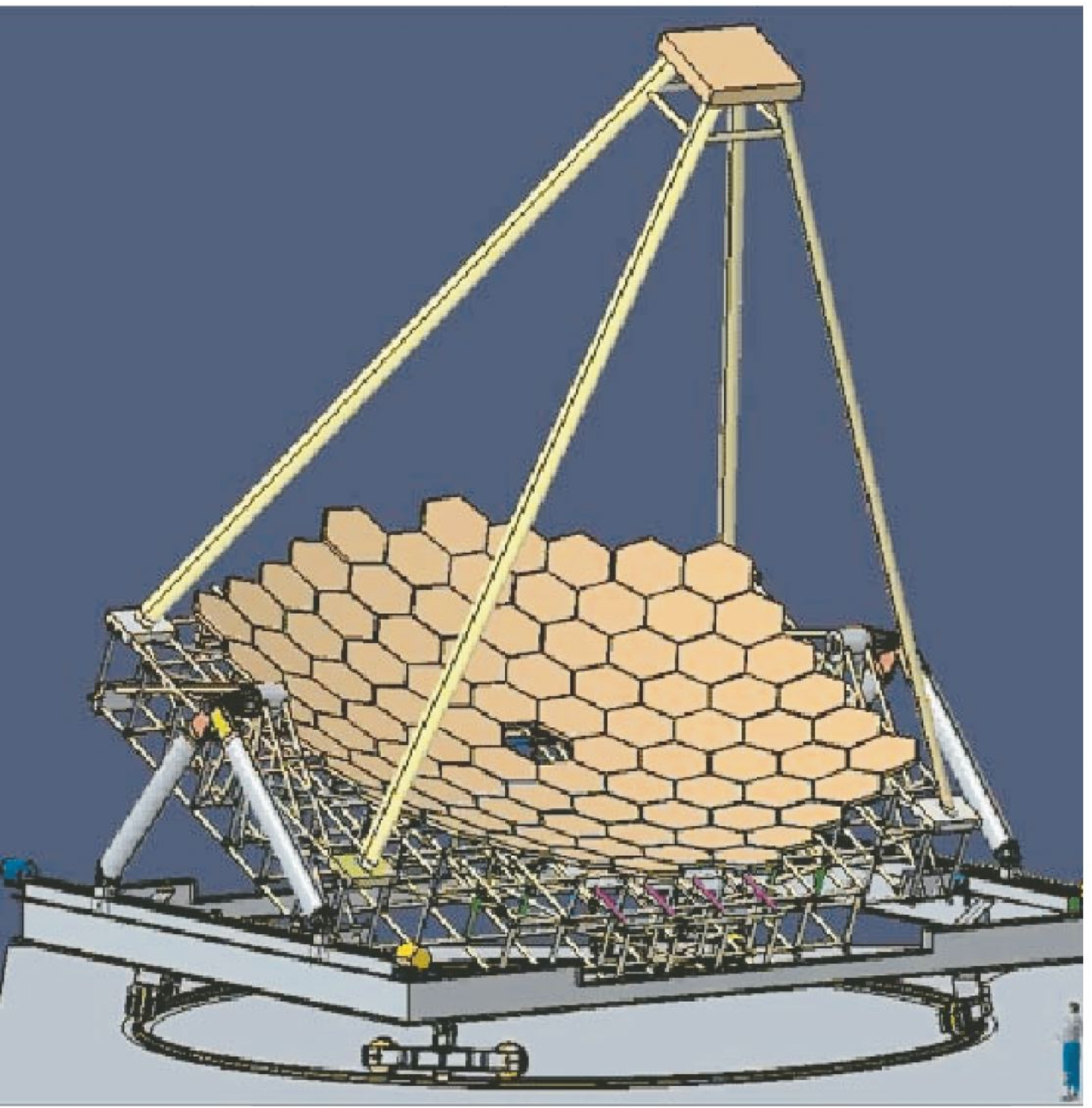}
\caption{\small Alternative alt-azimuth mounts, eliminating the elevation towers,
as studied for H.E.S.S. (left) and CTA (right).}
\label{fig:fliptel}
\end{figure}

Another unconventional mounting scheme is a lift-up mirror carried on a circular rail,
which eliminates the elevation
towers and, at least in some dish support schemes, allows the reduction of
bending torques on the dish due to the
camera support system. A conceptual design for such a scheme was worked out for H.E.S.S.
(see fig.~\ref{fig:fliptel} left), but again did not offer cost advantages.
With a different elevation mechanism, this support scheme
has been considered for the medium-sized CTA telescope (fig.~\ref{fig:fliptel} right).
A drawback of such systems
is that the centre of gravity moves as the telescope's elevation is changed,
requiring significantly increased drive power
compared to balanced systems, where the drives only have to counteract friction,
inertia, and certain wind loads.
 
\begin{figure}[hbtp]
\centering
\epsfig{height=0.4\linewidth,file=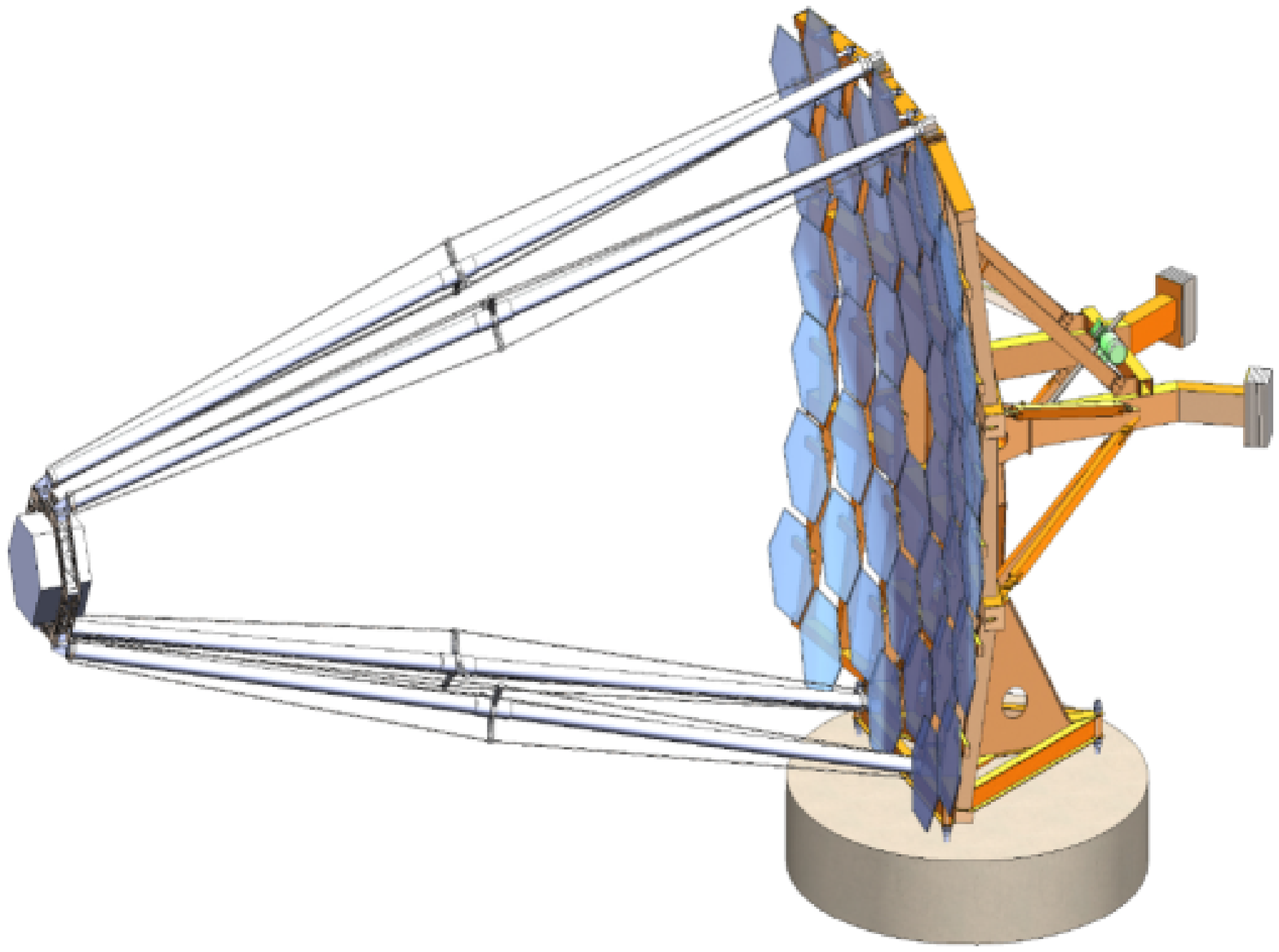} \hfill
\epsfig{height=0.4\linewidth,file=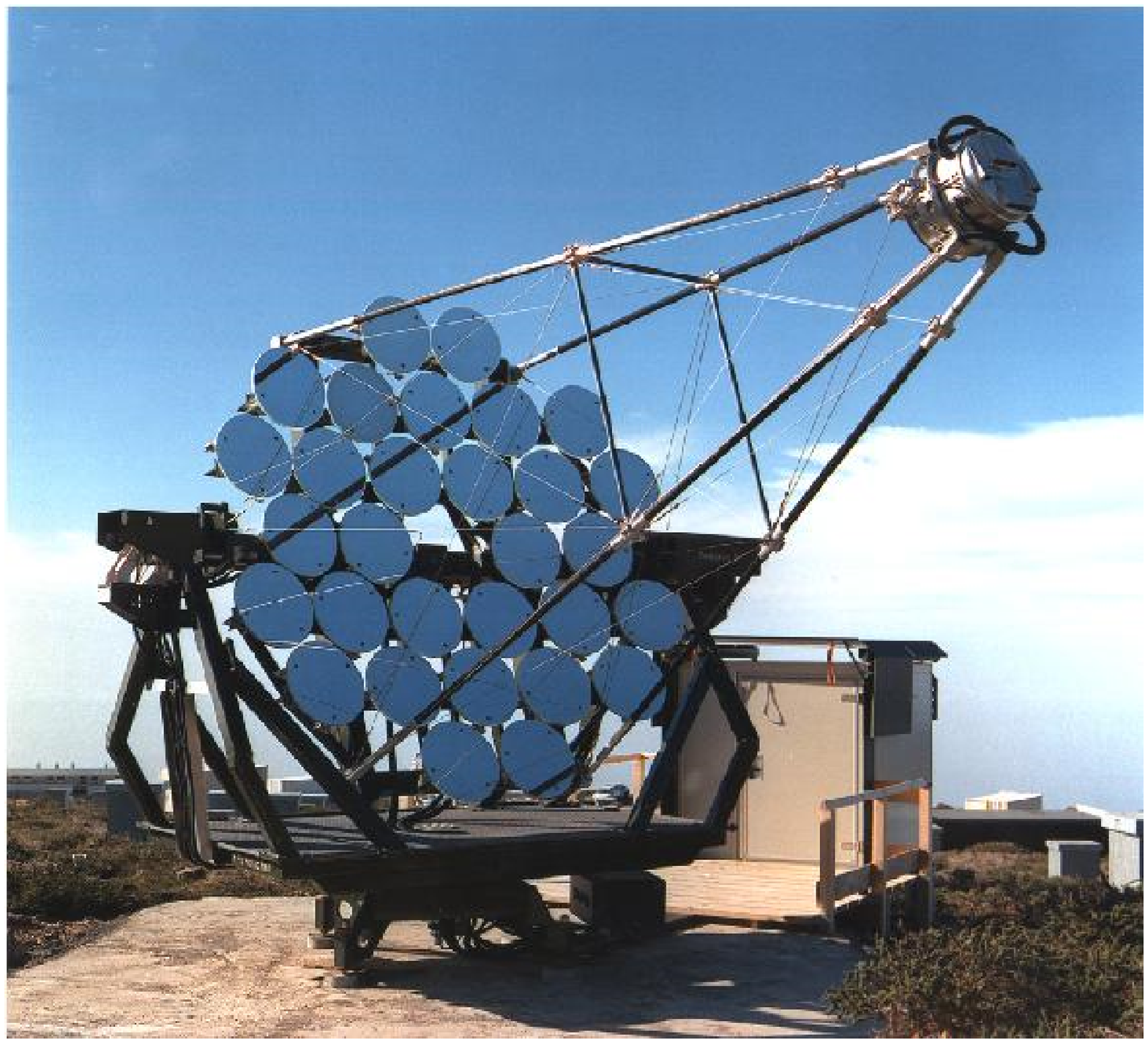}
\caption{\small Two options for the SST mount: a central positioner (left) or a HEGRA-type support (right).}
\label{fig:sst1}
\end{figure}

For the LST, only a rail design, as used by
H.E.S.S. and MAGIC, appears feasible. This is also a possible solution for the MST,
although here a central positioner is a viable option. The solution chosen for
the mount has significant influence on the dish design. When a rail mount is used, the dish is
supported either at its circumference, requiring a stiff dish envelope, or via an extra
elevation cradle as used in the H.E.S.S. II telescope. With a central positioner,
the dish is supported from its centre, and loads at the periphery of the dish must be minimised.
For the SST with its reduced weight and loads,
it appears cost effective to use a central positioner type mount as illustrated in
fig.~\ref{fig:sst1} (left) or to support the telescope by elevation towers but
replace the rail by a central azimuth bearing as is used in the HEGRA telescopes
(fig.~\ref{fig:sst1}, right).
 
Various types of drive systems are implemented in current telescopes.
The experience gained with these will inform the CTA designs.
Some central positioners can be purchased as commercial units and others are under development with
industrial partners. The main challenge
is the large torque that must be transmitted by a rather compact unit, resulting in high forces on
gears and bearings. Dual counter-acting drive units are unavoidable to compensate for play.
For rail-based mounts, azimuth drive systems are used,
e.g. friction drives (H.E.S.S. I), multiple driven wheels (H.E.S.S. II) and
rack-and-pinion drives, implemented using a chain (MAGIC).
%One of the latter two solutions is currently favoured.
For the elevation drive of the LST, a rack-and-pinion system is being considered,
again with the option of using a chain. For the SST and the MST, directly driven elevation
axes are an option.
 
Commercial servo systems will be used to control the drive motors, with multiple feedback loops:
for example, H.E.S.S. II e.g., uses an inner feedback loop to control motor speed and/or torque,
implemented in the servo controller,
an intermediate fast software-based feedback loop implemented in a local controller to control axis
motion and to balance multiple drive motors acting on an axis, and
an outer slower software-based feedback loop for absolute positioning and tracking,
based on absolute shaft encoders. Relatively low-cost
encoders provide a precision of $\le$10". At this level, pointing precision
is usually dominated by deformations of the dish and of the camera support,
causing deviations of the effective optical axis from the nominal pointing
monitored by the encoders.
 
Pointing can be corrected by a combination of lookup-based corrections of
elastic deformations, star guider CCD cameras monitoring the actual orientation of the dish, and
CCD cameras monitoring the position and orientation of the focal plane instrumentation
relative to the dish axis.
Using a combination of such measures allows an (off-line) pointing accuracy of about 10" to be achieved.
 
\subsubsection{Dish Structure and Camera Support}
 \label{sec:camsupp}
 
\begin{figure}[hbtp]
\centering
\epsfig{height=0.31\linewidth,file=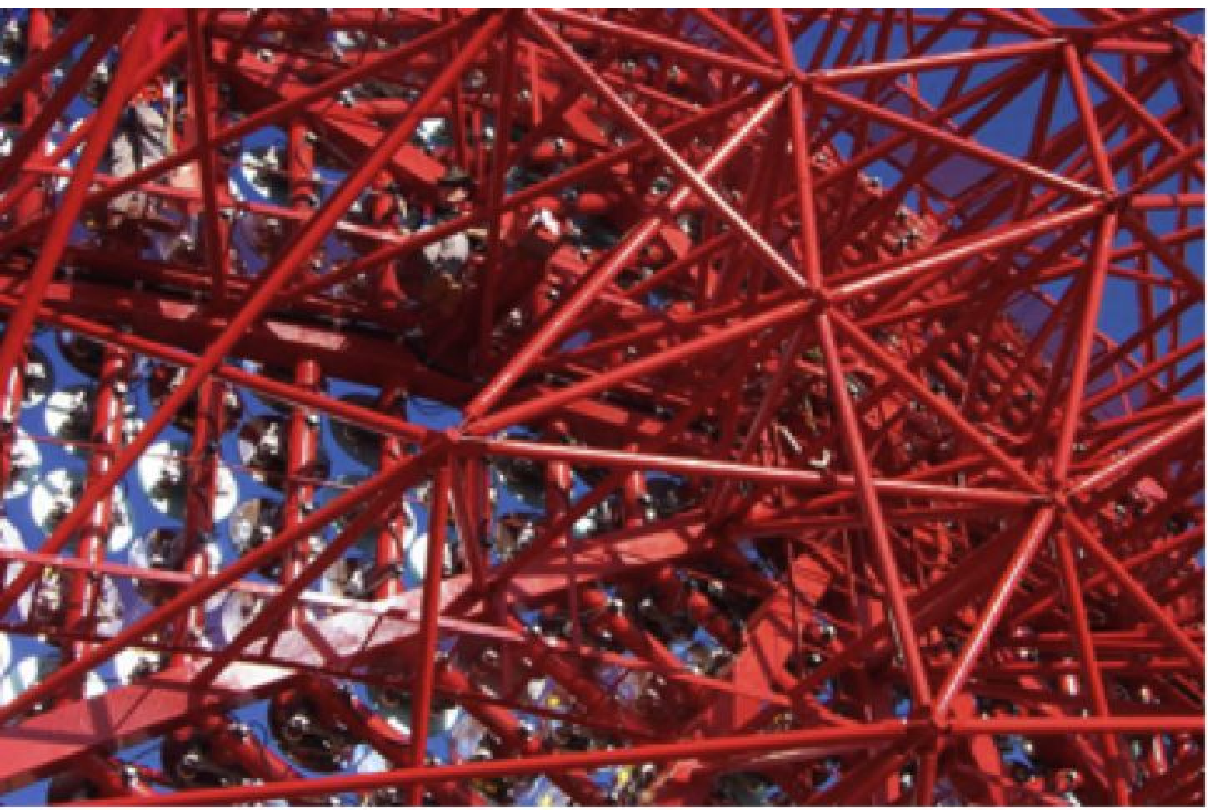}\hfill
\epsfig{height=0.31\linewidth,file=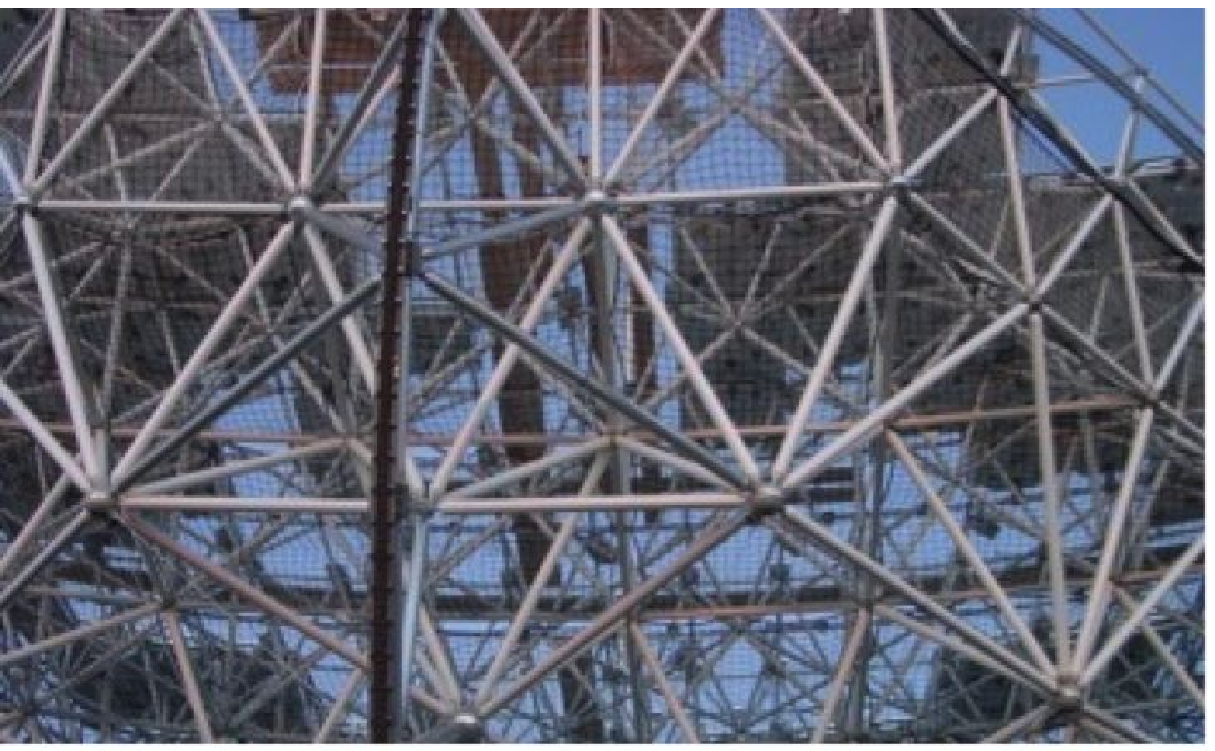}
\caption{\small Examples of the space-frame construction. H.E.S.S. steel
  space-frame (left) and the MAGIC three-layers CFRP space-frame (right).}
\label{fig:space_frame}
\end{figure}

\begin{figure}[hbtp]
\centering
\epsfig{width=0.65\linewidth,file=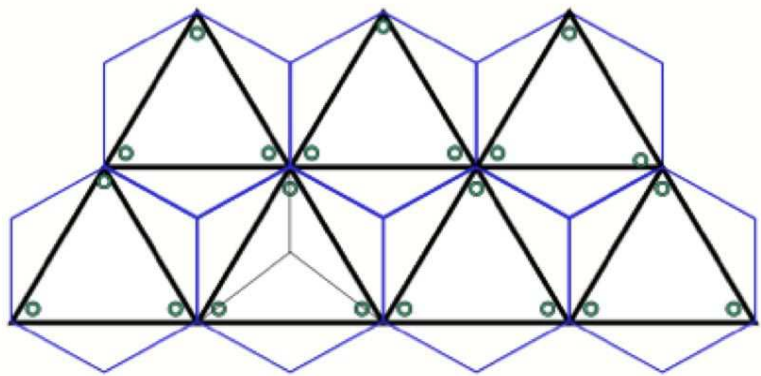}
\caption{\small Sketch of the triangular space frame top layer
  with hexagonal mirror elements (blue lines). The mirror supports points
  (green circle) are fixed close to the space frame corners.}
\label{fig:frame}
\end{figure}

The dish structures of the LST that is currently planned, has a space frame similar to that used
in different variants in the H.E.S.S., MAGIC and VERITAS telescopes (fig. \ref{fig:space_frame}).
A designs with only a minimal space frame is favoured for the dish of the MST.
Another option is
a relatively coarse space frame with an additional structure to provide
mirror attachment points. Alternatively, one can use a highly resolved space frame, based e.g. on
tetrahedron structures, where each mirror support point forms a node of the space frame
(fig. \ref{fig:frame}). The final choice will depend on
structural stability, cost and efficiency of production.
Stiffness requirements will depend on whether active mirror alignment
is employed to partly compensate for dish deformations.
This option is particularly interesting for the LST.\\
 
\noindent
{\bf Construction materials}
 
\noindent
The materials primarily used for the telescope structures are steel,
aluminium and, more recently, carbon fibre reinforced plastic (CFRP). All have their
advantages and drawbacks, particularly when building many telescopes at
remote sites:
 
\begin{description}
\item[Steel:] steel is the most commonly used material for
  past constructions, such as H.E.S.S. and VERITAS. It is generally the cheapest material, but results in
  rather heavy constructions. Nearly everywhere in the world expertise in steel fabrication and
  construction can be found.
\item[Aluminium:] Aluminium is less heavy than steel and has a higher
  specific Young's modulus, but it has the largest thermal expansion of
  all three materials considered here.
\item[CFRP:] CFRP is the strongest of the three materials and
  has the lowest weight, but it is the most expensive. It undergoes very little
  thermal expansion and is better as regards oscillation damping than the other
  materials, but connecting different elements is more difficult.
  This drawback might be overcome by an appropriate design, for example by use of composite-composite
  instead of metal-composite connections.
  CFRP is used in the MAGIC telescopes, to minimise their weight and moment of
  inertia to allow the maximum possible slewing speed.
\end{description}
 
\subsubsection{Current Baseline Designs}
 \label{sec:baseline}
 
\begin{figure}[hbtp]
\centering
\epsfig{width=0.53\textwidth,file=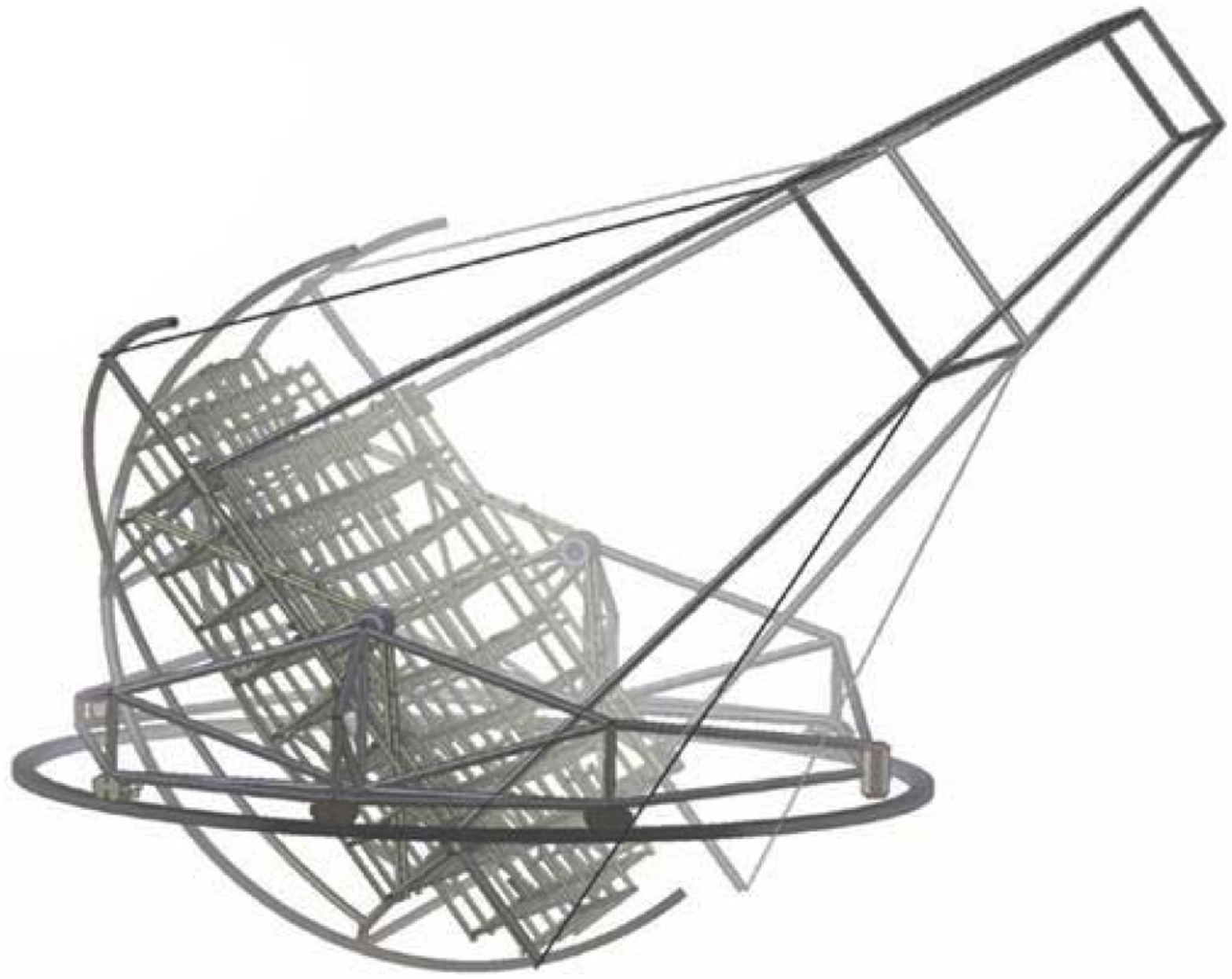} \hfill
\epsfig{width=0.46\textwidth,file=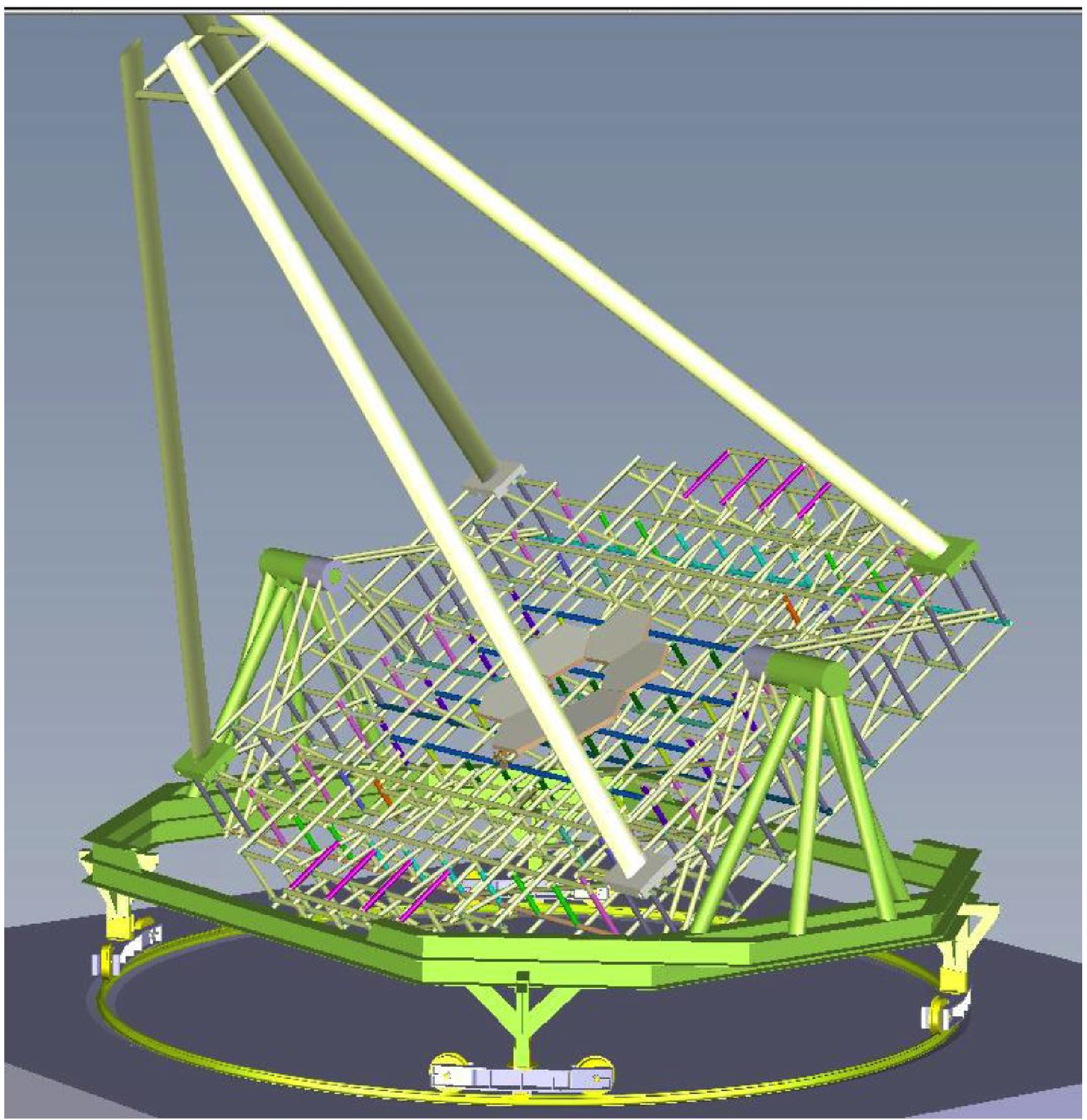}
\caption{\small {\bf left:}
Putting the telescope into a pit reduces the height of the telescope. \newline
{\bf right:} A CFRP dish on a steel mount.
In both cases the dish is held at the edge and the
azimuthal movement is realised by rails.}
\label{fig:zadig}
\end{figure}

\begin{figure}[t]
\centering
\epsfig{width=0.5\textwidth,file=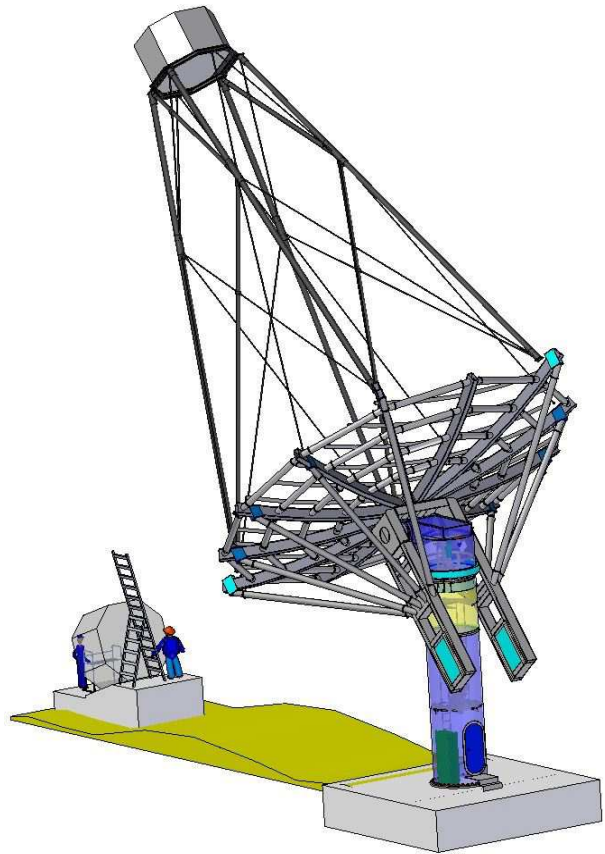}
\caption{\small This design makes use of a positioner for the movement around the
azimuthal and elevation axis.}
\label{fig:ANLDESY}
\end{figure}

For the MST and LST, the mechanically most complicated and costly structures,
as well as for the SST, the following designs have emerged as baseline options
(with other options still being pursued in parallel):
 
{\bf MST:} The general belief within the consortium is that the MST will become
the workhorse of the CTA observatory. This implies that quite a number of
telescopes will be built. Simplicity, robustness, reliability and the ease to maintenance are
therefore particularly important features. This led to the decision to build an
early prototype.
MC studies suggest that an $f/d$ of around 1.4 and a FoV of about 8$^\circ$ is required.
Three groups within the consortium have developed their designs
(Figs. \ref{fig:zadig}, \ref{fig:ANLDESY}).
%During the development phase
%the three designs converged significantly in their basic concepts.
 
The main idea in the first design was to have the elevation axis close to ground level.
This solution saves on the construction of elevation towers, but at the expense of a pit into which
the lower half of the dish disappears when the telescope is parked with the camera at
ground level (fig. \ref{fig:zadig} left).
The same team is working on a design that decouples the dish movement from the camera elevation.
The second design was based on a light and stiff dish, which consists solely of CFRP and is designed
in a way that avoids CFRP joints to metal (fig. \ref{fig:zadig} right). This design allows easy access
to the camera and mirrors. For the elevation, two options were foreseen, a lift-up
system and a more conventional swing-like mount.
 
The third design started from a mirror layout and a structural analysis. Two design
options were considered: one has similarity with the H.E.S.S. I telescopes,
the other with VERITAS. The second option with the central positioner
has been worked out in more detail as this design simplifies of construction and reduces costs
substantially (fig. \ref{fig:ANLDESY}).
A discussion between the three different design groups has started and
has led so far to the use of the CFRP camera structure of the second design in the third design.
All three designs are judged to be technically feasible,
as a consequence of which the costs will be the major criterion of choice.
After the decision on the design, a prototype will be constructed, probably next
to an institute and not at the experimental site. The main aim of this prototype will
be the optimisation and simplification of the instrument with respect to
construction and maintenance.
 
\begin{figure}[hbtp]
\centering
\epsfig{width=0.7\textwidth, file=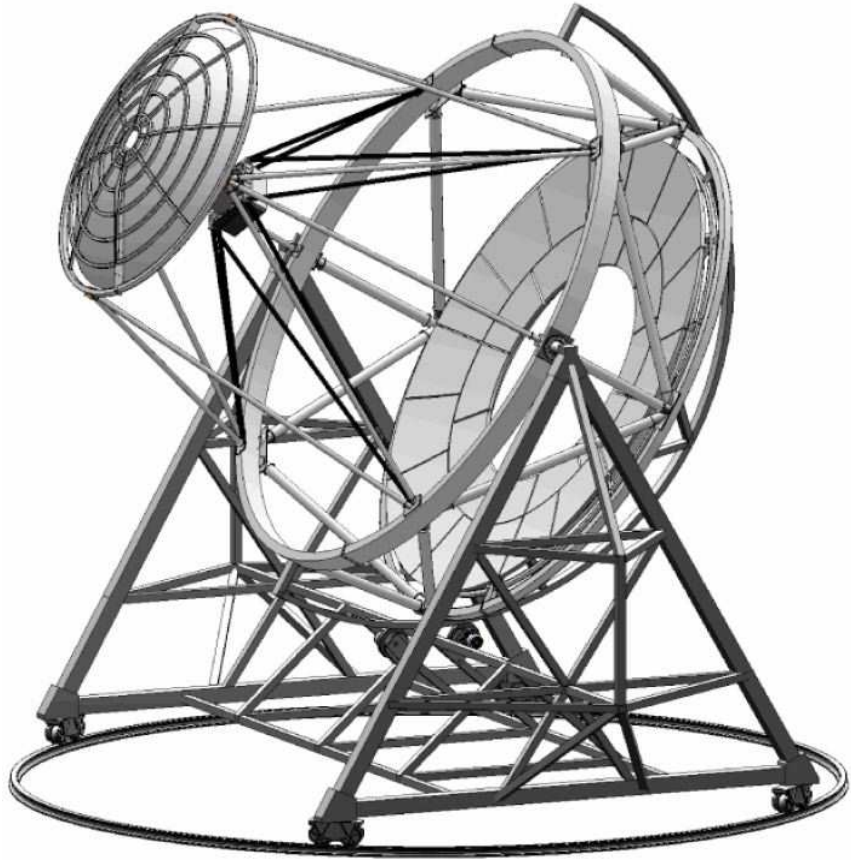}
\caption{\small Model of an AGIS Schwarzschild-Couder telescope and its
two-mirror aplanatic optical system. (from ref. \cite{agis_sims})}
\label{fig:agis_tel}
\end{figure}

In parallel with the prototyping of the single-mirror MST, the design of a
Schwarzschild-Couder telescope for AGIS has progressed (see fig. \ref{fig:agis_tel})
and work towards prototyping of components
and ultimately a full MST-SC prototype is underway in the US.
 
\begin{figure}[hbtp]
\centering
\epsfig{width=0.8\textwidth, file=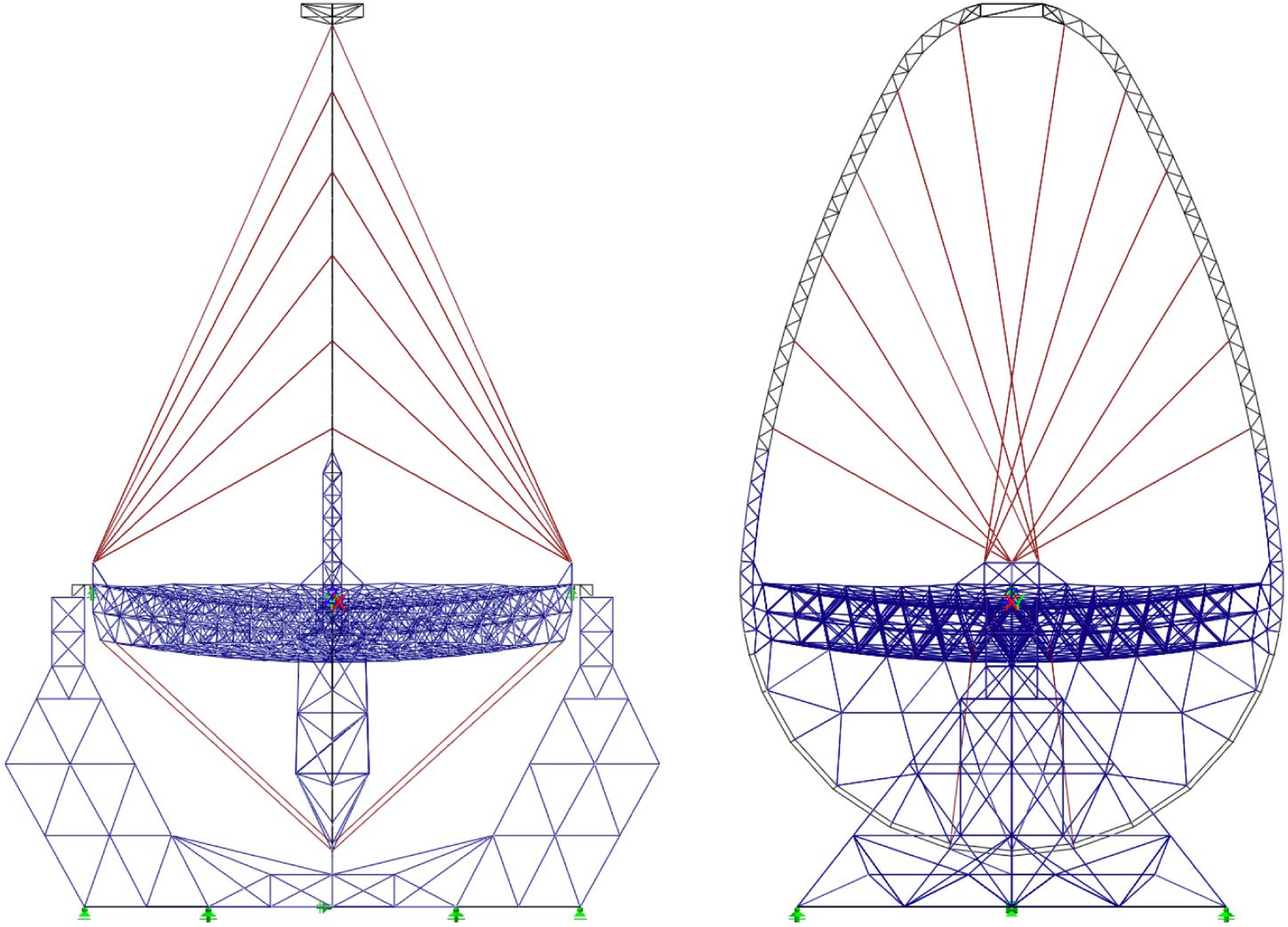}\\
\epsfig{width=0.8\textwidth, file=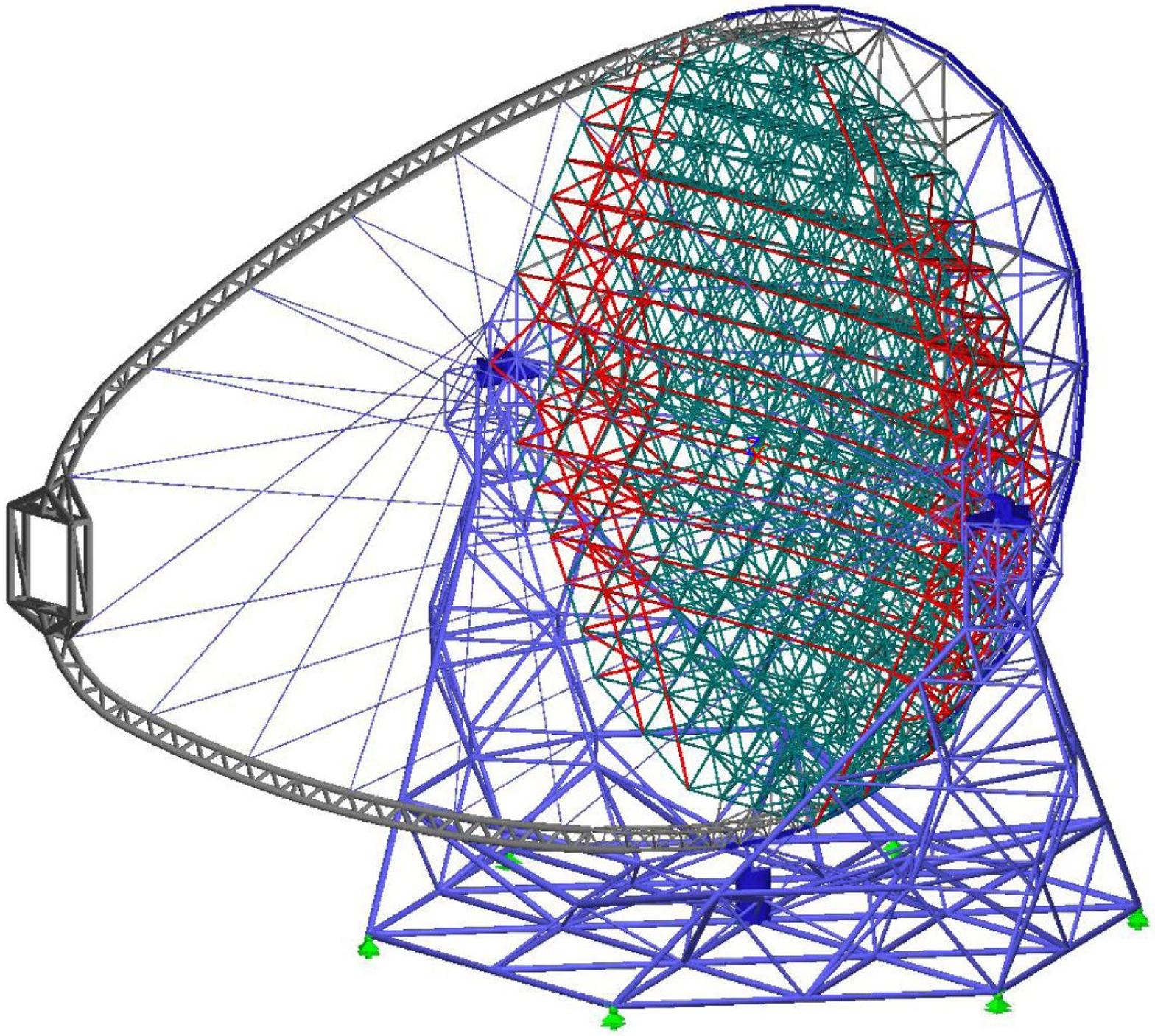}
\caption{\small Conceptual layout of the LST. The dish has a diameter of 23 m.}
\label{fig:LSTbaseline}
\end{figure}

{\bf LST:} For the LST, the current baseline design consists of
a parabolic dish of 23 m diameter with $f/d = 1.2$ constructed
using carbon fibre structure (an enlarged derivative of the proven MAGIC design).
The goal is to keep the total weight around 50\,t
(fig.~\ref{fig:LSTbaseline}). The dish uses a 3 or 4 layer
space frame, based on triangular elements, with hexagonal mirrors supported
from some of the nodes of the space frame.
The dish is supported by an alt-azimuth mount moving on 6 bogeys along a circular rail.
 
\begin{figure}[hbtp]
\centering
\epsfig{height=0.36\textwidth, file=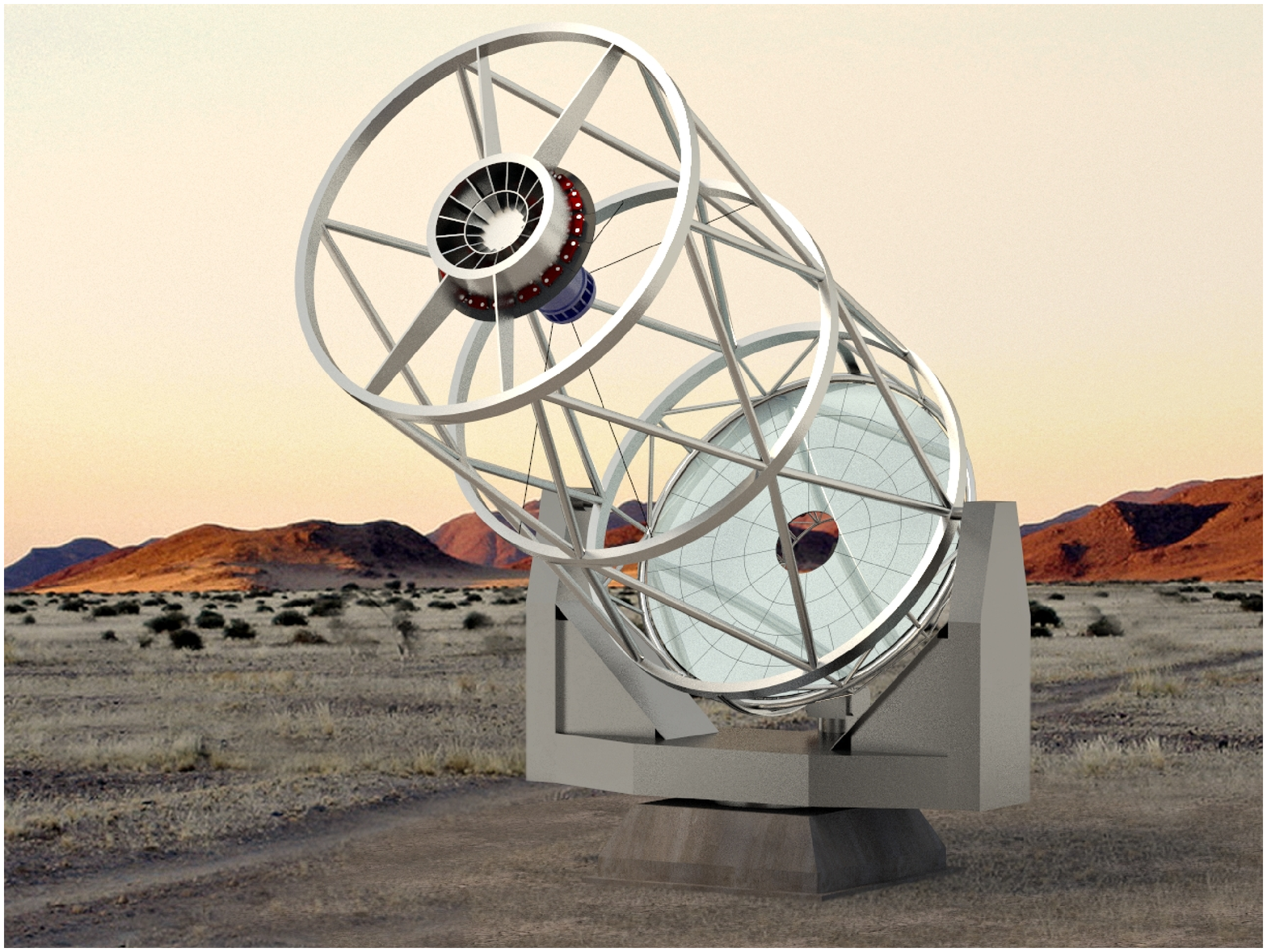}\hfill
\epsfig{height=0.36\textwidth,file=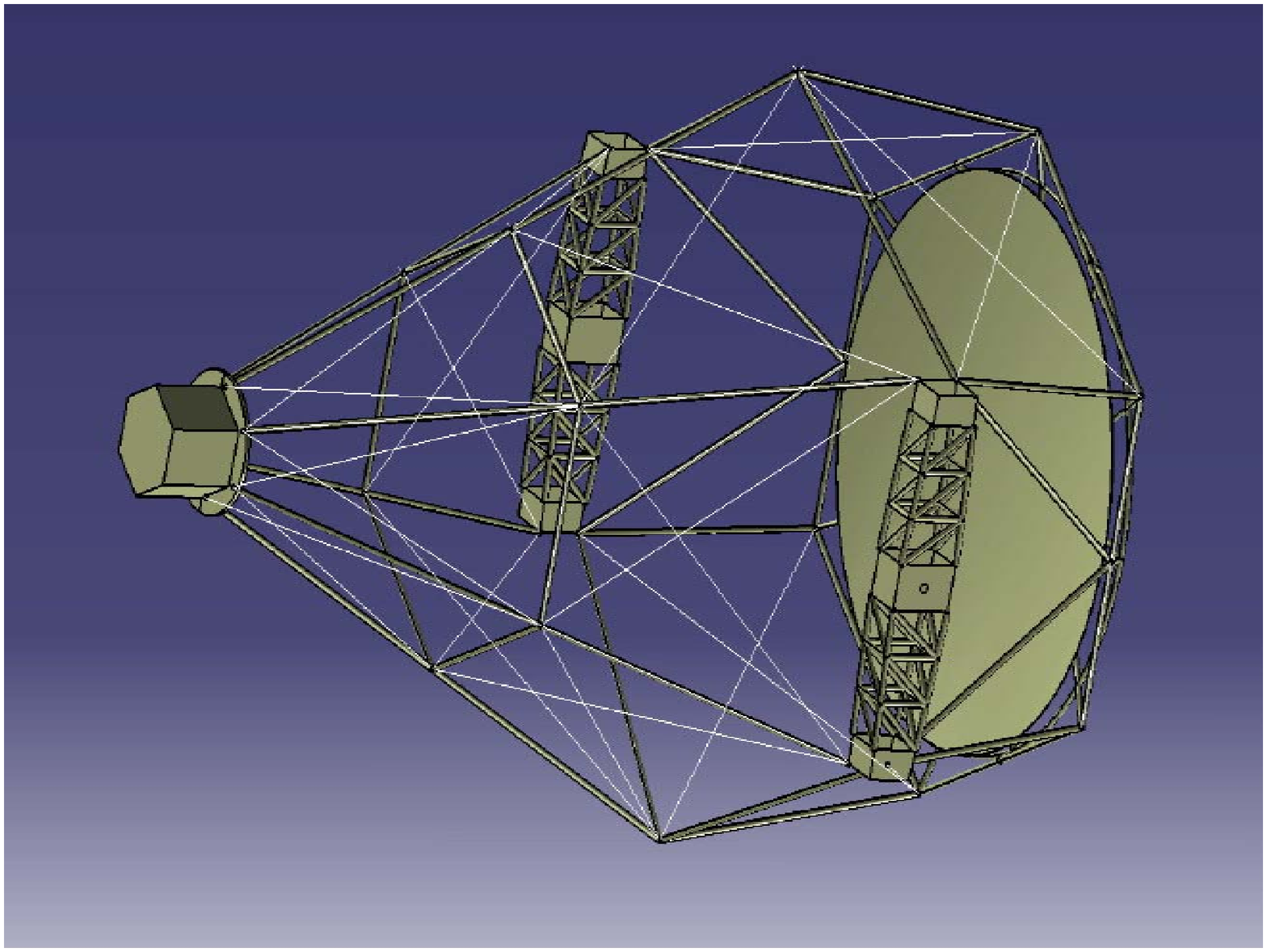}
\caption{\small
Conceptual layouts of a small telescope:
{\bf left:}  two-mirror system
{\bf right:} conventional one-mirror system. The dish is held at the edge and the
azimuth movement is realised by a central bearing.}
\label{fig:SST}
\centering
\end{figure}

{\bf SST:} For the SST, the mechanical design is less complex and several
therefore timescales are somewhat more relaxed.
Several options are still under study. The large FoV that is essential for the SST
results, for single-mirror designs, in a relatively large camera with high costs.
In comparison, the structure of the small telescope is cheap.
This large misbalance makes it sensible to investigate an
SST with a secondary optics which can potentially significantly reduce the
camera cost, at the price of a more expensive mechanical structure.
Whether this results is an overall saving is currently investigated.
A possible design of a two-mirror system is shown in fig. ~\ref{fig:SST} (left).
The design of a 6 m conventional
telescope is pursued in parallel (fig.~\ref{fig:SST}, right).
The costs of these two fundamentally different concepts are now being evaluated.
The result will determine which SST design will be selected.
 
%\input{Mirror.tex}
%%%%% HEADER: Mirror.tex
%%%%%%%%%%%%%%%%%%%%%%%%%%%%%%%%%%%%%%%%%%%%%%%%%%%%%%%%%%%%%%%%%%%
 
\subsection{Telescope Optics and Mirror Facets}
 \label{sec:optics}
 
\subsubsection{Telescope Optics}
 \label{sec:telopt}
 
\begin{figure}[hbtp]
\centering
\epsfig{height=0.5\linewidth,file=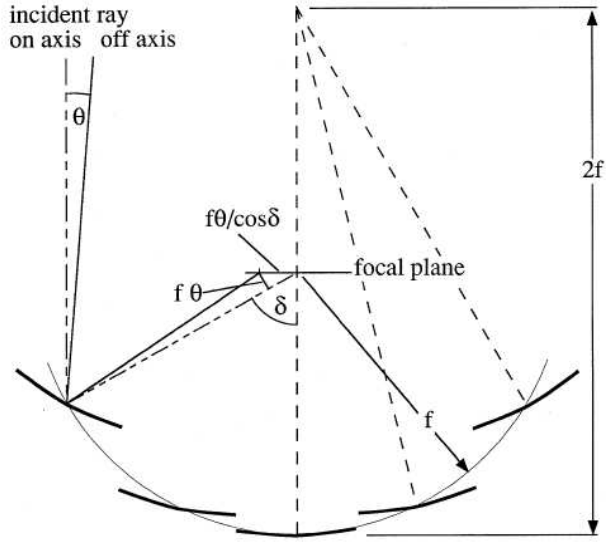}
\caption{\small Davies-Cotton mirror optics, with mirror facets of focal length $f$ arranged on a
sphere of radius $f$.}
\label{fig:DC}
\end{figure}

The reflector of each telescope images the Cherenkov light emitted by the air showers onto
the pixels of the photon detection system.
Apart from the total reflective area, which determines the amount of light that can be collected,
the important parameters of the reflector system are:
\begin{description}
\item[The point spread function.]
The PSF quantifies how well the reflector concentrates light from a point source.
The RMS width of the PSF should be less than half the pixel diameter for 40\% containment
if centred on a pixel,
(for a Gaussian PSF), or better than 1/3 of the pixel diameter  for 68\% containment.
\item[The time dispersion.] Different light paths through the telescope
results in a dispersion in the arrival time of photons on the camera,
which should not exceed the intrinsic width of about 3\,ns of
the Cherenkov light pulse from a gamma-ray shower.
\end{description}
The reflector is usually segmented into individual mirrors. For the optics layout,
most current instruments use
either a parabolic reflector, which minimises time dispersion, or a Davies-Cotton design
\cite{Ref_Mir_DaviesCotton},
where mirror facets of focal length $f$ (and hence radius of curvature $2f$) are arranged on a sphere of
radius $f$ (see fig.~\ref{fig:DC}), and which provides improved off-axis imaging.
At the large field angles required for imaging Cherenkov telescopes,
single-mirror designs suffer from significant optical aberrations with a resulting increase in PSF.
Dual-mirror designs
can provide significantly improved imaging, at the expense of a more complex telescope design
\cite{vassiliev}.
 
For a parabolic reflector of diameter $d$, focal length $f$ and focal ratio $F=f/d$,
the RMS width of the PSF can be approximated by \cite{schliesser}
$$
\sigma^2_\zeta = \frac{1}{512} \frac{\delta^2}{F^4} + \frac{1}{16} \frac{\delta^4}{F^2}
\qquad {\rm and} \qquad
\sigma^2_\eta  = \frac{1}{1536} \frac{\delta^2}{F^4}
$$
where $\delta$ is the field angle and $\sigma_\zeta$ and $\sigma_\eta$ are the widths of the PSF in the
radial and azimuthal directions, respectively. The spot size is always larger in the radial direction,
mostly due to the non-Gaussian tails of the PSF. For a parabolic reflector, the two spot dimensions differ
by a factor of more than 1.7, resulting in systematic distortions of Cherenkov images for
off-axis sources.
 
For a Davies-Cotton reflector with a planar focal surface, the corresponding expressions
are \cite{vassiliev}
$$
\sigma^2_\zeta = \frac{1}{1024} \frac{\delta^2}{F^4} \left(1 - \frac{1}{4 F^2}\right)
+ \frac{1}{256} \frac{\delta^4}{F^2} \left(4 + \frac{35}{6 F^2}\right)
\qquad {\rm and} \qquad
\sigma^2_\eta  = \frac{1}{1536} \frac{\delta^2}{F^4} \left(\frac{10}{9} + \frac{9}{32 F^6}\right).
$$
The difference between the radial and azimuthal spot sizes is less pronounced in this case,
typically around 20\%.
The Davies-Cotton design results in a flat distribution of photon arrival times, with a maximum time
difference of $D/(8F \cdot c)$, and an RMS time dispersion $\sigma_t = d/(16\sqrt{3}F \cdot c) \approx
0.12 d/F$\,ns/m.
 
Usually, the first term in the expansions for the PSF dominates, resulting in a roughly linear increase of the PSF
with the field angle $\delta$, and a quadratic dependence on $F$.
For typical parameter
values, $\sigma_\zeta$ is 20-30\% smaller for the Davies-Cotton design than for a parabolic mirror,
whereas $\sigma_\eta$ values are similar.
 
\begin{figure}[hbtp]
\centering
\epsfig{height=0.36\linewidth,file=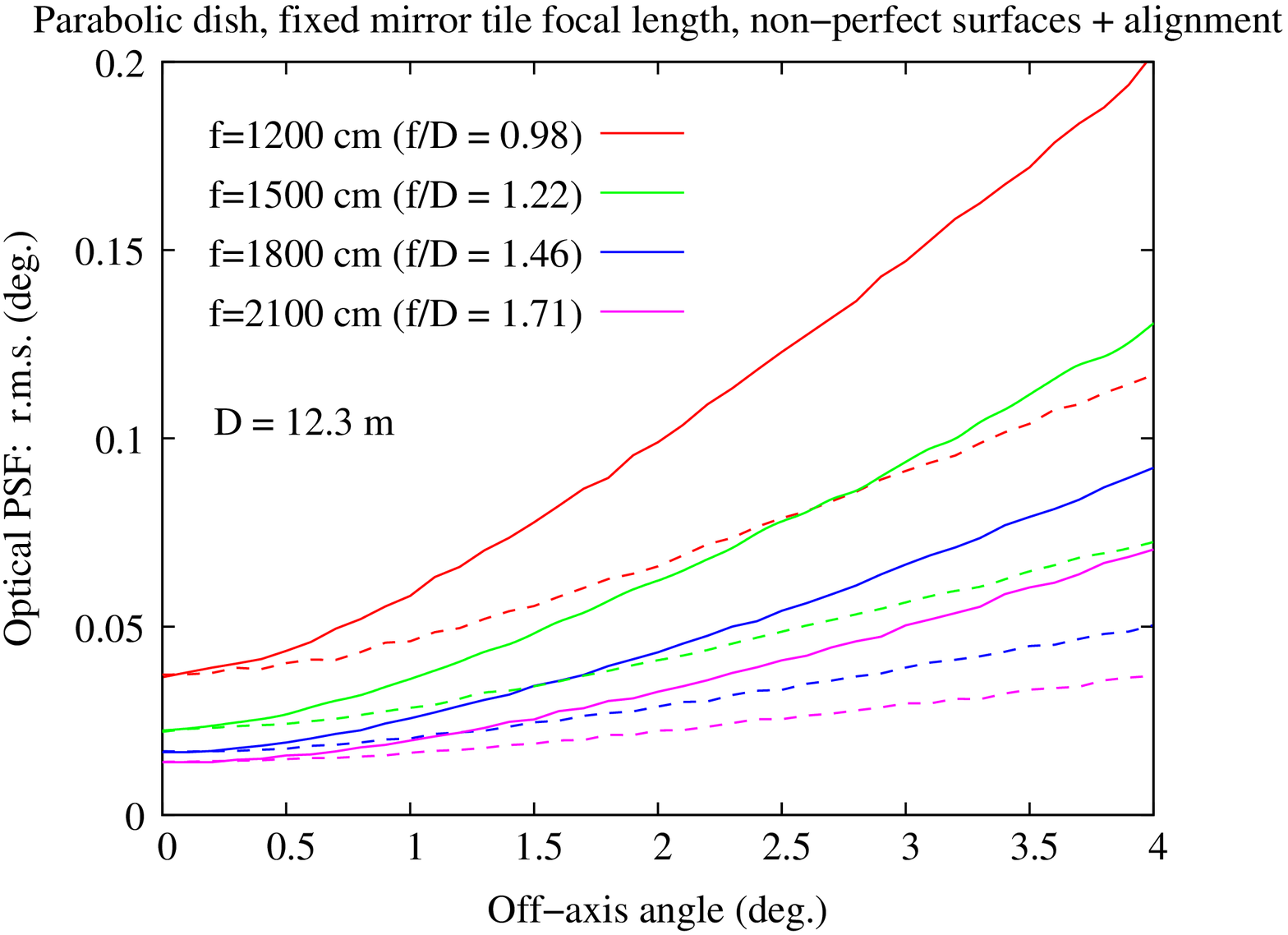} \hfill
\epsfig{height=0.36\linewidth,file=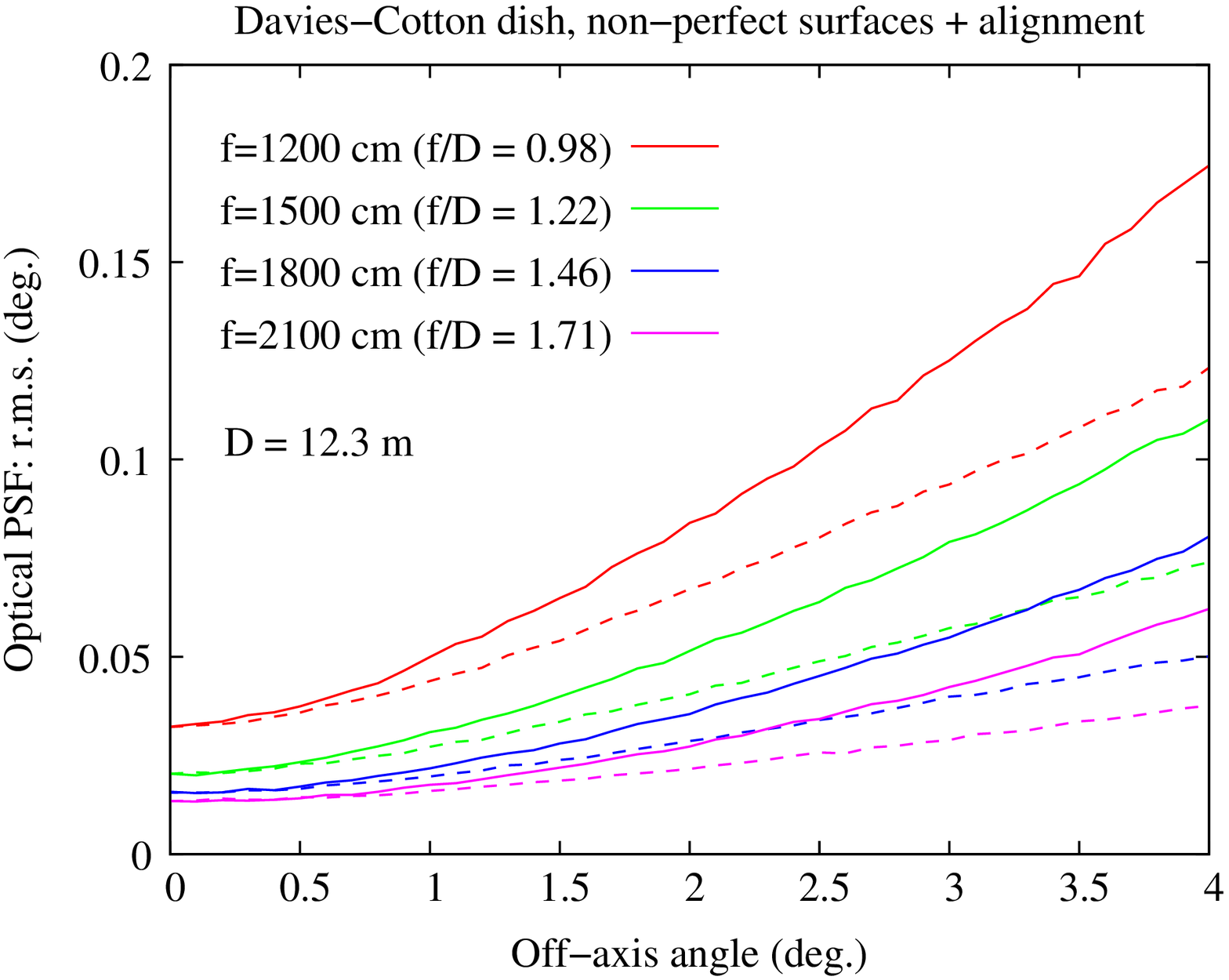}
\caption{\small PSF (RMS) as a function of field angle,
for a parabolic dish of different
$f/d$ (left) and for a Davies-Cotton dish (right).
Full lines represent the radial component of the PSF, dashed
lines the transverse component.}
\label{fig:optics}
\end{figure}

%Fig.~\ref{fig:optics} illustrates how the PSF varies across the field of view, for different values
%of $F$.
The expressions given above assume perfect shapes of the mirror facets, and
very small facets for the Davies-Cotton design.
In real applications, individual mirror facets will have an intrinsic spot size,
which to a first approximation must be added quadratically to the PSFs given above.
%Similarly, imperfections in the alignment of facets introduce a constant additional term, but actuator-based
%alignment schemes have been demonstrated to keep this term at an uncritical
%level \cite{hessalignment,magicalignment}.
Parabolic mirrors
can be constructed using spherical facets with focal lengths that are adjusted in 2-3 steps, rather
than varying continuously according to their radial position.  The optimal radii
$r_1$ and $r_2$ for aspherical mirrors
at a distance $R$ from the optical axis of a parabolic dish of focal length $f$ are
$$
\frac{r_1}{2f} = \sqrt{1+\frac{R^2}{4f^2}} \approx 1 + \frac{R^2}{8f^2}
{\rm \qquad and \qquad}
\frac{r_2}{2f} = \sqrt{\left(1+\frac{R^2}{4f^2}\right)^3} \approx 1 + \frac{3R^2}{8f^2} \quad .
$$
Use of spherical facets will cause a typical
contribution to the spot size of order $(d/8f)^2$, equivalent to that caused by the typical spread of
1\% in facet focal length.
Effects on the PSF are hence modest. The same
holds for the influence of the facet size in the Davies-Cotton layout, as long as the number of
facets is still large.
Fig.~\ref{fig:optics} illustrates how the PSF varies across the field of
view, for different values of $f/d$, based on a realistic
Monte Carlo simulation, including the effects of the PSF of the
individual mirrors, the alignment inaccuracy, and the use of
spherical mirror facets for the parabolic reflector.
 
For the SST and MST, among single-reflector designs
a Davies-Cotton geometry provides the best imaging over a large field of view. For the
LST only a parabolic dish is possible due to the large time dispersion a Davies-Cotton design would introduce.
To achieve a PSF of 3' over a $7^\circ$ field of view, an $F$ value of about 1.5 is required.
 
\begin{figure}[hbtp]
\centering
\epsfig{width=\linewidth,file=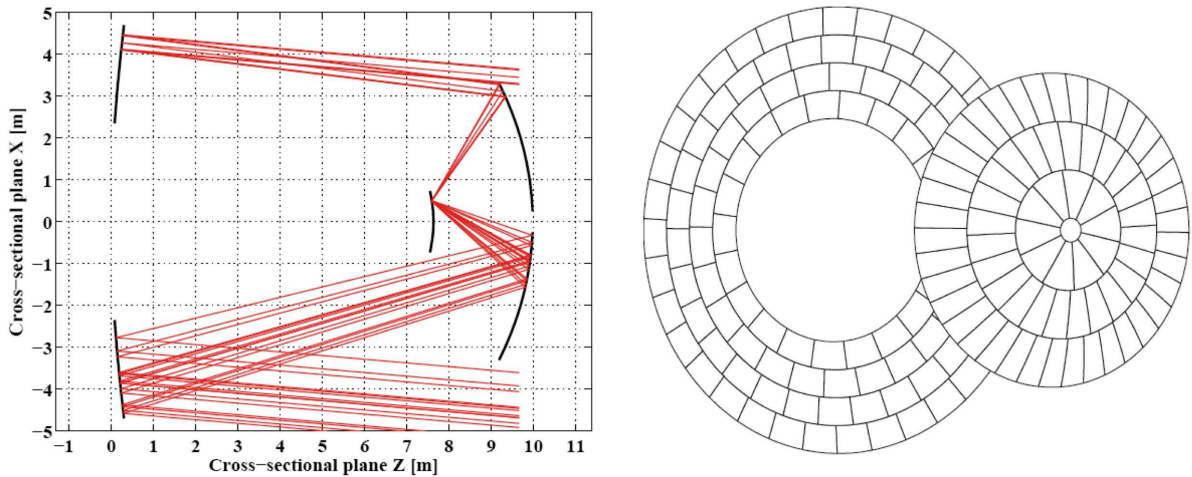}
\caption{\small Dual-reflector optics design for Cherenkov telescopes providing an
improved PSF over a large field
of view combined with a small plate scale \cite{vassiliev}.}
\label{fig:dual_mirror}
\end{figure}

Dual-mirror telescopes have so far not been used in Cherenkov astronomy, but obviously allow improved
compensation of optical errors over a wide field of view. In \cite{vassiliev} dual-reflector designs
are discussed in depth, with particular emphasis on the Schwarzschild-Couder design which combines a small plate
scale (adapted to the use of multi-anode PMTs as photo-sensors) with a 3' PSF across a $5^\circ$ radius field
of view (see fig.~\ref{fig:dual_mirror}).
Compared to single-reflector designs, where the camera has to be supported at a large
distance $F\cdot d$ from the dish, the dual-reflector design is quite compact. Drawbacks include the fact that
non-spherical mirrors are needed, which are more difficult to fabricate, and that the tolerances on the relative
alignment of optical elements are rather tight. Also, the large secondary reflector results in
significant shadowing of the primary reflector.
CTA's US collaborators, together with some European groups,
plan to build a Schwarzschild-Couder telescope of 12\,m.
While current CTA designs are based on single-reflector telescopes,
a dual-reflector construction could be adopted in particular for the SST or the MST,
should the developments prove promising.
 
To realise the PSFs given above, obviously the orientation of the mirror facets has to be stable to
a fraction of the
PSF under varying dish orientations, temperatures, temperature gradients, and wind loads.
Due to the reflection, orientation errors enter with a factor of 2 into the PSF. The facet
orientation can be stabilised either by using a rigid dish, or by
active compensation of dish deformations. For example, the mechanical structure of the H.E.S.S. telescopes
is designed to keep the  facet orientation stable to within 0.14\,mrad (0.5') RMS over the elevation range
45-90$^\circ$ and the operational range for  wind loads and temperatures
\cite{Ref_Mir_HessOpticsI}. In MAGIC, an active mirror alignment system compensates for dish
deformations \cite{Ref_Mir_Biland}.
The initial alignment of the mirror facets, as well as the calibration of active systems, is usually
carried out using images of bright stars
and has been demonstrated to have a precision well below the typical 3' PSF
(e.g. \cite{hess_optics2}).
 
Of additional interest is the precision with which the real dish shape needs to approximate the
ideal shape. Use of straight beam segments to approximate a curved dish may simplify production
considerably. Two effects matter: an otherwise ideal facet
displaced by $\delta z$ along the optical axis will generate a spot of angular diameter
$\Delta \zeta = d_{\rm facet} \delta z / f^2$ where $d_{\rm facet}$ is the facet diameter. The
corresponding RMS is $\sigma_\zeta = \Delta \zeta /4$. Typically, facets have a PSF
of 1\,mrad diameter or better. Limiting additional contributions due to imperfect facet placement
to 1\,mrad, which implies that they matter only near the centre of the field of view, where
the facet PSF dominates imaging, one finds $\delta z < 10^{-3} f^2/d_{\rm facet}$, or 0.2\,m
for $f = 15$\,m and
$d_{\rm facet} = 1$\,m. If the focal distance is wrong for a given facet, the spot location
for off-axis rays will also be shifted by $\Delta \zeta = \delta \Delta z / F$,
which should again be small compared to the spot size, for typically requiring
$\Delta z < 0.1$\,m.
Another limit comes from the time dispersion introduced by this deviation,
which is $\Delta T = 2 \Delta z/c$, implying that $\Delta z$ should not exceed 0.1-0.2\,m.
In summary, mirror placement along the optical axis should be within 10\,cm of the nominal
position for the MST.

\subsubsection{Mirror Facets}
 \label{sec:facets}
 
\noindent
Because of its large size, the reflector of a Cherenkov telescope is composed of many
individual mirror facets.  It is therefore important to balance the ease and cost of production
techniques against the required optical precision.
In total, CTA will need of the order of $10^4$\,m$^2$ of mirror area,
an order of magnitude more than current instruments.
As the telescopes are required to observe the Cherenkov light emitted from the many particle tracks of
an extensive air shower,  the necessary optical precision of the mirror system
is relatively relaxed. Focusing
can be worse than is required for mirrors for optical astronomy
by about two orders of magnitude, and the distance of the mirror facets to the focal plane
needs to be correct only to within a few cm, as opposed to the sub-wavelength precision
needed in optical astronomy.
 
The mirror facets for CTA will probably have a hexagonal shape and dimensions of 1-2.5\,m$^2$.
Large mirror facets have the advantage that they
reduce the number of facets on a dish and the number of support
points and alignment elements required.
On the other hand, in particular for Davies-Cotton optics, the
optical performance worsens as mirror facets
become larger. Also, the choice of manufacturing technologies
becomes rather limited. For these reasons,
the current baseline for the MST is to use hexagonal mirrors
of 1.2\,m (flat-to-flat) diameter.
Performance criteria for facets are equivalent to those for
current instruments as regards the spot size, the reflectance
and requirements on the long-term durability. The reflected
light should largely be contained in a 1 mrad diameter area,
the reflectance in the 300-600\,nm range
must exceed 80\%,
and facets must be robust against ageing when exposed to the
environment at the chosen site for several years.
Spherical facets are in most cases a sufficiently accurate approximation.
For a parabolic dish, a variation
of facet focal length with distance from the dish centre may be considered,
although gains are modest for a dish with
relatively large $f/d$.
 
Several
technologies for the production of mirror facets for Cherenkov
experiments were used in the past, or are under development at
present. These can be divided into two classes:
technologies using grinding/polishing or milling of
individual mirrors, as used for most current instruments; and
replication
techniques, where mirrors are manufactured using a mould or template,
which has obvious advantages for mass production.
 
\begin{figure}[hbtp]
\centering
\epsfig{width=0.4\linewidth,file=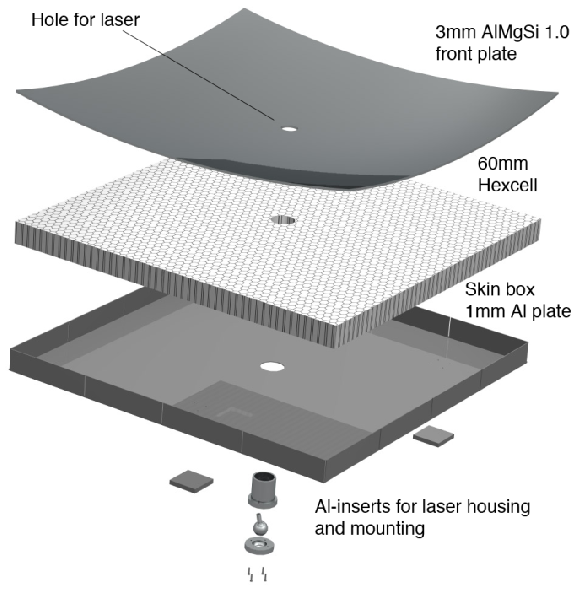} \\
\epsfig{width=0.49\linewidth,file=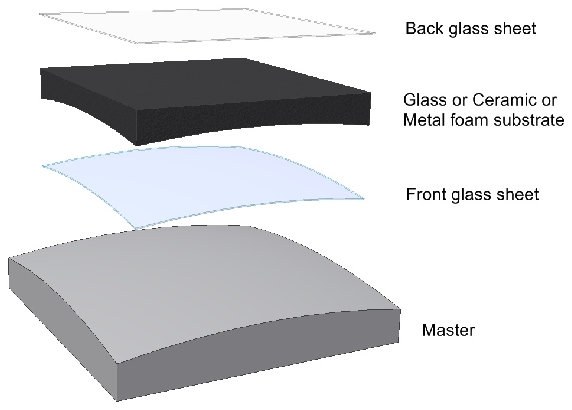} \hfill
\epsfig{width=0.49\linewidth,file=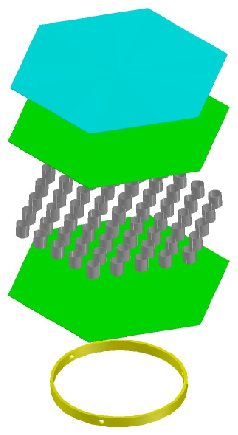} \\
\epsfig{width=0.49\linewidth,file=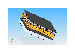} \hfill
\epsfig{width=0.3\linewidth,file=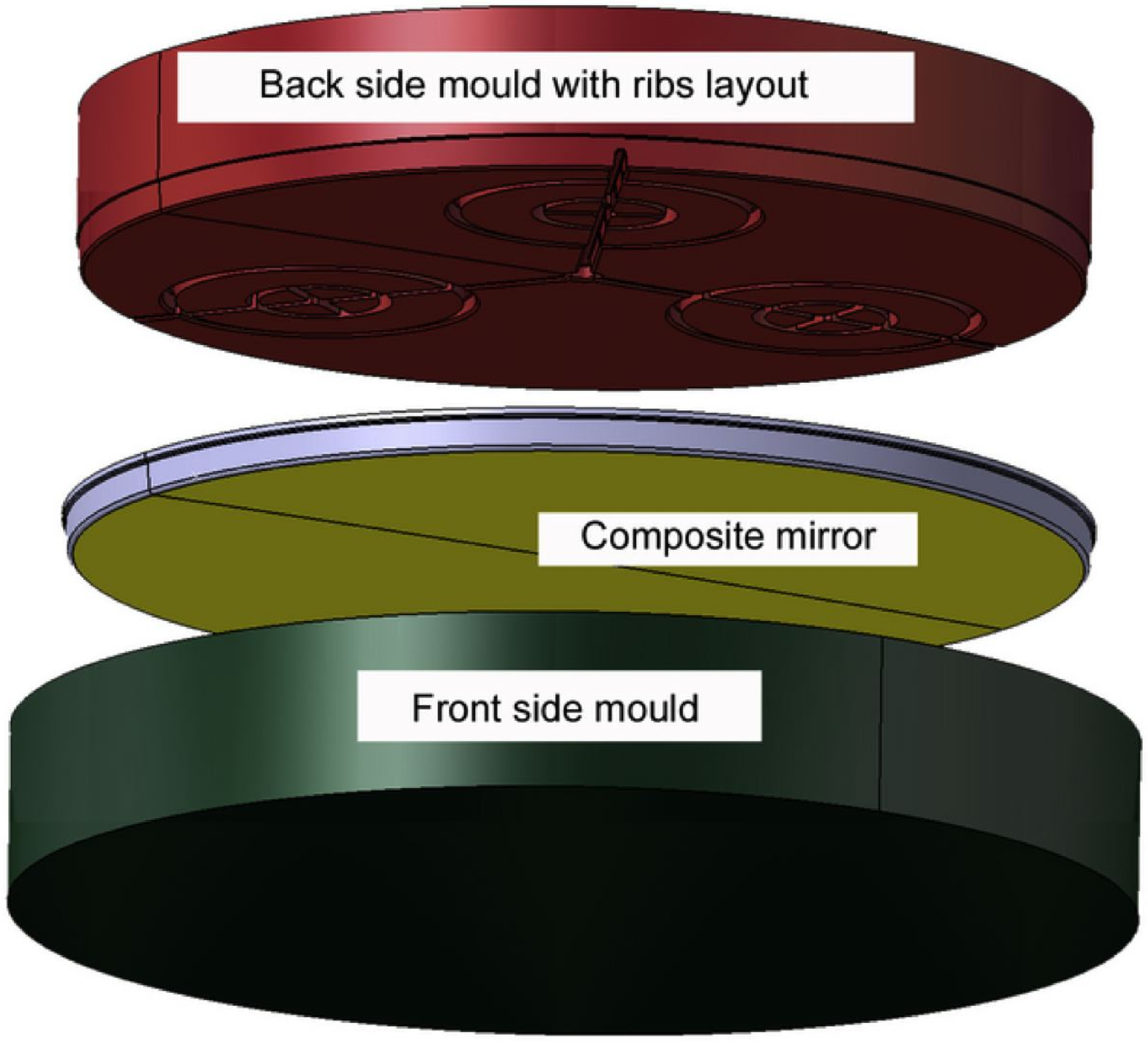} \hfill
\caption{\small Various mirror types under consideration for CTA:
{\bf top:} Diamond-milled aluminium honeycomb mirrors.
{\bf middle left:} Cold slumped glass-foam sandwich mirrors.
{\bf middle right:} Open fibre-reinforced plastics mirror (carbon fibre or glass fibre).
{\bf lower left:} Carbon-fibre composite mirror with CFRP honeycomb.
{\bf lower right:} Carbon-fibre composite mirror produced with SMC technology.}
\label{fig:mirrortypes}
\end{figure}

Facet types produced using
grinding or milling techniques include:

\begin{description}
\item[Glass mirrors] which have been the standard solutions for many past and present
Cherenkov telescope (e.g. HEGRA, CAT, H.E.S.S., VERITAS). The mirrors were produced
from machined and polished
glass blanks that were front-coated in vacuum with aluminium and some
weather-resistant transparent protection layer, such as vacuum deposited SiO$_2$ (HEGRA, CAT, H.E.S.S.),
or alternatively Al$_2$O$_3$ applied by
anodisation (VERITAS). These mirrors exhibit high reflectivity and good PSFs and
there is extensive production experience. Drawbacks are their fragility and weight, in
particular if facets of $\ge$1\,m$^2$ are considered. Their front-side coating shows relatively
fast ageing and degradation when exposed permanently to the wind and weather. A
typical degradation of the reflectance of around 5\% per year is observed
for a single $\sim100$\,nm SiO$2$ protection layer. Production and handling of thin (few cm) and large
(1\,m$^2$) facets is non-trivial.
\item[Diamond-milled aluminium mirrors] are used in the
MAGIC telescopes \cite{Ref_Mir_Doro}; these light-weight
mirrors are composed of a sandwich of two thin aluminium layers, separated
by an aluminium hexcell
honeycomb structure that ensures rigidity, high temperature conductivity and
low weight (see fig.~\ref{fig:mirrortypes}). After a rough pre-milling
that ensures
approximately the right curvature of the aluminium surface, the mirror is
precisely machined using diamond-milling techniques. A thin layer of quartz
of $\sim100$\,nm thickness,
with some carbon admixture,  is plasma coated on the mirror surface
for protection against corrosion. Diamond-milled mirrors have proven more resistant
to ageing effects (reflectance loss of 1-2\% per year)
than mirrors with a thin reflective coatings on glass or other substrates,
presumably since the reflective
layer cannot be locally destroyed.
On the other hand, the initial reflectance of diamond-milled mirrors is
a few percent lower.
\end{description}

An ongoing development is the mass production of mirror
panels by means of replication technologies. These
are cost effective and can be
used to produce non-spherical and very
light-weight mirrors with good and reproducible optical quality.
Replication methods look to be
promising for the large-scale production of CTA mirror facets and will be
considered as the baseline design,
although long-term tests are still required.
Replica production methods include:
 
\begin{description}
\item[Cold slumped glass mirrors.]
The mirror panels are composed of two thin glass sheets (1-2 mm) glued
as to a suitable core material, giving a structure with the necessary rigidity.
Construction proceeds as fllows:
At room temperature, the front glass sheet is formed to the required
optical shape on a master by means of vacuum suction. The core material and the
second glass sheet are glued to it. After the curing of the glue, the
panel is released from the master, sealed and coated in the same way as a glass mirror.
Half of the mirrors of the MAGIC II telescope were produced with
this technology using an aluminium honeycomb Hexcell structure as core
material \cite{Ref_Mir_Pareschi}.
For CTA, other core materials are under investigation, such as various foams.
Especially promising is an all-glass closed-cell foam
that can be pre-machined to the required curvature
(see fig.~\ref{fig:mirrortypes}). Further investigation of the effects
of thermal insulation between the front and the back of the mirror
caused by foams is required.
 
\item[Aluminium foil mirrors.]
Aluminium
honeycomb sandwich mirrors with reflective
aluminium sheets of 1-2\,mm thickness
(made e.g. by the company Alanod) are also being studied in detail.
Their main limitation currently results
from the imperfect reflection properties of the aluminium foil.
\item[Fibre reinforced plastics mirrors.]
Several attempts are being made to use carbon- or glass-fibre
reinforced plastic
materials to produce light-weight mirror facets.
Three different technologies are currently under development for CTA:
(a) an open sandwich structure of glass-fibre or carbon-fibre
reinforced plastic, consisting of two flat plates and spacers
with either an
epoxy layer cast on one plate or a bent
thin glass sheet glued to it to form the mirror surface;
(b) a closed structure of two carbon-fibre reinforced plastic
plates bent to the required radius of curvature,
an intermediate pre-machined CFRP honeycomb for stability and
a thin glass sheet as reflecting surface; and
(c) a one-piece design using a compound containing carbon-fibre
and the high-temperature and high-pressure sheet moulding
(SMC) technology, which is frequently used in the automotive industry.
To form a smooth surface in the same production step an in-mould coating
technology is under investigation which would allow for production times
of the order of just a few minutes per substrate.
See fig.~\ref{fig:mirrortypes} for the different mirror types.
%Another replica technique used in particular in applications where thin or lightweight mirror facets
%are required uses carbon fibres embedded in expoxy as the main structural material, sometimes
%embedding a foam core. Two different replica techniques are under investigation for CTA: (a) composite
%mirrors with front and back side
%molded carbon fibre plates with a carbon fibre structure in between, and (a) mirrors where in a single mold
%the front surface and a rear supporting structure with ribs are formed. Like for glass surfaces,
%reflective and protection layers have to be added in separate production steps.
\end{description}
 
Since the mirrors are permanently exposed to the environment,
degradation of mirror reflectivity is a serious concern. In the case of
aluminium-coated mirrors, water can creep along the interface of the
glass and aluminium layer because the aluminium does not stick perfectly to
the glass surface and the protective layers often have pin holes.
In contrast, solid aluminium mirrors show
localised corrosion, which, even when deep, affects only a very
small fraction of the surface. Possible cures for the
glass mirrors could be intermediate layers improving adhesion, e.g.
of chromium or SiO, or more resistant protection layers, for example
multiple layers which reduce the probability of pin holes in the coating.
Multi-layer protective coatings could also be used to
enhance reflectivity in the relevant wavelength region.
These are under investigation as are purely dielectric coatings
without any aluminium, which consist of multiple layers with different
refractive indices. These latter can in principle provide reflectances of up to
98\% and would not suffer from the rather weak adhesion of aluminium to glass.
 
Another option to improve the mirror lifetime is to apply the reflective
coating to the protected back side
of a thin glass sheet, which could then be used in the replication techniques
described above. Disadvantages are transmission losses in the glass, the
requirement of a very uniform glass thickness, as the mould defines the shape
of the front side but the reflective layer is on the back,
and, in addition, icing problems due to radiation cooling of the front surface.
 
In summary, many different technologies for the production of
mirror facets are under
investigation. For several of them, large-scale production experience
exists already, others are in a development phase.
A challenge in mirror production will be to find the optimum compromise
between mirror lifetime and production costs.
Current production costs are
1650 \euro{}/m$^2$ for the 0.7 m$^2$ H.E.S.S. II glass mirrors, 2450 \euro{}/m$^2$
for the 1.0 m$^2$ MAGIC II milled-aluminium mirrors, and 2000 \euro{}/m$^2$
for the 1.0 m$^2$
MAGIC II cold-slumped glass mirrors.
The much larger
production scale of CTA and the use of optimised techniques is expected to result
in a significant reduction in cost,
in particular for the replication technologies.
Current baseline specifications for MST mirror facets are summarised in tab.~\ref{tab:mirrorspec}.
 
\begin{table}[htbp]
\centering
\begin{tabular}{|l|l|}
\hline
 shape & hexagonal \\
 size  & 1200 mm flat-to-flat \\
 type  & spherical (aspherical opt.)\\
 focal length & $\sim 16$\,m \\
 reflectance & $>$80\% between 300 and 600 nm \\
 spot size & $<$1\,mrad diameter (80\% containment)\\
\hline
\end{tabular}
\caption{\small Baseline specifications for mirror facets (MST)}
\label{tab:mirrorspec}
\end{table}
 
\subsubsection{Mirror Support and Alignment}
\label{sec:support}
 
To achieve design performance, mirror facets need to be aligned with a precision which is
about an order of magnitude better than the optical point spread function,
i.e. given the PSF requirement of $<$1\,mrad the alignment
precision needs to be
well below 0.1\,mrad or 100\,$\mu$m (assuming a typical 1\,m lever
arm between mirror support points). Various alignment methods
are in use for existing telescopes:
\begin{description}
\item[Manual alignment.] Using an appropriate adjustment mechanism, mirror facets are manually aligned
after mounting. For technical reasons, alignment is usually performed at or near
the stow position of the telescope.
Deformations in dish shape between the stow position and the average observation position
(at 60-70$^\circ$ elevation) can be compensated by
``misaligning'' mirrors by the appropriate amount in the stow position.
This scheme is used in the VERITAS telescopes \cite{Ref_Mir_Veritas}.
\item[Actuator-based alignment.] Initial alignment of mirrors is carried out by remote-con\-trol\-led
actuators, using the image of a star viewed on the camera lid by a CCD camera on the dish, and
implementing a feedback loop which moves all
facet spots to a common location. This scheme is employed by H.E.S.S. \cite{hess_optics2}.
\item[Active alignment.] Remote-controlled actuators are used not only for initial
alignment of facets, but also to compensate for deformations of the dish, in particular as
a function of elevation. If dish deformations are elastic,
reproducible and not very large compared to the point spread function, alignment corrections
can be based on a lookup table of actuator positions as a function of telescope pointing.
If deformations are large or inelastic,
a closed feedback loop can be implemented by actively monitoring facet pointing, using lasers
attached to each facet and
imaged onto a target in the focal plane. Active alignment is used by MAGIC \cite{Ref_Mir_Biland}.
\end{description}
Technically, the requirements for the actuator-based alignment and the active alignment
are very similar, the main
difference being that for active alignment of a significant fraction of facets need to be moved simultaneously or nearly simultaneously
as telescope pointing changes, requiring parallel rather than serial control of actuators
and a higher-capacity power supply.
Since manual alignment of the large number of CTA mirror facets is impractical,
certainly the medium-sized
and large telescopes will be equipped with actuators. The small and
medium sized telescopes will have mechanically stable dish structures which do
not necessarily require active control, but active (look-up table driven) mirror control could be
implemented to maintain optimum point spread function over the entire elevation range.
 
Desirable features for actuators include a movement range of at least 30\,mm
and a built-in relative or, better, absolute position encoder which allows
the actuator to be moved by an exact pre-defined amount. This is
particularly relevant for lookup-based corrections.
For active alignment, the positioning speed needs to be such, that
the changes in mirror alignment are performed within the time needed to
move the telescope to a different position. For actuator based alignment
it is sufficient to be able to perform an initial alignment within a few days and
possible re-alignments within a few hours.
When not moving, actuators should be self-blocking to avoid movements e.g. in the case of power failure.
The actuators need to perform reliably and without significant maintenance
over the expected lifetime of the CTA array of over 20\,years. (The 
mean time between failure (MBTF) should be 100 years.) 
 
\begin{figure}[hbtp]
\centering
\epsfig{width=0.9\linewidth,file=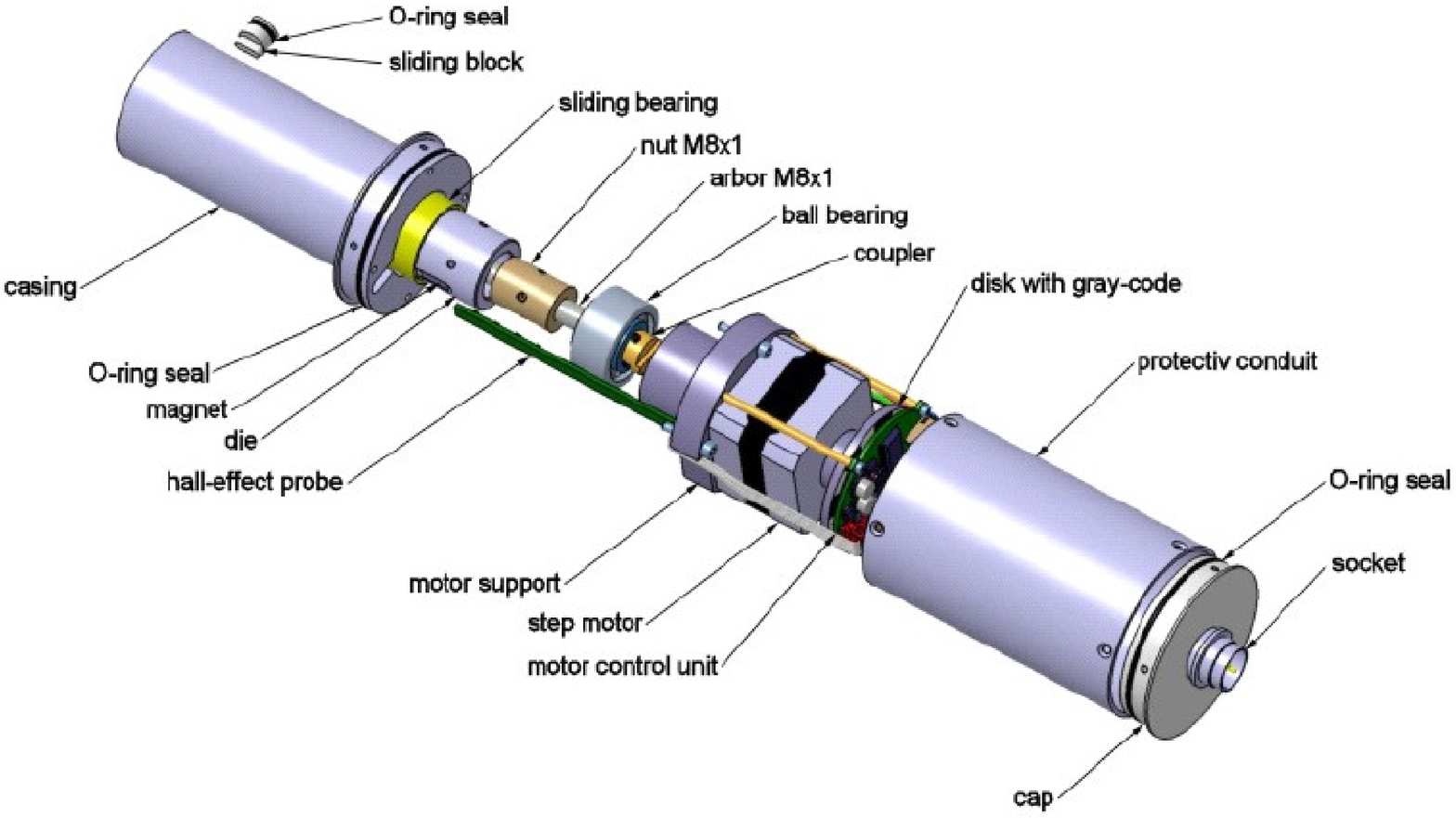}
\epsfig{width=0.7\linewidth,file=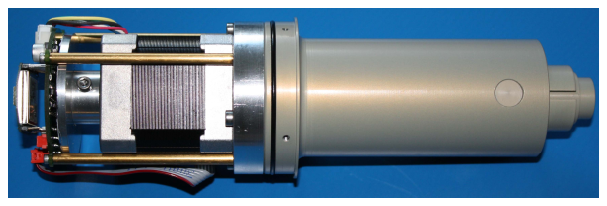}
\caption{\small Prototype mirror actuator based on a stepper-motor driven spindle, providing absolute position
encoding and a wireless control interface.}
\label{fig:actuator}
\end{figure}

Fig.~\ref{fig:actuator} shows a prototype actuator design based on a
spindle driven by a
stepper motor, with a combination of a digital Gray-code rotation encoder
and analogue signals from four Hall probes providing absolute position sensing.
The actuator is controlled by wireless communication using the Zigbee
industry standard, with each actuator identified by a unique (48 bit) code.
A broadcast mode is also available, which could be used to communicate the  current
elevation to all actuators allowing the controller to look up and apply the relevant individual
correction values.
 
\begin{figure}[p]
\centering
\epsfig{width=0.7\linewidth,file=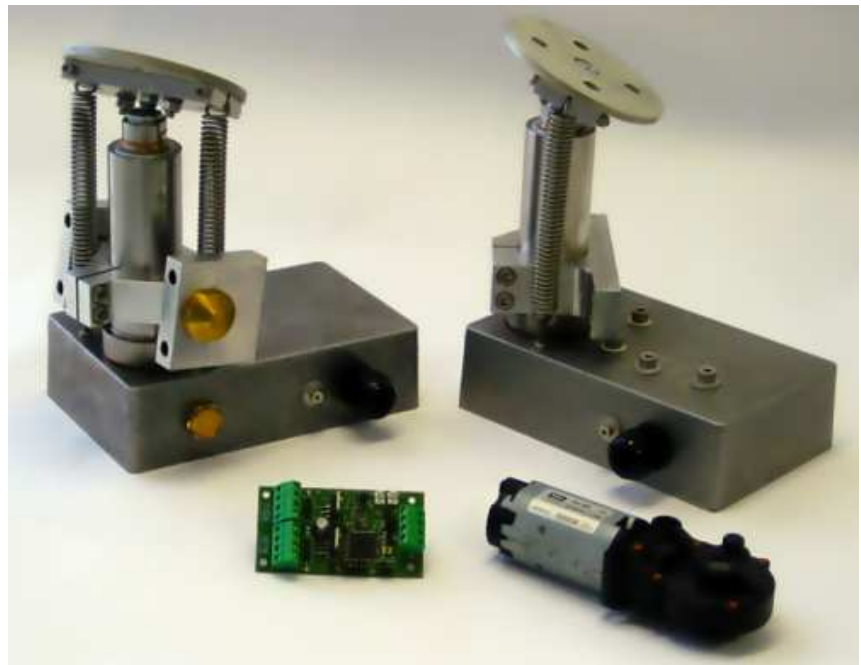}
\epsfig{width=0.7\linewidth,file=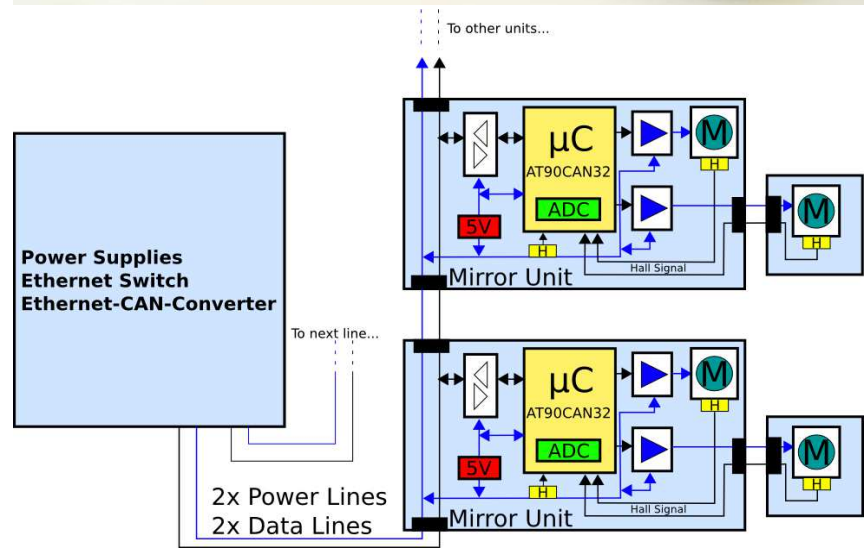}
\caption{\small Upper part: Prototype mirror control actuators, motor, and micro-controller
board for the solution based on relative position encoding.
Lower part: Electronics layout of the setup.}
\label{fig:amc_tuebingen}
\end{figure}

A second solution can be seen in fig.~\ref{fig:amc_tuebingen}.
The upper part of the figure shows the motor, two actuators, and the
micro-controller board of one mirror unit. This device uses servo motors with a Hall
sensor attached to the motor axis which makes possible relative positioning
of the actuator with high accuracy. The communication is based on CAN
(Controller Area Network), a multi-master broadcast serial bus standard
which is used in the automotive industry and other areas where there is demand for high
reliability. The communication of the telescope units with the control
computer is done via Ethernet. The electronics layout is depicted in
the lower part of the figure.
 
Mirror facets will be attached at three points, two equipped with
actuators and one universal joint. The facet
mounting scheme should allow the installation of the facets from the front,
without requiring access from the space-frame
side of the dish. This can be achieved by supporting mirrors at the
outer circumference, where attachment points
are easily accessible, or using screws or attachment bolts going
through the mirror.
Current baseline specifications for the mirror alignment system
are summarised in tab.~\ref{tab:actuatorspec}.
 
\begin{table}[p]
\centering
\begin{tabular}{|l|l|}
\hline
precision & $<$0.1 \, mrad \\
initial alignment & $<$ few days \\
re-alignment (actuator based alignment) & $<$ few hours \\
re-alignment (active alignment) & $<$ slewing time to new position \\
lifetime & $>$20 \, years \\
\hline
\end{tabular}
\caption{\small Baseline specifications for mirror alignment actuators}
\label{tab:actuatorspec}
\end{table}

%\input{Camera.tex}
%%%%% HEADER: Camera.tex
%%%%%%%%%%%%%%%%%%%%%%%%%%%%%%%%%%%%%%%%%%%%%%%%%%%%%
%%  streamlined and cut by Richard White.
 
%%%%%%%%%%%%%%%%%%%%%%%%%%%%%%%%%%%%%%%%%%%%%%%%%%%%%
\subsection{Photon Detection, Electronics, Triggering and Camera Integration}
\label{sec:camera}
%%%%%%%%%%%%%%%%%%%%%%%%%%%%%%%%%%%%%%%%%%%%%%%%%%%%%
 
The cameras developed for gamma-ray detections with current atmospheric Cherenkov
telescopes have reached the sensitivity required to perform detailed investigations
of many astrophysical sources. Further advancing  Cherenkov telescope performance requires,
in particular, that the energy range covered be extended, i.e. that
the gamma-ray energy threshold be reduced and detection capabilities be extended at high energies,
enhancing the flux sensitivity, and improving angular resolution and particle identification.
Lowering the threshold energy and
increasing the sensitivity of an IACT requires that more Cherenkov photons be collected and/or
that these are detected more efficiently.
The efficiency of the collection of Cherenkov photons and their conversion to photoelectrons in the
photo-sensor must therefore be improved;
the non-sensitive regions (dead areas) in the camera must be minimised, for example by
using light guides, the effective photon conversion efficiency
increased by exploiting
novel technical developments. Enlarging the energy range requires appropriate electronics
with a sufficiently large dynamic range.
 
Achieving the required performance necessitates the development and the production of electronics components
dedicated to CTA. Sophisticated application-specific integrated circuits (ASICs)
for equipping the front-end part of the readout chain
are under study. These have the advantage that they  minimise signal distortion, decrease the
power consumption and ultimately reduce the cost of the experiment considerably. Integrated readout
systems take advantage of the recent development of analogue memories for data buffering. An alternative
solution is a fully digital readout scheme. The amplified signal from the photon sensors is directly
digitised by an analogue-to-digital converter (ADC) and buffered in a deep memory. Readout and triggering
benefit from continuous data storage to avoid deadtime.
 
The integration of detectors and the associated electronics reduces the size of the apparatus and
embedded cameras have operational advantages, particularly at an isolated site and
given the number of about one hundred cameras that will be required for CTA.
The usage of complete spare cameras, rather than spare components for these cameras, can
significantly simplify the maintenance of the system.
The camera typically consists of a cylindrical structure built
completely from low mass components. It holds a matrix of photon sensor cells, carefully optimised to
make maximal use of the incoming light, and is fixed to the arms of the telescope in the focal
plane, above the dish of the telescope. For embedded cameras, the light
sensor, readout and trigger system, data acquisition system and power supply are integrated in a modular
mechanical structure. The only connections that enter the camera are the input power, the communication
network, and any central trigger cables.
Disadvantages are a heavy camera requiring considerable cooling power and a heavy camera support structure.
 
\subsubsection{Photon Detection}
\label{sec:photondetection}
 
The photon sensors most commonly used in IACTs are photomultipliers with alkali
photo cathodes and electron multipliers
based on a chain of dynodes. The technology is well-established, but is subject to continuous
development and improvement.
PMTs have established themselves as the best available low light level
sensors for ultra-fast processes. The relatively high peak quantum efficiency (QE) currently available
(up to 30\%), together with high gains of up to 10$^6$ and low noise, allow the reliable measurement even
of single photoelectrons. A dynamic range of about 5000 photoelectrons is obtainable with PMTs.
The PMTs convert impinging photons into a charge pulse of size measured in number of
photoelectrons. IACTs usually use PMTs with bialkali type photo-cathodes, as these provide
the highest QE. They are sensitive in the wavelength range of 300-600 nm
(200-600 nm if a PMT with a quartz window is used). The bialkali PMT sensitivity curve is well-matched
to the spectrum of Cherenkov light arriving at ground level from air showers. As a rule, one needs to
amplify this pulse in order to match the sensitivity of the data acquisition (DAQ) electronics.
However, new photon detectors are under study and the CTA cameras must be designed
to allow their integration if their performance and cost provide significant advantages over PMTs. \\
 
\begin{figure}[b]
\centering
\epsfig{width=\textwidth, file=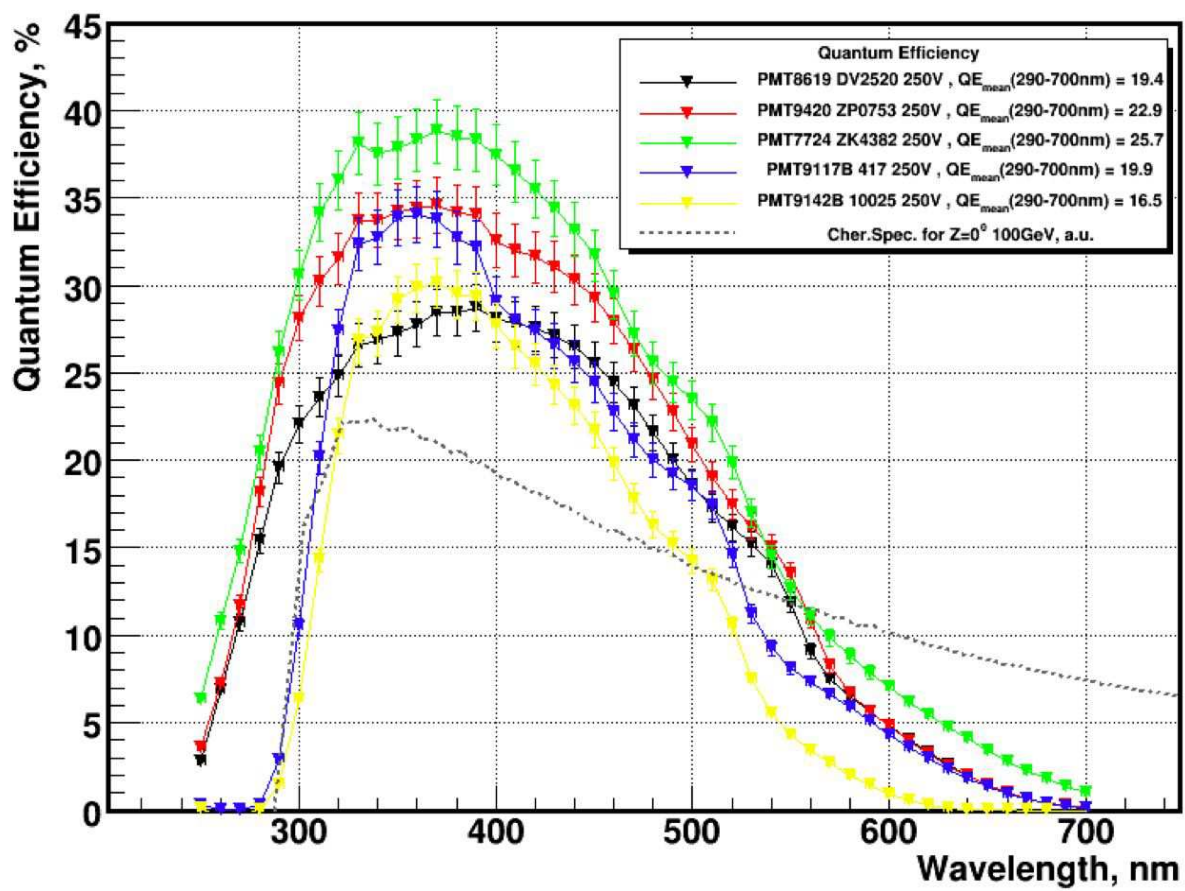}
\caption{\small
Spectral response of several types of super bi-alkali PMTs from Hamamatsu (green, red and black)
and Electron Tubes Enterprises (yellow and blue), compared to the spectrum of Cherenkov light
produced by vertical 100 GeV gamma rays on the ground (grey, dashed),
convoluted with the standard atmospheric transmission for the observation height of
2200 m a.s.l.. The numbers in the inset give the convolution of the
QE curve of a given PMT with the dashed line.}
\label{fig:spectral_response}
\end{figure}

%-----------------------------------------------------------------------------------------------------
\noindent {\bf Criteria for Photo-detectors:}
\begin{description}
 
\item[Spectral sensitivity:] The spectrum of Cherenkov light is cut off below 300\,nm, due to atmospheric
transmission effects, and falls off as $1/\lambda^2$ towards longer wavelengths\footnote{Due to 
Rayleigh and Mie scattering the actual spectrum at the camera is considerably flatter (see fig. 
\ref{fig:spectral_response}). At large zenith angles the UV part of the spectrum might no longer be detectable.}.  
Candidate photo-detectors
should be matched to the peak in this spectrum at around 350\,nm. At large wavelengths, beyond about 550\,nm,
the signal-to-noise ratio becomes increasingly unfavourable due of the increasing intensity of the
night sky background in this region. Above $\sim$650 nm strong emission lines are present in the
Cherenkov spectrum, originating from the rotational levels of (OH) groups. It is therefore desirable
but not essential to measure up to wavelengths of about 600-650\,nm.
(The more accurately the absolute charge in an image is measured, the better the absolute calibration.)
 
\item[Sensor area:] Currently favoured pixel sizes are around $0.1^\circ$ for the LST, $0.18^\circ$
for the MST, and $0.25^\circ$ for the SST. For conventional telescope designs (single mirror optics,
with Davies-Cotton or parabolic reflectors), these angular sizes translate to linear dimensions of
40\,mm, 50\,mm and 35\,mm,
respectively. If a secondary-optics design is used for the SST, a size of $0.2^\circ$ represents
around 6\,mm. For the secondary-optics design for the MST, a smaller angular pixel size of $0.07^\circ$
equates to the same physical size of 6\,mm. Light-collecting Winston cones in front of any sensor
reduce the required sensor size by a factor of 3 to 4 compared to the pixel size
and can decrease the amount of dead space between pixels.
 
\item[Sensor uniformity:] Sensor non-uniformities below $\sim$10\% are tolerable. Larger
non-uniformities should be avoided as they introduce an additional variable component in the light
collection and thus increase the variance of the output signal.
 
\item[Dynamic range and linearity:] Sensors should be able to detect single photons and provide a
dynamic range of up to 5000 photo-electrons, with linearity deviations below a few per cent. Non-linearities
can be tolerated if they can be accurately corrected for in the calibration procedure.
 
\item[Temporal response:] The time dispersion of Cherenkov photons across a camera image depends
on the energy of the primary gamma ray. At low energies, the dispersion is only few
nanoseconds.  Matched short signal integration windows are used to minimise the noise.
The photo-sensor must not significantly lengthen the time structure
of a Cherenkov light pulse. It is desirable
to determine the pulse arrival times with sub-nanosecond precision for sufficiently large light pulses.
 
\item[Lifetime:] Sensors will detect photons from the night-sky background at a typical rate of
about 100 to 200\,MHz for the telescopes with large collection areas (MST and LST). If operation
is attempted
when the moon is up, this rate can increase by an order of magnitude. Sensors should have a lifetime
of 10\,years for an annual exposure of up to $\sim$2000\,hours.
This can be achieved using PMTs with only 6 to 8 dynodes, operated at a gain of 30000 to 50000,
followed by a fast AC-coupled preamplifier.
 
\item [Rate of spurious signals:] Spurious signals from photo-detectors can result in an increase of
trigger rates and a degradation of trigger thresholds. This is a particular issue for photomultiplier sensors where
residual gas atoms in the tubes are ionised by impinging electrons. The resulting afterpulses, produced by
positively charged heavy ions bombarding the photo-cathode, may have large amplitude and long
delays relative to the primary electron. Photomultipliers should be selected with an afterpulse
probability below ($\sim10^{-4}-10^{-5}$).
 
\item[Operational characteristics:] To ensure efficient and reliable operation of the systems,
sensors should show good short- and medium-term stability, and only gradual ageing, if any. Sensors
should be able to survive high illumination levels.
 
\item[Cross-talk:] Although cross-talk for photomultipliers is very low, it may be an issue for
alternative sensor solutions such as silicon photomultipliers or multi-anode photomultipliers.
Crosstalk between adjacent pixels must be kept $\le 1$\%.
 
\item[Cost and manufacturing considerations:] In total, the CTA consortium is intending to use
$\sim 10^5$ sensor channels. Thus, the photo-detectors comprise a major fraction of the total
capital cost of the project and any innovations which allow their cost to be reduced should
be carefully considered. One important criterion is that the manufacturer/supplier
must be able to provide the necessary number
of sensors to the required specification with an acceptable and reliably known lead time.
 
\end{description}
% \clearpage
%-----------------------------------------------------------------------------------------------
\noindent {\bf Candidate Photo-detectors} \\
 
The baseline photo-detector for CTA is the PMT. However, there may be alternative solutions that
reach maturity on approximately the right timescale for CTA construction. Modular cameras for the
LST, MST and SST are therefore desirable to allow the exchange of photo-detectors without major
alterations to the trigger and readout electronics chain. In the case of a secondary optics design
of the MST or SST, conventional PMTs are not available in the appropriate physical size, and
therefore the choice of a secondary optics telescope design would depend heavily on the availability
of alternative photo-detectors, such as those presented here.
\smallskip
 
\noindent {\it Baseline Solution - Photomultipliers}: The spectral sensitivity of conventional
PMTs, see fig.~\ref{fig:spectral_response}, with their falling sensitivity at
large wavelengths, provides a reasonably good match to the spectrum of Cherenkov light on the ground.
The baseline solution for CTA is to use PMTs with enhanced quantum efficiency compared to those
currently used in H.E.S.S., for example. Such tubes are becoming commercially available and offer
$\sim$50\% advantage in photon detection efficiency over conventional PMTs.
\smallskip
 
\noindent {\it Silicon Photomultipliers (SiPMs)}: (known also as MPPCs, GAPDs and
Micro-channel APDs) are novel light sensors that are rapidly reaching maturity. The more recent SiPMs consist of
single pixels which contain several hundred to thousands cells, coupled to a single output,
Each cell is operated in Geiger mode. An arriving photon can trigger the
cell, after which that cells suffers significant deadtime, but leaves the surrounding cells
ready to collect other arriving photons. The
photon-counting dynamic range is comparable to the number of cells. Silicon photo-sensors
could provide higher photon detection efficiencies than the latest PMTs at lower cost and without
the requirement for high-voltage. However, silicon sensors typically require cooling to reduce the dark count to a manageable
level and also suffer from optical cross-talk and are not as well matched to the Cherenkov light spectrum
as PMTs. They therefore require further improvement and commercialisation. However, depending on
the time scale and cost of such a development, SiPMs could be considered as a candidate sensor for
replacing the PMTs or, alternatively, as an upgrade path for all telescope sizes. They are of particular
interest for the SST secondary optics option, where their physical size is better suited to the plate
scale of the telescope.
\smallskip
 
\noindent {\it Multi-Anode Photomultipliers:} MAPMTs
provide multiple pixels in a compact package, with properties similar to monolithic PMTs. Such
devices offer individual pixel sizes of the order of 6 mm, suitable for secondary optics schemes.
Enhanced quantum efficiency versions with up to 64 channels are now available. The suitability of
MAPMTs must be assessed, and properties such as the uniformity, cross-talk, dynamic range and detection
efficiency are currently under investigation. \\
 
%----------------------------------------------------------------------------------------------
\noindent{\bf Associated Systems}
\smallskip
 
\begin{figure}[hbtp]
\centering
\epsfig{height=0.39\textwidth,file=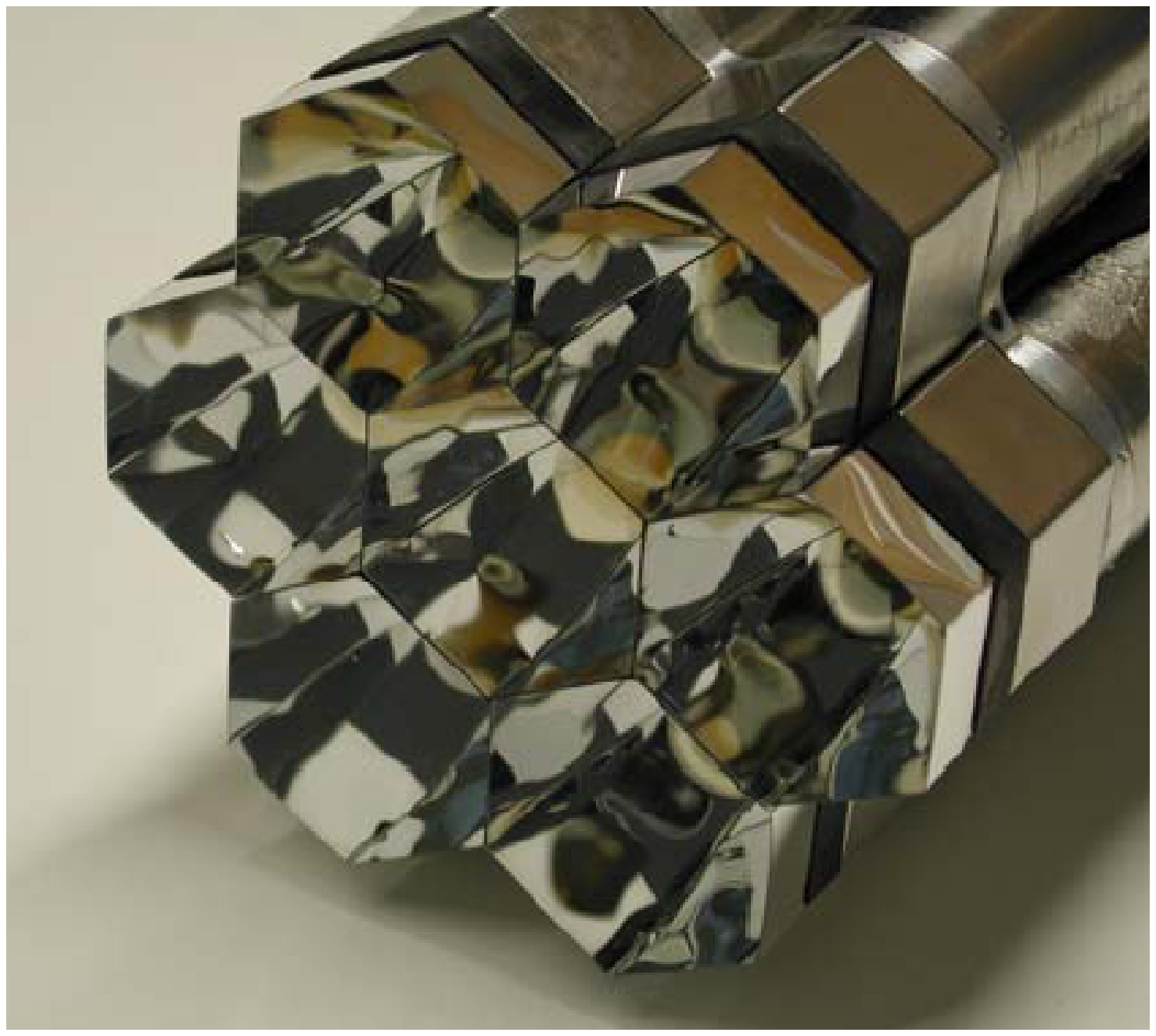} \hfill
\epsfig{height=0.39\textwidth,file=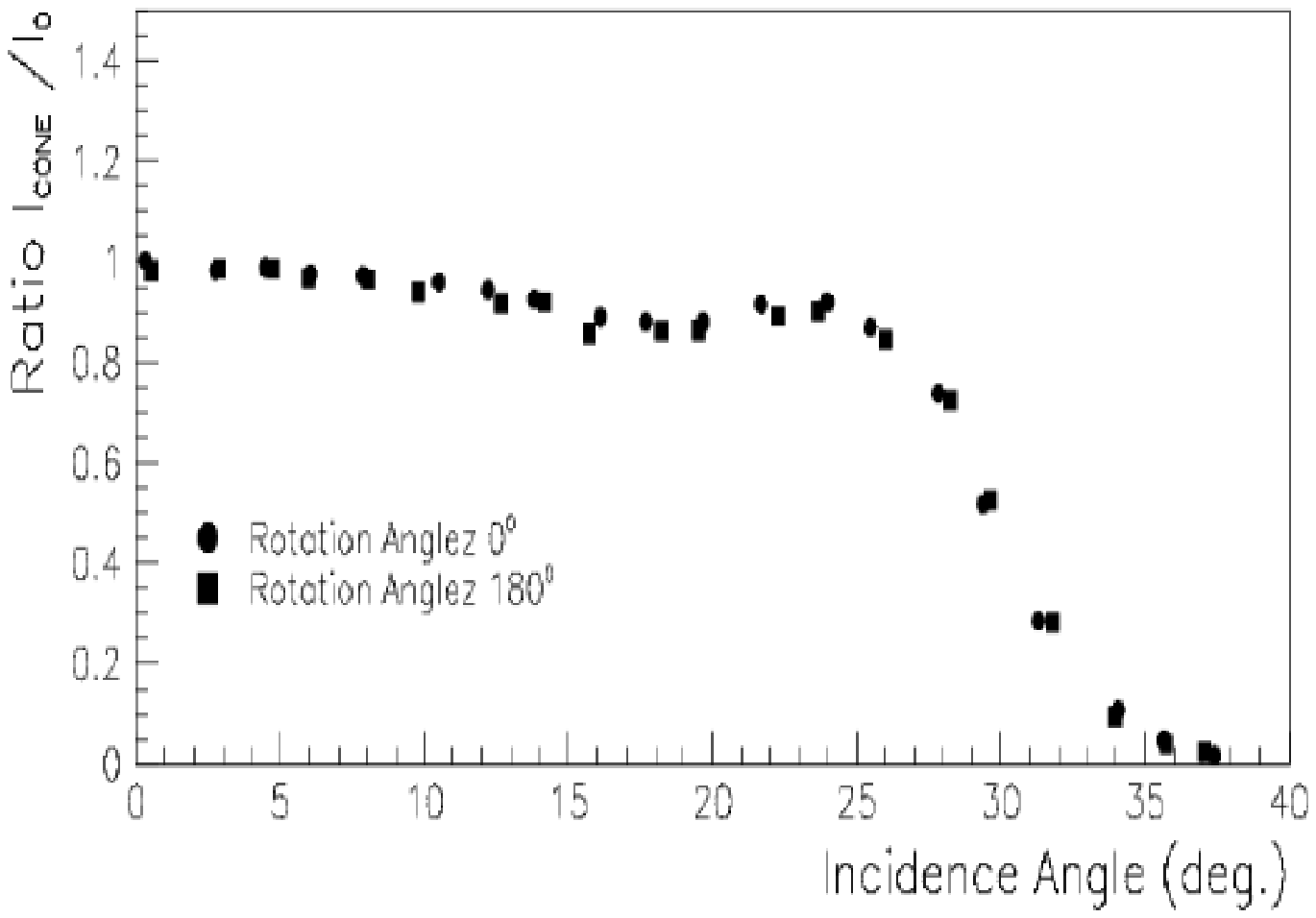}
\caption{\small {\bf Left:} PMT pixel cluster with light funnels.
{\bf Right:} Angular response of a typical light funnel,
normalised to the on-axis response.}
\label{fig:cone}
\end{figure}

\noindent {\it Light-collecting Winston cones}: Winston cones placed in front of any sensor
could reduce the
required sensor size by a factor of 3 to 4 (see fig. \ref{fig:cone}).
However, compression is limited by Liouville's theorem, which states
that the phase-space volume of an ensemble of photons
%$\int n^2 \cos \theta\,dS\,d\Omega$
is conserved.
Lightcones can minimise the dead space between pixels
and reduce the amount of stray light from the night sky impinging on the sensors at large incidence angles.
Fig.~\ref{fig:cone} illustrates the typical angular response of a light funnel.
Current lightcones have a net transmission of about 80\%. Improved cones may
allow increased performance at modest cost.
\smallskip
 
\noindent {\it Plexiglas input window}: To avoid the deposition of dust on the photo-detectors and
lightcones (if used), a Plexiglas window could be utilised to seal the camera. The transmission
losses of $\sim$8\% for a 3\,mm thick GS 2458 window may be considered as well-justified because of the
absence of deterioration of the light throughput on long time scales. The use of a sealed camera
and plexiglas window must be investigated for each telescope size.
For the LSTs and  MSTs, sealing the camera does not significantly
increase its total cost.
For the SST, a sealed system may represent a significant proportion of the
cost for the camera, but give advantages in maintenance and long-term performance.
\smallskip
 
\noindent {\it HV supply}: PMTs and MAPMTs need to be provided with a stable and adjustable
high voltage supply. The first dynodes are often supplied through a passive divider chain,
the last using
an active divider to provide more power, improving the dynamic range and allowing stabilisation.
The HV system also needs to provide a current-limiter or over-current
trip circuit for protection in case of excessive illumination of the PMTs, due to bright stars,
moon shine or, even worse, daylight. Several options are under study for CTA:
a) Cockcroft-Walton type, b)
transistor-based active divider type and c) one central power supply providing individually
attenuated voltages to different channels.
 
\subsubsection{Electronics}
\label{sec:electronics}
 
%---------------------------------------------------------------------------------
\noindent {\bf Signal Recording Electronics}
\smallskip
 
Air-shower induced photo-sensor signals have a pulse width of a few ns, superimposed on a random
night sky background with typical rates of some 10 MHz to more than 100 MHz, depending on mirror
size and pixel size (which is therefore different for the LST, MST and SST). Optimum capture of
air-shower signals implies high bandwidth and short integration times. Ideally, the dynamic range
and noise should be such that single photoelectron signals are resolved, and signals of a few
thousand photoelectrons are captured without truncation. The recording electronics must delay or store
the signals whilst a trigger is generated, indicating that the event is to be captured and
read out. The generation of a trigger signal could take from 0.1 to a few $\mu$s within a
single telescope, depending on the complexity of the trigger scheme, and $\ge$10 $\mu$s  if
trigger signals between several telescopes are combined.
 
Advances in signal recording and processing provide the possibility of recording a range of
signal parameters, from the integrated charge, to the full pulse shape over a fixed time window. Whilst
it is not yet clear that the full pulse shape is needed, it is desirable to record at least a few
parameters of the pulse shape rather than just the integrated charge. In this way, absolute timing
information would be available, allowing improved background rejection and
adaptive integration windows.
Increasing the bandwidth of the signal recording system will allow improved timing and shorter
integration gates, resulting in reduced levels of night sky background under the signal. However,
as the bandwidth of the system is increased, so is the cost. Whilst such an approach may be justified
for the LST, where night sky background is high, the Cherenkov pulses are very fast and the number of
telescopes is low, this is not necessarily the case for the SST, where the night sky background is low,
the Cherenkov pulses are not as fast and any cost savings could be used to build more telescopes.
The bandwidth of the electronics chain for a given telescope size should be motivated by examining
its consequences for the array sensitivity and energy threshold through Monte Carlo simulations.
Currently, there is no clear answer as to the optimum choice for any telescope size, and the signal
sampling frequencies under discussion range from a few 100 MSample/s to $\sim$2 GSample/s.
 
Two techniques for signal recording and processing are in use in existing IACT arrays,
These are
based around Flash Analogue-to-Digital Converters (FADCs) and analogue sampling memories, and form
the basis for the CTA development:
 
\smallskip
\begin{figure}[hbtp]
\begin{center}
\epsfig{width=0.7\textwidth, file=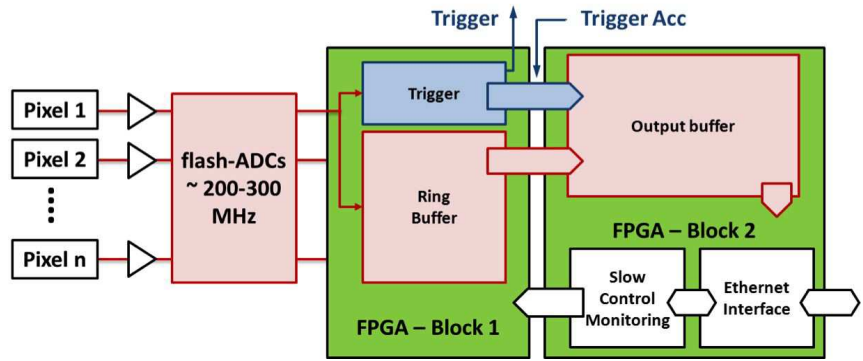}
\end{center}
\caption{FADC based recording systems with purely digital trigger acting on the digitised data.}
\label{fig:elecxyz}
\end{figure}
\noindent {\it Flash Analogue-to-Digital Converters}: FADCs digitise
the photon-sensor signals at rates of a few 100 MSample/s to a few GSample/s, writing the output into
a digital ring buffer, often realised as a very large scale integration gate array
which also provides control logic
and digital readout. The modest cost of digital buffers allows large trigger latency; delays
of tens of microseconds can be realised. However, the dynamic range of FADCs is limited and
typically no more than 8 to 10 bits are available, requiring either parallel conversion with
different gains or dynamic gain switching, as used in the 500 MSample/s, 8-bit VERITAS
FADC system  \cite{veritas_fadc}. The rather high cost of the fastest FADCs has led to
the development of systems in which several channels are time-multiplexed onto one ADC, as used
in the MAGIC 2 GSample/s, 10-bit FADC system.
%\cite{astro-ph07092363}.
In principle,
FADC based recording systems allow the use of a purely digital trigger, acting on the digitised
data in the ring buffer, to select air shower events. Such a system is sketched
in fig. \ref{fig:elecxyz}.
None of the systems implemented so far uses this approach. Instead,
parallel analogue trigger circuitry is used, adding not insignificant complexity to the electronics
layout. The steadily increasing power of VLSI gate arrays may soon make
digital trigger processors an attractive and feasible option.
 
As well as being expensive, FADCs suitable for IACTs are traditionally also bulky and power
hungry, negating the possibility of integrating the readout electronics into the camera and
requiring the transmission of analogue signals over many tens of meters to a counting house.
However, the recent development of low-power, low-cost FADCs in recent years imply this situation
may be changing, at least for modest-speed FADCs. In response to this, a 250
MSample/s system, named FlashCam, is under development for CTA. Monte Carlo
simulations have shown that, at least for the MST and SST, 250 MSample/s is a sufficiently fast sampling rate
to allow correct pulse shape reconstruction.
Hardware prototyping is under way to confirm this simulated result. The sensitivity of
the complete array with such a readout system must still be assessed.
 
\smallskip
\noindent {\it Analogue Sampling Memories}: Analogue sampling memories consist of banks of
switched capacitors which are used in turn to record the signal shape. The maximum recording depth is given
by the sampling time multiplied by the number of storage capacitors, which ranges from 128 to
a few 1000, implying at most a few microseconds of trigger latency. Trigger signals are
derived using additional analogue trigger circuits.
 
\begin{figure}[hbtp]
\begin{center}
\epsfig{width=0.7\textwidth, file=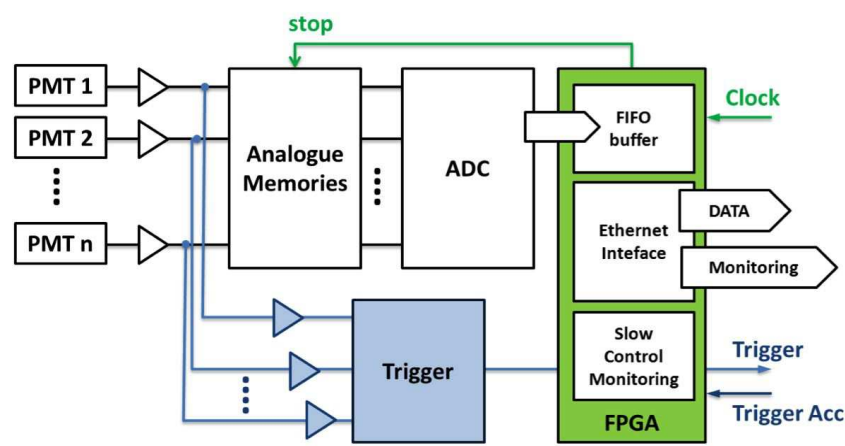}
\end{center}
\caption{Analogue memory based recording systems. The analogue trigger is formed in parallel
to the data shaping and buffering.}
\label{fig:elecxxx}
\end{figure}
Current ASIC implementations stop the recording of signals after a camera trigger and initiate the
digitisation of the charge stored on a selected range of capacitors, thereby introducing front-end
deadtime of few microseconds. The signal is then converted to a digital format using an ADC and
can be stored in a local Field-Programmable Gate Array (FPGA) before transfer (see fig. \ref{fig:elecxxx}).
The ADC is typically used to
digitise the pulse integrated over a time window, and therefore can have a
sampling frequency an order of magnitude lower
than those considered in the FADC readout scheme. Additional information,
such as the pulse width and arrival time, can also be stored, which is highly desirable.
A First-in, First-out (FIFO)
memory between the digital conversion and the FPGA can be used to smooth the distribution of
arrival times of events to reduce fluctuations in the data acquisition rate.
 
The dynamic range of analogue samplers is up to 12 bits. As with FADC-based systems,
parallel channels with different gains or non-linear input stages can be employed to record
a larger dynamic range of Cherenkov signals. Examples of such systems include the HESS I
readout system, which is based on the ARS ASIC \cite{ars},
the HESS II readout, based on the further
developed Swift Analogue Memory (SAM)
ASIC with significantly reduced readout deadtime \cite{sam}, and the MAGIC II
readout system, using the Domino Ring Sampler (DRS) ASIC \cite{drs}.
Several analogue sampling based schemes are
under development for CTA, including a project based on the next-generation version of the
SAM chip, termed NECTAr, a DRS4 based project called Dragon and a project based on the Target
ASIC originally intended for AGIS. The main parameters of some of these ASICs are summarised
in tab.~\ref{tab:asics}.
\begin{table}[htbp]
\begin{center}
\small
\begin{tabular}{|l|c|c|c|c|c|}
\hline
 & ARS & SAM & NECTAR & DRS4 & TARGET\\
\hline
GSamples/s             & 1 & 1 - 3.2 & 1 - 3.2 & 0.2 - 6  & 1 \\
Channels/chip          & 5 & 2 & TBD & 8 &16\\
Samples/channel        & 128 & 256 & $>$1024 & 1024 &16000\\
Analogue bandwidth (MHz) & 80 & $>$400 & $300$ & 950 & 380\\
Dynamic range (bits)     & 9 - 10 & 12 & $\ge$12  & 11 - 12 & 9 - 10\\
Integrated trigger disc. & no & no & no & no & yes\\
Integrated ADC           & no & no & yes & no & yes \\
Integrated digital control logic & (PLL) & yes & yes & (PLL) & yes \\
Typ. readout latency ($\mu$s) & 60 & $<$2 & $<$2 & $<$3 & $<$3 \\
Power cons. (mW) & 500 & 300 & 150 - 300 & 65 & 10 \\
Status/use & HESS I & HESS II & design & MAGIC II & design \\
\hline
\end{tabular}
\normalsize
\caption{\small Characteristics of switched-capacitor signal-recording ASICs.}
\label{tab:asics}
\end{center}
\end{table}
 
\begin{figure}[htbp]
\begin{center}
  \epsfig{width=0.7\textwidth,file=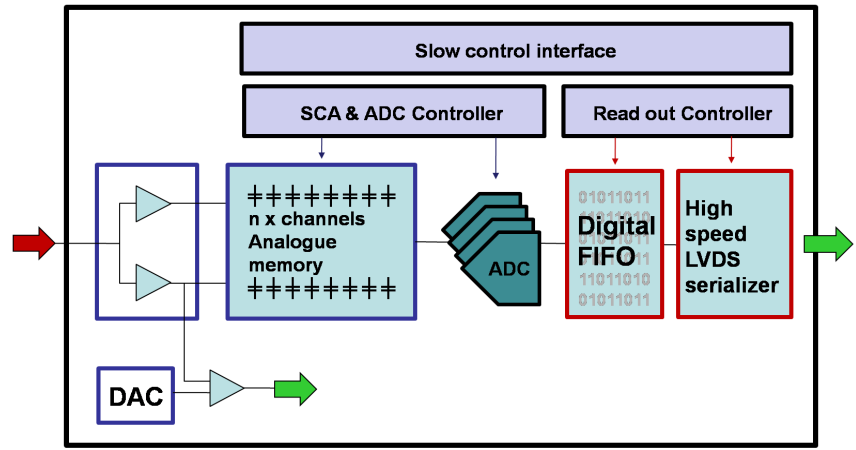}
  \caption{\small High-level integrated analogue sampling ASIC.
  The single ASIC amplifies, stores,
  and digitises the analogue signal, and buffers the digital data before sending them to
  the central camera recording system.}
 \label{fig_sampler_asic}
\end{center}
\end{figure}

While FADC systems may ultimately offer somewhat superior performance, analogue
samplers could allow lower cost, in particular if much of the auxiliary circuitry
surrounding and supporting the sampler ASIC, such as pixel trigger circuits, ADCs,
digital buffer and readout controllers can be integrated into a single multi-channel
ASIC (fig. ~\ref{fig_sampler_asic}). This is analogous to the readout for silicon
strip sensors, where single readout ASICs typically accommodate 128 channels and
where the cost per channel is at the level of a few \euro{}.
\begin{figure}[hbtp]
\begin{center}
 \epsfig{width=0.7\textwidth,file=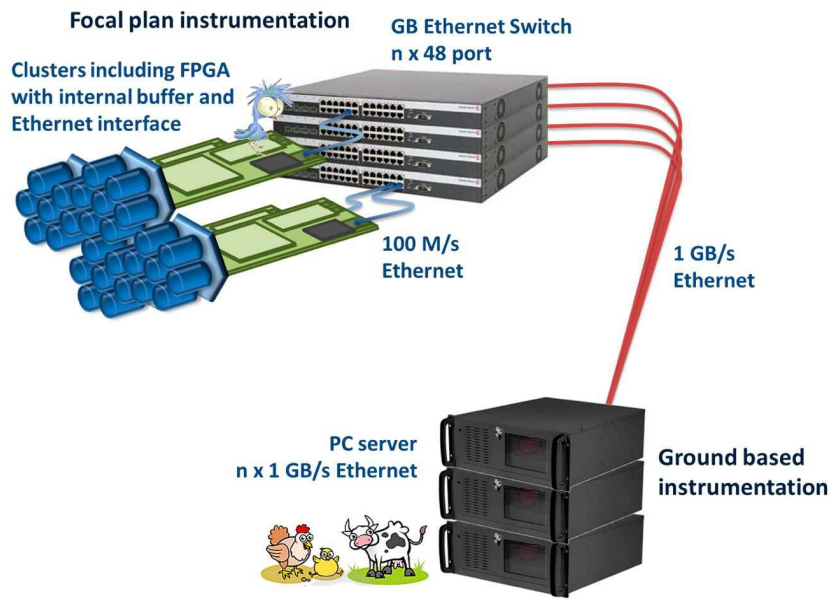}
  \caption{\small Possible scheme for an Ethernet-based front-end to back-end
  readout. A group of pixels with their ADCs is controlled by a dedicated FPGA. The same FPGA can
  be used to buffer the data and to transmit them through a dedicated Ethernet network to a camera
  computer (PC Server), which buffers the data in its RAM and preprocesses events before sending
  them to an event building farm.}
\label{fig:readout}
\end{center}
\end{figure}
At the current stage of CTA electronics design, analogue samplers and FADCs will
be pursued in parallel. Existing ASICs such as the SAM or DSR4 are probably adequate for
use in CTA. MC simulations should help to decide if dual-gain channels are needed,
which would imply significantly
increasing electronics cost.
A specific development effort is also aimed at producing nonlinear
input stages providing signal compression.\\

%-----------------------------------------------------------------------
\noindent{\bf Readout Electronics}
\smallskip
 
Readout of digitised data has so far either relied on custom-built bus systems to collect data from electronics
units covering the camera focal plane (such as the ``drawers'' of the H.E.S.S. telescopes), or
has located the digitisation electronics in commercial VME or PCI crate systems.
As a flexible and cost-effective alternative, the use of commercial
Ethernet systems has recently been explored  \cite{astro-ph08120762}, using normal switches to
buffer data sent via a low-level Ethernet protocol (fig. \ref{fig:readout}).
A low-cost
front-end gate array emulates the Ethernet interface. Data transfer is asynchronous, with
buffering in the front-end gate array, eliminating a source of deadtime. To enable synchronisation,
events are tagged at the front end with an event marker. In tests with 20 sender nodes,
transmitting via a switch to a receiver PC, loss-free transmission of more than  $10^{10}$
packets with a data rate of more than 80 MByte/s was achieved. Current up-to-date servers
can operate with 2$\times$4 GBit interfaces and cope with the resulting data flow.
It is therefore expected that
loss-free transmission of the front-end data, even of a 2000 pixel camera operating at
data rates of 600 MByte/sec, should not be a problem. Nevertheless, various forms of zero
suppression could be implemented in the front end, reducing data rates by up to
an order of magnitude. Since the Ethernet system operates in full-duplex mode, it can also be
used for the control and parameterisation of the front-end components, such as HV supplies,
and to set parameters for triggering, digitisation etc. It would not be necessary to design a separate
command bus, as employed in most current cameras.
 
\subsubsection{Triggering}
\label{sec:triggering}
 
%------------------------------------------------------------------------------------
\noindent{\bf Triggering the Telescopes}
\smallskip
 
Arrays of Cherenkov telescopes typically employ multi-level trigger schemes to keep the
rate of random triggers from the night sky background low. At the first level, signals
from individual pixels are discriminated above a threshold. These pixel-level signals are
input to a second level, topological trigger. The topological trigger is used to identify
concentrations of Cherenkov signals in local regions of the camera, via patten recognition
or a sum of first-level triggers, to form a telescope-level trigger. A third,
array-level trigger, is formed by combining trigger information from several telescopes.
 
The trigger chain within a telescope may follow a digital, or analogue path. In H.E.S.S.,
Magic and VERITAS, analogue schemes are used, but for CTA several approaches for both options
are under investigation. A digital scheme would require the continuous digitisation
(with one or more bits) of the signal coming from the PMTs. Components that look
for coincidences from digitised signals with a predefined timing are commercially available.
In a digital scheme, the trigger is very flexible and almost any algorithm can
be implemented, even a posteriori.
Trigger algorithms and parameter settings for each camera can easily be adapted
for each telescope type, array configuration and for the physics programme
(e.g. energy range). Both sector and topological trigger concepts can be implemented in a
digital trigger system.
The information provided by a digital trigger is essentially ``screenshots'' of the
camera every given time slice. Even if only one or a few bits are used to encode
the trigger information, it may be worthwhile to add this information to the data stream.
In the extreme case of a digital trigger with sufficient resolution,
only the digital stream could be used, as is proposed for the FlashCam development.
On the other hand, the digitisation frequencies available at reasonable cost may
yield worse rejection of random triggers from the night sky background
compared to an analogue approach.
 
\begin{figure}[hbtp]
\begin{center}
 \epsfig{width=0.7\textwidth,file=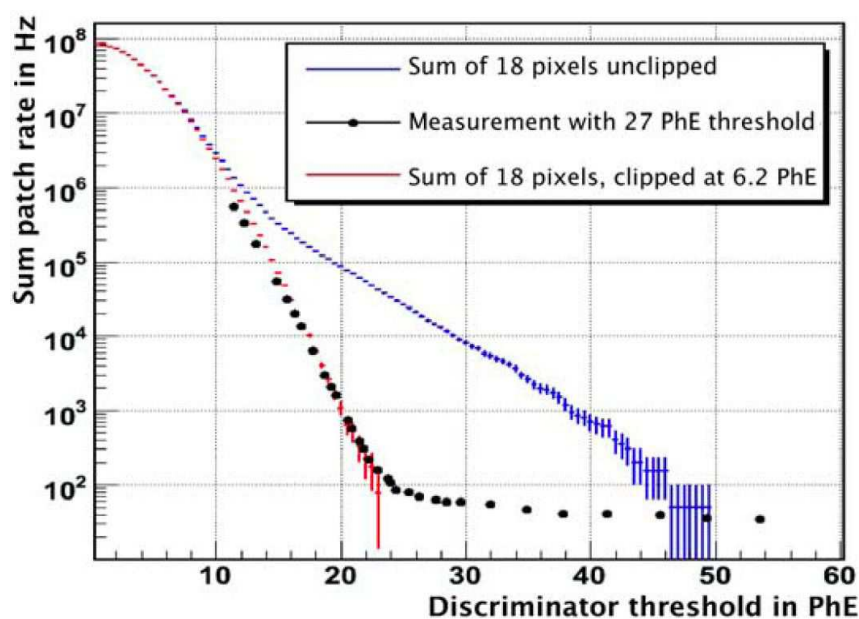}
  \caption{\small Trigger rate (in Hz) against the discriminator threshold (in photo electrons).
  NSB dominates at low thresholds.  Without clipping
  afterpulses largely dominate the rate (blue),
  while they are effectively eliminated by clipping (red).
The black points beyond 25 photo-electrons are due to cosmic showers.}
\label{fig:trigyyy}
\end{center}
\end{figure}

The telescope trigger is traditionally formed by looking for a number of pixels above threshold,
or a number of neighbouring pixels above threshold, within the camera. This is typically
implemented by dividing the camera up into sectors, which must overlap to provide a uniform
trigger efficiency across the camera. By requiring several pixels to trigger at once, random fluctuations
due to the night sky background and PMT afterpulses are greatly reduced.
Alternative schemes are also under investigation.
These include a sum trigger,
which can lead to a significant reduction of the
trigger threshold \cite{CrabPulsar}. In the sum trigger, the
analogue or digital sum of all pixels in a cluster is formed and a threshold is set to
initiate a trigger. It is necessary to clip pixel signals before summing to prevent
large afterpulses triggering a cluster (see fig. \ref{fig:trigyyy}).
All these approaches can be implemented in both an analogue or a digital path.
 
The size of camera sectors and their overlaps have implications for the threshold and detection
efficiency. Given the different goals, they may differ for the LST, MST and SST. Schemes for
all telescope sizes are under investigation using Monte Carlo simulations. The shape and dimensions
of the mechanical clusters put limitations on possible camera sectors and their overlap,
but these are only of second order.
The default reaction to a trigger is the read out of the entire camera.
Also an autonomous-cluster trigger is under study, however. This allows sections of
the camera to form trigger decisions
independently and these are read out autonomously. In the high-energy range, observations at low elevation
produce images that propagate in time through the camera. The propagation time can be much higher
than the usual integration window of the Cherenkov signal acquisition. The autonomous readout would
allow the recording of the time slot when the signal is in each section of the camera, following
the propagation of the image through the camera. For low energy showers, the shower image covers
a small region of the camera and it can be useful to read out only a part of the focal plane
to save bandwidth on the network and to lower the deadtime of the system.
 
The earlier a trigger system enters a purely digital level, the more easily and reliably it can be simulated.
Schemes which rely on the addition of very fast analogue or digital signals are potentially more
powerful, but could be sensitive to details of the pulse shapes and to the transition-time dispersion
between different PMTs, requiring a pre-selection of PMTs with similar transition times or the
implementation of matched delays to compensate for intrinsic differences.
 
Note that no signal recording scheme rules out the use of a given triggering scheme,
but the use of FADCs to record the signal would allow implementation of a
digital trigger based on the already digitised signals, hence reducing the cost
and complexity of the system.\\
 
%----------------------------------------------------------------------------------
\noindent{\bf Triggering the Array}
\smallskip
 
Current array trigger schemes for systems of Cherenkov telescopes
\cite{hess_trigger} %astro-ph0709.4438
provide asynchronous trigger decisions, delaying telescope trigger signals by an appropriate amount to
compensate for the time differences when the Cherenkov light reaches the telescopes, and
scanning trigger signals for pre-programmed patterns of telescope coincidences. The time to
reach a trigger decision and to propagate it back to the telescopes is about 1 $\mu$s or
more. While FADC based readout systems can buffer signals for this period, analogue-sampling
ASICs will usually not provide sufficient memory depth, and require the halting of waveform
sampling after a telescope trigger, while awaiting a third-level telescope coincidence trigger.
The resulting deadtime of a few $\mu$s limits telescope (second-level) trigger rates to
some 10\,kHz, which does not represent a serious limitation. The latest
analogue-sampling ASICs allow digitisation of stored signals on time scales of 2 to 3 $\mu$s
(see tab. \ref{tab:asics}), comparable to the array trigger latency. In this case, a new
option for the triggering of the array becomes possible: pixel signals are readout and digitised after
each telescope trigger, and are stored in digital memory, tagged with an event number.
Given that data are buffered and that buffers can easily be made large, restrictions on
array trigger latency are greatly relaxed (with GByte memory, about 1 s of data
can be buffered) and one can implement a software-based asynchronous trigger. With
each local trigger, an absolute timestamp is captured for the event with an accuracy of
the order of 1 ns and transmitted to the camera CPU. This computer collects the time stamps
and possibly additional trigger information for each event, e.g. pixel trigger patterns,
and transmits them every 10-100 ms via standard Ethernet using TCP/IP to a dedicated
central trigger computer. The central computer receives all time stamps from all telescopes
and uses this information to test for time coincidences of the events and to derive the
telescope system trigger. In addition, the time and trigger information can be used
to obtain a first estimate of the core position and shower direction. Following
the central trigger decision, the central trigger CPU sends the information to the
corresponding telescopes about which of the buffered events are to be rejected and
which fulfil the system trigger condition and should be pre-processed in the camera
CPU and transmitted for further stereoscopic processing. Assuming a local trigger rate
of 10 kHz and that about 100 Byte of trigger information are generated from each
telescope, the central trigger computer needs to handle up to 100 MByte/sec in a 100
telescope system, which can be readily be done with today's technology. In such a trigger
scheme, the central trigger decision is software-based, but the ``hard'' timing
from the camera trigger decision is used. It is therefore scalable, fully flexible and all types
of sub-systems can be served in parallel. At the same time it uses the shortest possible
coincidence gates and provides an optimum suppression of accidental coincidences.
 
\subsubsection{Camera Integration}
\label{sec:cameraintegration}
 
Signal transmission from the photo-sensors to the recording electronics represents a critical
design issue if the electronics is located far from the photo-sensors. Conventional cables
limit bandwidth, are bulky and difficult to route across telescope bearings, and are costly.
MAGIC uses optical signal transmission,
%\cite{magic_opttrans},
circumventing the first two problems, at considerable expense. H.E.S.S. avoids signal
transmission altogether by combining 16 photo-sensors and their associated electronics in
``drawers'', requiring only power and Ethernet connection to the camera \cite{hess_camera},
but limiting flexibility as regards upgrades of individual components.
 
At least for the SST and MST, which are produced in significant quantity
and where costs of the electronics is a decisive factor, the most effective solution seems to be
to combine
photo-sensors and electronics in the camera body. The design should allow easy swapping
of the camera for a spare unit, allowing convenient maintenance and repair of faulty cameras at a central
facility. However, over the expected lifetime of CTA,
upgrades at least of the photo-sensors are likely. The same may be true for the trigger and
data recording systems, where novel networking components may allow transmission of significantly
larger amounts of digital data than is currently possible. A viable option could therefore be, rather
than combining photo-sensors and electronics in a single mechanical unit, to build a photo-sensor
plane with short connections to electronics units, which in turn feed a trigger system via a
flexible interface (fig.~\ref{fig_camera}). For ease of mechanical assembly, both photo-sensors
and electronics will be packaged into multi-channel units.
 
\begin{figure}[htbp]
\begin{center}
 \epsfig{file=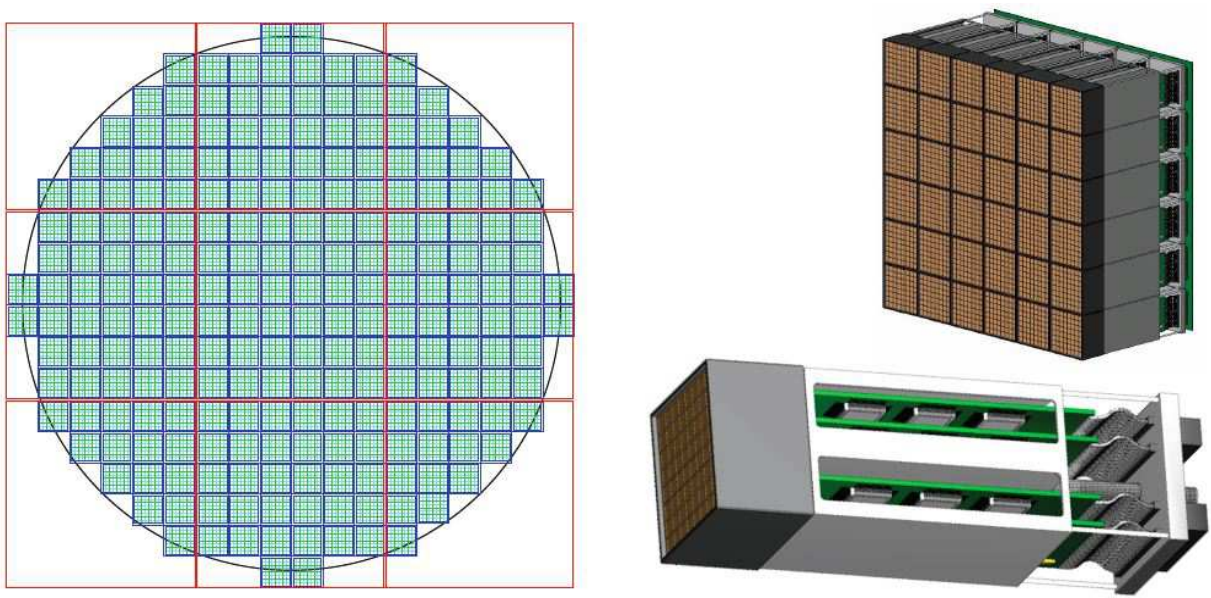, width=\textwidth}
 \caption{\label{fig:agis_pixelunit}
 \small
 Instrumentation of a 50 cm diameter camera for a dual mirror
 telescope from 64-pixel multi-anode PMTs. One pixel is about 6$\times$6 mm$^2$.}
\end{center}
\end{figure}

\begin{figure}[htbp]
\begin{center}
 \epsfig{file=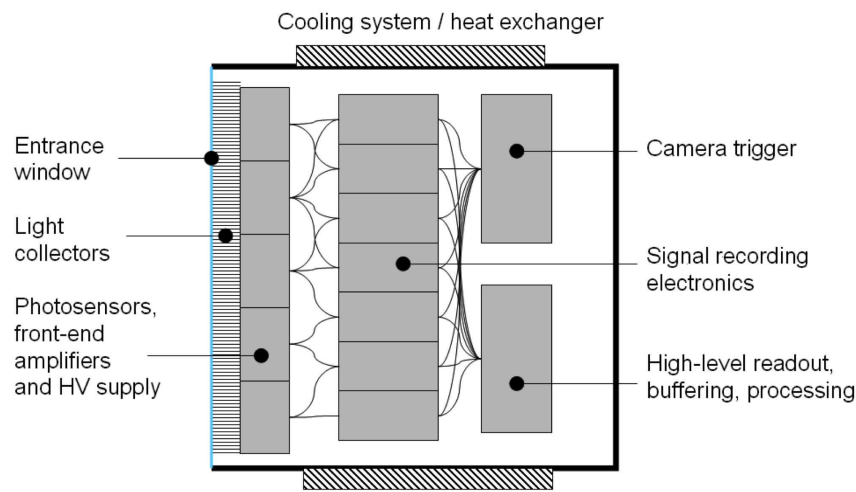, width=0.7\textwidth}
 \caption{\label{fig_camera} \small Concept for packaging of the electronic
 contained in the camera.}
\end{center}
\end{figure}

Dual-mirror solutions, such as the Schwarzschild-Couder telescopes,
require much smaller cameras and can therefore
utilise cheap multi-anode photo-sensors. Fig. \ref{fig:agis_pixelunit}
shows a possible solution considered for AGIS, using  64-pixel multi-anode PMTs
\cite{agis_pixelunit}.

Mechanical packaging of the entire camera and sealing against the environment is crucial for
stable performance. In its daytime configuration with closed camera lid, the camera body should
be reasonably waterproof. Dust penetrating the camera and deposited on connectors and on optical
components is a serious issue. To protect the photo-sensors and the light-collecting funnels and
allow for easy cleaning, an optical entrance window made of near-UV transparent material is
desirable, even if this induces a modest light loss due to reflection. While larger-scale integration
should reduce power consumption compared to current systems, a camera will nevertheless consume
kilowatts of power and must be cooled. Air cooling requires high-quality filtering of the airflow into
the camera. Closed-circuit cooling systems, involving internal circulation of a cooling medium
and appropriate heat exchangers
improve long-term reliability, but add cost and weight.
 
%\input{Atac.tex}
%%%%% HEADER: Atac.tex
%%%%%%%%%%%%%%%%%%%%%%%%%%%%%%%%%%%%%%%%%%%%%%%%%%%%%%%%%%%%%%%%%%%%%%%%%%%%%%%%
 
\subsection{Calibration and Atmospheric Monitoring}
\label{sec:atac}
 
The higher sensitivity of CTA means good gamma-ray statistics for many
sources. Therefore, the instrument's systematic uncertainties may
limit the accuracy of the measurements. The atmosphere is an
integral part of an IACT and so monitoring and correcting for atmospheric
inhomogeneity must be addressed in addition to the detailed calibration and
monitoring of the response and characteristics of the telescopes. Work is
ongoing to address both issues, as well as their interplay, with the goal of
characterising the systematic uncertainties to an unprecedented level.
 
Already teams of world-experts have gathered to develop state-of-the-art instrumentation for
atmospheric monitoring and the associated science for CTA.
These teams are actively participating in the corresponding CTA work package (ATAC).
 
\subsubsection{Telescope Calibration}
\label{sec:telcal}
\begin{figure*}[th]
  \centering
  \epsfig{height=4in,file=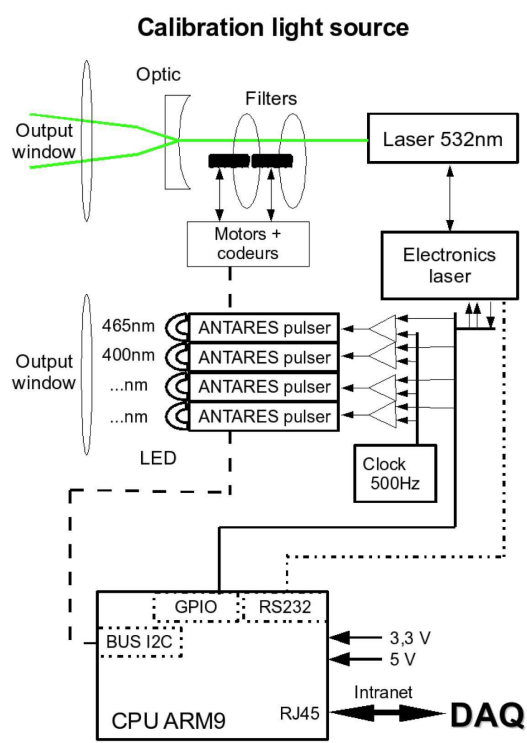}
  \caption{\small Layout of the H.E.S.S. II flatfielding and single photoelectron device.
  For a large array of telescopes, it is likely that the laser will be replaced by LEDs and that
  the mechanical filter wheel will be replaced by an electronic system.}
  \label{fig1}
 \end{figure*}

The calibration of the CTA telescopes has two distinct aspects. Firstly, the absolute gain of
the system must be determined. Secondly, the pointing accuracy of the telescope must be measured.
The necessity of the precise measurement of the gain of each electronic channel for CTA requires
the development of a single and reliable calibration device which can measure the
flatfielding coefficients and the ratio between a single photoelectron and the number of digital
counts recorded. This development will add to existing experience in building calibration devices,
for example the H.E.S.S. II flatfielding system as shown in fig. \ref{fig1}.
Overall, absolute calibration is achieved by reconstructing the rings generated by local muons.
A special pre-scaled single-telescope trigger could be implemented to enhance the rate at which these are
recorded.

In the development of the calibration apparatus, many challenges must be
addressed. The first is the difficulty of uniformly illuminating large,
wide field-of-view cameras. This problem is twofold: firstly, diffusers must be able to
present a uniform signal to the edge of the field of view; secondly, the
pixels across the field of view must uniformly accept the diffused signal on
their photo-cathodes. This second aspect can be difficult to achieve when
reflective lightcones on the camera edge have a different
acceptance to those in the centre to a close-by, centrally diffused light
source.
The use of different colour light sources would allow the
quantification of any differences and/or changes in the quantum efficiency of
pixels.
An additional challenge concerns the measurement of the single photo-electron
response. Current telescope systems measure this either in-situ with a low light
background level, or indirectly using photon statistics \cite{VERITASSPE}.
A comparison of these two methods allows the study of their associated systematic errors
and the choice for the best system for CTA.\\
 
The requirements of the telescope pointing measurement are somewhat simpler, but vitally important.
Here, a system of two CCD cameras mounted on each telescope is envisaged. The first measures the
position of the night-sky relative to the telescope dish and the second the position of the
telescope camera relative to the dish. In combination, the system allows the astronomical pointing
of the telescope to be assessed accurately.
 
\subsubsection{Atmospheric Monitoring}
\label{sec:atmon}
\begin{figure}[htbp]
  \centering
  \epsfig{width=0.7\textwidth,file=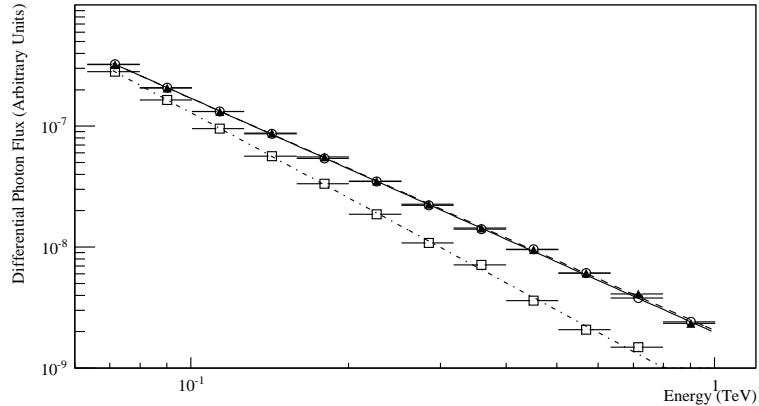}
  \caption{\small Recovering spectral information for non-ideal observing conditions. From a full simulation
  database a randomly sampled spectrum of $10^5$ events with spectral slope of  E$^{-2.3}$ is drawn.
  These events are then reconstructed using simulation-based look-up tables
  which give the reconstructed
  energy as a function of the camera image brightness
  and the reconstructed distance to the shower.
  For different atmospheric conditions (described in tab. \ref{tab1}), a reconstructed spectrum
  is derived.  The  open circles show the reconstructed differential spectrum for case 1, the open
  squares for case 2 and the closed triangles for case 3. By incorporating Lidar data into the
  reconstruction (case 3) a corrected spectrum can be recovered with approximately the
  same normalisation and slope as for a clear night sky (case 1).}
  \label{fig2}
\end{figure}

The calibration of the CTA telescopes is one critical calibration and monitoring task,
a second is the monitoring of the atmosphere which forms part of the detector.
This is where the particle shower is initiated
by the incident gamma-ray and the medium through which the Cherenkov photons must travel.
The estimation of the energy
of an individual gamma-ray is based on the calorimetric energy deposited in the atmosphere,
which in turn is measured
via Cherenkov photon emission. Therefore, any change in atmospheric quality can affect the signal
detected. To investigate this effect, a set of benchmark simulations of a 97-telescope array design were
initiated to test the performance of an array of imaging Cherenkov telescopes under the presence of
varying atmospheric conditions. Simulations were produced for a clear atmosphere and an
atmosphere with a significant layer of low-level dust, as derived from measurements taken with
a 355 nm single-scattering Lidar deployed on the Namibian Highlands. These  show that, if
unaccounted for, the changing atmospheric quality produces a significant shift in the
reconstructed gamma-ray spectrum. This can be seen in fig. \ref{fig2}.
 
\begin{table}[hbtp]
   \centering
  \begin{tabular}{|c|c|c|}
  \hline
  Case   &  Simulation Derived From & Lookup Derived From \\
   \hline
    1 & Clear Database& Clear Database \\
    2 & Dusty Database& Clear Database \\
    3 & Dusty Database&Dusty Database \\
    \hline
  \end{tabular}
  \caption{\small The combination of look-up tables (as derived from simulations)
 and simulated spectra produced to derive the effect of the atmosphere on
 reconstructed spectra illustrated in fig. \ref{fig2}.}
  \label{tab1}
 \end{table}
Many current Cherenkov telescope arrays have in-situ single-scattering Lidars. This type of Lidar
possesses a strong and variable systematic error in the derived transmission, up to approximately
50-60\% \cite{Lidar}. After discussions with members of the Pierre Auger Observatory (PAO) and
other atmospheric scientists, CTA has decided to adopt the Raman Lidar technique as the tool of
choice for accurately probing atmospheric quality. Below and around shower maximum, it is believed
that this technique will reduce the systematic error in derived transmission to approximately
5\% \cite{Lidar}. Therefore, Raman Lidars are currently under development, which will be installed
at the sites of some existing Cherenkov telescopes in order to test their efficacy in ground-based gamma-ray analysis.
If successful, these atmospheric monitoring systems will allow CTA to significantly reduce the
systematic error in energy measurements and derived source fluxes.
% and hence increase its active lifetime.

%\input{QA.tex}
%%%%% HEADER: QA.tex
%% QA %%%%%%%%%%%%%%%%%%%%%%%%%%%%%%%%%%%%%%%%%%%%%%%%%%%%%%%%%%%%%%%%%
 
\subsection{Quality Assurance}
 \label{sec:QA}
 
Since the design study phase, the CTA project has included a
work package named ``Quality Assurance and Risk assessment''.
The objective of this WP is to implement a uniform approach to risk
analysis in the design, commissioning and operation of the telescopes and of
the facility, and for quality assurance of the telescope components and of the
assembly procedures.
 
``Risks'' are any features that can be a threat to the success of the project.
They can have negative effects on the cost, schedule and technical performance of CTA.
The aim of project risk management is to identify, assess, reduce,
accept (where necessary), and control project risks in a systematic and
cost-effective manner, taking into account technical and programming
constraints.
 
Quality Assurance ensures a satisfactory level of quality for all
steps of the design study. This level of quality is guaranteed by
correct implementation of the pre-defined quality criteria and the
participation of all the project actors.
 
Including quality assurance and risk assessment from the very start of
the design study phase will have a positive effect on the building
schedule and cost of CTA.
 
The objective of the design study is to develop telescopes
which will be produced in series during the building phase, so the
study will be done in partnership with industry. Quality assurance and
risk assessment will ensure that the project will have good
traceability and a good control of risks from the outset.
 
%%``Quality assurance and risk assessment'' WP organization:
 
This WP is managed by a coordinator
who defines standards and quality methods for the
project.
To ensure the implementation of quality in the project
laboratories, ``Local Quality Correspondents'' (LQCs) will be identified
and trained. These people will dedicate part of their time to quality
issues, proportional to their laboratory participation in the overall
project.
 
The main tasks of the WP participants are:
\begin{itemize}
\item To define the quality insurance organisation (the roles of the participants)
\item To ensure that quality control and risk analysis procedures are
defined and applied uniformly across the project to ensure high quality
and reliability of hardware and software
\item To ensure that the risk analysis, including dependability
(reliability, availability, maintenance and safety) is defined based on
the technical configuration proposed
\item To ensure support and expertise to implement the quality system and
associated tools across the project
\item To verify the coherence of the procedures and protocols in order to
approve them for subsequent release and use
\item To verify the application of the quality procedures across the project
\item To identify and reduce technical and management risks
\end{itemize}
 
% Major interfaces:
Quality assurance and risk assessment concern the whole project. Thus, the members of
``Quality assurance and risk assessment'' will have active
links to all work packages, to the project management and to all laboratories involved in
building parts of CTA.

\clearpage
 
%%
%\input{Site.tex}
%%%%% HEADER: Site.tex
%%%%%%%%%%%%%%%%%%%%%%%%%%%%%%%%%%%%%%%%%%%%%%%%%%%%%%%%%%%%%%%%%%%
 
\section{CTA Site Selection}
 \label{sec:site}
 
Selection of sites for CTA is obviously crucial for achieving optimum performance and science
output. Criteria for site selection include, among others, geographical conditions, observational
and environmental conditions and questions of logistics, accessibility, availability, stability
of the host region, and local support:
 
\begin{description}
\item[Geographical conditions:] For best sky coverage, the latitude of the sites should be around
$30^\circ$ north and south, respectively. The sites have to provide a reasonably flat
area of about 1\,km$^2$ (north) and at least 10\,km$^2$ (south). Optimum overall performance is obtained
for site altitudes between about 1500 and 4000\,m. Even higher altitudes allow further reduction of
the energy threshold \cite{5at5} at the expense of performance at medium and high energies and might
be considered for the northern array. Desirable is also a low component of geomagnetic field
parallel to the surface, since such fields deflect air shower particles.
\item[Observational conditions:] Obviously, the fraction of clear nights should be high.
For good sites, this fraction is well above 60\%, reaching up to 80\% for the very best sites.
Artificial light pollution
must be well below the natural level of night sky background, which excludes sites
within some tens of km of major population centres.
Atmospheric transparency should be good, implying dry locations with
low amounts of aerosols and dust in the atmosphere.
\item[Environmental conditions:] Environment and climate influences both the operational efficiency
and the survival conditions of the instrument. Wind speeds above 10\,m/s may impact observations;
peak wind speeds, which may range from below 100\,km/h to beyond 200\,km/h depending on the site, have
a major impact on telescope structure and cost. Sand storms and hail represent a major danger for
unprotected mirror surfaces. Snow and ice prevent observations and will influence
instrument costs, e.g. by making heating systems necessary and requiring increased structural stability.
Seismic activity will similarly increase requirements on telescope structures and buildings.
\item[Infrastructure and logistics:] A well-developed infrastructure, e.g. as a result of already existing observatories,
is an advantage. Connection to the power grid and high speed internet access are mandatory.
There should be good access to the site, i.e. nearby airports for air travel to/from Europe and elsewhere,
and local access roads. A major population centre with technical and commercial infrastructure within
convenient travel distance is desirable.
\item[Other criteria:] These include availability of the site for construction,
guarantees for long-term operation and access,
political stability of the host region,
safety of personnel, both during travel and stay, and
availability of local administrative, technical and funding support as well as
possibilities for scientific cooperation with local groups.
\end{description}
For both the observational and the environmental conditions, a long-term (multi-year)
data record is required to allow dependable decisions o be made.
While archival remote-sensing data can provide
some information, well-explored sites with existing installations and good records are favoured. It is
unlikely that any site is optimal in all respects, so the different criteria will have to be balanced
against each other.
Reliable and efficient operation of the observatory should be a key criterion.
 
Site evaluation includes a number of different approaches, at different stages
of progress for a candidate site:
\begin{itemize}
\item Use of remote-sensing archival data and local archival data to evaluate observing conditions and
environmental conditions.
\item Site visits and information gathering by local collaborating groups on logistics aspects.
\item Dedicated CTA measurements; since long-term measurements are excluded, this approach is useful
only for those quantities where short campaigns can provide meaningful results,
such as the determination of natural and artificial night-sky brightness.
\end{itemize}
\begin{figure}[htbp]
\centering
  \epsfig{width=0.8\textwidth, file=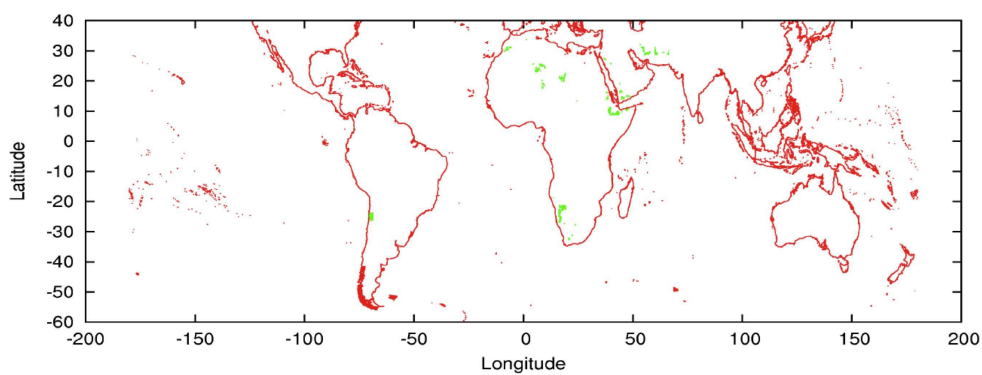}
  \caption{\label{fig_site1} \small Green areas indicate sites above 1500\,m a.s.l.,
  which offer sufficiently flat areas,
  minimal artificial background light
  and an average cloud cover of $<$40\%, selected on the basis of topological and satellite data.}
\end{figure}
 
A first preselection can look for sufficiently large and flat areas above
1500\,m a.s.l. (based on a topological model of the Earth \cite{topomodel}),
with the requirement that the artificial background light is minimal (as determined from satellite images
\cite{lightpollution}),
and that average cloud coverage is less than 40\% (as provided by the
International Satellite Cloud Climatology Project, ISCCP, based on the analysis of satellite
data \cite{isccp}).
The resulting map (fig.~\ref{fig_site1}) shows very few locations matching these basic criteria,
among them the well-known sites in Chile and Namibia. However, while the ISCCP data has the advantage of
covering the whole planet, the resolution is relatively coarse and sites with very local conditions
(such as mountain tops) may deviate significantly from the ``pixel'' average. Also, daytime and nighttime
cloud cover will usually be different. Only the latter is relevant for Cherenkov astronomy.
For identification of potential observatory sites, special algorithms and high-resolution data have been
provided by Erasmus \cite{erasmus} for the identification of potential observatory sites,
but only for selected areas, such as the Chilean sites,
the Indian site at Hanle or the Yanbajing site in Tibet.
Similar searches are conducted using MODIS and ISCCP maps
%\cite{sitenote}
as well as the recently released ESO application FriOwl that provides access to an
extensive database of information from the last 40 years \cite{friowl}.
 
Based on these preliminary evaluations, potentially interesting sites
have been selected at which
detailed studies will be conducted in the coming months.\\
 
\noindent Northern site candidates are:
% with details given in Table~\ref{table_sites}:
\begin{description}
\item[Canary Islands La Palma and Tenerife:] These are well-known and well-explored
observatory sites at about $26^\circ$\,N, about 2400 m a.s.l., with the Observatorio del Roque de
los Muchachos on La Palma and the Observatorio del Teide on Tenerife.
\item[Hanle in India, in the Western Himalayas:] This high-altitude site ($33^\circ$\,N, 4500\,m a.s.l.)
hosts a small observatory and an array of Cherenkov instruments which is deployed by Indian groups.
%%Access to the site is difficult.
\item[San Pedro Martir, Baja California:] Well-established astronomical site that hosts already
two observatories run by UNAM (Universidad Autonoma de Mexico). It is situated at about
$31^\circ$\,N, at 2800 m a.s.l..
\end{description}
 
\noindent Southern site candidates are:
\begin{description}
\item[Khomas Highland of Namibia:] This is a well-known astronomical site,
at 1800 m a.s.l. and $23^\circ$\,S, and is the
home of the H.E.S.S. instrument. The region offers a range of suitable, large and flat areas.
\item[Chilean sites:] Chile is home to some of the World's premier optical observatories.
However, availability
of sufficiently large cites near these locations is limited. A possible site is north of
La Silla at $29^\circ$\,S and 2400 m a.s.l. Another potential site is near Cerro Paranal,
with even better observing conditions, but no sufficiently flat area in this region
has been identified so far.
\item[El Leoncito Reserve in Argentina:] This site is at $32^\circ$\,S and 2600 m a.s.l. and hosts
the El Leoncito Astronomical Observatory.
% but its atmospheric conditions are somewhat
% inferior to the best Chilean sites.
\item[Puna Highland in Argentina:] The region offers some large sites at 3700 m a.s.l..
with sky quality equivalent to the best Chilean ones. These sites have good access to
a railway line.
\end{description}
 
The final decision among otherwise identical sites may rely on considerations such as financial
or in-kind contributions by the host regions. It is likely that an inter-governmental agreement will
be required to assure long-term availability of the site,
as well as guaranteed access and free transfer
of data. At the same level, issues such as import taxes, value added tax and fees etc. should
be addressed. Such agreements exist, for H.E.S.S., Auger and other observatories operated
by international collaborations.

\clearpage
 
%%
%\input{Outlook.tex}
%%%%% HEADER: Outlook.tex
%% OLD HEADER: Concept.tex
%%%%%%%%%%%%%%%%%%%%%%%%%%%%%%%%%%%%%%%%%%%%%%%%%%%%%%%%%%%%%%%%%%%
 
\section{Outlook}
\label{sec:outlook}
 
The Cherenkov Telescope Array was conceived back in 2005,
and was then promoted by members of the HESS and MAGIC collaborations.
It was soon apparent that
a gamma-ray observatory could be designed with existing technologies
that was much more powerful than any of the existing facilities.
An improvement of a factor of 10 in sensitivity around a TeV and
an extension of the energy range from a few tens of GeV to $>$100 TeV,
well beyond the currently accessible range,
was achievable with an array of a large number ($\approx$ 100) of
differently sized telescopes.
 
With the results from current Cherenkov telescopes
pouring in, it became obvious that with such an instrument
a vast number of sources of very different types
could be discovered and studied with unprecedented precision.
Answers to long-standing questions in a number of
science areas seemed possible. The extent and the diversity of the science case was,
and is, stunning (see sec. \ref{sec:phys}).
CTA would truly be the first large open observatory for
astronomy of the extreme universe beyond the GeV range.
 
Not surprisingly, many scientists were attracted to
CTA and its science grew rapidly, as did the number of supporters
who now form
a large international collaboration which is
investigating how best to realise the project.
CTA has received consistently excellent reviews and
high rankings in Science Roadmaps in Europe and across the world.
CTA is an acknowledged ESFRI project, features high on the roadmaps of future projects
of ApPEC, ASPERA and ASTRONET and has been well received by national funding agencies.
The potential of CTA is well recognised outside Europe,
with the USA, Japan, India, Brazil and Argentina, and other countries
contributing significantly.
The US Decadal Survey endorsed a strong US participation in CTA as
one of the four most important ground based initiatives in the next ten years.
 
Since 2006, and specifically in a 4-year design study, it has been shown that CTA,
with observatories in the northern and southern hemisphere,
can be built to achieve its goal performance, at an investment cost in the range of
150 M\euro{}, which is a modest price for an installation of such scientific potential.
 
CTA has recently received substantial funding
from the European Community, for preparing for construction and operation, and
from national funding agencies, for development and prototyping.
There is much excitement amongst all participants, and the wider science community,
about the prospects that CTA will soon from design to reality.
 
% original
%  This report is an account of the main design work performed so far and
%  a solid basis for the prototyping and construction phases ahead.
% original
  In this report, an account of the main design work performed so far is presented,
  which constitutes a solid basis for the prototyping and construction phases that lie ahead.
% compromise
%  In this report the concepts and results of the main CTA design work have been presented.
%  They constitute a solid basis for the prototyping and construction phases ahead.
% proposal manel
%  The materials which have been presented and discussed in this document are a selected collection
%  of the Design Concepts work performed so far
%  and constitutes
%  a solid basis for the prototyping and construction phases ahead.
The Preparatory Phase (3 years) and the subsequent construction phase
(2013-2018) will pose many challenges.
But CTA is a well-organised international collaboration of
25 countries and $>$600 scientists with extensive expertise
in all relevant areas. Its members are eager and ready to
tackle the problems that lie ahead.
 
This effort is well worth it, as CTA will provide a huge science return
in astrophysics, particle physics, cosmology and fundamental physics,
and lead to a bright future for ground-based gamma ray astronomy.

\section*{Acknowledgements}

We gratefully acknowledge financial support from
the following agencies and organisations:
Ministerio de Ciencia, Tecnolog\'ia e Innovaci\'on Productiva (MinCyT), 
Comisi\'on Nacional de Energ\'ia At\'omica (CNEA) and Consejo Nacional 
de Investigaciones Cient\'ificas y T\'ecnicas (CONICET) Argentina;
State Committee of Science of Armenia;
Ministry for Research, 
CNRS-INSU, CNRS-IN2P3 and CEA, France;
Max Planck Society, BMBF, DESY, Helmholtz Association, Germany;
MIUR, Italy;
Netherlands Research School for Astronomy (NOVA), 
Netherlands Organization for Scientific Research (NWO);
Ministry of Science and Higher 
Education and the National Centre for Research and Development, Poland;
MICINN support through the National R+D+I, CDTI funding
plans and the CPAN and MultiDark Consolider-Ingenio 2010 programme, Spain.
Swedish Research Council, Royal Swedish Academy of Sciences financed, Sweden;
Swiss National Science Foundation (SNSF), Switzerland;
Leverhulme Trust, Royal Society, Science and Technologies Facilities Council, Durham University, UK;
National Science Foundation, Department of Energy,
Argonne National Laboratory,
University of California,
University of Chicago,
Iowa State University, 
Institute for Nuclear and Particle Astrophysics (INPAC-MRPI program),  
Washington University McDonnell Center for the Space Sciences, USA

\clearpage
 
%% refs
%\input{refs.tex}
%%%%% HEADER: refs.tex
%%%%%%%%%%%%%%%%%%%%%%%%%%%%%%%%%%%%%%%%%%%%%%%%%%%%%%%%%%%%%%%%%%%

\end{document}